\documentclass[twoside,12pt]{article}
\usepackage{epsfig}
\usepackage{graphicx}% Include figure files
\usepackage{bm}% bold math
\usepackage{amsmath}
\usepackage{amssymb}
\numberwithin{equation}{section}

\def\Journal#1#2#3#4{{#1} {#2} (#4) #3 }

\def\NPA{{\em Nucl. Phys.} A}

\def\PRO{{\em Prog. Theor. Phys.}}
\def\NPB{{\em Nucl. Phys.} B}

\def\PLB{{\em Phys. Lett.} B}

\def\PL{{\em Phys. Lett.}}
\def\PRL{\em Phys. Rev. Lett.}
\def\PREV{\em Phys. Rev.}
\def\PREP{\em Phys. Rep.}
\def\PRA{{\em Phys. Rev.} A}
\def\PRD{{\em Phys. Rev.} D}
\def\PRC{{\em Phys. Rev.} C}

\def\ANNP{\em Ann. Phys. (N.Y.)}

\def\JPG{{\em J. Phys.} G}

\newcommand{\be}{\begin{equation}}
\newcommand{\ee}{\end{equation}}
\newcommand{\bea}{\begin{eqnarray}}
\newcommand{\eea}{\end{eqnarray}}

\def\mib#1{\hbox{\boldmath $#1$}}
\def\mbf#1{{\mib #1}}
\def\eq#1{Eq.\ (\ref{#1})}

\def\lesssim{\begin{array}{c} < \\[-3mm] \sim \\ \end{array}}

\def\bS{{\mbf S}}

\def\bL{{\mbf L}}
\def\bP{{\mbf P}}
\def\bR{{\mbf R}}
\def\bS{{\mbf S}}

\def\bX{{\mbf X}}

\def\ba{{\mbf a}}
\def\bb{{\mbf b}}
\def\be{{\mbf e}}
\def\bk{{\mbf k}}
\def\bn{{\mbf n}}
\def\bp{{\mbf p}}
\def\bq{{\mbf q}}
\def\br{{\mbf r}}

\def\bx{{\mbf x}}

\def\bfsigma{{\mbf \sigma}}

\def\CA{{\cal A}}

\def\CM{{\cal M}}
\def\CO{{\cal O}}
\def\CP{{\cal P}}

\def\CT{{\cal T}}

\def\CY{{\cal Y}}

\def\SM3{\Sigma N (3/2)}
\def\SN1{\Sigma N (1/2)}

\def\Sp{\Sigma^- p}

\def\TS1{\hbox{}^3S_1}
\def\TD1{\hbox{}^3D_1}
\def\aagp{{\alpha \alpha^\prime}}
\def\tint{{\rm int}}
\def\tlab{{\rm lab}}

\def\CVwe{{\cal V}(\omega, \varepsilon)}

\def\VRGM{V^{\rm RGM}(\varepsilon)}
\def\VRGMA{V^{\rm RGM}_\alpha(\varepsilon_\alpha)}
\def\VRGMB{V^{\rm RGM}_\beta(\varepsilon_\beta)}
\def\VRGMC{V^{\rm RGM}_\gamma(\varepsilon_\gamma)}

\def\TTE{\widetilde{T}^{(3)}_\alpha(E, \varepsilon_\alpha)}
\begin{document}

\title{ \vspace{1cm}
Baryon-baryon interactions in the $SU_6$ quark model \\
and their applications to light nuclear systems}
\author{
Y.\ Fujiwara,$^{1}$ Y.\ Suzuki,$^{2}$ C.\ Nakamoto,$^{3}$ \\
$\hbox{}^{1}$ Department of Physics, Kyoto University, 
Kyoto 606-8502, Japan \\
$\hbox{}^{2}$ Department of Physics, and Graduate School
of Science and Technology, \\
Niigata University, Niigata 950-2181, Japan \\
$\hbox{}^{3}$ Suzuka College of Technology, Suzuka 510-0294, Japan
}%

\maketitle

\begin{abstract}
Interactions between the octet-baryons ($B_8$) in the
spin-flavor $SU_6$ quark model are investigated
in a unified coupled-channels framework
of the resonating-group method (RGM).
The interaction Hamiltonian for quarks consists of the phenomenological
confinement potential, the color Fermi-Breit interaction
with explicit flavor-symmetry breaking (FSB),
and effective-meson exchange potentials of scalar-,
pseudoscalar- and vector-meson types.
The model parameters are determined to reproduce
the properties of the nucleon-nucleon ($NN$) system
and the low-energy cross section data
for the hyperon-nucleon interactions.
Mainly due to the introduction of the vector mesons,
the $NN$ phase shifts at non-relativistic energies
up to $T_{\rm lab}=350~\hbox{MeV}$ are greatly improved
in comparison with the previous quark-model $NN$ interactions.
The deuteron properties and the low-energy observables
of the $B_8 B_8$ interactions, including the inelastic
capture ratio at rest for the $\Sigma^- p$ scattering,
are examined in the particle basis
with the pion-Coulomb correction.
The nuclear saturation properties and
the single-particle (s.p.) potentials of $B_8$ in nuclear medium
are examined through the $G$-matrix calculations, using
the quark-exchange kernel.
The $\Sigma$ s.p.~potential is
weakly repulsive in symmetric nuclear matter.
The s.p.~spin-orbit strength for $\Lambda$ is very small,
due to the strong antisymmetric spin-orbit force
generated from the Fermi-Breit interaction.
The qualitative behavior of the $B_8B_8$ interactions is  
systematically understood
by 1) the spin-flavor $SU_6$ symmetry of $B_8$,
2) the special role of the pion exchange,
and 3) the FSB of the underlying quark Hamiltonian.
In particular, the $B_8 B_8$ interaction becomes
less attractive according to the increase of strangeness,
implying that there exists no $B_8 B_8$ di-baryon bound state
except for the deuteron.
The strong $\Lambda N$--$\Sigma N$ coupling results from
the important tensor component of the one-pion exchange.
The $\Lambda \Lambda$--$\Xi N$--$\Sigma \Sigma$ coupling
in the strangeness $S=-2$ and isospin $I=0$ channel
is relatively weak, since this coupling is caused by
the strangeness exchange.
The $B_8 B_8$ interactions are then
applied to some of few-baryon systems
and light $\Lambda$-hypernuclei
in a three-cluster Faddeev formalism using two-cluster RGM kernels.
An application to the three-nucleon system
shows that the quark-model $NN$ interaction
can give a sufficient triton binding energy with
little room for the three-nucleon force.
The hypertriton Faddeev calculation
indicates that the attraction of the $\Lambda N$ interaction
in the $\hbox{}^1S_0$ state is only slightly more attractive
than that in the $\hbox{}^3S_1$ state.
In the application to the $\alpha \alpha \Lambda$ system,
the energy spectrum of $\hbox{}^9_\Lambda \hbox{Be}$ is
well reproduced using the $\alpha \alpha$ RGM kernel.
The very small spin-orbit splitting
of the $\hbox{}^9_\Lambda \hbox{Be}$ excited states
is also discussed.
In the $\Lambda \Lambda \alpha$ Faddeev calculation,
the NAGARA event for $\hbox{}^{\ \,6}_{\Lambda \Lambda}\hbox{He}$ is
found to be consistent with the quark-model $\Lambda \Lambda$ interaction.
\end{abstract}

%\eject
%\tableofcontents

\bigskip

\begin{center}
Table of Contents
\end{center}

\begin{enumerate}
\setlength{\itemsep}{0mm}
\item[1.] Introduction
\item[2.] Formulation
\begin{enumerate}
\setlength{\itemsep}{0mm}
\item[2.1] RGM formalism for quark clusters
\item[2.2] Effective-meson exchange potentials
\item[2.3] The Lippmann-Schwinger formalism
for $(3q)$-$(3q)$ RGM and the $G$-matrix equation
\item[2.4] Three-cluster Faddeev formalism using two-cluster RGM 
kernels
\end{enumerate}
\item[3.] Results and discussions
\begin{enumerate}
\setlength{\itemsep}{0mm}
\item[3.1] Two-nucleon system
\item[3.2] Hyperon-nucleon interactions
\begin{enumerate}
\setlength{\itemsep}{0mm}
\item[3.2.1] The spin-flavor $SU_6$ symmetry
in the $B_8 B_8$ interactions
\item[3.2.2] $\Sigma^+ p$ system
\item[3.2.3] $\Lambda N$ system
\item[3.2.4] $\Sigma^- p$ system
\item[3.2.5] The inelastic capture ratio at rest in
the low-energy $\Sigma^- p$ scattering
\item[3.2.6] $YN$ cross sections
\item[3.2.7] $G$-matrix calculations
\end{enumerate}
\item[3.3] Interactions for other octet-baryons
\begin{enumerate}
\setlength{\itemsep}{0mm}
\item[3.3.1] $S=-2$ systems
\item[3.3.2] $S=-3$ systems
\item[3.3.3] $S=-4$ systems
\end{enumerate}
\item[3.4] Triton and hypertriton Faddeev calculations
\begin{enumerate}
\setlength{\itemsep}{0mm}
\item[3.4.1] Three-nucleon bound state
\item[3.4.2] The hypertriton
\end{enumerate}
\item[3.5] Application to $\hbox{}_\Lambda^9 \hbox{Be}$ and
$\hbox{}^{\ \,6}_{\Lambda \Lambda} \hbox{He}$ systems
\begin{enumerate}
\setlength{\itemsep}{0mm}
\item[3.5.1] The $\alpha \alpha \Lambda$ system
for $\hbox{}^9_{\Lambda}\hbox{Be}$
\item[3.5.2] The $\Lambda \Lambda \alpha$ system 
for $\hbox{}^{\ \,6}_{\Lambda \Lambda} \hbox{He}$
\end{enumerate}
\end{enumerate}
\item[4.] Summary
\end{enumerate}

\section{Introduction}

Since the discovery of quantum chromodynamics (QCD), many
efforts have been made to understand the structure of
hadrons and hadron-hadron interactions based on this
fundamental theory of the strong interaction. 
See, e.g., a review article \cite{VA05} for these efforts. 
Such studies are of crucial importance in understanding 
many-body systems of nuclei and hadronic matter. 
A direct derivation of these elements from QCD, however, 
involves tremendous difficulties 
such as the quark confinement and multi-gluon effects in the
low-energy phenomena. This has motivated a study using 
other effective models 
which take account of some important QCD characteristics
like the color degree of freedom, one-gluon exchange, etc., 
even though they may not be rigorously derived from QCD. 

In this review article, we will focus on the 
interactions between the octet-baryons ($B_8$), 
studied in the non-relativistic quark model (QM),
and their applications to light nuclear systems.  
The QM study of the baryon-baryon interaction
started from applying the three-quark ($3q$) model
for the nucleon ($N$) to the $(3q)$-$(3q)$ system, in order to 
obtain a microscopic understanding of the short-range
repulsion of the nucleon-nucleon ($NN$) interaction.
Almost 30 years ago Liberman \cite{LI77} first
calculated the quark-exchange kernel of the phenomenological
confinement potential and 
the $(\bfsigma_i \cdot \bfsigma_j)(\lambda^C_i \cdot \lambda^C_j)$-type
color-magnetic interaction originating from 
the Fermi-Breit (FB) interaction.
The latter was suggested by De R{\'u}jula, Georgi
and Glashow \cite{RU75} as a possible candidate
of the quark-quark ($qq$) residual interaction
based on the asymptotic freedom of QCD.
The effect of antisymmetrization among quarks
is most straightforwardly
taken into account in the framework of 
a microscopic nuclear cluster model, the
resonating-group method (RGM). 
The RGM equation for the $(3q)$-$(3q)$ system
was solved by Oka and Yazaki \cite{OK80} not only for
the $NN$ interaction, but also for the $\Delta \Delta$ interaction.
Harvey \cite{HA80} pointed out the appearance
of the so-called hidden-color 
components in the many-quark systems. 
The nature of the repulsive core of the $NN$ interaction
was studied by Faessler, Fernandez, L{\"u}beck
and Shimizu \cite{FA82} in the framework of RGM,
including $NN$, $\Delta \Delta$ and the hidden-color
channels.
These early studies were followed by a number of works
\cite{SU84,WS84,WO86,OSY87,SH89},
which repeatedly confirmed
that the medium- and long-range effects, usually described
by meson-exchange potentials, 
are absent in the QM and must be incorporated 
to obtain a realistic description of the baryon-baryon interaction.
The first such attempt for the $NN$ interaction
was carried out \cite{OK83} by supplementing
the well-established one-pion exchange
potential (OPEP) and a phenomenological medium-range central potential
in the Schr{\"o}dinger-type equation equivalent to the RGM equation.
The deuteron properties were investigated in relation with
the compact $(6q)$-bag configuration \cite{YA86}.
The most successful calculation of the $NN$ interaction
in this approach was given in Ref.~\cite{TA89}.
The authors introduced the spin-spin and non-central terms
of the OPEP, assuming that the pions directly couple
with the quarks in the $(3q)$ core, and thereby yielding
an effective-meson exchange potential (EMEP) at the baryon level.
An alternative approach is to assume OPEP between the quarks
in the RGM formalism and
to calculate the quark-exchange kernel explicitly \cite{SH84}.
This program was carried out by the T{\"u}bingen group~\cite{ST88,BR90}, 
who extended the model of the $NN$ interaction
to the hyperon-nucleon ($YN$) and
hyperon-hyperon ($YY$) interactions as well.
Their approach includes a pseudoscalar (PS) meson
exchange between the quarks as well as
a phenomenological $\sigma$-like potential at the baryon level.
A fully microscopic calculation including both PS-
and $\sigma$-meson exchange potentials at the 
quark level was undertaken by the Salamanca
group \cite{FE93,VA94} for the $NN$ interaction
in the framework of the chiral QM.
This model was further extended by the Beijing group
\cite{ZH94,QI95,YU95,SH99}
to a chiral $SU_3$ model
with strangeness degree of freedom,
and applied to the $YN$ and $YY$ interactions.
In the meantime, meson-exchange models have also made a 
progress in the description of the medium-range attraction
of the $YN$ interaction. 
The new J{\"u}lich $YN$ model \cite{HA05} has discussed
the contributions of the scalar-isoscalar ($\sigma$)
and vector-isovector ($\rho$) exchange channels, constrained by
a microscopic model of correlated $\pi \pi$ and $K {\bar K}$ exchanges.
The extended Nijmegen soft-core model ESC04 \cite{ESC04a,ESC04b} has
introduced two PS-meson exchange and meson-pair exchange,
in addition to the standard one-boson exchange
and short-range diffractive exchange potentials.

Our viewpoint for applying the RGM formalism
to the baryon-baryon interactions is that the $SU_6$ QM with
some realistic modifications like peripheral
mesonic or $(q{\bar q})$ effects can essentially reproduce
the whole low-energy hadronic phenomena.
For the low-energy phenomena related to the $S$-wave
ground-state hadrons, the predictions with the FB interaction
of the one-gluon exchange have no apparent disagreement with experiment
except for some trivial cases.
Though the FB spin-orbit ($LS$) interaction is believed not to be
favorable for the very small energy splitting
of the negative-parity nucleon excited states, known as 
the ``missing $LS$ force problem'' of the $P$-wave baryons,
it was argued in Ref.~\cite{FU93} that
this problem can be resolved by a proper treatment of
a compact $(3q)$ structure embedded
in the $(3q)(q{\bar q})$ continua.
In other papers \cite{SHT84,NA93}, we showed that
the same FB $LS$ force has almost right order of magnitude
compatible with the experimental information
on the $NN$ and $YN$ spin-orbit interaction.
Fujiwara and Hecht discussed in~\cite{FU85,FU87} the off-shell
FB interaction in the non-relativistic framework
and incorporated the meson-exchange effects
to the $(3q)$-$(3q)$ RGM by explicitly
introducing $(q{\bar q})$ and $(q{\bar q})^2$ excitations.
As the main feature of the $NN$ interaction was reproduced in 
the extended QM without introducing any extra
meson parameters, it may be considered the first step toward
the conventional meson-exchange description through a
possible mechanism of $(q{\bar q})$ exchanges between nucleons. 
An extension of the model to include the antiquark degree
of freedom is also made to describe other hadronic interactions
such as the pion-pion and pion-nucleon scatterings as well as
the proton-antiproton annihilation processes
\cite{FA81,OK85,BU89}.
%\cite{FA81-BU89}.

We note that to apply the RGM formalism to the $(3q)$-$(3q)$ system
is not straightforward, since there are some new features
arising from the intrinsically relativistic nature of the system.
The standard non-relativistic RGM formulation should be 
augmented with the following features: 
1) the full FB interaction with flavor-symmetry breaking (FSB),
including the momentum-dependent Breit retardation term
and the Galilean non-invariant $LS$ term, 
2) a $(0s)^3$ harmonic-oscillator (h.o.) 
wave function with a common size
parameter $b$ for all the $(3q)$ systems,
3) a perturbative treatment of the FSB, in terms of
the pure $SU_6$ QM wave functions with the [3]-symmetric
spatial function,
and 4) correct reduced masses and threshold energies
in the RGM formalism.
See Refs.~\cite{NA95,RGMFa} for the background of these extensions.

The effective $qq$ interaction in our model 
consists of a phenomenological quark-confining potential, 
the one-gluon exchange FB interaction,
and other terms generating EMEP's from various meson-exchange mechanisms.
Various versions of our QM for the baryon-baryon interactions emerge
from several possibilities of 
how to incorporate these EMEP's in the RGM equation.
A simplest method is to use effective local potentials 
$V^{\hbox{eff}}$ at the baryon level, which are determined
from the existing one-boson exchange potentials (OBEP)
with some adjustable parameters.
From the correspondence between the RGM formalism
for composite particles and the
Schr{\"o}dinger equation for structureless particles,
the relative-motion wave function should be
renormalized with the square root of the normalization
kernel $\sqrt{N}$. A nice point of this
prescription is that we only need the normalization kernel to
generate the exchange terms.
Our earliest version of the QM, {\bf RGM-F} \cite{NA95,RGMFa,RGMFb},
uses this prescription for EMEP's, where 
$V^{\hbox{eff}}$ is assumed to be proportional to the scalar- (S-) meson
and PS-meson exchange potentials of the Nijmegen model-F \cite{NA79}.
More specifically, the central force
of the S-meson nonet and the tensor force
of the $\pi$ and $K$ mesons in the Nijmegen model-F are used.
With only two adjustable parameters determined in the $NN$ sector,
RGM-F can reproduce reasonably well all the low-energy
cross section data of the $YN$ systems.

In the next step, we have improved the way of introducing
the EMEP's in two respects.
One is to calculate the spin-flavor factors exactly
at the quark level and the other is to include
the spin-spin terms originating from all the PS mesons.
We have shown in Ref.~\cite{PRL} that 
the available $NN$ and $YN$ data can simultaneously be reproduced 
in the standard $(3q)$-$(3q)$ formulation,
if one assumes the full PS- and S-meson nonet
exchanges at the quark level and properly introduces
the FSB in the quark sector.
An explicit evaluation of quark-exchange kernels for the EMEP leads 
to a strong constraint on the flavor dependence
of each $NN$ and $YN$ channel.
The $SU_3$ relation of the coupling constants emerges as a natural
consequence of the $SU_6$ QM. For S mesons,
one of the $SU_3$ parameters, $F/(F\!+\!D)$ ratio, turns out
to take the $SU_6$ value of a purely electric type.
Under this restriction, the low-energy $NN$, $\Sigma^+p$
and $\Lambda p$ total cross sections cannot be reproduced consistently
with a unique mixing angle of the flavor-singlet and octet S mesons.
We resolve this problem in two ways; one is to change
the mixing angle only for $\Sigma^+ p$ channel
and the other is to employ the same approximation as RGM-F, changing
the $F/(F\!+\!D)$ ratio, solely for the isoscalar S mesons,
$\epsilon$ and $S^*$.
These models are called {\bf FSS} and {\bf RGM-H} \cite{PRL,FSS,SCAT},
respectively. Since their predictions are
not much different except for the roles of
the $LS^{(-)}$ force in the $\Lambda N$--$\Sigma N(I=1/2)$
coupled-channels (CC) system, the result with FSS will be discussed 
in this article.

In the most recent model called {\bf fss2} \cite{fss2,B8B8},
we have also included the following interaction pieces mediated 
by S and vector (V) mesons: $LS$, quadratic spin-orbit ($QLS$),
and momentum-dependent Bryan-Scott terms.
The introduction of these terms to the EMEP's is primarily
motivated by the insufficient description
of the $NN$ experimental data with our previous models.
Due to these improvements, the $NN$ phase shifts
at non-relativistic energies up to $T_{\rm lab}=350~\hbox{MeV}$
have attained an accuracy almost comparable to that of OBEP's.
The model parameters are searched in the isospin basis
to fit the $NN$ phase-shift analysis
with $J \leq 2$ and $T_{\rm lab} \leq 350$ MeV,
the deuteron binding energy, the $\hbox{}^1S_0$ $NN$ scattering
length, and the low-energy $YN$ total cross section data.
The model fss2 reproduces the $YN$ scattering data
and the essential features of the $\Lambda N$--$\Sigma N$ coupling
equally well with our previous models.
Another important progress for the QM baryon-baryon interaction
has come from a technique to solve the RGM equation
in the momentum representation. Here we solve the $T$-matrix
equation of the RGM kernel, for which all the Born terms
are derived analytically. The partial wave decomposition
of the Born kernel is carried out numerically
in the Gauss-Legendre integral quadrature.
This method, called Lippmann-Schwinger RGM (LS-RGM) \cite{LSRGM},
enables us not only to obtain very stable and accurate results
for the $S$-matrix over a wide energy range up to a few GeV,
but also to apply our QM baryon-baryon interactions
to physically interesting problems such as $G$-matrix
and three-cluster Faddeev calculations.
The single-particle (s.p.) potentials of
the $B_8$ ($N$, $\Lambda$ and $\Sigma$) and
the strength of the s.p.~spin-orbit potential are predicted
through the $G$-matrix calculations \cite{GMAT,SPLS}.

We have carried out coupled-channels RGM (CCRGM) calculations both
in the isospin basis and in the physical particle basis.
The former is convenient to show concisely the characteristics
of the $B_8 B_8$ interactions, but the latter is necessary
to make a realistic evaluation of cross sections for a comparison
with experimental data. The latter is also necessary
for the study of the charge dependence of the $NN$ interaction
and the charge-symmetry breaking (CSB) of the $\Lambda N$ interaction.
The charge-independence breaking (CIB) is partially explained
by the so-called pion-Coulomb correction \cite{MA01},
which implies 1) the small mass difference
of the neutron and the proton, 2) the mass difference of the
charged pions and the neutral pion, and 3) the Coulomb effect.
For the CSB of the $\Lambda N$ interaction,
the early version of the Nijmegen potential \cite{NA73} already
focused on this CSB in the OBEP including the pion-Coulomb correction
and the correct threshold energies
of the $\Lambda N$--$\Sigma N$ coupling in the particle basis.
We always include the pion-Coulomb correction when the particle basis
is used for the calculation.
For the Coulomb effect, we calculate the full exchange kernel
without any approximation.
The pion-Coulomb correction and the correct treatment
of the threshold energies are found to be
very important to analyze the low-energy observables 
in the $\Sigma N$--$\Lambda N$ CC problem.

In the next section, we briefly describe the formulation
of the $(3q)$-$(3q)$ RGM and its application to 
the $G$-matrix and Faddeev calculations.
These formulations are made in the momentum representation.
The essential points to be emphasized include the energy dependence
of the RGM kernel and the existence of the Pauli-forbidden state.  
The treatment of the Coulomb force in the
momentum representation is also given.
Section 3 presents results and discussions.
We start with the two-nucleon properties
and proceed to the $YN$ interaction, 
the $G$-matrix calculations, the $YY$ interaction
and the other $B_8 B_8$ interactions including
the $\Xi N$, $\Xi Y$ and $\Xi \Xi$ interactions.
The importance of the spin-flavor $SU_6$ symmetry
in our QM framework is stressed through a systematic analysis
of the phase-shift behavior and total cross sections
in some particular channels. 
In Sec.~3.4, our QM baryon-baryon interactions,
FSS and fss2, are applied to the calculations of the triton
and hypertriton binding energies in the three-cluster Faddeev formalism.  
The application to other light hypernuclei,
$\hbox{}^9_\Lambda \hbox{Be}$ and
$\hbox{}^{\ \,6}_{\Lambda \Lambda} \hbox{He}$, is given in Sec.~3.5.
The spin-orbit splitting of $\hbox{}^9_\Lambda \hbox{Be}$ excited
states is also examined by the Faddeev and $G$-matrix calculations.
A summary is drawn in Sec.~4.

\section{Formulation}

\subsection{RGM formalism for quark clusters}

The RGM is a microscopic theory
which describes the dynamical motion of clusters, on the basis
of two- (and three-, $\cdots$) body interactions of constituent
particles. For the application to baryon-baryon interactions
in the QM, the $B_8$'s are assumed to be simple
$s$-shell clusters composed of the $(0s)^3$ h.o. wave functions
with a common size parameter $b$.  
They have spin-flavor and color degrees of freedom as well,
for which we follow the standard non-relativistic $SU_6$
description. 
We use the Elliott notation to specify the $SU_3$ quantum
numbers of $B_8$, and write $B_8=B_{(11)a}$ with
$a \equiv Y I$ (hypercharge $Y$, isospin $I$) to denote
a flavor function $N$ ($YI=1\,\frac{1}{2}$), $\Lambda$ ($YI=0\,0$),
$\Sigma$ ($YI=0\,1$) or $\Xi$ ($YI=-1\,\frac{1}{2}$).
We also use the notation $\alpha=a I_z$ to denote
a specific member of the isospin multiplet for each baryon. 
Then the QM description of the $B_8$ is achieved
with the correspondence
\begin{eqnarray}
B_{(11)\alpha} \sim
W^{[3]}_{\frac{1}{2}(11)\alpha}(123)
&=&\sum_{S(\lambda\mu)=0\,(01),\;1\,(20)}\frac{1}{\sqrt{2}}\;
[\;[\;w_{\frac{1}{2}}(1)w_{\frac{1}{2}}(2)\;]_{S}\;
w_{\frac{1}{2}}(3)\;]_{\frac{1}{2}} \nonumber \\
& &\times\;[\;[\;F_{(10)}(1)F_{(10)}(2)\;]_{(\lambda\mu)}\;
F_{(10)}(3)\;]_{(11)\alpha}\ ,
\label{form1}
\end{eqnarray}
because the spin-flavor part of $B_8$ is totally symmetric.
Here $w_{\frac{1}{2}}$ is the spin wave function 
and $F_{(10)}$ the flavor wave function of a single quark.
The QM wave functions of the $B_8$ are expressed as
\begin{eqnarray}
\phi_{B_8}(123)=\phi^{({\rm orb})}(123)
~W^{[3]}_{1/2(11)\alpha}(123)~C(123)\ ,
\label{form2}
\end{eqnarray}
where $C(123)$ is the color-singlet wave function.
The orbital wave function, $\phi^{({\rm orb})}(123)$,
is an intrinsic part of $B_8$ and contains no dependence
on the c.m. coordinate $\bX_G=(\bx_1+\bx_2+\bx_3)/3$.
The orbital functions for the $(3q)$-clusters
are thus assumed to be flavor-independent and are taken to be the
same for all the $B_8$'s.

The RGM wave function for the $(3q)$-$(3q)$ system
can be expressed as
\begin{equation}
\Psi=\sum_\alpha \CA^\prime
\left\{ \phi_\alpha \chi_\alpha(\bR) \right\}\ \ ,
\label{form3}
\end{equation}
where the channel wave function $\phi_\alpha=\phi^{\rm (orb)}
\xi^{SF}_\alpha \xi^C$ is composed of the orbital part 
of the internal
wave function $\phi^{\rm (orb)}=\phi^{\rm (orb)}(123)\phi^{\rm (orb)}(456)$,
the isospin-coupled basis $\xi^{SF}_\alpha$ of the spin-flavor
$SU_6$ wave functions and $\xi^C=C(123)C(456)$.
A detailed definition of $\xi^{SF}_\alpha$ with a set of specific
two-baryon quantum numbers $\alpha$ will be given below.
The antisymmetrization operator $\CA^\prime$ in \eq{form3}
makes $\Psi$ totally antisymmetric under the exchange
of any quark pairs and can be reduced to
the form $\CA^\prime \rightarrow (1/2)(1-9P_{36})(1-P_0)$ with
the core-exchange operator of the two $(3q)$-clusters
$P_0=P_{14}P_{25}P_{36}$.

We now explain the basis $\xi^{SF}_\alpha$.
In the general $B_8 B_8$ interaction, there appear several
new features which are absent in the $NN$ interaction.
These are mainly related to the rich flavor contents
of the $B_8 B_8$ system constrained by the spin-flavor $SU_6$
symmetry and the generalized Pauli principle. Since these features
are common in both OBEP and QM descriptions,
we will use the baryon's degree of freedom as much as possible.
Here we use the isospin-coupled basis for presentation, but
the transformation to the physical particle basis is
straightforward.
The $SU_3$-coupled basis of two $B_8$'s is defined by
\begin{eqnarray}
\left[ B_{(11)} B_{(11)} \right]_{(\lambda \mu)\alpha; \rho}
& = & \left[ B_{(11)}(1)\,B_{(11)}(2)
\right]_{(\lambda \mu)\alpha; \rho} \nonumber \\
& = & \sum_{a_1, a_2} \langle (11)a_1 (11)a_2 \parallel
(\lambda \mu)a \rangle_\rho
\left[ B_{(11)a_1}(1)\,B_{(11)a_2}(2) \right]_{II_z}\ ,
\label{form4}
\end{eqnarray}
where the last brackets denote the isospin coupling
and $\rho$ denotes a possible multiplicity label of
the $SU_3$ coupling, $(11)\times (11)
\rightarrow (22)+(30)+(03)+(11)_s+(11)_a+(00)$
($\rho=1$ for $(11)_a$ and $\rho=2$ for $(11)_s$).
We assume that the first $B_{(11)}$ is always for
particle 1 and the second for particle 2.
A set of the internal quantum numbers, $\alpha$ ($\alpha=a I_z$), 
stand for the hypercharge $Y=Y_1+Y_2$, the total isospin $I$ 
and its $z$-component $I_z$, which are all conserved quantities 
if we assume the isospin symmetry and the charge conservation 
$Q/e = Y/2 + I_z$. Only when the particle-basis is used, the isospin
$I$ is no longer ``exactly'' conserved.
The flavor-symmetry phase $\CP$ of the two-baryon system
is determined from the symmetry
of the $SU_3$ Clebsch-Gordan coefficients through
\begin{equation}
P^F_{12} \left[ B_{(11)} B_{(11)} \right]_{(\lambda \mu)\alpha; \rho}
=\CP \left[ B_{(11)} B_{(11)}
\right]_{(\lambda \mu)\alpha; \rho}\ ,
\label{form5}
\end{equation}
with
\begin{eqnarray}
\CP = \left\{ \begin{array}{c}
(-1)^{\lambda+\mu} \quad \hbox{in multiplicity-free case}\ , \\
(-1)^\rho \qquad \hbox{for} \quad (\lambda \mu)=(11)\ . \\
\end{array} \right.
\label{form6}
\end{eqnarray}
Here $P^F_{12}=(\lambda^F_1 \cdot \lambda^F_2)/2+1/3$ is the 
flavor-exchange operator with $\lambda^F_i$ ($i=1,~2$) being
the Gell-Mann matrices for the flavor degree of freedom.
It is important to maintain the definite flavor symmetry
even in the isospin-coupled basis and in the particle basis,
since the flavor symmetry is
related to the total spin and the parity through the generalized
Pauli principle;
\begin{equation}
(-1)^{L+S} \CP=1\ ,
\label{form7}
\end{equation}
where $L$ is the relative orbital angular momentum.
For the $NN$ system with $Y=2$, the total isospin $I$ is
good enough to specify $\CP$; i.e., $\CP=(-1)^{1-I}$, so that
\eq{form7} becomes the well-known rule $(-1)^{L+S+I}=-1$.
More generally, the flavor-exchange symmetry $\CP$ is
actually redundant for the systems of identical particles
with $a_1=a_2$ and $I_1=I_2$, since it is uniquely determined
as $\CP=(-1)^{2I_1-I}$.
Let us introduce a simplified notation $B_i \equiv B_{(11)a_i}$
($i=1,\ldots,4$ in general) with $a_i=Y_i I_i$. Then the symmetrized
isospin basis is defined through
\begin{eqnarray}
\left[ B_1 B_2 \right]^\CP_{I I_z}
& = & \frac{1}{\sqrt{2(1+\delta_{B_1 B_2})}}
\left\{ \left[ B_1 B_2 \right]_{II_z}+\CP (-1)^{I_1+I_2-I}
\left[ B_2 B_1 \right]_{II_z} \right\} \nonumber \\
& = & \sqrt{\frac{2}{1+\delta_{B_1 B_2}}}
\sum_{(\lambda \mu) \rho \in \CP}
\langle (11)a_1 (11)a_2 \parallel (\lambda \mu)a; \rho \rangle
\left[ B_{(11)} B_{(11)} \right]_{(\lambda \mu)\alpha; \rho}\ ,
\label{form8}
\end{eqnarray}
where the sum is over $(\lambda \mu)\rho$ compatible
with \eq{form6}.
As is apparent from the construction,
the basis in \eq{form8} is naturally
an eigenstate of the flavor-exchange operator $P^F_{12}$ with
the eigenvalue $\CP$.
It is now straightforward to incorporate the spin degree of freedom
in the above formulation and extend the argument
to the spin-flavor $SU_6$ wave functions of the $(3q)$-$(3q)$ system.
In the QM, the isospin basis with a definite flavor-exchange symmetry
is given by
\begin{eqnarray}
\xi^{SF}_\alpha
& = & \frac{1}{\sqrt
{2(1+\delta_{a_1 a_2})}}
\biggl\{ \,\left[ W^{[3]}_{\frac{1}{2}(11)a_1}(123)
\,W^{[3]}_{\frac{1}{2}(11)a_2}(456) \right]_{SS_zII_z} \nonumber \\
& + & \CP (-1)^{I_1+I_2-I}
\left[ W^{[3]}_{\frac{1}{2}(11)a_2}(123)
\,W^{[3]}_{\frac{1}{2}(11)a_1}(456) \right]_{SS_zII_z} \biggr\}\ .
\label{form9}
\end{eqnarray}
The subscript $\alpha$ in \eq{form9} specifies
a set of quantum numbers of the channel wave function;
\begin{equation}
\alpha \equiv \left[\frac{1}{2}(11)\,a_1 \frac{1}{2}(11)a_2 \right]
SS_z YII_z; \CP\ .
\label{form10}
\end{equation}
In the particle basis, we use the notation
$B_{\alpha_i}=B_{(11)\alpha_i}$ and define
\begin{eqnarray}
\left(B_{\alpha_1} B_{\alpha_2}\right)^\CP
& = & \frac{1}{\sqrt{2(1+\delta_{\alpha_1, \alpha_2})}}
\left(B_{\alpha_1} B_{\alpha_2}+\CP B_{\alpha_2} B_{\alpha_1}
\right) \nonumber \\
& = & \sqrt{\frac{1+\delta_{a_1, a_2}}{1+\delta_{\alpha_1, \alpha_2}}}
\sum_{II_z} \langle I_1 I_{1z} I_2 I_{2z} | I I_z \rangle
\left[B_1 B_2\right]^\CP_{II_z}\ .
\label{form11}
\end{eqnarray}
The channel quantum number $\alpha$ in \eq{form10} in this case
is specified by  
$\alpha \equiv \left[\frac{1}{2}(11)\,\alpha_1
\frac{1}{2}(11)\alpha_2 \right]SS_z; \CP$ with
$Y=Y_1+Y_2$ and $I_z=I_{1z}+I_{2z}$.
For the system of identical particles with $\alpha_1=\alpha_2$,
the flavor-exchange symmetry is again redundant and $\CP=1$.
The relationship between the isospin basis and the particle basis
in \eq{form11} is inversely written as
\begin{eqnarray}
\left[B_1 B_2\right]^\CP_{II_z}
=\sum_{I_{1z} I_{2z}}
\sqrt{\frac{1+\delta_{\alpha_1, \alpha_2}}
{1+\delta_{a_1, a_2}}}
\langle I_1 I_{1z} I_2 I_{2z} | I I_z \rangle
\left(B_{\alpha_1} B_{\alpha_2} \right)^\CP \ .
\label{form12}
\end{eqnarray}
The bases in Eqs. (\ref{form9}) and (\ref{form11}) are
the eigenstates of the core exchange
operator $P^{SF}_0=P^{SF}_{14}P^{SF}_{25}P^{SF}_{36}$
with the eigenvalue $(-1)^{1-S}\CP$,
where $P^{SF}_{ij}=P^\sigma_{ij}P^F_{ij}$ exchanges
the spin-flavor coordinates.

The QM Hamiltonian consists of the rest-mass plus non-relativistic
kinetic-energy term, the quadratic confinement potential,
the full FB interaction with explicit quark-mass dependence,
and the S-, PS-, and V-meson exchange potentials
acting between quarks:
\begin{eqnarray}
H=\sum^6_{i=1} \left( m_ic^2+\frac{\bp^2_i}{2m_i}-T_G \right)
+\sum^6_{i<j} \left( U^{\rm Cf}_{ij}+U^{\rm FB}_{ij}
+\sum_\beta U^{{\rm S}\beta}_{ij}
+\sum_\beta U^{{\rm PS}\beta}_{ij}
+\sum_\beta U^{{\rm V}\beta}_{ij} \right).
\label{form13}
\end{eqnarray}
Here $U^{\rm Cf}_{ij}=-(\lambda^C_i \cdot \lambda^C_j)\,a_c\,r^2$ with
$r= | \br | = | \bx_i - \bx_j |$ is
the confinement potential of quadratic power law, which gives
a vanishing contribution to the interaction in the present formalism.
The $qq$ FB interaction $U^{\rm FB}_{ij}$ is composed
of the following pieces \cite{SU84,SP87}:
\begin{equation}
U^{\rm FB}_{ij}=U^{CC}_{ij}+U^{MC}_{ij}+U^{GC}_{ij}
+U^{sLS}_{ij}+U^{aLS}_{ij}+U^{T}_{ij} \ \ ,
\label{form14}
\end{equation}
where the superscript $(CC)$ stands for the color-Coulombic
or $(\lambda^C_i \cdot \lambda^C_j)/r$ piece,
$(MC)$ for the momentum-dependent Breit retardation term
or $(\lambda^C_i \cdot \lambda^C_j)\Big\{
({\bp}_i \cdot {\bp}_j) + {\br}({\br} \cdot {\bp}_i)
\cdot {\bp}_j/r^2
\Big\} / (m_i m_j r)$  piece,
$(GC)$ for the combined color-delta and color-magnetic
or $(\lambda^C_i \cdot \lambda^C_j)$
$\Big\{ 1/(2m^2_i) + 1/(2m^2_j) + 2/(3m_im_j)
(\bfsigma\hbox{}_i \cdot \bfsigma\hbox{}_j)\Big\}
\delta({\br})$ piece,
$(sLS)$ for the symmetric $LS$, $(aLS)$ for the
antisymmetric $LS$, and $(T)$ for the tensor term.
It is important to note that all the contributions from the FB
interaction are generated from the quark-exchange diagrams,
since we assume color-singlet cluster wave functions.
The EMEP's, $U^{{\rm S}\beta}_{ij}$,
$U^{{\rm PS}\beta}_{ij}$, and $U^{{\rm V}\beta}_{ij}$ will be
discussed in the next section.
When the calculations are made in the particle basis,
the Coulomb force is also introduced at the quark level.

The RGM equation for the parity-projected relative
wave function $\chi^\pi_\alpha(\bR)$ is derived
from the variational
principle $\langle\delta\Psi|E-H|\Psi\rangle=0$,
and it reads as \cite{SH89,FSS}
\begin{equation}
\left[~\varepsilon_\alpha + \frac{\hbar^2}{2\mu_\alpha}
\left(\frac{\partial}{\partial {\bR}} \right)^2~\right]
\chi^\pi_\alpha({\bR})=\sum_{\alpha^\prime} \int d {\bR}^\prime
~\CM_\aagp({\bR}, {\bR}^\prime; E)
~\chi^\pi_{\alpha^\prime}({\bR}^\prime)\ ,
\label{form15}
\end{equation}
where $\CM_{\alpha \alpha^\prime}({\bR}, {\bR}^\prime; E)$ is
composed of various pieces of the interaction kernels
as well as the direct potentials of EMEP:
\begin{eqnarray}
\CM_\aagp({\bR}, {\bR}^\prime; E)
= \delta({\bR}-{\bR}^\prime)
\sum_\beta \sum_\Omega V^{\Omega\beta}
_{{\rm D}\,\alpha \alpha^\prime}({\bR})
+ \sum_\Omega \CM^{\Omega}_\aagp ({\bR}, {\bR}^\prime)
-\varepsilon_\alpha~\CM^N_\aagp({\bR}, {\bR}^\prime)\ .
\label{form16}
\end{eqnarray}
The relative energy $\varepsilon_\alpha$ in channel $\alpha$ is
related to the total energy $E$ of the system
in the c.m. system through $\varepsilon_\alpha=E-E^{\tint}_a$,
where $E^{\tint}_a=E^{\tint}_{a_1}+E^{\tint}_{a_2}$ with $a=a_1a_2$.
In \eq{form16} the sum over $\Omega$ in the direct term
implies various contributions of interaction types
for the meson-exchange potentials, while $\beta$ specifies
the meson species. 
These EMEP interaction types are classified to $\Omega
=CN,~SS,~TN,~LS$, and $QLS$, which are the spin-independent central,
spin-spin, tensor, $LS$, and quadratic $LS$ terms, respectively.
On the other hand, $\Omega$ for the exchange
kernel $\CM_{\alpha\alpha'}^{\Omega}({\bR}, {\bR}^\prime)$ involves
not only the exchange kinetic-energy ($K$) term but also
various pieces ($CC$, $MC$, $GC$, $sLS$, $aLS$, $T$) of the FB interaction,
as well as the EMEP contributions corresponding to the direct term.
The structure of the EMEP contributions is rather involved and contains
some ambiguities in the way of introducing them even starting
from the meson-exchange potentials at the quark level,
as will be discussed in the next section.
Here we only refer to a general procedure to solve
the RGM equation (\ref{form15}) and a couple of important features 
of the present RGM framework using solely $qq$ interactions.

A simple way to solve the CCRGM
equation (\ref{form15}) is the variational method
developed by Kamimura \cite{KA77}.
Although this technique gives accurate results
up to about $T_{\rm lab}=300$ MeV,
it seems almost inaccessible to higher energies
due to the rapid oscillation of the relative wave functions.
We alternatively use a method
to solve the CCRGM equation in the momentum representation,
namely, an LS-RGM equation \cite{LSRGM}.
The idea of solving the RGM equation in the momentum representation
is not new \cite{KU93}.
It is also used by Salamanca group in the study of the 
QM baryon-baryon interactions \cite{SAL00}.
In this method all the necessary Born
amplitudes (or the Born kernels) for the
quark-exchange kernels are analytically derived
by using a transformation formula,
which is specifically developed for momentum-dependent
two-body interactions acting between quarks \cite{LSRGM}.
The partial-wave decomposition of the Born kernel is
carried out numerically in the Gauss-Legendre integral
quadrature. The LS-RGM equation is then solved
by using the techniques developed by Noyes \cite{NO65} and
Kowalski \cite{KO65}.
Although this method requires more CPU time than the variational
method, it gives very stable and accurate results
in a wide energy range.
Since we first calculate the Born amplitudes of the RGM kernel,
it is almost straightforward to proceed
to the $G$-matrix calculations \cite{GMAT}, as well as many
Faddeev calculations in the three-cluster Faddeev formalism. 
The transformation to the RGM equation (\ref{form15}) is easily
carried out by simple Fourier transformations.

A nice property of the present RGM formalism is that
the internal-energy contribution is already subtracted
in the exchange kernel. Namely,
$\CM^{(\Omega)}_{\alpha \alpha^\prime}
(\bR, \bR^\prime)$ in \eq{form16}
for the central components $\Omega=K,~CC,~MC$, $GC$, $CN$, and $SS$
is defined through its corresponding original exchange kernel
$\CM^{(\Omega) exch}_{\alpha \alpha^\prime}(\bR, \bR^\prime)$:
\begin{equation}
\CM^{(\Omega)}_{\alpha \alpha^\prime}(\bR, \bR^\prime)
=\CM^{(\Omega) exch}_{\alpha \alpha^\prime}(\bR, \bR^\prime)
-\left( E^{(\Omega)}_{a_1}+E^{(\Omega)}_{a_2} \right)
~\CM^{N}_{\alpha \alpha^\prime}(\bR, \bR^\prime)\ \ ,
\label{form17}
\end{equation}
where $E_a^{(\Omega)}$ denotes the $\Omega$-term contribution to the
internal energy of the $B_8$ specified by the flavor label $a=a_1 a_2$.
Owing to this subtraction, the mass term of the kinetic-energy
operator in \eq{form13} as well as the confinement potential
with $\Omega=\hbox{Cf}$ exactly cancels between the first
and the second terms on the right-hand side of \eq{form17}.
We consider this feature one of the advantages
of the RGM formalism, because the present QM produces
those results which are independent
of the strength of the confinement potential and are insensitive
to the details of the confinement phenomenology.

We should note that the original Hamiltonian (\ref{form13})
in general contains the derivative operators
$\partial/\partial \bx_i$ and $\partial/\partial \bx_j$,
which cannot be reduced to the derivative of the
relative coordinate $\br=\bx_i-\bx_j$ between two quarks.
The appearance of these Galilean non-invariant terms,
like the momentum-dependent Breit retardation term $U^{MC}_{ij}$
and the antisymmetric $LS$ term $U^{aLS}_{ij}$, is
a direct consequence of the more strict Lorentz invariance
at the relativistic level, and their RGM kernel should be explicitly
evaluated in the total c.m. system \cite{SHT84,NA95}.
We include the $U^{MC}_{ij}$ term and take account of its
contribution to the relative kinetic-energy term explicitly.
Even with this procedure, the calculated reduced mass
of the $\Sigma N$ system is degenerate
with that of the $\Lambda N$ system.
This is one of the limitations of the non-relativistic QM,
in which the inertia masses (of $\Sigma$ and $\Lambda$ in
the present case) are not always reproduced correctly.
In order to use the correct reduced masses
in the CCRGM equation,
we make the following replacement only for the exchange
kernels $\CM^{(K)}_{\alpha \alpha^\prime} (\bR, \bR^\prime)$ and
$\CM^{(MC)}_{\alpha \alpha^\prime} (\bR, \bR^\prime)$:
\begin{eqnarray}
\CM^{(\Omega)}_{\alpha \alpha^\prime} (\bR, \bR^\prime)
\rightarrow
\widetilde{\CM}^{(\Omega)}_{\alpha \alpha^\prime} (\bR, \bR^\prime)
=\frac{\mu_{\alpha_0}}{\mu^{exp}_{\alpha_0}}
~\CM^{(\Omega)}_{\alpha \alpha^\prime} (\bR, \bR^\prime)
\qquad \Omega=K \quad \hbox{and} \quad MC\ ,
\label{form18}
\end{eqnarray}
where $\alpha_0$ is the incident channel. (This prescription 
is slightly different from that of Refs.~\cite{RGMFb,FSS}.)
For such a case that the exact Pauli-forbidden state exists
in the CC system, a modification is necessary as discussed
in Sec.~4 of Ref.~\cite{GRGM}.
The essential idea is as follows.
Let us express the CCRGM equation (\ref{form15}) as
\begin{eqnarray}
(\varepsilon - H_0-V^{\rm RGM}) \chi =0 \ ,
\label{form19}
\end{eqnarray}
with $\varepsilon=E-E_{\rm int}$ and $V^{\rm RGM}=V_{\rm D}
+G+\varepsilon K$.
Here $V^{\rm RGM}$ represents the RGM kernel in \eq{form16}
with $V_{\rm D}$ the direct potential, $G$ the exchange kernel
$\sum_\Omega \CM^{\Omega}_\aagp ({\bR}, {\bR}^\prime)$,
and $K$ the exchange normalization kernel 
$-\CM^N_\aagp({\bR}, {\bR}^\prime)$.
The essential feature of the RGM equation (\ref{form19}) is
the existence of the trivial
solution $\chi=u$, which satisfies $Ku=u$.
If we use the projection operator on the Pauli-allowed space,
$\Lambda=1-|u\rangle \langle u|$, \eq{form19} can
equivalently be written as
\begin{equation}
\Lambda (\varepsilon - H_0-V^{\rm RGM} ) \Lambda \chi = 0\ .
\label{form20}
\end{equation}
In this form, we can safely replace the calculated reduced mass
$\mu_\alpha$ with the empirical mass $\mu^{\rm exp}_\alpha$.
Quite obviously, this procedure is accurate only when the
calculated mass is a good approximation for the empirical mass. 
The same procedure is also applied when we formulate the CCRGM
equation in the particle basis. The present constituent QM does not
reproduce the small mass difference
of the isospin multiplets exactly even if the FSB
of the quark masses and the Coulomb interaction
at the quark level are properly introduced.
We therefore solve the modified equation
\begin{equation}
\Lambda (\varepsilon^{\rm exp} - H^{\rm exp}_0-V^{\rm RGM})
\Lambda \chi = 0\ ,
\label{form21}
\end{equation}
instead of \eq{form20},
where $\varepsilon^{\rm exp}$ and $H^{\rm exp}_0$ are
defined by 
\begin{equation}
\varepsilon^{\rm exp}=E-E^{\rm exp}_{\rm int}\ ,\qquad
H^{\rm exp}_0=-\frac{\hbar^2}{2 \mu^{\rm exp}}\left(
\frac{\partial}{\partial \bR}\right)^2
=\frac{\mu}{\mu^{\rm exp}}H_0\ .
\label{form22}
\end{equation}
This procedure to use the empirical reduced masses and threshold
energies in the Pauli-allowed model space has a wide applicability
to remedy the inflexibility of the RGM framework.
If we restore \eq{form22} to the original Sch{\"o}dinger-type
equation (\ref{form19}), we obtain an additional correction term
to the RGM kernel, which is supposed to be small enough for
practical calculations, but necessary to preserve the exact
treatment of the Pauli principle:
\begin{equation}
(\varepsilon^{\rm exp}-H^{\rm exp}_0-V^{\rm RGM}-\Delta G) \chi=0\ .
\label{form23}
\end{equation}
Here $\Delta G$ is given by
\begin{equation}
\Delta G=\Lambda \left( \Delta E_{\rm int}+\Delta H_0 \right)
\Lambda - \left(\Delta E_{\rm int}+\Delta H_0 \right)
\label{form24}
\end{equation}
with
\begin{equation}
\Delta E_{\rm int}=E^{\rm exp}_{\rm int}-E_{\rm int}\ ,\qquad
\Delta H_0=H^{\rm exp}_0-H_0
=\left(\frac{\mu}{\mu^{\rm exp}}-1\right) H_0\ .
\label{form25}
\end{equation}
A detailed example is given in Ref.~\cite{GRGM}
for the $\Lambda N$--$\Sigma N (I=1/2)$ CCRGM,
which involves a Pauli-forbidden state
in the $(11)_s$ $SU_3$ representation
for the $\hbox{}^1S_0$ state \cite{RGMFb}.
With this procedure we can employ
the empirical reduced masses and threshold energies
without impairing the major role of the Pauli principle.

\subsection{Effective-meson exchange potentials}

In this subsection, we outline a procedure to
derive the RGM kernel, starting from a general form of 
the $qq$ interaction, and discuss appropriate 
forms of the EMEP's used in the model FSS and fss2.
Refer to the original papers \cite{KI94,SU83,FSS,fss2,LSRGM} for
detailed calculations.

\begin{table}[b]
\caption{
The coefficients $\alpha^\Omega$, the corresponding
spatial functions $u^\Omega(\bk, \bp_1, \bp_2)$,
and the spin-flavor operators $w^{\Omega}$
for the Fermi-Breit interaction.
The quark pair $i,~j=1,~2$ is assumed.
For the non-central terms, the angular-momentum
coupling $U^T=\alpha^T 3 \sqrt{10} \left[ u^T w^T \right]^{(0)} w^C$ etc.
should be taken.
}
\label{corr}
%\bigskip
\begin{center}
\setlength{\tabcolsep}{1mm}
\renewcommand{\arraystretch}{1.1}
\begin{tabular}{cccc}
\hline
$\Omega$ & $\alpha^\Omega/\alpha_S$ & $u^{\Omega}$
& $w^{\Omega}$ \\
\hline
$CC$  & 1 & $\frac{4\pi}{k^2}$ & 1 \\
$MC$  & $\left(\frac{1}{m_{ud}}\right)^2$
 & $\frac{4\pi}{k^2}\left[\frac{1}{k^2}\left(\bk \cdot \bp_1\right)
\left(\bk \cdot \bp_2\right)-\left(\bp_1 \cdot \bp_2\right)\right]$
& $\frac{\left(m_{ud}\right)^2}{m_1 m_2}$ \\
$GC$ & $-\pi\left(\frac{1}{m_{ud}}\right)^2$ & 1 &
$\frac{1}{2}\left[ \left(\frac{m_{ud}}{m_1}\right)^2
+\left(\frac{m_{ud}}{m_2}\right)^2\right]
+\frac{2}{3}\frac{\left(m_{ud}\right)^2}{m_1 m_2}\left(\bfsigma_1
\cdot \bfsigma_2 \right)$ \\
$sLS$ & $\frac{3}{4}\left(\frac{1}{m_{ud}}\right)^2$
& $\frac{4\pi}{k^2} i \left[ \bk, \frac{1}{2}(\bp_1-\bp_2)\right]$
& $\frac{1}{3}\left(\frac{m_{ud}}{m_1}\right)^2 \bfsigma_1
+\frac{1}{3}\left(\frac{m_{ud}}{m_2}\right)^2 \bfsigma_2
+\frac{2}{3}\frac{\left(m_{ud}\right)^2}{m_1 m_2} \left(
\bfsigma_1+\bfsigma_2\right)$ \\
$aLS$ & $-\frac{1}{8}\left(\frac{1}{m_{ud}}\right)^2$
& $\frac{4\pi}{k^2} i \left[ \bk, \bp_1+\bp_2\right]$
& $-\left(\frac{m_{ud}}{m_1}\right)^2 \bfsigma_1
+\left(\frac{m_{ud}}{m_2}\right)^2 \bfsigma_2
+\frac{2\left(m_{ud}\right)^2}{m_1 m_2} \left(
\bfsigma_1-\bfsigma_2\right)$ \\
$T$ & $\frac{1}{12}\left(\frac{1}{m_{ud}}\right)^2$
& $\frac{4\pi}{k^2} \CY_{2\mu}(\bk)$
& $\frac{\left(m_{ud}\right)^2}{m_1 m_2}
\left[\bfsigma_1 \bfsigma_2 \right]^{(2)}_\mu$ \\
\hline
\end{tabular}
\end{center}
\end{table}

The systematic evaluation of the quark-exchange kernel
is carried out for a two-body
interaction
\begin{eqnarray}
U^\Omega_{ij}=\alpha^\Omega w^{\Omega}_{ij}
w^C_{ij}
~u^{\Omega}_{ij}\ ,
\label{form26}
\end{eqnarray}
where $\alpha^\Omega$ is a simple numerical factor
including a single quark-gluon coupling constants
$\alpha_S=g^2_{qG}/(4\pi)$ and various
types of quark-meson coupling constants $g_{qq\beta}$,
$f_{qq\beta}$, etc.,
$w^{\Omega}_{ij}$ represents the spin-flavor
part, $w^C_{ij}$ the color part
and $u^{\Omega}_{ij}$ the spatial part.
The color part is $w^C_{ij}=(1/4)(\lambda^C_1 \cdot \lambda^C_2)$
for the FB interaction and $w^C_{ij}=1$ for EMEP's.
In Table \ref{corr}, $\alpha^\Omega$, $w^\Omega_{ij}$
and $u^\Omega_{ij}$ are
given for the FB interaction \cite{SU84}, while 
those for EMEP's in Tables VII and VIII of Ref.~\cite{fss2}.
Four different types of the spin-flavor
factors $\Omega=C,~SS,~T,~LS$ are required for the most
general EMEP up to the V mesons; $w^C_{ij}=1$,
$w^{SS}_{ij}=(\bfsigma_i\cdot \bfsigma_j)$,
$w^T_{ij}=[\bfsigma_i \bfsigma_j]^{(2)}_\mu$,
and $w^{LS}_{ij}=(\bfsigma_i+\bfsigma_j)/2$.
For the flavor-octet mesons, these spin operators
should be multiplied with $(\lambda^F_i \cdot \lambda^F_j)$. 
The non-central factors are defined by the reduced matrix elements
for the tensor operators of rank 1 and 2. For example, the
tensor operator is expressed as
\begin{eqnarray}
S_{12}(\bk, \bk)=3 \left(\bfsigma_i \cdot \bk \right)
\left(\bfsigma_j \cdot \bk \right) 
-\left(\bfsigma_i \cdot \bfsigma_j\right) \bk^2
=3\sqrt{10}\left[\,\left[ \bfsigma_i \bfsigma_j\right]^{(2)}
\CY_2(\bk)\,\right]^{(0)} \ \ ,
\label{form27}
\end{eqnarray}
with $\CY_{2\mu}(\bk)=\sqrt{4\pi/15}\,\bk^2 Y_{2\mu}(\widehat{\bk})$.
The Born kernel for the two-body interaction (\ref{form26}) is
calculated by
\begin{equation}
M(\bq_f, \bq_i)=\langle\,e^{i \bq_f \cdot \bR}~\phi\,\vert
\sum_{i<j}^6 U_{ij} \CA^\prime \,\vert
\,e^{i \bq_{\,i}\cdot \bR}~\phi \rangle
=\sum_{x \CT} X_{x \CT}
~M_{x \CT}(\bq_f, \bq_i)\ \ .
\label{form28}
\end{equation}
where $\phi=\phi^{\rm (orb)}\,\xi$ with $\xi=\xi^{SF} \xi^C$ is
the h.o. internal wave function of the $(3q)$-$(3q)$ system
and the superscript $\Omega$ for specifying the interaction piece,
as well as the channel subscript $\alpha$,
are omitted for simplicity.
In \eq{form28}, the spatial part of the Born kernel is
expressed as
\begin{equation}
M_{x \CT}(\bq_f, \bq_i)
=\langle\,z_x~e^{i \bq_f \cdot \bR}\,\phi^{\rm (orb)}\,\vert
\,u_{ij}\,\vert
\,e^{i \bq_{\,i}\cdot \bR}\,\phi^{\rm (orb)} \rangle
\quad \hbox{with} \quad (i,j) \in \CT\ \ .
\label{form29}
\end{equation}
It is specified by the number of exchanged quarks, $x=0,~1$,
and various interaction types, $\CT$.
More specifically, $x=0$ with $z_0=1$ corresponds
to the direct term (EMEP's only)
and $x=1$ with $z_1=P_{36}$ the (one-quark) exchange term.
The interaction types, $\CT=E,
~S,~S^\prime,~D_+$ and $D_-$, correspond
to some specific $(i,j)$ pairs of quarks \cite{KI94}.
The spin-flavor-color factor $X_{x \CT}$ is
defined through
\begin{eqnarray}
X_{x \CT} & = & C_x \langle z_x~\xi\,|\,\sum_{i<j}^\CT w_{ij}
w^C_{ij}\,|\,\xi \rangle \nonumber \\
& = & \left\{
\begin{array}{c}
X^C_{0 \CT} \langle~\xi^{SF}\,|\,\sum_{i<j}^\CT w_{ij}
\,|\,\xi^{SF}\,\rangle \\ [3mm]
(-9) X^C_{1 \CT} \langle~P_{36}~\xi^{SF}\,|
\,\sum_{i<j}^\CT w_{ij}
\,|\,\xi^{SF}\,\rangle \\
\end{array} \right.
\quad \hbox{for} \quad
x=\left\{ \begin{array}{c}
0 \\ [3mm]
1 \\ \end{array} \right.,
\label{form30}
\end{eqnarray}
where $C_0=1$ and $C_1=-9$.
The color-factors $X^C_{x \CT}=\langle z_x~\xi^C\,|
\,w^C_{ij}\,|\,\xi^C \rangle$ for each $(i,j) \in \CT$ are
given as follows. For the quark sector
with $w^C_{ij}=(1/4)(\lambda^C_i \cdot \lambda^C_j)$,
$X^C_{0E}=-(2/3)$,
$X^C_{0D_+}=0$, and $X^C_{1E}=X^C_{0S}=X^C_{0S^\prime}=-(2/9)$,
$X^C_{1D_+}=1/9$, $X^C_{1D_-}=4/9$.
For the EMEP sector with $w^C_{ij}=1$, $X^C_{0 \CT}=1$ and
$X^C_{1\CT}=1/3$.
The spin-flavor-color factor for the exchange normalization kernel
is defined by $X_N=(-3) \langle P_{36}~\xi^{SF}\,|
\,\xi^{SF}\,\rangle$.
We also need $X_K=24 \langle P_{36}~\xi^{SF}\,|\,Y_6-Y_5\,|
\,\xi^{SF}\,\rangle$ with the hypercharge operator $Y_i$ for the exchange
kinetic-energy kernel of the $YN$ systems.

To calculate the matrix elements in \eq{form30},
it is convenient first to calculate three-quark matrix elements
of various operators with respect to the [3]-symmetric and
[21]-symmetric $SU_6$ wave functions. We need a small number of
operators for this calculation, which are always expressed
as composite operators of the two $SU_6$ unit vectors \cite{SP87,FU92}.
After representing the spin-flavor factors in $(3q)$-$(3q)$ system
as operator products of these composite operators,
we can evaluate the $(3q)$-$(3q)$ matrix elements of
these operator products, using a simple formula which combines
the $(3q)$ matrix elements in various ways.
The advantage of this method is that it is directly related
to the straightforward separation of spin-isospin dependence
of various non-central forces appearing
in the $B_8 B_8$ interactions.
The final results of some spin-flavor factors are already published
in our previous papers.
In the quark sector, the operator form of the spin-flavor
factor is expressed as $X^\Omega_\CT$, which is essentially
$X^\Omega_{1\CT}$ in \eq{form30} but has a slightly
different numerical factor.
The spin-flavor-color factors in the quark sector
for the $NN$ system are given in Eq.\,(B.6) of Ref.~\cite{FU87}.
We should note that $X_\CT$ there is
actually $X^{GC}_\CT$ for $\CT=D_+$ and $D_-$,
and $X^{GC}_S=X^{GC}_{S^\prime}=-X_E$.
(Note that $X_E$ is not $X^{GC}_E$.)
Furthermore, $X^{MC}_{D_-}=X^{MC}_{D_+}=X_N$, 
$X^{MC}_S=X^{MC}_{S^\prime}=-2 X_N$, $X^{MC}_E=-X_N$,
and $X_K=0$ for $NN$.
The spin-orbit and tensor factors for $YN$ systems
are found in Appendix C of Ref.~\cite{FU97}.
The one-quark exchange ($x=1$) central
factors $(X^\Omega_\CT)_{B_3 N B_1 N}$ in the $YN$ system
with $Y=\Lambda$, $\Sigma$ and $\Xi$ are
given in Appendix B.3 of Ref.~\cite{LSRGM}.

We define the Born kernel for the exchange
kernel $\CM_\aagp({\bR}, {\bR}^\prime; E)$ in \eq{form16} as
\begin{eqnarray}
& & M^{\rm B}_\aagp (\bq_f, \bq_i; E)
=\langle\,e^{i \bq_f \cdot {\bR}}\,\vert
\,\CM_\aagp ({\bR}, {\bR}^\prime; E)
\,\vert\,e^{i \bq_{\,i}\cdot {{\bR}}^\prime} \rangle \nonumber \\
& & =\sum_\beta \sum_\Omega M_{{\rm D}\,\aagp}^{\Omega\beta}(\bq_f, \bq_i)
\,\CO^\Omega(\bq_f, \bq_i)
+\sum_\Omega M_\aagp^\Omega(\bq_f, \bq_i)\,\CO^\Omega(\bq_f, \bq_i)
-\varepsilon_\alpha~M_{\aagp}^N(\bq_f, \bq_i)\ \ .
\label{form32}
\end{eqnarray}
Each component of the Born kernel is given
in terms of the transferred momentum $\bk=\bq_f-\bq_i$ and
the local momentum $\bq=(\bq_f+\bq_i)/2$.
In \eq{form32} the space-spin
invariants $\CO^\Omega=\CO^\Omega(\bq_f, \bq_i)$ are given
by $\CO^{central}=1$ and
\begin{eqnarray}
& & \CO^{LS} = i \bn \cdot \bS\ ,\quad
\CO^{LS^{(-)}} = i \bn \cdot \bS^{(-)}\ ,\quad 
\CO^{LS^{(-)}\sigma} = i \bn \cdot \bS^{(-)}\,P_\sigma\ ,
\nonumber \\
& & \hbox{with} \quad \bn=[ \bq_i \times \bq_f ]\ ,\quad 
\bS=\frac{1}{2}(\bfsigma_1+\bfsigma_2)\ ,\quad
\bS^{(-)}=\frac{1}{2}(\bfsigma_1-\bfsigma_2)\ ,\quad
P_\sigma=\frac{1+\bfsigma_1 \cdot \bfsigma_2}{2}\ . \qquad
\label{form33}
\end{eqnarray} 
For the tensor and $QLS$ parts, it is convenient to take four
natural operators defined by
\begin{equation}
\CO^{T} = S_{12}(\bk, \bk)\ ,\quad
\CO^{T^\prime} = S_{12}(\bq, \bq)\ ,\quad 
\CO^{T^{\prime \prime}} = S_{12}(\bk, \bq)\ ,\quad
\CO^{QLS} = S_{12}(\bn, \bn)\ ,
\label{form34}
\end{equation}
where $S_{12}(\ba, \bb)=(3/2) [\,({\bfsigma}_1 \cdot \ba)
(\bfsigma_2 \cdot \bb)+(\bfsigma_2 \cdot \ba)
(\bfsigma_1 \cdot \bb)\,]
-(\bfsigma_1 \cdot \bfsigma_2)(\ba \cdot \bb)$.
The calculation of \eq{form32} is most easily carried out
by using a special type of the transformation formula,
developed in Appendix A of Ref.~\cite{LSRGM} for the systems
of two $(0s)^3$-clusters.
This formula is particularly useful to calculate
the Born kernel from the momentum-dependent two-body
interaction. For the direct term of the static two-body
interaction, this formula is reduced to the simple
multiplication of the Gaussian form factor
$F(\bk)^2=e^{-(bk)^2/3}$ to the two-body interaction
in the momentum representation.
For the exchange term, the exchange normalization kernel
is factored out. The formula is especially simple for the
Gaussian interaction, but is also applicable to the Yukawa
functions preserving the factorization property. 
The resultant Born kernel is expressed
using generalized Dawson's integrals $\widetilde{h}_n(x)$ and
modified Yukawa functions $\widetilde{{\cal Y}}_\alpha(x),
~\widetilde{{\cal Z}}_\alpha(x)$ etc., which are given
by the error function of the imaginary argument.
The momentum dependence of the two-body interaction appears
as the polynomial terms in $\bq$ up to the second order.
The direct Born
kernel $M_{{\rm D}\,\aagp}^{\Omega\beta}(\bq_f, \bq_i)$ in
the EMEP sector is shown below.
The exchange Born kernel $M_{\aagp}^{(\Omega)}(\bq_f, \bq_i)$
is given in Appendix B of Ref.~\cite{LSRGM} for the FB interaction
and in Appendix A of Ref.~\cite{fss2} for the EMEP's.

The EMEP's at the quark level are most easily formulated
in the momentum representation, by using the second-order
perturbation theory with respect to the quark-baryon vertices.
We employ the following $qq$ interaction,
which is obtained through the non-relativistic reduction
of the one-boson exchange amplitudes
in the parameter  $\gamma=(m/2m_{ud})$ (where $m$ is the
exchanged meson mass and $m_{ud}$ is the up-down quark mass):

\begin{eqnarray}
U^{\rm S}(\bq_f, \bq_i) & = & gg^\dagger \frac{4\pi}{\bk^2+m^2}
\left(~-1+\frac{\bq^2}{2m_{ud}^2}
-\frac{1}{2m_{ud}^2}i\bn \cdot \bS \right)\ , \nonumber \\ [3mm]
U^{\rm PS}(\bq_f, \bq_i) & = & -ff^\dagger \frac{1}{m_{\pi^+}^2}
\frac{4\pi}{\bk^2+m^2} \left[~(\bfsigma_i \cdot \bk)
(\bfsigma_j \cdot \bk)-(1-c_\delta) (m^2+ \bk^2)
\frac{1}{3}(\bfsigma_i \cdot \bfsigma_j)~\right]
\ , \nonumber \\ [3mm]
U^{\rm V}(\bq_f, \bq_i) & = & \frac{4\pi}{\bk^2+m^2} \left\{
f^e {f^e}^\dagger \left(~1+\frac{3\bq^2}{2m_{ud}^2}\right)
-f^m {f^m}^\dagger \frac{2}{(m_{ud} m)^2}
\Bigl[~(\bfsigma_i \cdot \bn)(\bfsigma_j \cdot \bn)
\right. \nonumber \\ 
& & - \left. (1-c_{qss})
\frac{1}{3}\bn^2 (\bfsigma_i \cdot \bfsigma_j)~\Bigr]
-\left(f^m {f^e}^\dagger + f^e {f^m}^\dagger \right)
\frac{2}{m_{ud} m} i\bn \cdot \bS~\right\}\ .
\label{new1}
\end{eqnarray}
Here the quark-meson coupling constants are expressed
in the operator form in the flavor space \cite{FU92,FU97}.
For example, the product of the
two different coupling-constant operators $g$ and $f$ are
expressed as
\begin{eqnarray}
gf^\dagger = \left\{
\begin{array}{c}
g_1 f_1 \\
g_8 f_8 \Sigma_a \lambda_{a i} \lambda_{a j} \\
\end{array}
\right.
~\hbox{for}
~\left\{
\begin{array}{c}
\hbox{singlet~mesons}\\
\hbox{octet~mesons}\\
\end{array}
\right. \ .
\label{new2}
\end{eqnarray}
The meson mixing between
the flavor-singlet and octet mesons is very important, implying
\begin{eqnarray}
f_{\eta^\prime}=f_1 \cos \theta+f_8 \sin \theta~\lambda_8 \quad,
\qquad f_\eta=-f_1 \sin \theta+f_8 \cos \theta~\lambda_8 \ ,
\label{new3}
\end{eqnarray}
instead of $f_1$ and $f_8 \lambda_8$ in \eq{new2} for the PS mesons.
A similar transformation is also applied to the S- and V-meson
coupling constants.
The $SU_3$ parameters of the EMEP coupling constants
are therefore $f_1$, $f_8$ and $\theta$.
The S-meson exchange EMEP in \eq{new1} involves not only
the attractive leading term, but also the
momentum-dependent $\bq^2$ term and the $LS$ term.  
In the PS-meson exchange operator,
the parameter $c_\delta$ is introduced only for the one-pion
exchange, in order to reduce the very strong effect
of the delta-function type contact term involved in the
spin-spin interaction. The case $c_\delta=1$ corresponds to 
the full expression, while $c_\delta=0$ corresponds to
the case with no spin-spin contact term.
The V-meson exchange potential is composed of
the electric-type term, the magnetic-type term
and the cross term. In the electric term, the central
force generated by the $\omega$-meson exchange potential
is usually most important, and it also includes the $\bq^2$-type
momentum-dependent term.
As to the introduction of the V-meson EMEP to the QM,
the problem of double counting is addressed especially in the
strong short-range repulsion from the $\omega$ meson \cite{YA90}.
We will not discuss this problem here, but avoid this double counting
for the short-range repulsion and the $LS$ force,
by simply choosing appropriate coupling
constants for V mesons, i.e., $f^{\rm Ve}_1,~f^{\rm Vm}_1 \sim 1$,
and $f^{\rm Ve}_8=0$ (see Table \ref{table1}).
The magnetic term is usually important
for the isovector $\rho$ meson, and yields the spin-spin,
tensor and $QLS$ terms in the standard OBEP.
The choice in \eq{new1} is to keep only the $QLS$ term with the
spin-spin term proportional to $\bL^2$, the reason for which is
discussed below. Finally, the cross term between the electric
and magnetic coupling constants leads to
the $LS$ force for the $qq$ interaction.
The antisymmetric $LS$ ($LS^{(-)}$) force
with $\bS=(\bfsigma_1-\bfsigma_2)/2$ is not generated from EMEP,
because the flavor operator in \eq{new2} is the Gell-Mann matrix
and also because the mass difference between the up-down
and strange quarks is ignored in \eq{new1}.

We should keep in mind that these EMEP's are by no means
the consequences of the real meson-exchange
processes taking place between the quarks. First of all,
the static approximation used to derive the meson-exchange
potentials between the quarks is not permissible,
since the masses of S mesons and V mesons are more than
twice as heavy as the quark mass $m_{ud}=$ 300 - 400 $\hbox{MeV}/c^2$.
Since the parameter $\gamma$ is not small,
the non-relativistic reduction is not justified.
Also, the very strong S-meson central attraction
is just a replacement
of the real processes of the $2\pi$ exchange,
the $\pi \rho$ exchange, the $\Delta$ excitations and so forth.
The V mesons are supposed to behave as composite particles
of the $(q{\bar q})$ pairs.
Furthermore, the choice of terms in \eq{new1} is quite {\em ad hoc} and
phenomenological. We should consider \eq{new1} an effective
interaction to simulate those residual interactions
between the quarks which are not taken into account by the
FB interaction.
The full expression of the Born kernel \eq{form32} is lengthy,
but the direct term from the EMEP's is simple enough to
show the main characteristics of the EMEP's introduced
in the present model:
\begin{eqnarray}
& & M^{\rm S}_{\rm D}(\bq_f, \bq_i) = g^2 \frac{4\pi}{\bk^2+m^2}
e^{-\frac{1}{3}(bk)^2} \left\{X^C_{0D_+}
\left[-1+\frac{1}{2(3m_{ud})^2}
\left(\bq^2+\frac{9}{2b^2} \right)\right]
% \right. \nonumber \\
% & & \left.
-\frac{3}{2(3m_{ud})^2}X^{LS}_{0D_+}
i\bn \cdot \bS \right. \nonumber \\
& & \ \hspace{0mm} 
\left. -\frac{3}{2(3m_{ud})^2}X^{LS^{(-)}}_{0D_+}
i\bn \cdot \bS^{(-)}~\right\}\ , \nonumber \\
& & M^{\rm PS}_{\rm D}(\bq_f, \bq_i) = -f^2 \frac{1}{m_{\pi^+}^2}
\frac{4\pi}{\bk^2+m^2}e^{-\frac{1}{3}(bk)^2} X^T_{0D_+}
\left[~(\bfsigma_1 \cdot \bk)
(\bfsigma_2 \cdot \bk) -(1-c_\delta)(m^2+\bk^2)\frac{1}{3}
(\bfsigma_1 \cdot \bfsigma_2)\right]\ , \nonumber \\
& & M^{\rm V}_{\rm D}(\bq_f, \bq_i) = \frac{4\pi}{\bk^2+m^2}
e^{-\frac{1}{3}
(bk)^2} \left\{~(f^e)^2~X^C_{0D_+}\left[~1+\frac{3}{2(3m_{ud})^2}
\left(\bq^2+\frac{9}{2b^2} \right)
~\right] \right. \nonumber \\
& & \ \hspace{0mm} - (f^m)^2 \frac{2}{(3m_{ud} m)^2} X^T_{0D_+}
\left[~(\bfsigma_1 \cdot \bn)(\bfsigma_2 \cdot \bn)
-(1-c_{qss})\left(\frac{\bn^2}{3}+\frac{\bk^2}{b^2}\right)
(\bfsigma_1 \cdot \bfsigma_2)
\right. \nonumber \\ [0mm]
& & \ \hspace{0mm} \left. \left. +\frac{3}{2b^2}[\bfsigma_1 \times \bk]
\cdot [\bfsigma_2 \times \bk]~\right]
- 2f^m f^e \frac{2}{3m_{ud} m} X^{LS}_{0D_+}
i\bn \cdot \bS
- 2f^m f^e \frac{2}{3m_{ud} m} X^{LS^{(-)}}_{0D_+}
i\bn \cdot \bS^{(-)}~\right\}\ .
\label{new4}
\end{eqnarray}
Here $X^\Omega_{0D_+}$ represents the spin-flavor factors
related to the spin-flavor operators in \eq{new1}.
Besides the Gaussian factor $\exp \{-(bk)^2/3 \}$ related
to the $(0s)^3$-cluster wave functions,
the finite size effect of the baryons also appears as the constant
zero-point oscillation terms accompanied with the $\bq^2$ terms,
which appear in the S- and V-meson contributions.
For the $QLS$ force, the same effect appears as the tensor force
having the form $[\bfsigma_1 \times \bk]\cdot
[\bfsigma_2 \times \bk]$. The magnitude of this term is about
one third, compared to the original tensor term
appearing at the level of the $qq$ interaction.
The advantage of using the $QLS$ force in \eq{new1},
instead of the tensor force, is that we can avoid
the $\pi$-$\rho$ cancellation of the tensor force
for the coupling term of the $S$ and $D$ waves.
The $\varepsilon_1$ parameter of the $NN$ interaction is very
sensitive to this coupling strength.

We have four QM parameters; the h.o. size
parameter $b$ for the $(3q)$-clusters, the up-down quark
mass $m_{ud}$, the strength of the quark-gluon
coupling constant $\alpha_S$, and the mass ratio
of the strange to up-down quarks $\lambda=(m_s/m_{ud})$.
A reasonable range of the values for these parameters
in the present framework is $b=0.5$ - 0.6 fm,
$m_{ud}=300$ - 400 MeV/$c^2$, $\alpha_S \approx 2$,
and $\lambda=1.2$ - 1.7.
Note that we are dealing with the constituent QM with
explicit mesonic degree of freedom.
The size of the system determined from the $(3q)$ wave function
with $b$ (the rms radius of the $(3q)$ system
is equal to $b$) is related to the quark distribution,
which determines the range in which the effect
of the FB interaction plays an essential role
through the quark-exchange kernel.
The internal energies of the clusters
should be calculated from the same Hamiltonian as used in the
two-baryon system, and contain not only the quark contribution
but also various EMEP contributions.
The value of $\alpha_S$ is naturally correlated with $b$, $m_{ud}$,
and other EMEP parameters.
This implies that $\alpha_S$ is merely a parameter in our framework,
and has very little to do with the real quark-gluon coupling
constant of QCD.
The quark and EMEP contributions
to the baryon mass differences
between $N$ and $\Delta$ and between $\Lambda$ and $\Sigma$,
calculated in the isospin basis, is tabulated
in Table III of Ref.~\cite{fss2}.

For the EMEP part, we have three parameters $f_1$, $f_8$,
and $\theta$ for each of the S, PS, Ve (vector-electric)
and Vm (vector-magnetic) terms.
The $F/(F+D)$ ratio is no longer a free parameter in the present
framework, but takes the pure $SU_6$ values $\alpha_e=1$ (for S and Ve)
and $\alpha_m=2/5$ (for PS and Vm).
It is convenient to use the coupling constants
at the baryon level, in order to compare our result
with the predictions by other OBEP models.
These are related to the coupling constants at the quark level
used in Eqs.\ (\ref{new1}) and (\ref{new4}) through
a simple relationship
\begin{eqnarray}
& & f^{\rm S}_1=3 g_1 \quad, \qquad f^{\rm S}_8=g_8 \quad,
\qquad f^{\rm PS}_1=f_1 \quad,
\qquad f^{\rm PS}_8=\frac{5}{3} f_8 \ ,\nonumber \\
& & f^{\rm Ve}_1=3 f^e_1 \quad, \qquad f^{\rm Ve}_8=f^e_8 \quad,
\qquad f^{\rm Vm}_1=f^m_1 \quad,
\qquad f^{\rm Vm}_8=\frac{5}{3} f^m_8 \ .
\label{new5}
\end{eqnarray}
Through this replacement, the leading term for each meson
in \eq{new4} precisely coincides with
that of the OBEP with Gaussian form factors.
In the present framework,
the S-meson masses for $\epsilon$,
$S^*$, $\delta$ and $\kappa$, are also considered free parameters
within some appropriate ranges,
expected from some real mesons,
$f_0(600)$ or $\sigma$, $f_0(980)$, $a_0(980)$
and $K^*_0(800)$ or $\kappa$, respectively.
We further introduce three parameters to improve the fit of
the $NN$ phase shifts to the empirical data \cite{fss2},
$c_\delta$ the strength factor for the delta-function type
spin-spin contact term of the one-pion exchange potential (OPEP),
$c_{qss}$ the strength factor for the spin-spin term
of the $QLS$ force, and $c_{qT}$ the strength factor for the
tensor term of the FB interaction.

\begin{table}[htb]
%\bigskip
\caption{
Quark-model parameters, $SU_3$ parameters of the EMEP's, S-meson
masses, and some reduction factors $c_\delta$ etc. for the models
fss2 and FSS.
The $\rho$ meson in fss2 is treated in the
two-pole approximation, for which $m_1$ ($\beta_1$)
and $m_2$ ($\beta_2$) are shown below the table.
}
\label{table1}
\begin{center}
\renewcommand{\arraystretch}{1.1}
\setlength{\tabcolsep}{4mm}
\begin{tabular}{ccccc}
%\begin{tabular}{
%@{\hspace{1.0cm}}c@{\hspace{1.0cm}}c
%@{\hspace{1.0cm}}c@{\hspace{1.0cm}}c
%@{\hspace{1.0cm}}c@{\hspace{1.0cm}}}
\hline
 &  $b$ (fm) & $m_{ud}$ (MeV/$c^2$)
 & $\alpha_S$ & $\lambda=m_s/m_{ud}$ \\
\hline
fss2 & 0.5562 & 400 & 1.9759 & 1.5512 \\
FSS    & 0.616 & 360 & 2.1742 & 1.526 \\
\hline
 & $f^{\rm S}_1$ & $f^{\rm S}_8$ & $\theta^{\rm S}$ &
$\theta^{\rm S}_4$ $^{\,1)}$ \\
\hline
fss2 & 3.48002 & 0.94459 & $33.3295^\circ$ & $55.826^\circ$ \\
FSS    & 2.89138 & 1.07509 & $27.78^\circ$ & $65^\circ$ \\
\hline
 & $f^{\rm PS}_1$ & $f^{\rm PS}_8$ & $\theta^{\rm PS}$ & \\
\hline
fss2 & $-$    & 0.26748 & $-$ & (no $\eta, \eta^\prime$) \\
FSS    & 0.21426 & 0.26994 & $-23^\circ$ &  \\
\hline
 & $f^{\rm Ve}_1$ & $f^{\rm Ve}_8$ & $f^{\rm Vm}_1$
 & $f^{\rm Vm}_8$ $^{\,2)}$ \\
\hline
fss2 & 1.050 & 0 & 1.000 & 2.577 \\
\hline
(MeV/$c^2$) & $m_\epsilon$ & $m_{S^*}$ & $m_\delta$
 & $m_\kappa$ \\
\hline
fss2 & 800 & 1250 & $846^{\,3)}$ & 936 \\
FSS    & 800 & 1250 & 970 & 1145 \\
\hline
 & $c_\delta$ & $c_{qss}$ & ${c_{qT}}^{\ 5)}$ &  \\
\hline
fss2 &  $0.4756^{\,4)}$ & 0.6352 & 3.139 & \\
FSS & 0.381 & $-$ & $-$ & \\
\hline
\end{tabular}
\end{center}
\begin{enumerate}
\setlength{\itemsep}{0mm}
\item[1)] $\theta^{\rm S}_4$ is used only for $\Sigma N (I=3/2)$.
\item[2)] $\theta^{\rm V}=35.264^\circ$ (ideal mixing) and
two-pole $\rho$ meson with $m_1$ ($\beta_1$) = 664.56 $\hbox{MeV}/c^2$
(0.34687) and $m_2$ ($\beta_2$) = 912.772 $\hbox{MeV}/c^2$ (0.48747)
\protect\cite{ST94} are used.
\item[3)] For the $NN$ system, $m_\delta=720~\hbox{MeV}/c^2$ is used.
\item[4)] Only for $\pi$, otherwise 1.
\item[5)] The enhancement factor for the Fermi-Breit tensor term.
\end{enumerate}
\end{table}

We determine these parameters so as to fit
the phase shift analysis SP99 \cite{SAID} for the $np$ scattering
with the partial waves $J \leq 2$ and
the incident energies $T_{\rm lab} \leq 350~\hbox{MeV}$,
under the constraint that the deuteron binding energy
and the $\hbox{}^1S_0$ $NN$ scattering length are reproduced.
Some parameters are very sensitive to reproduce the available data
for the low-energy $YN$ total cross sections.
In the parameter search, we minimize the $\chi^2$ value defined as
\begin{eqnarray}
\sqrt{\chi^2}=\left\{\frac{1}{N}\sum_{i=1}^N
\left( \delta_i^{\rm cal}-\delta_i^{\rm exp}\right)^2\right\}
^{\frac{1}{2}}\ ,
\label{new6}
\end{eqnarray}
where no experimental error bars are employed because
they are not given in the energy-dependent solution of the phase-shift
analysis. Here the sum over $i=1$ - $N$ covers
various angular momenta and energies,
and the mixing parameters, $\varepsilon_1$ and $\varepsilon_2$,
in units of degree.
The value $\sqrt{\chi^2}$ gives a measure
for the deviation of the calculated
phase shifts from the empirical values.
The result of the parameter search is shown in Table \ref{table1}.
The parameters of the previous model FSS are also
shown for comparison. The resultant $\chi^2$ values
are $\sqrt{\chi^2} \approx 3^\circ$ in FSS and
$\sqrt{\chi^2}=0.59^\circ$ in fss2 for the $np$ scattering,
when the energies $T_{\rm lab}=25$, 50, 100, 150, 200, 300 MeV
and the partial waves up to $J=2$ are employed.

\subsection{The Lippmann-Schwinger formalism
for ${\protect\mbf (3q)}$-${\protect\mbf (3q)}$ RGM
and the ${\protect\mbf G}$-matrix equation}

As mentioned in the Introduction, the RGM equation (\ref{form15}) is
solved in the momentum representation.
The basic LS-RGM equation is given by
\begin{eqnarray}
T_{\gamma \alpha}(\bp, \bq; E)
=V_{\gamma \alpha}(\bp, \bq; E)
+\sum_\beta \frac{1}{(2\pi)^3} \int d \bk
~V_{\gamma \beta}(\bp, \bk; E)
\frac{2 \mu_\beta}{\hbar^2}
\frac{1}{k_\beta^2-k^2+i \varepsilon}
~T_{\beta \alpha}(\bk, \bq; E)\ ,
\label{fm7}
\end{eqnarray}
where the quasipotential $V_{\gamma \beta}(\bp, \bq; E)$ is
calculated from the basic Born kernel (\ref{form32}) through
\begin{equation}
V_{\gamma \alpha}(\bk, \bk^\prime; E)
=\frac{1}{2} \left[~M^{\rm B}_{\gamma \alpha}(\bk, \bk^\prime; E)
+(-1)^{S_\alpha} \CP_\alpha
~M^{\rm B}_{\gamma \alpha}(\bk, -\bk^\prime; E)~\right]\ .
\label{fm8}
\end{equation}
Using the partial wave decomposition of the LS-RGM equation \cite{LSRGM},
we introduce the definition
\begin{eqnarray}
T_{\gamma \alpha}(\bk, \bk^\prime; E)
=\sum^{\qquad \prime}_{JM\ell \ell^\prime}
4\pi~T^J_{\gamma S^\prime \ell^\prime,\alpha S \ell}(k, k^\prime; E)
\sum_{m m^\prime} \langle \ell^\prime m^\prime
S^\prime S_z^\prime | JM \rangle
\langle \ell m S S_z | JM \rangle
\,Y_{\ell^\prime m^\prime}(\widehat{\bk})
\,Y_{\ell m}^*(\widehat{\bk}^\prime)\ ,
\label{fm21}
\end{eqnarray}
where $\gamma=\left[1/2(11)\,c_1, 1/2(11)c_2 \right]$
$S^\prime S^\prime_z YII_z; \CP^\prime$.
The prime on the summation symbol indicates 
that $\ell$ and $\ell^\prime$ are limited to such values that satisfy
the condition $(-1)^\ell=(-1)^S\CP$ and $(-1)^{\ell^\prime}
=(-1)^{S^\prime}\CP^\prime$.
We write the spin quantum
numbers $S$ and $S^\prime$ of the $T$-matrix 
explicitly for the later convenience,
although these are already included in the definition
of $\alpha$ and $\gamma$.
The LS-RGM equation is reduced to
\begin{eqnarray}
T^J_{\gamma S^\prime \ell^\prime, \alpha S \ell}(p, q; E)
& = & V^J_{\gamma S^\prime \ell^\prime, \alpha S \ell}(p, q; E)
+\sum_{\beta S^{\prime \prime} \ell^{\prime \prime}}^{\qquad \prime}
\frac{4\pi}{(2\pi)^3} \int^{\infty}_{0} k^2\,d\,k 
V^{J}_{\gamma S^\prime \ell^\prime,
\beta S^{\prime \prime}
\ell^{\prime \prime}}(p, k; E)
\nonumber \\
& & \times
\frac{2 \mu_\beta}{\hbar^2}
\frac{1}{k_\beta^2-k^2+i \varepsilon}
~T^J_{\beta S^{\prime \prime} \ell^{\prime \prime}, \alpha S \ell}
(k, q; E)\ ,
\label{fm22}
\end{eqnarray}
where $E=E_\beta^{\rm int}+(\hbar^2/2 \mu_\beta) k_\beta^2$
and $V^J_{\gamma S^\prime \ell^\prime,
\alpha S \ell}(p, q; E)$ is the partial-wave decomposition
of $V_{\gamma \alpha}(\bp, \bq; E)$.
We use the Gauss-Legendre 20-point quadrature formula
to carry out the numerical integration.
The Lippmann-Schwinger equation (\ref{fm22}) involves a pole
at $k=k_\beta$ in the Green function.
A proper treatment of such a singularity for positive energies
is well known. Here we use the technique developed by
Noyes \cite{NO65} and Kowalski \cite{KO65},
and separate the momentum region of $k$ (and also $p$ and $q$) into
two parts. After eliminating the singularity,
we carry out the integral over $0 \leq k \leq k_\beta$ by
the Gauss-Legendre 15-point quadrature formula,
and the integral over $k_\beta \leq k < \infty$ by
using the Gauss-Legendre 20-point quadrature formula
through the mapping, $k=k_\beta+\tan\,[(\pi/4)(1+x)]$.

Once we obtain the phase shift parameters for all the partial
waves, it is straightforward to calculate the scattering
amplitude, which is defined by the on-shell $T$-matrix through
\begin{equation}
M_{\gamma \alpha}(\bq_f, \bq_i; E)
=-\frac{\sqrt{\mu_\gamma \mu_\alpha q_f
q_i}}{2\pi \hbar^2}
T_{\gamma \alpha}(\bq_f, \bq_i; E)\ .
\label{fm17}
\end{equation}
Here $\bq_f$ and $\bq_i$ are
related to the total energy $E$ by the on-shell condition
$E=E_\gamma^{\rm int}+\varepsilon_\gamma=E_\alpha^{\rm int}
+\varepsilon_\alpha$ with $\varepsilon_\gamma=(\hbar^2/2\mu_\gamma){q_f}^2$
and $\varepsilon_\alpha=(\hbar^2/2\mu_\alpha){q_i}^2$.
For the calculation of the polarization observables,
it is customary to write the scattering amplitude
\eq{fm17} in the invariant form with respect to
the space-spin operators. In the $YN$ scattering,
the three basic vectors, $\bk=\bq_f-\bq_i$,
$\bq=(\bq_f+\bq_i)/2$ and $\bn=[\bq_i \times \bq_f]
=[\bq \times \bk]$ are not mutually orthogonal.
We therefore use $\bk=\bq_f-\bq_i$, $\bn=[\bq \times \bk]$ and
$\bP=[\bk \times \bn]=\bk^2 \bq-(\bk \cdot \bq)\bk$ to define
the invariant amplitudes:   
\begin{eqnarray}
& & M(\bq_f, \bq_i; E) = g_0+h_0~i(\bfsigma_1+\bfsigma_2)
\cdot \widehat{\bn}
+h_-~i(\bfsigma_1-\bfsigma_2)\cdot \widehat{\bn}
+ h_n~(\bfsigma_1\cdot \widehat{\bn})
\cdot (\bfsigma_2\cdot \widehat{\bn}) \nonumber \\
& & +h_k~(\bfsigma_1\cdot \widehat{\bk})
\cdot (\bfsigma_2\cdot \widehat{\bk})
+h_P~(\bfsigma_1\cdot \widehat{\bP})
\cdot (\bfsigma_2\cdot \widehat{\bP})
+ f_+ \left\{(\bfsigma_1\cdot \widehat{\bk})
(\bfsigma_2\cdot \widehat{\bP})+(\bfsigma_1\cdot \widehat{\bP})
(\bfsigma_2\cdot \widehat{\bk})\right\}
\nonumber \\
& & + f_- \left\{(\bfsigma_1\cdot \widehat{\bk})
(\bfsigma_2\cdot \widehat{\bP})-(\bfsigma_1\cdot \widehat{\bP})
(\bfsigma_2\cdot \widehat{\bk})\right\}\ .
\label{in1}
\end{eqnarray}
The eight invariant amplitudes, $g_0, \cdots, f_-$, are complex
functions of the total energy $E$ and the scattering
angle, $\cos \theta=(\widehat{\bq}_i \cdot \widehat{\bq}_f)$.
Three of the eight invariant amplitudes,
$h_-$, $f_+$ and $f_-$, do not appear for the $NN$ scattering
due to the identity of two particles ($h_-$ and $f_-$) and
the time-reversal invariance ($f_-$ and $f_+$).
These three terms correspond to
the non-central forces characteristic in the $YN$ scattering;
i.e., $h_-$ corresponds to $LS^{(-)}$,
$f_+$ to $S_{12}(\br, \bp)$, and $f_-$ to $LS^{(-)}\sigma$,
respectively \cite{FU97}.
In particular, the antisymmetric $LS$ interactions,
$LS^{(-)}$ and $LS^{(-)}\sigma$,
involve the spin change between 0 and 1, together with
the transition of the flavor-exchange symmetry $\CP \neq \CP'$.
In the $NN$ scattering this process is not allowed,
since the flavor-exchange symmetry is uniquely specified
by the conserved isospin: $\CP=(-1)^{1-I}$.
On the contrary, these interactions are in general all possible
in the $YN$ scattering, which gives an intriguing interplay
of non-central forces.

The invariant amplitudes are expressed by the $S$-matrix elements
through the partial-wave decomposition of 
the scattering amplitude \eq{fm17}.
Expressing the invariant amplitudes as
\begin{eqnarray}
M_{\gamma \alpha}(\bq_f, \bq_i; E)
=\sum_{JM\ell \ell^\prime}^{\qquad \prime}
4\pi~R^J_{\gamma S^\prime \ell^\prime,\alpha S \ell}
%\nonumber \\
%& & \times 
\sum_{m m^\prime} \langle \ell^\prime m^\prime
S^\prime S_z^\prime | JM \rangle
\langle \ell m S S_z | JM \rangle
\,Y_{\ell^\prime m^\prime}(\widehat{\bq}_f)
\,Y_{\ell m}^*(\widehat{\bq}_i)\ ,
\label{in2}
\end{eqnarray}
we obtain the partial-wave component $R^J_{\gamma S^\prime \ell^\prime,
\alpha S \ell}=(1/2i)(S^J_{\gamma S^\prime \ell^\prime,
\alpha S \ell}-\delta_{\gamma, \alpha} \delta_{S^\prime, S}
\delta_{\ell^\prime, \ell})$ as follows:
\begin{equation}
R^J_{\gamma S^\prime \ell^\prime, \alpha S \ell}
=-\frac{\sqrt{\mu_\gamma \mu_\alpha q_f q_i}}{2\pi \hbar^2}
~T^J_{\gamma S^\prime \ell^\prime, \alpha S \ell}(q_f, q_i; E)\ .
\label{in3}
\end{equation}
The formulas given in Appendix D of Ref.~\cite{LSRGM} are
then used to reconstruct the invariant amplitudes by the solution
of the LS-RGM equation (\ref{fm22}).
All the scattering observables are expressed
in terms of these invariant amplitudes.
For example, the differential cross section and the
polarization of the scattered particle are
given by
\begin{eqnarray}
\frac{d \sigma}{d \Omega} & \equiv & \sigma_0(\theta)
= |g_0|^2 + |h_0+h_-|^2+|h_0-h_-|^2
+|h_n|^2+|h_k|^2+|h_P|^2
+|f_++f_-|^2+|f_+-f_-|^2\ , \nonumber \\ [2mm]
P(\theta) & = & 2~\hbox{Im} \left[ g_0(h_0+h_-)^*+
h_n (h_0-h_-)^*\right]
%\nonumber \\
%& & 
+2~\hbox{Im} \left[ h_k(f_+-f_-)^*-h_P(f_++f_-)^* \right]\ \ .
\label{in4}
\end{eqnarray}

The LS-RGM equation (\ref{fm7}) is readily
extended to the $G$-matrix equation by a replacement
of the free propagator with the ratio of the angle-averaged
Pauli operator and the energy denominator:
\begin{eqnarray}
G_{\gamma \alpha}(\bp, \bq; K, \omega)
=V_{\gamma \alpha}(\bp, \bq; E)
+\sum_\beta \frac{1}{(2\pi)^3} \int d~\bk
~V_{\gamma \beta}(\bp, \bk; E)
\frac{Q_\beta(k, K)}{e_\beta(k, K; \omega)}
~G_{\beta \alpha}(\bk, \bq; K, \omega)\ \ .
\label{fm9}
\end{eqnarray}
A detailed description of this formalism
is given in Ref.~\cite{GMAT}. The formula to calculate the
Scheerbaum factor \cite{SC76} for the s.p.~spin-orbit potential
from the $G$-matrix solution is
given in Ref.~\cite{SPLS}. The energy dependence of the
quasipotential $V_{\gamma \alpha}(\bp, \bq; E)$ in \eq{fm9} is
dealt with as follows. The total energy of the two
interacting particles in the nuclear medium is not conserved.
Since we only need the diagonal $G$-matrices to calculate
s.p.~potentials and the nuclear-matter properties in the
lowest-order Brueckner theory, we simply use
$\varepsilon_\gamma=E_a^{\rm int} -E_c^{\rm int}
+(\hbar^2/2\mu_\alpha)q^2$ both
in $V_{\gamma \alpha}(\bp, \bq; E)$ and $V_{\gamma \beta}
(\bp, \bk; E)$ in \eq{fm9}.
The meaning and the adequacy of this procedure are discussed
in Ref.~\cite{GRGM} in a simple model.

Calculations in the isospin basis is sufficient to obtain
the main characteristics of the baryon-baryon interactions.
For more detailed investigations of the CSB etc.,
we also solve the LS-RGM equation (\ref{fm22}) in the particle basis
to incorporate the pion-Coulomb correction.
We use the empirical baryon masses and evaluate spin-flavor factors
for the charged pions and the neutral pion separately.
The other spin-flavor factors for heavier mesons and the FB interaction
are generated from those in the isospin basis
through the simple isospin relations.
The Coulomb force is introduced at the quark level
by using the quark charges. The exchange Coulomb kernel has the
same structure as the color-Coulombic term of the FB interaction.
All the necessary expressions for the spatial integrals are
given in Ref.~\cite{fss2}.

The standard technique by Vincent and
Phatak \cite{VP74} is employed to solve the Lippmann-Schwinger
equation with the Coulomb force in the momentum representation.
This technique requires the introduction of a cut-off radius $R_C$ for
the Coulomb interaction. In the RGM formalism, we have to introduce
this cut-off at the quark level,
in order to avoid the violation of the Pauli principle.
The two-body Coulomb force assumed in the present
calculation is therefore written as
\begin{eqnarray}
U^{CL}_{ij}=Q_i Q_j e^2 \frac{1}{r_{ij}} \Theta(R_C-r_{ij})\ ,
\label{pa1}
\end{eqnarray}
where $Q_i$ is the charge of the quark $i$ in units of $e$,
and $\Theta$ the Heaviside step function.
%~Q_j
%=2/3$ for the up quark and $-1/3$ for the down and strange quarks.
The Coulomb contribution to the internal energies becomes zero
for the positive charged particles like the proton and $\Sigma^+$.
The direct Coulomb potential is given by
\begin{eqnarray}
V_{\rm D}(R)=Z_1 Z_2 e^2 \frac{1}{r}\left\{
\hbox{erf}\left(\beta R\right)
-\frac{1}{2}\left[\hbox{erf}\left(\beta (R+R_C)\right)
+\hbox{erf}\left(\beta (R-R_C)\right)\right]\right\} \ ,
\label{pa4}
\end{eqnarray}
where $Z_i e$ ($i=1,~2$) is the charge of each baryon,
$\beta=\sqrt{3}/(2b)$, and
$\hbox{erf}\,(x)=(2/\sqrt{\pi})\int^x_0 e^{-t^2} dt$ stands
for the error function.
The exchange Coulomb kernel is also slightly modified
from the pure Coulomb kernel.
Note that, even in the $np$ and $nn$ systems, we have
small contributions from the Coulomb interaction through
the exchange Coulomb kernel.
The value $R_C$ should be sufficiently large
to be free from any nuclear effect beyond $R_C$.
Then the final $S$-matrix is calculated from the condition
that the wave function obtained by solving
the Lippmann-Schwinger equation with the cut-off Coulomb
force is smoothly connected to
the asymptotic Coulomb wave function.
In principle, this procedure is correct only when
the direct Coulomb potential has a sharp cut-off. 
We have developed a new method \cite{apfb05} which 
can be applied even when the cut-off is not sharp
like \eq{pa4}.
In this method, we introduce two extra radii $R_{\rm in}$
and $R_{\rm out}$, which satisfy $R_{\rm in} < R_C < R_{\rm out}$.
We choose $R_{\rm out}$ such that the direct
Coulomb potential \eq{pa4} becomes negligibly small
for $r > R_{\rm out}$.
On the other hand, $R_{\rm in}$ is selected such that 
the Coulomb wave functions are approximately local
solutions of \eq{pa4}, avoiding the effects of the nuclear
force and the smooth transition of the cut-off effect.
We solve a local-potential problem of \eq{pa4}
from $R_{\rm out}$ to $R_{\rm in}$, and prepare two
independent modified plane
waves, $\widetilde{u}_\ell(k, R_{\rm in})$
and $\widetilde{v}_\ell(k, R_{\rm in})$.
The standard procedure by Vincent and Phatak can now be
applied to the $S$-matrix of the LS-RGM equation
by using these $\widetilde{u}_\ell(k, R_{\rm in})$,
$\widetilde{v}_\ell(k, R_{\rm in})$, in order to obtain
the nuclear $S$-matrix with the Coulomb effect.
We find that this procedure gives quite accurate result
within the difference of less than 0.1 degree,
unless the incident energies are extremely small.
For the practical calculations, we have chosen
$R_{\rm in}$ - $R_C$ - $R_{\rm out}$=6 - 8 - 12 fm.

For the $YN$ interaction, more consideration is required
for the treatment of the threshold energies.
Here we use the non-relativistic expressions for the threshold energies
in the calculations with the isospin basis, while in those with the
particle basis the relativistic kinematics is employed.
We note that the mass difference of $\Sigma^-$ and $\Sigma^+$ is
about 8 MeV and is fairly large. The $\Lambda p$ system has the total
charge $Z=1$ and the $\Sigma^- p$ system $Z=0$. 
In the $YN$ systems, the direct Coulomb term exists only
in the $\Sigma^+ p$ channel and the $\Sigma^- p$ channel.
In calculating the EMEP contribution
to the masses of the $\Sigma$ isospin multiplet,
we find a large cancellation between the neutral and charged
pion contributions.
In general, the pion-Coulomb correction is not sufficient
to reproduce the empirical mass difference of various $B_8$
isospin multiplets.
The disagreement of the calculated threshold energies
with the empirical values is a common feature of
any microscopic models. Fortunately, we have
a nice method to remedy this flaw without violating
the Pauli principle. As discussed in Sect. 2.1 and in Ref.~\cite{GRGM},
we only need to add a small correction term $\Delta G$ to
the exchange kernel, in order to use the empirical threshold energies.
The same technique is also applied to the reduced mass corrections.

\subsection{Three-cluster Faddeev formalism using two-cluster RGM 
kernels}

To apply realistic QM baryon-baryon interactions to few-baryon systems
such as the hypertriton, we have to find a basic
three-cluster equation formulated using
microscopic two-cluster quark-exchange kernel of the RGM.
This is non-trivial not only because
the quark-exchange kernel is non-local and energy dependent,
but also because RGM equations sometimes involve redundant components
due to the effect of the antisymmetrization, the Pauli-forbidden states. 
Here we propose a simple three-cluster equation,
which is similar to the three-cluster orthogonality condition
model (OCM) \cite{HO74}, but employs the two-cluster RGM kernel
as the interaction potential \cite{TRGM}.
The three-cluster Pauli-allowed space is constructed
by the orthogonality of the total wave functions
to the pairwise Pauli-forbidden states. Although this definition
of the three-cluster Pauli-allowed space is not exactly
equivalent to the standard definition
given by the three-cluster normalization kernel,
this condition of the orthogonality is essential
to achieve the equivalence of the
proposed three-cluster equation and the Faddeev equations which
employ a singularity-free $T$-matrix derived
from the RGM kernel (the RGM $T$-matrix) \cite{TRGM}.

Let us first consider a simple system composed
of three identical spinless clusters,
interacting via the two-cluster RGM kernel
$\VRGM=V_{\rm D}+G+\varepsilon K$.
We assume that there is only one Pauli-forbidden state
$|u\rangle$ for each pair of the clusters,
satisfying $K|u\rangle=|u\rangle$.
Using the equivalence of \eq{form19} to \eq{form20},
we can separate $\VRGM$ into two distinct parts as
\begin{eqnarray}
\VRGM=V(\varepsilon)+v(\varepsilon)\ ,
\label{fad3}
\end{eqnarray}
where
\begin{eqnarray}
& & V(\varepsilon)=(\varepsilon-H_0)
-\Lambda (\varepsilon-H_0) \Lambda
=\varepsilon |u\rangle \langle u|
+\Lambda H_0 \Lambda-H_0\ ,\nonumber \\
& & v(\varepsilon)=\Lambda \VRGM \Lambda
=\Lambda \left( V_{\rm D}+G+\varepsilon K\right) \Lambda\ .
\label{fad4}
\end{eqnarray}
Note that $\Lambda V(\varepsilon) \Lambda=0$
and $\Lambda v(\varepsilon) \Lambda=v(\varepsilon)$; namely,
$v(\varepsilon)$ represents the net effect of $\VRGM$ in
the Pauli-allowed space.
Using these properties, we can express \eq{form19} 
or \eq{form20} as
\begin{eqnarray}
\Lambda \left[\,\varepsilon-H_0-v(\varepsilon)\,\right]
\Lambda \chi=0\ .
\label{fad5}
\end{eqnarray}
The separation of $\VRGM$ in \eq{fad3} enables us to deal with the
energy dependence of the exchange RGM kernel independently
in the Pauli-forbidden space and in the allowed space.
Let us generalize \eq{fad3} to
\begin{eqnarray}
\CVwe=V(\omega)+v(\varepsilon)\ .
\label{fad6}
\end{eqnarray}
We will see that the energy dependence involved in $V(\omega)$ can be
eliminated by the orthogonality condition
to the Pauli-forbidden state.
Since the direct application of the $T$-matrix formalism
to \eq{form19} leads to a singular off-shell $T$-matrix \cite{GRGM},
we first consider the subsidiary equation
\begin{eqnarray}
\left[\,\omega -H_0-\VRGM\,\right] \chi=0\ ,
\label{fad7}
\end{eqnarray}
with $\omega \neq \varepsilon$,
and extract a singularity-free
off-shell $T$-matrix starting from the standard $T$-matrix
formulation for the ``potential'' $\VRGM$.
A formal solution of the $T$-matrix equation
\begin{eqnarray}
T(\omega,\varepsilon)=\VRGM+\VRGM G^{(+)}_0(\omega)
T(\omega,\varepsilon)
\label{fad8}
\end{eqnarray}
with $G^{(+)}_0(\omega)=1/(\omega-H_0+i0)$ is given by
\begin{eqnarray}
T(\omega,\varepsilon) & = & \widetilde{T}(\omega, \varepsilon)
+(\omega-H_0)|u\rangle \frac{1}{\omega-\varepsilon}
\langle u|(\omega-H_0) \ ,\nonumber \\
\widetilde{T}(\omega, \varepsilon) & = & T_v(\omega, \varepsilon)
-\left[1+T_v(\omega, \varepsilon)
G^{(+)}_0(\omega)\right]|u \rangle
\frac{1}{\langle u|G^{(+)}_v(\omega, \varepsilon)|u \rangle}
\langle u| \left[1+G^{(+)}_0(\omega)
T_v(\omega, \varepsilon)\right] , \qquad
\label{fad9}
\end{eqnarray}
where $T_v(\omega, \varepsilon)$ is defined by
\begin{eqnarray}
T_v(\omega, \varepsilon)
=v(\varepsilon)+v(\varepsilon) G^{(+)}_0(\omega)
T_v(\omega, \varepsilon) \ .
\label{fad10}
\end{eqnarray}
The basic relationship, necessary for deriving the Faddeev equations
of composite particles, is 
\begin{eqnarray}
G^{(+)}_0(\omega) T(\omega,\varepsilon)
=G^{(+)}_0(\omega) \widetilde{T}(\omega, \varepsilon)
+|u\rangle \frac{1}{\omega-\varepsilon}
\langle u|(\omega -H_0)\ ,
\label{fad20}
\end{eqnarray}
with
\begin{eqnarray}
G^{(+)}_0(\omega) \widetilde{T}(\omega, \varepsilon)
=G^{(+)}_\Lambda(\omega, \varepsilon) \CVwe
-|u\rangle \langle u| \ .
\label{fad21}
\end{eqnarray}
Here $G^{(+)}_{\Lambda}(\omega, \varepsilon)$ is given by
\begin{eqnarray}
G^{(+)}_{\Lambda}(\omega, \varepsilon)
=G^{(+)}_v(\omega, \varepsilon)
-G^{(+)}_v(\omega, \varepsilon)|u\rangle
\frac{1}{\langle u|G^{(+)}_v(\omega, \varepsilon)|u \rangle}
\langle u|G^{(+)}_v(\omega, \varepsilon)\ ,
\label{fad12}
\end{eqnarray}
with $G^{(+)}_v(\omega, \varepsilon)
=1/(\omega-H_0-v(\varepsilon)+i0)$.
The full $T$-matrix $T(\omega, \varepsilon)$ in \eq{fad9} is
singular at $\varepsilon=\omega$,
while $\widetilde{T}(\omega, \varepsilon)$ does not
have such a singularity.
For $\varepsilon \neq \omega$, $T(\omega, \varepsilon)$ satisfies
the relationship
\begin{eqnarray}
\langle u|G^{(+)}_0(\omega) T(\omega,\varepsilon)|\phi \rangle
=\langle \phi | T(\omega,\varepsilon) G^{(+)}_0(\omega)
|u \rangle=0
\label{fad22}
\end{eqnarray}
for the plane wave solution $|\phi \rangle$ with
the energy $\varepsilon$,
i.e., $(\varepsilon-H_0)|\phi \rangle=0$.
This is a direct consequence of more general relationship
\begin{eqnarray}
\langle u| \left[\,1+G^{(+)}_0(\omega)
\widetilde{T}(\omega, \varepsilon)
\,\right]=0 \ \ ,\qquad \left[\, 1+\widetilde{T}(\omega, \varepsilon)
G^{(+)}_0(\omega)\,\right] |u \rangle=0 \ ,
\label{fad23}
\end{eqnarray}
which follows from \eq{fad21}.

The three-body equation with the pairwise
orthogonality conditions is expressed as
\begin{eqnarray}
P \left[\,E-H_0-\VRGMA-\VRGMB-\VRGMC\,\right] P \Psi=0\ ,
\label{three1}
\end{eqnarray}
where $H_0$ is the free three-body kinetic-energy operator
and $\VRGMA$ stands for the RGM kernel \eq{fad3} for
the $\alpha$-pair, etc.
The operator $P$ projects on the Pauli-allowed space
with [3] symmetry \cite{HO74}.
More precisely, we solve the eigenvalue problem
\begin{eqnarray}
\sum_\alpha|u_\alpha \rangle \langle u_\alpha
| \Psi_\lambda \rangle
=\lambda~| \Psi_\lambda \rangle
\label{three2}
\end{eqnarray}
in the model space with [3] symmetry, $|\Psi_\lambda \rangle \in [3]$,
and define $P$ as
%a projection operator onto the space spanned
%by eigenvectors with the eigenvalue $\lambda=0$:
%
\begin{eqnarray}
P=\sum_{\lambda=0}|\Psi_\lambda \rangle \langle \Psi_\lambda |\ .
\label{three3}
\end{eqnarray}
The two-cluster relative energy $\varepsilon_\alpha$ in the
three-cluster system is self-consistently determined through
\begin{eqnarray}
\varepsilon_\alpha=\langle P \Psi|
\,h_\alpha+\VRGMA\,|P \Psi \rangle\ ,
\label{three1-2}
\end{eqnarray}
using the normalized three-cluster wave
function $P \Psi$ with $\langle P \Psi| P \Psi \rangle=1$.
Here $h_\alpha$ is the free kinetic-energy operator
for the $\alpha$-pair.
Expressing $P \Psi=\psi_\alpha+\psi_\beta+\psi_\gamma$ with
the Faddeev components $\psi_\alpha$ etc., we can derive
the Faddeev equations \cite{TRGM}
\begin{eqnarray}
\psi_\alpha=G^{(+)}_0(E) \TTE (\psi_\beta+\psi_\gamma)\ .
\label{three18}
\end{eqnarray}
Note that $\TTE$ is essentially the non-singular part of the
two-cluster RGM $T$-matrix \eq{fad9}:
\begin{eqnarray}
\TTE=\widetilde{T}_\alpha(E-h_{\bar \alpha},
\varepsilon_\alpha)\ ,
\label{three19}
\end{eqnarray}
and that the solution of \eq{three18} automatically satisfies
$\langle u_\alpha| \psi_\alpha+\psi_\beta+\psi_\gamma\rangle=0$ due
to \eq{fad23}.
Since we can also derive \eq{three1} from \eq{three18},
these two equations are completely equivalent.

This three-cluster Faddeev formalism is applied
in Ref.~\cite{TRGM} to a toy model composed of  
three di-neutrons ($3 d^\prime$) and a more realistic model
composed of three $\alpha$-clusters.
The di-neutron clusters are described with $(0s)^2$ h.o. wave functions
with $\nu=0.12~\hbox{fm}^{-2}$ and a Serber-type Gaussian interaction is
assumed between two neutrons.
The Pauli-forbidden state between the two di-neutrons is
an $N=0$ $(0s)$ state, where $N$ is the total h.o. quanta
for the relative motion. We adjusted the strength of the
two neutron interaction such that $2d^\prime$ has
a zero-energy bound state, and obtained a $3 d^\prime$ bound state
of about $E_{3d^\prime} \approx -0.5$ MeV.
Using the self-consistent procedure for the energy
parameter $\varepsilon$ in the $d^\prime d^\prime$ RGM kernel,
we have obtained a complete agreement between this Faddeev
calculation and a variational calculation in terms of
the translationally invariant h.o. basis.

In the $3\alpha$ system, the structure
of the $2\alpha$ normalization kernel $K$ is not that simple.
The eigenvalue $\gamma_N$ for $K$ is given
by $\gamma_N=2^{2-N}-3\delta_{N, 0}$ for the even $N$.
In the relative $S$-wave we have two Pauli-forbidden
states, $(0s)$ and $(1s)$, while in the $D$-wave only
one $(0d)$ h.o. state is forbidden.
The relative motion between the two $\alpha$-clusters
starts from $N=4$ h.o. quanta.
The large value $\gamma_4=1/4$ in the Pauli-allowed space
makes the self-consistent procedure through \eq{three1-2} very important.
Because of this rather involved structure
of the $2\alpha$ Pauli-forbidden states,
the Faddeev equations of the $3\alpha$ system
allow the existence of the redundant or trivial solutions in the Faddeev
equation (\ref{three18}) \cite{RED}.
These redundant solutions are generated from solving
the overlap eigenvalue equation for the rearrangement
\begin{eqnarray}
\langle u|S|uf^\tau \rangle=\tau |f^\tau \rangle\ ,
\label{eq4}
\end{eqnarray}
where $S$ is the permutation operator $S=(123)+(123)^2$ and
$|uf^\tau\rangle=|u\rangle |f^\tau\rangle$ is a product of
two functions corresponding to the two momentum Jacobi-coordinate
vectors, $\bk$ and $\bq$, respectively.
The eigenstate with $\tau=-1$ gives a Faddeev component
$\psi^\tau_0=G_0 |uf^\tau \rangle$, which gives a vanishing
total wave function $P \Psi=(1+S)\psi^\tau=0$ but satisfies
the Faddeev equation (\ref{three18}) owing to the orthogonality
relations of the RGM $T$-matrix \eq{fad23}.
In the $3\alpha$ system with the total angular momentum $L=0$,
we have two such solutions with $\tau=-1$.  
In spite of this complexity, we can eliminate these redundant
components by slightly modifying the Faddeev equation in \eq{three18}.
For details, Ref.~\cite{RED} should be referred to.
Comparison between the Faddeev calculation and
the variational calculation in the h.o. basis was carried out
in the same condition, incorporating the cut-off
Coulomb force with $R_C=10$ fm.
For the two-body effective interaction,
we use the three-range Minnesota
force \cite{TH77} with the Majorana exchange mixture $u=0.94687$.
The h.o. width parameter, $\nu=0.257~\hbox{fm}^{-2}$,
is assumed for the $(0s)^4$ $\alpha$-clusters.
The $\alpha \alpha$ phase shifts are nicely reproduced
in the $\alpha \alpha$ RGM calculation,
using this effective $NN$ force.
We have obtained $E_{3\alpha}=-9.592$ MeV,
$\varepsilon=13.481$ MeV, and $c_{(04)}=0.971$,
by taking the partial waves up to $\lambda=8$ in the
Faddeev calculation \cite{BE9L}.
Here $c_{(04)}$ is the amplitude of the lowest shell-model component
with the $SU_3$ (04) representation.
The corresponding variational
calculation using up to $N=60$ h.o. quanta yields
$E_{3\alpha}=-9.594$ MeV,
$\varepsilon=13.480$ MeV, and $c_{(04)}=0.971$,
which implies that the difference is only 1 - 2 keV.
We have also compared the present Faddeev calculations
of the $3\alpha$ system with the fully microscopic
$3\alpha$ RGM or GCM calculations, using various $NN$ effective
forces \cite{post}. 
We find that the present $3\alpha$ ground-state energies
are only 1.5 - 1.8 MeV higher than those
of the microscopic calculations.
This implies that the genuine three-cluster exchange effect,
which is neglected in the present three-cluster formalism,
is attractive in nature, and is not repulsive claimed
in many semi-microscopic $3\alpha$ models.
Oryu {\em et al.} carried out $3\alpha$ Faddeev calculation
using $2\alpha$ RGM kernel \cite{OR94}.
They obtained very large binding energy, $-E_{3\alpha}=28.2$ MeV
with no Coulomb force.
Since the effect of the Coulomb force is at most 6 MeV,
this $3\alpha$ binding energy is extremely large.
This may result from that they did not treat
the $\varepsilon K$ term in the RGM kernel properly,
and that the Pauli-forbidden components are not
exactly eliminated.

The present three-cluster Faddeev formalism is also
applied to the three-cluster OCM in Ref.~\cite{OCMFAD}.
For this application, we only need
to replace $\VRGM$ in \eq{fad3} with
\begin{eqnarray}
V^{\rm OCM}(\varepsilon)=\varepsilon |u\rangle \langle u|
+\Lambda \left(h_0+V^{\rm eff}\right)\Lambda-h_0\ ,
\label{eq15}
\end{eqnarray}
where $h_0$ is the free two-cluster kinetic-energy operator
and $V^{\rm eff}$ is an appropriate effective potential
which reproduces the properties of the two-cluster system.
If the Pauli-forbidden state $|u\rangle$ is a real
bound state of $V^{\rm eff}$, satisfying
$\left( \varepsilon_B-h_0-V^{\rm eff} \right) |u_B\rangle=0$,
\eq{eq15} is further simplified to
\begin{eqnarray}
V^{\rm OCM}(\varepsilon)=V^{\rm eff}
+(\varepsilon-\varepsilon_B)|u_B\rangle \langle u_B|\ .
\label{eq17}
\end{eqnarray}
In these cases, the corresponding OCM $T$-matrix
$\widetilde{T}(\omega, \varepsilon)$ defined in \eq{fad9}
becomes $\varepsilon$-independent.
The Faddeev equation takes the same form as \eq{three18},
but no self-consistency condition like \eq{three1-2} is
necessary. By this procedure, we can prove that the solutions
of the pairwise three-cluster OCM equation (\ref{three1})
(with $\VRGM$ replaced by $V^{\rm eff}$)
are achieved by Kukulin's method of orthogonalizing
pseudopotentials \cite{KU78}.
Here again, some complexity arises for the $3\alpha$ system,
when the deep $\alpha \alpha$ potential by Buck, Friedrich
and Wheatly \cite{BF77} is employed with the bound-state
solutions for the Pauli-forbidden states.
In this case, one of the redundant solutions
with the dominant [21] $N=4$ (40) component does not become
completely redundant, but becomes an almost redundant
Faddeev component with the eigenvalue $\tau \approx -0.999$
in \eq{eq4}. Such a solution allows a large admixture
of the [3]-symmetric $3\alpha$ component given by
\begin{equation}
\phi^{[3]}_\tau=\frac{(1+S)|uf^\tau\rangle}{\sqrt{3(1+\tau)}}\ ,
\label{eq18}
\end{equation}
which has a large overlap with the shell-model like
compact configuration, $|[3]~N=8~(04)\rangle$.
The ground-state solution of the modified Faddeev equations
or similar equations (see, for example, Eq.\,(22) of Ref.~\cite{AMRED})
gives a large binding energy with this dominant configuration,
and a small admixture of the redundant components will never
be completely eliminated.
On the other hand, if the Kukulin's method
of orthogonalizing pseudopotentials is used to eliminate
the redundant components completely, one would also
repel such a component as $\phi^{[3]}_\tau$, resulting
in a very small binding energy of the $3\alpha$ ground
state \cite{tur01,TB03,DE03}.
Refer to Ref.~\cite{AMRED} for details. 

\section{Results and discussions}

\subsection{Two-nucleon system}

\begin{figure}[t]
\begin{center}
\epsfxsize=0.70\textwidth
\epsffile{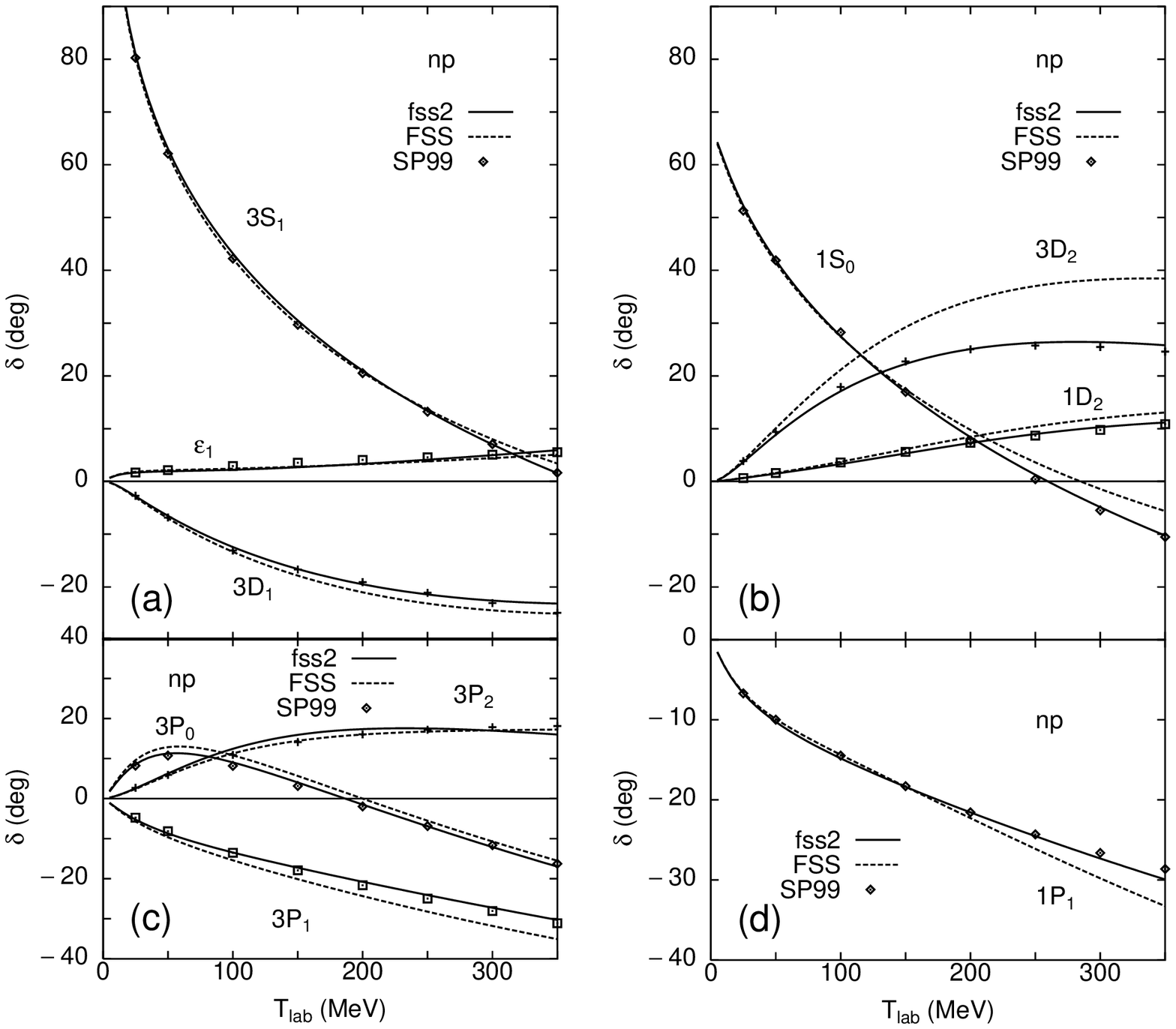}
%\vspace{-40mm}
\epsfxsize=0.70\textwidth
\epsffile{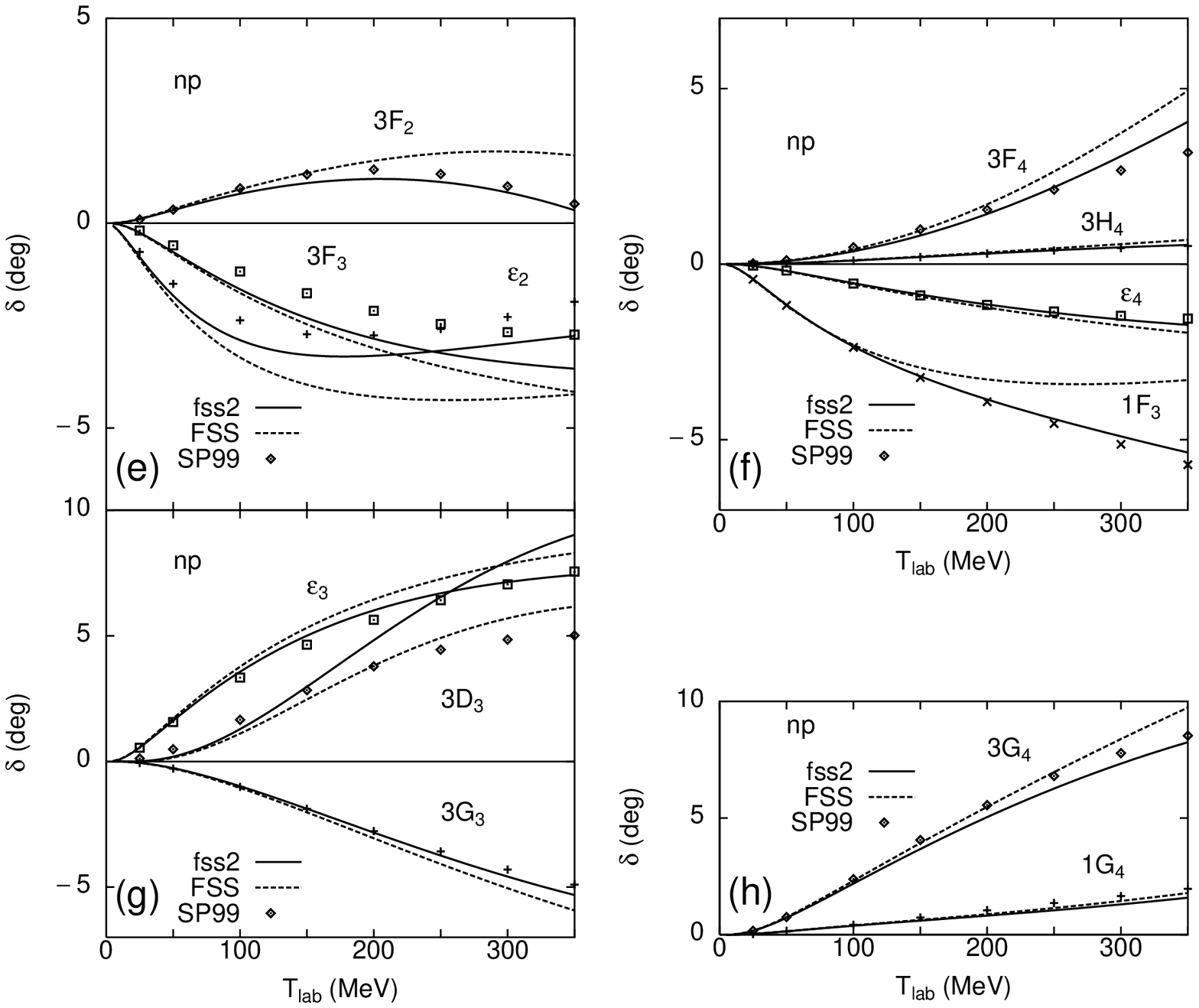}
\caption{
Calculated $np$ phase shifts by fss2
in the isospin basis, compared with the
phase-shift analysis SP99 by Arndt {\em et al.} \protect\cite{SAID}
The dotted curves indicate the results given by FSS.
}
\label{npphase}
\end{center}
\end{figure}

As already discussed, the reliability of the $YN$ and $YY$ interaction
model is judged from whether or not the $NN$ data are
reproduced in such a common framework that can be extended to
the $YN$ and $YY$ systems.
A motivation to upgrade the model FSS \cite{FSS} to fss2 \cite{fss2} is
in a better description of the $NN$ data by including
the important effects of the V mesons and the momentum-dependent
higher-order terms of the S- and V-meson central forces. 
For example, the $\hbox{}^3D_2$ phase shift in FSS is
more attractive than experiment by $10^\circ$ around $T_{\rm lab}=
300~\hbox{MeV}$, because the one-pion tensor force is too strong. 
The strong one-pion tensor force is partially weakened
in the standard OBEP's by the $\rho$-meson tensor force.
We use the $QLS$ force mainly from the $\rho$-meson exchange
as a phenomenological ingredient of this cancellation.
Furthermore, some phase shifts
of other partial waves deviate from the empirical ones
by a couple of degrees.  
Another improvement is required for the central attraction.
The $G$-matrix calculation using the quark-exchange
kernel explicitly \cite{GMAT} shows that the energy-independent
attraction, dominated by the $\epsilon$-meson exchange, is unrealistic,
since in our previous models the s.p.~potentials
in symmetric nuclear matter show a strongly attractive behavior
in the momentum region $5~\hbox{-}~20~\hbox{fm}^{-1}$.
This flaw is removed in Ref.~\cite{LSRGM} by introducing
the momentum-dependent higher-order term
of the flavor-singlet S-meson exchange potential,
which was first carefully examined by Bryan and Scott \cite{BR67}.
In the higher energy region, the $LS$ term of the S mesons
also makes an appreciable contribution.

\begin{figure}[t]
\begin{center}
\begin{minipage}{0.45\textwidth}
\epsfxsize=1.0\textwidth
\epsffile{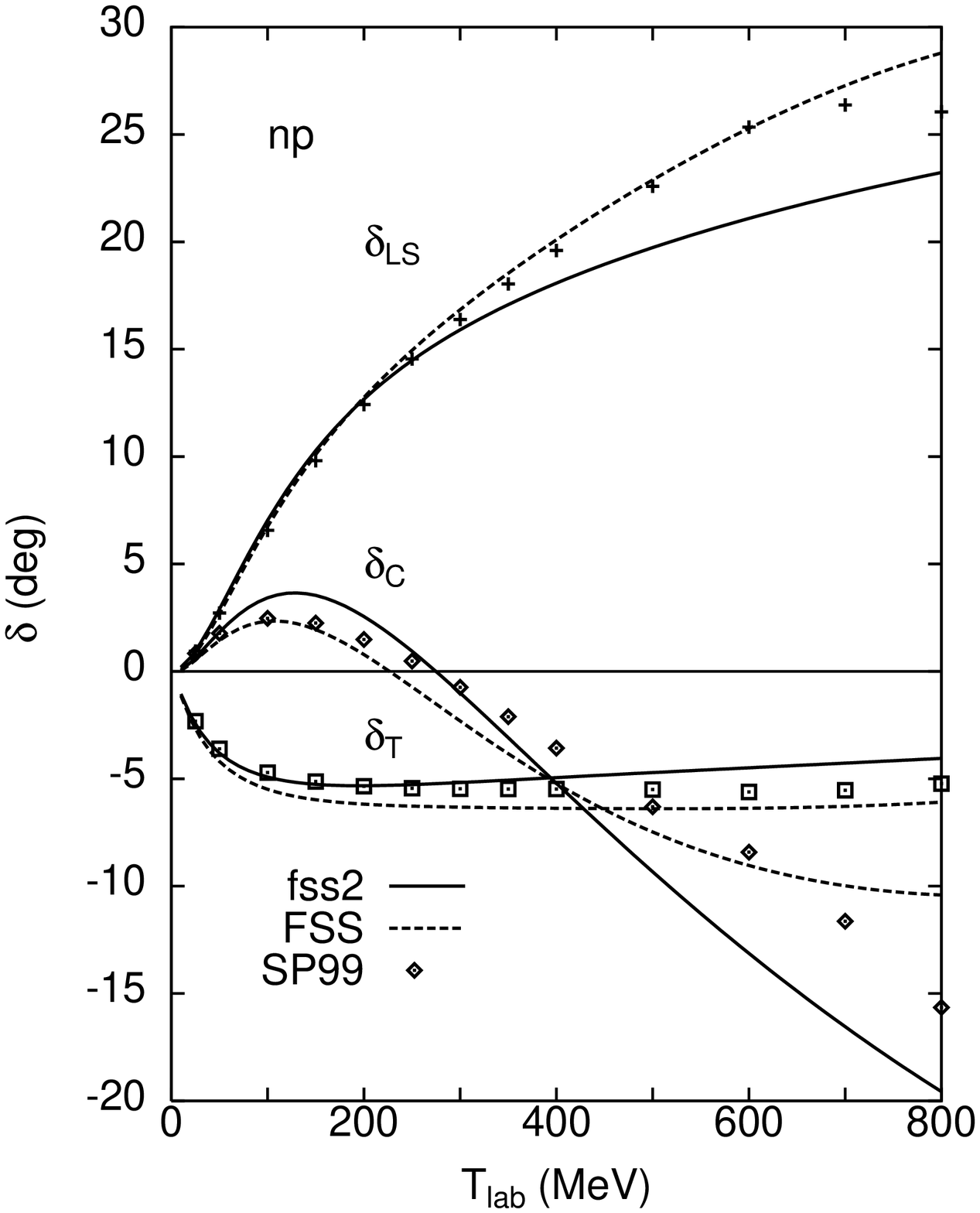}
\end{minipage}~%
\qquad
%\hfill~%
\begin{minipage}{0.47\textwidth}
\caption{Decomposition of the $\hbox{}^3P_J$ phase shifts
for the $np$ scattering
to the central ($\delta_C$), $LS$ ($\delta_{LS}$)
and tensor ($\delta_T$) components
up to the energy $T_{\rm lab}=800$ MeV.
The results given by fss2 (solid curves) and FSS (dashed curves)
are compared with the decomposition of the empirical
phase shifts SP99 \protect\cite{SAID}.
}
\label{phcom}
\end{minipage}~%
\end{center}
\end{figure}

Figures \ref{npphase}(a) - \ref{npphase}(h) compare
the $np$ phase shifts
and the mixing angles $\varepsilon_J$ predicted by fss2 with
those of SP99 by Arndt {\em et al.} \cite{SAID}.
For comparison, the FSS results are also shown
with the dotted curves.
The phase-shift parameters are shown for the partial waves
up to $J=4$ with the energies less than 350 MeV.
More information is obtained from the QMPACK homepage \cite{QMPACK}.
For the non-relativistic energies less than $T_{\rm lab}=300$ MeV,
the $NN$ phase-shift parameters are almost perfectly reproduced,
except for the $\hbox{}^3D_3$ state.
The $\sqrt{\chi^2}$ values in \eq{new6} are calculated using
the phase-shift analysis SP99 \cite{SAID} and PWA93 \cite{ST93} of
the Nijmegen group. 
The fss2 gives $\sqrt{\chi^2}= 0.59^\circ$ and $0.60^\circ$ for SP99
and PWA93, respectively.
These are compared to the $\chi^2$ values
of CD-Bonn \cite{MA01} and ESC04 \cite{ESC04b};
$\sqrt{\chi^2}=0.52^\circ$ ($0.16^\circ$)
and $1.16^\circ$ ($1.01^\circ$) for CD-Bonn
and ESC04, respectively, when SP99 (PWA93) is used.

\begin{figure}[htb]
\begin{center}
\epsfxsize=0.78\textwidth
\epsffile{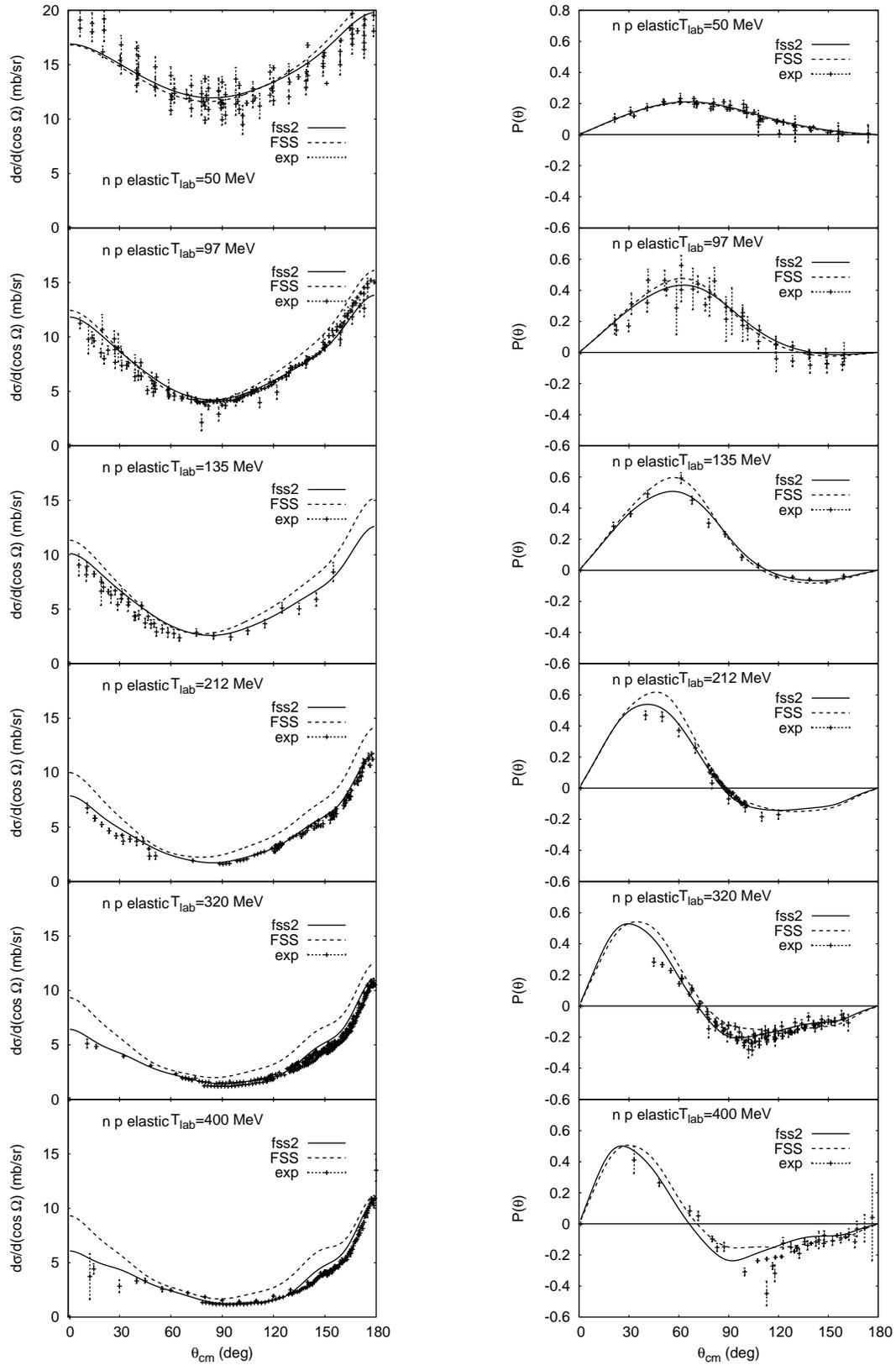}
%\vspace{-40mm}
\caption{
Calculated $np$ differential cross sections
and the polarization compared with
the experiment \protect\cite{SAID}.
Calculation is performed using fss2 (solid curves) and
FSS (dashed curves) with the full pion-Coulomb correction included.
}
\label{npdif}
\end{center}
\end{figure}

For energies higher than 300 MeV, the inelastic process
of the pion production takes place, and the effect of the
$\Delta N$ channel becomes important.
The inelasticity parameters determined from the phase shift analysis,
however, show that they are still rather small even up to the 
energies $T_{\rm lab}=800$ MeV. We have therefore examined 
the results of the single-channel $NN$ calculation up to 800 MeV
in Ref.~\cite{fss2}, in order to investigate a possible effect
of the $\Delta N$ channel coupling.
Up to this energy, the $\hbox{}^3D_2$ phase shift is well
reproduced by the $QLS$ component. Even in the other partial waves,
the improvement of the phase-shift parameters is usually
achieved. This includes 1) $\hbox{}^3P_0$, $\hbox{}^3P_1$,
and $\hbox{}^3G_4$ phase shifts, 2) $\hbox{}^3S_1$,
$\hbox{}^1S_0$, $\hbox{}^1P_1$, $\hbox{}^1F_3$,
and $\hbox{}^3H_4$ phase shifts at higher
energies $T_{\rm lab}=400$ - 800 MeV,
and 3) some improvement in $\hbox{}^3F_2$ phase shift
and $\varepsilon_2$ mixing parameter.
On the other hand, $\hbox{}^3P_2$ and $\hbox{}^3D_3$ phase
shifts turn out worse and $\hbox{}^3F_4$ phase shift
is not much improved.
The problem of fss2 lies in the $\hbox{}^3P_2$ and
$\hbox{}^3D_3$ phase shifts at the intermediate and
higher energies $T_{\rm lab}=300$ - 800 MeV.
The empirical $\hbox{}^3P_2$ phase shift gradually decreases
if we ignore the weak dispersion-like behavior.
Our result, however, decreases too rapidly.
Our $\hbox{}^3D_3$ phase shift is too attractive
by $4^\circ$ - $6^\circ$.
The disagreement of the $\hbox{}^3D_3$ phase shift
and the deviation of the $\hbox{}^3D_1$ phase shift
at higher energies suggests that our description
of the central, tensor and $LS$ forces
in the $\hbox{}^3E$ states requires further improvement.
The insufficiency in the $\hbox{}^3O$ partial
waves is probably related to the imbalance of the central force
and the $LS$ force in the short-range region.
The decomposition of the $\hbox{}^3P_J$ phase shifts
to the central, $LS$ and tensor components, shown in
Fig.~\ref{phcom}, implies that the $\hbox{}^3O$ central
force is too repulsive at higher
energies $T_{\rm lab} \geq 400$ - 500 MeV.
We find that whenever the discrepancy
of the phase shifts between the calculation
and the experiment is large, the inelasticity parameters
are also very large. In particular, the inelasticity
parameters of the $\hbox{}^3P_2$, $\hbox{}^1D_2$,
and $\hbox{}^3F_3$ states rise very rapidly as the energy increases,
and reach more than $20^\circ$ at $T_{\rm lab}=800$ MeV.
The elastic phase shift for each of these states
shows a dispersion-like resonance behavior in the energy
range from 500 MeV to 800 MeV. These are the well-known
di-baryon resonances directly related
to the $\Delta N$ threshold in the isospin $I=1$ channel.
The present single-channel calculation is not
capable of describing these resonances,
unlike the $\hbox{}^1S_0 (NN)$--$\hbox{}^5D_0 (N \Delta)$
CCRGM calculation in Ref.~\cite{VA95}.

\begin{figure}[htb]
\begin{center}
\epsfxsize=0.78\textwidth
\epsffile{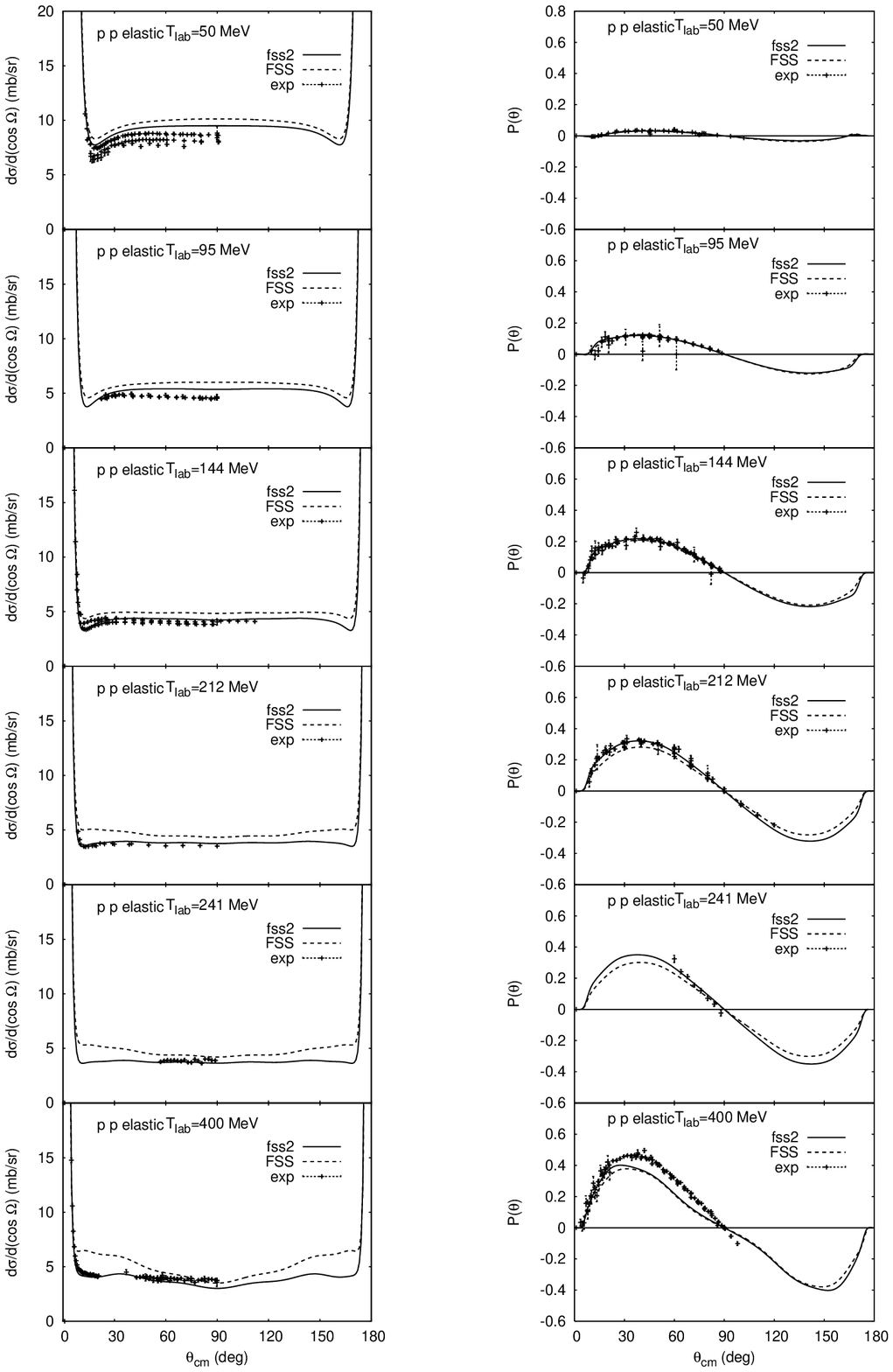}
%\vspace{-40mm}
\caption{
The same as Fig.~\protect\ref{npdif} but
for the $pp$ differential cross sections and the polarization.}
\label{ppdif}
\end{center}
\end{figure}

\begin{figure}[t]
\begin{center}
\epsfxsize=0.78\textwidth
\epsffile{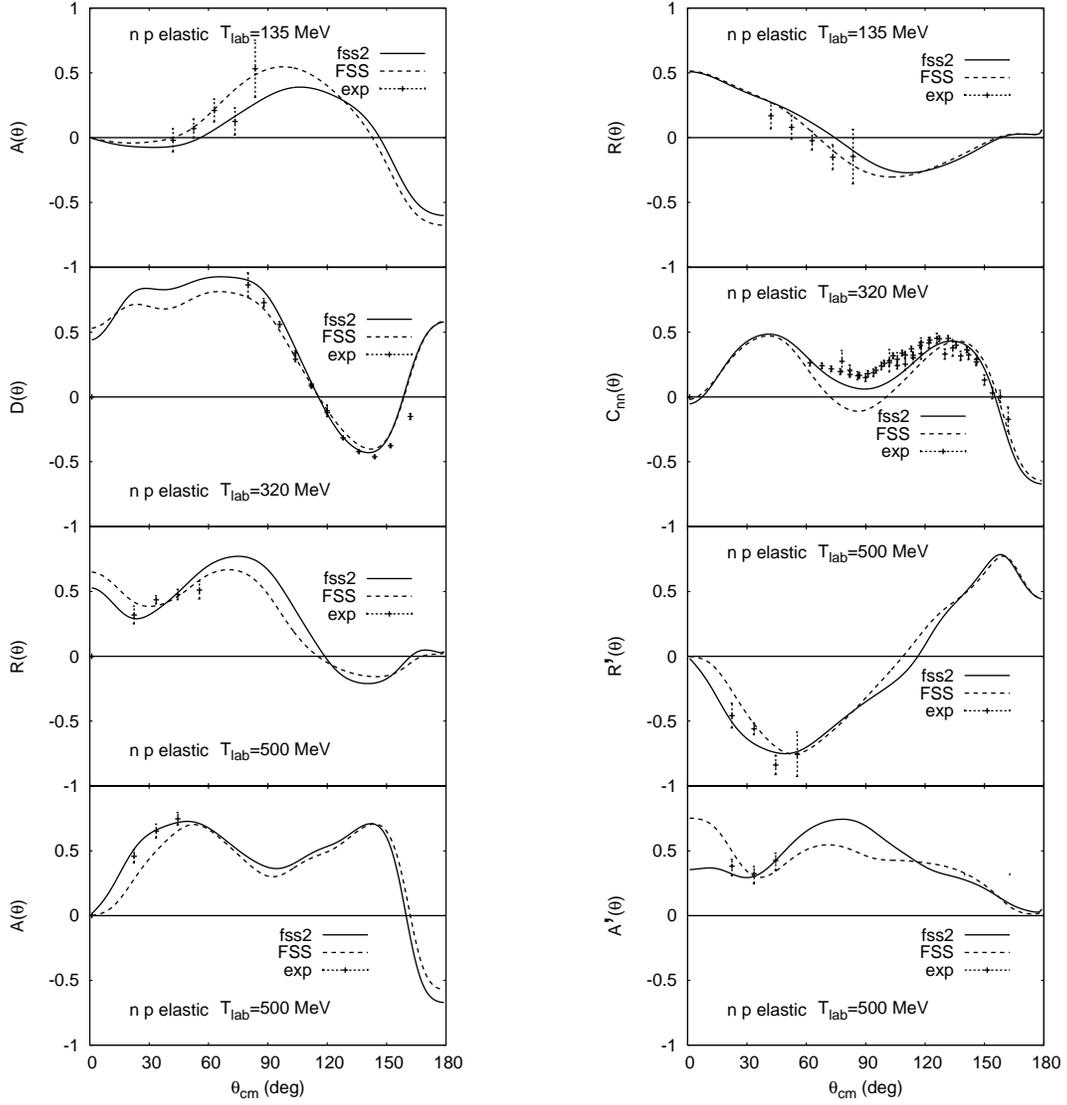}
%\vspace{-40mm}
\caption{
The same as Fig.~\protect\ref{npdif} but for the other
spin observables.}
\label{np_obs}
\end{center}
\end{figure}

\begin{figure}[htb]
\begin{center}
\epsfxsize=0.7\textwidth
\epsffile{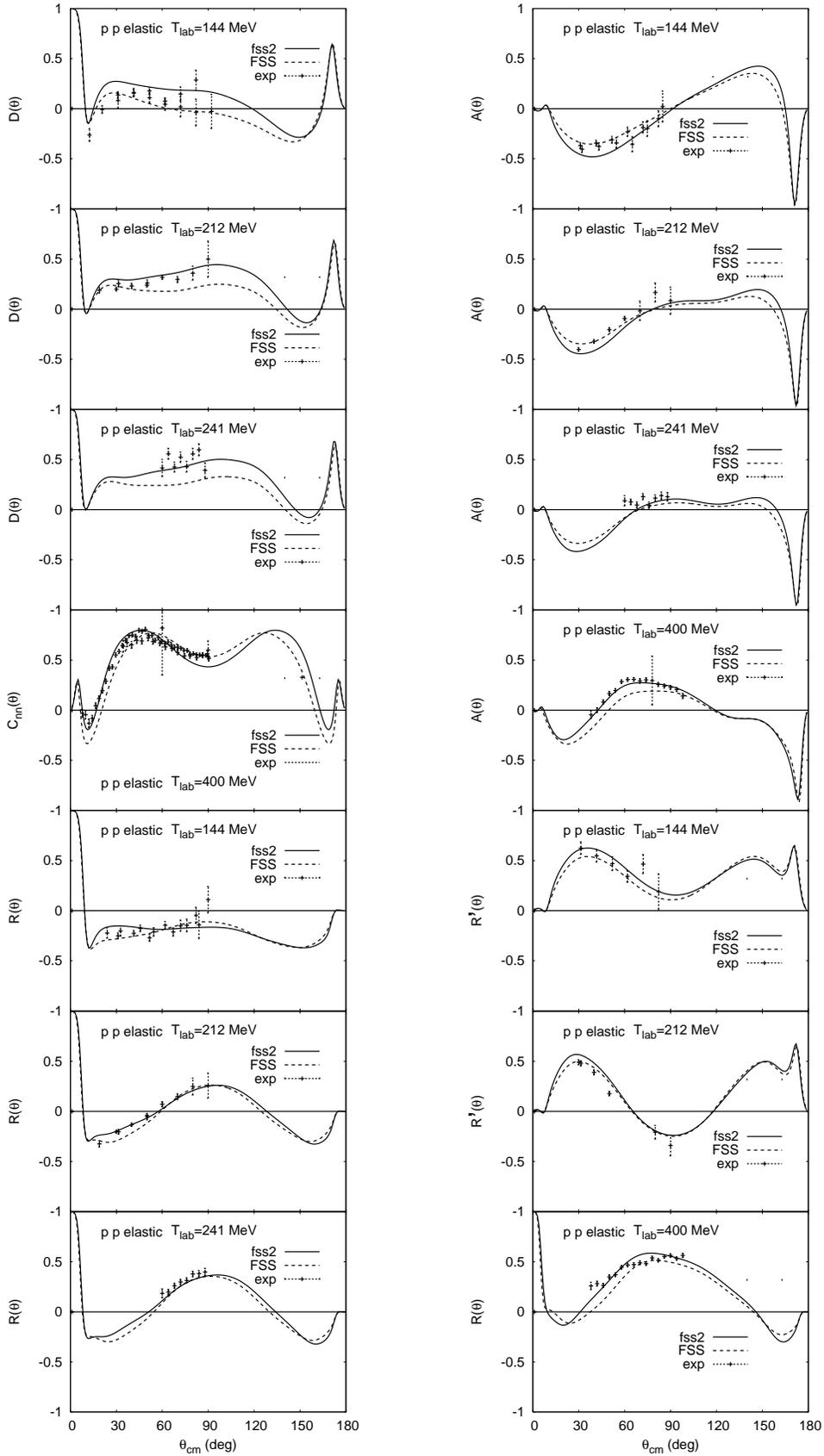}
%\vspace{-40mm}
\caption{
The same as Fig.~\protect\ref{ppdif} but
for the other spin observables.}
\label{pp_obs}
\end{center}
\end{figure}

Figure \ref{npdif} compares with experiment \cite{SAID} the
fss2 (solid curves) and FSS (dashed curves) predictions of the differential
cross sections ($d\sigma/d\Omega$) and
the polarizations ($P(\theta)$) for the elastic $np$ scattering.
The same observables for the elastic $pp$ scattering
are plotted in Fig.~\ref{ppdif}, and some other
observables like the depolarization ($D(\theta)$),
the vector analyzing powers ($A(\theta)$, $A^\prime(\theta)$,
$R(\theta)$), etc., in Figs.~\ref{np_obs} and \ref{pp_obs}.
These observables are obtained in the particle basis with the full Coulomb
force incorporated.
We find some improvement in the differential cross sections.
First, the previous FSS overestimation
of the $np$ differential cross sections at the forward angle
at $T_{\rm lab} \geq 100$ MeV is corrected. Secondly, the bump
structure of the $np$ differential cross sections
around $\theta_{\rm cm}=130^\circ$ at energies $T_{\rm lab}
=300$ - 800 MeV has disappeared. 
The overestimation of the $pp$ differential cross sections
at $\theta_{\rm cm}=10^\circ$ - $30^\circ$ at energies $T_{\rm lab}
=140$ - 400 MeV is improved.
However, the essential difficulties of FSS, namely the
oscillatory behavior of the $np$ polarization
around $\theta_{\rm cm}=110^\circ$ and
that of the $pp$ polarization around the symmetric
angle $\theta_{\rm cm}=90^\circ$ for
higher energies $T_{\rm lab}\geq 400$ MeV are not resolved.
Furthermore, the $pp$ differential cross sections show
a deep dip at angles $\theta_{\rm cm}
\leq 30^\circ$ and $\geq 150^\circ$ for $T_{\rm lab}\geq 500$ MeV.
The low-energy $pp$ cross sections
at $\theta_{\rm cm}=90^\circ$ for $T_{\rm lab}\leq 100$ MeV are
still overestimated.

In order to find a possible reason for the unfavorable
oscillations of our polarizations, we compared
in Fig.~3 of Ref.~\cite{fss2} the five independent $pp$ invariant
amplitudes at the highest energy $T_{\rm lab}=800$ MeV,
predicted by fss2 and the phase-shift analysis SP99.
They are composed of the real and imaginary parts
of $g_0$ (spin-independent central), $h_0$ ($LS$),
$h_n$ ( $(\bfsigma_1\cdot \widehat{\bn})
(\bfsigma_2\cdot \widehat{\bn})$-type tensor),
$h_k$ ( $(\bfsigma_1\cdot \widehat{\bk})
(\bfsigma_2\cdot \widehat{\bk})$-type tensor),
and $h_P$ ( $(\bfsigma_1\cdot \widehat{\bP})
(\bfsigma_2\cdot \widehat{\bP})$-type tensor)
invariant amplitudes in \eq{in1}.
The result by SP99 was calculated
using only the real parts of the empirical
phase-shift parameters.
Recalling that the polarization is given by
the cross term contribution of the central, $LS$, and tensor
invariant amplitudes (i.e., $P(\theta)=2~\hbox{Im}
\left[(g_0+h_n)(h_0)^*\right]$ from \eq{in4} since
$h_{-}=f_\pm=0$ for $NN$),
we find that the disagreement
in $\hbox{Im}~h_n$ and $\hbox{Re}~h_0$ is most serious.
Since the oscillatory behavior of $\hbox{Im}~h_n$ in SP99
also appears in $\hbox{Im}~h_k$ and $\hbox{Im}~h_P$,
it is possible that this is an oscillation caused by
the $NN$--$\Delta N$ channel coupling through
the one-pion spin-spin and tensor forces.
From the figure, we can also expect the underestimation
of the differential cross sections
at $\theta_{\rm cm}\leq 30^\circ$ in this energy region.
Namely, the imaginary part of $g_0$ is too small
and the real part of $g_0$ is strongly reduced in fss2.

\begin{figure}[t]
\begin{center}
\begin{minipage}{0.48\textwidth}
\epsfxsize=1.0\textwidth
\epsffile{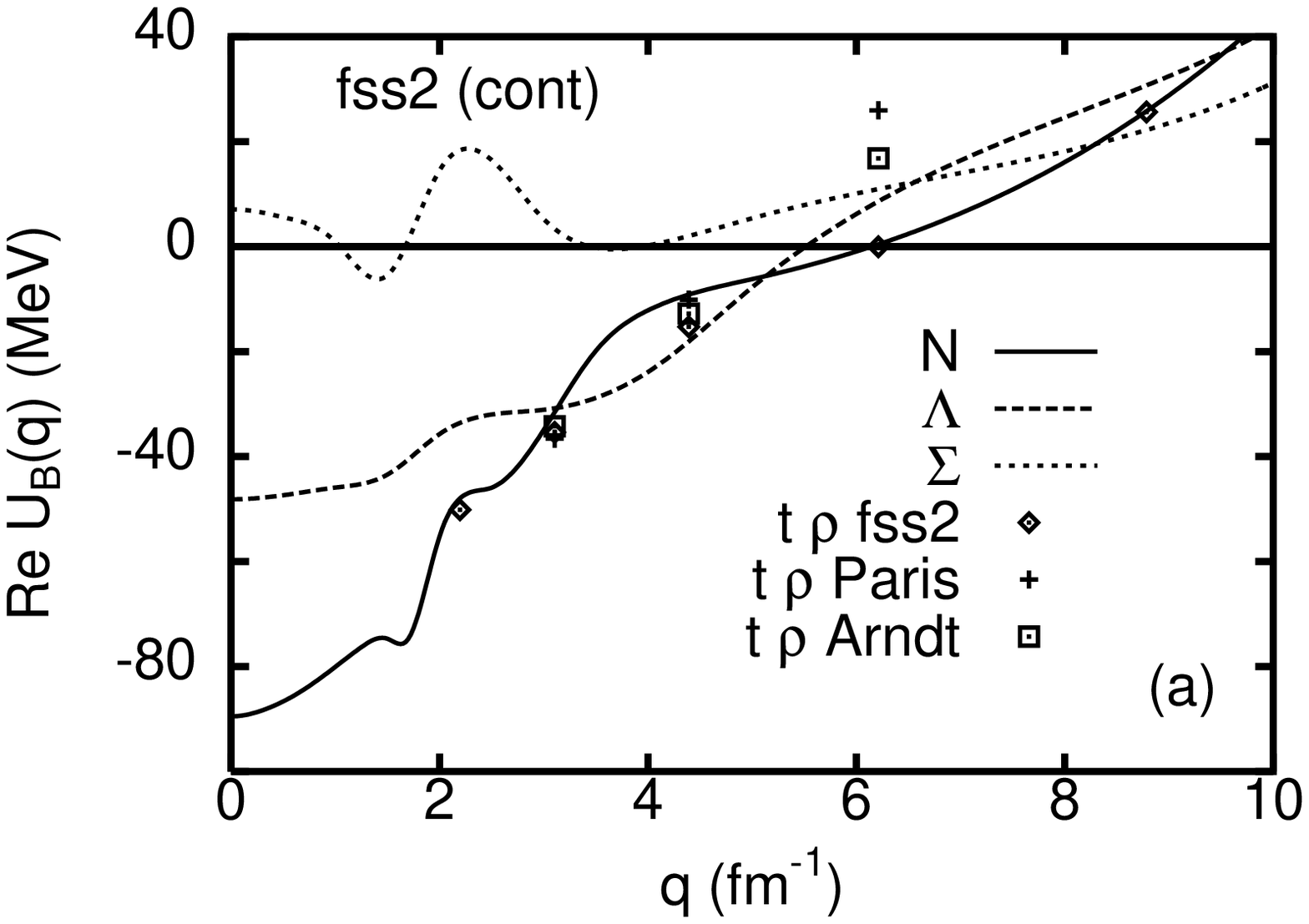}
\end{minipage}~%
\qquad
%\hfill~%
\begin{minipage}{0.48\textwidth}
\epsfxsize=1.0\textwidth
\epsffile{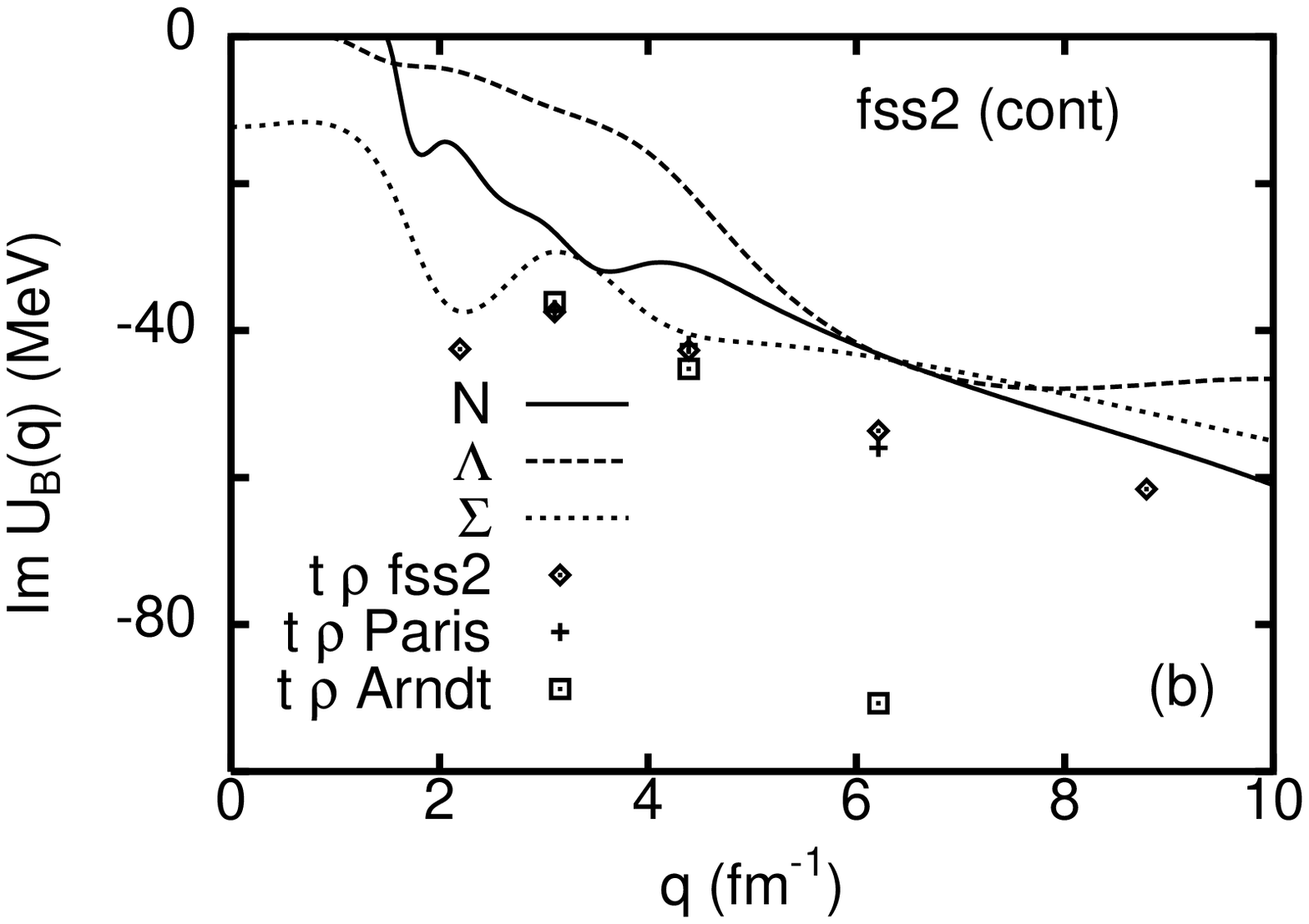}
\end{minipage}
\bigskip
\caption{
The momentum dependence of the s.p.~potentials $U_B(q)$
predicted by the $G$-matrix calculation of fss2
in the continuous prescription for intermediate spectra:
(a) The real part, $\hbox{Re}~U_B(q)$.
(b) The imaginary part, $\hbox{Im}~U_B(q)$.
The nucleon s.p.~potentials obtained
by the $t^{\rm eff}\rho$ prescription are also shown
for fss2, the Paris potential \protect\cite{PARI}
and the empirical phase shifts SP99 \protect\cite{SAID}.
The momentum points selected correspond to $T_{\rm lab}=100$,
200, 400, 800, and 1600 MeV for the $NN$ scattering.
The partial waves up to $J=8$ are included in fss2 and
the Paris potential, and $J=7$ in SP99.
}
\label{trho}
\end{center}
\end{figure}

Another application of the invariant amplitudes is
the $t^{\rm eff}\rho$ prescription for calculating
the s.p.~potentials of the nucleons and hyperons
in nuclear matter. It is discussed in Ref.~\cite{LSRGM} that
the s.p.~potentials predicted by
the model FSS in the $G$-matrix calculation
show fairly strong attractive behavior in the momentum
interval $q=5\,\hbox{--}\,20~\hbox{fm}^{-1}$ for all the baryons.
In particular, $U_N(q)$ in the continuous prescription
becomes almost $-80$ MeV at $q=10~\hbox{fm}^{-1}$.  
This momentum interval corresponds to the incident-energy
range $T_{\tlab}=500~\hbox{MeV}\,\hbox{--}\,8~\hbox{GeV}$
in the $NN$ scattering.
The $t^{\rm eff}\rho$ prescription is a convenient way to
evaluate the s.p.~potentials in the asymptotic momentum
region in terms of the spin-independent invariant amplitude
at the forward angle $g_0(\theta=0)$.
Since the present model fss2 incorporates the momentum-dependent
Bryan-Scott term, the asymptotic behavior of
the s.p.~potentials in the large momentum region is improved.
We can see this in Fig.~\ref{trho}, where the s.p.~potentials 
of $N$, $\Lambda$, and $\Sigma$ calculated in the $G$-matrix
approach are shown in the momentum
range $q=0$ - 10 $\hbox{fm}^{-1}$.
Figures \ref{trho}(a) and \ref{trho}(b) show
the real and imaginary parts of $U_B(q)$,
obtained in the continuous choice for intermediate spectra.
The solid curves for the nucleon s.p.~potential are compared with
the results by the $t^{\rm eff}\rho$ prescription
with respect to the $T$-matrices of fss2,
the Paris potential \cite{PARI},
and the empirical phase shifts SP99 \cite{SAID}.
The partial waves up to $J\leq 8$ are included in fss2 and
the Paris potential, and $J\leq 7$ in SP99.
The momentum points calculated correspond
to the energies $T_{\rm lab}=100$, 200, 400, 800, and 1,600 MeV.
We find that the real part of $U_N(q)$ nicely
reproduces the result of the $G$-matrix calculation
even at such a low energy as $T_{\rm lab}=100$ MeV.
On the other hand, the imaginary part by
the $t^{\rm eff}\rho$ prescription usually overestimates the
exact result especially at lower energies.

\begin{figure}[b]
\begin{center}
\begin{minipage}{0.45\textwidth}
\epsfxsize=0.9\textwidth
\epsffile{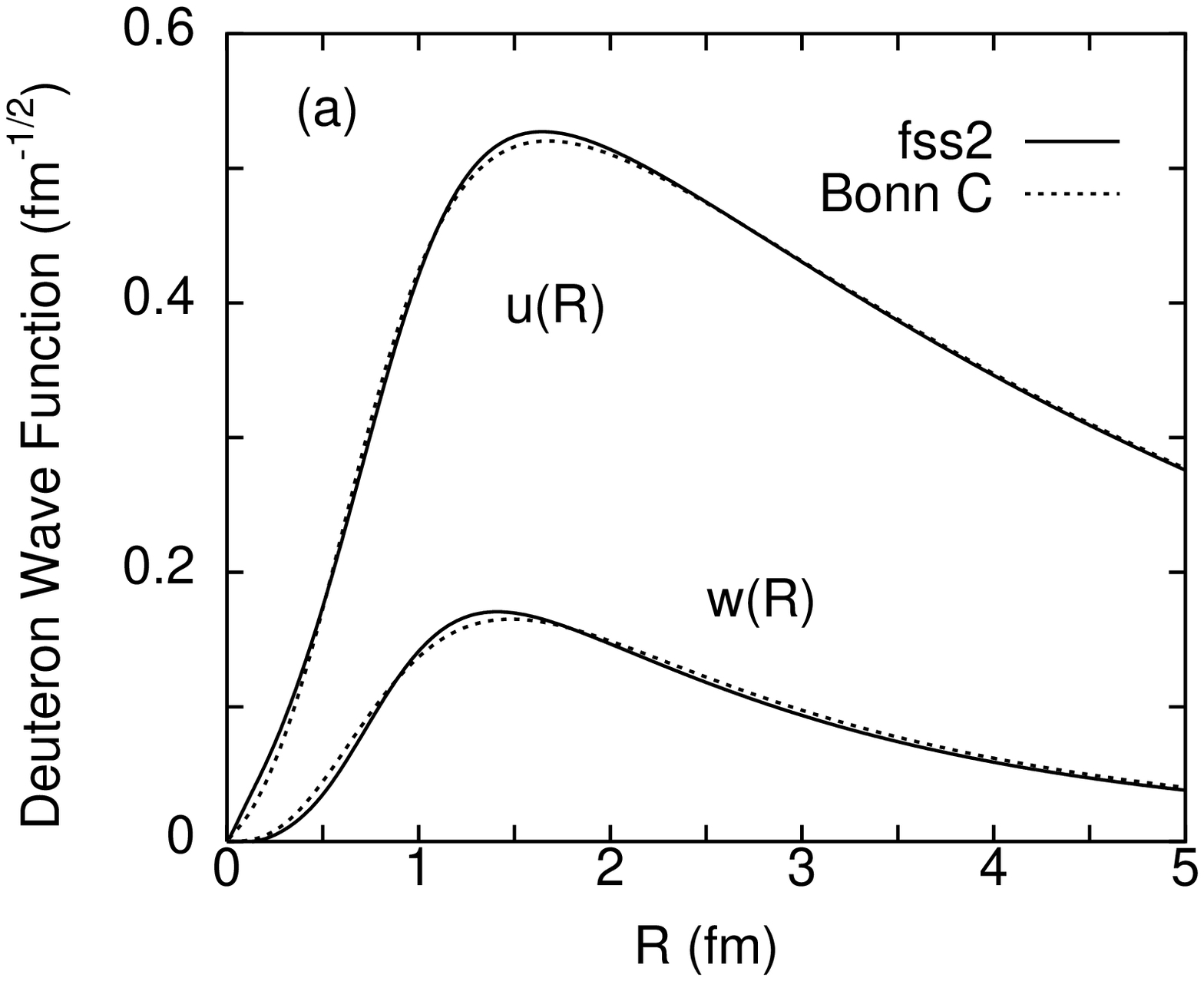}
\end{minipage}~%
\qquad
%\hfill~%
\begin{minipage}{0.45\textwidth}
\epsfxsize=0.9\textwidth
\epsffile{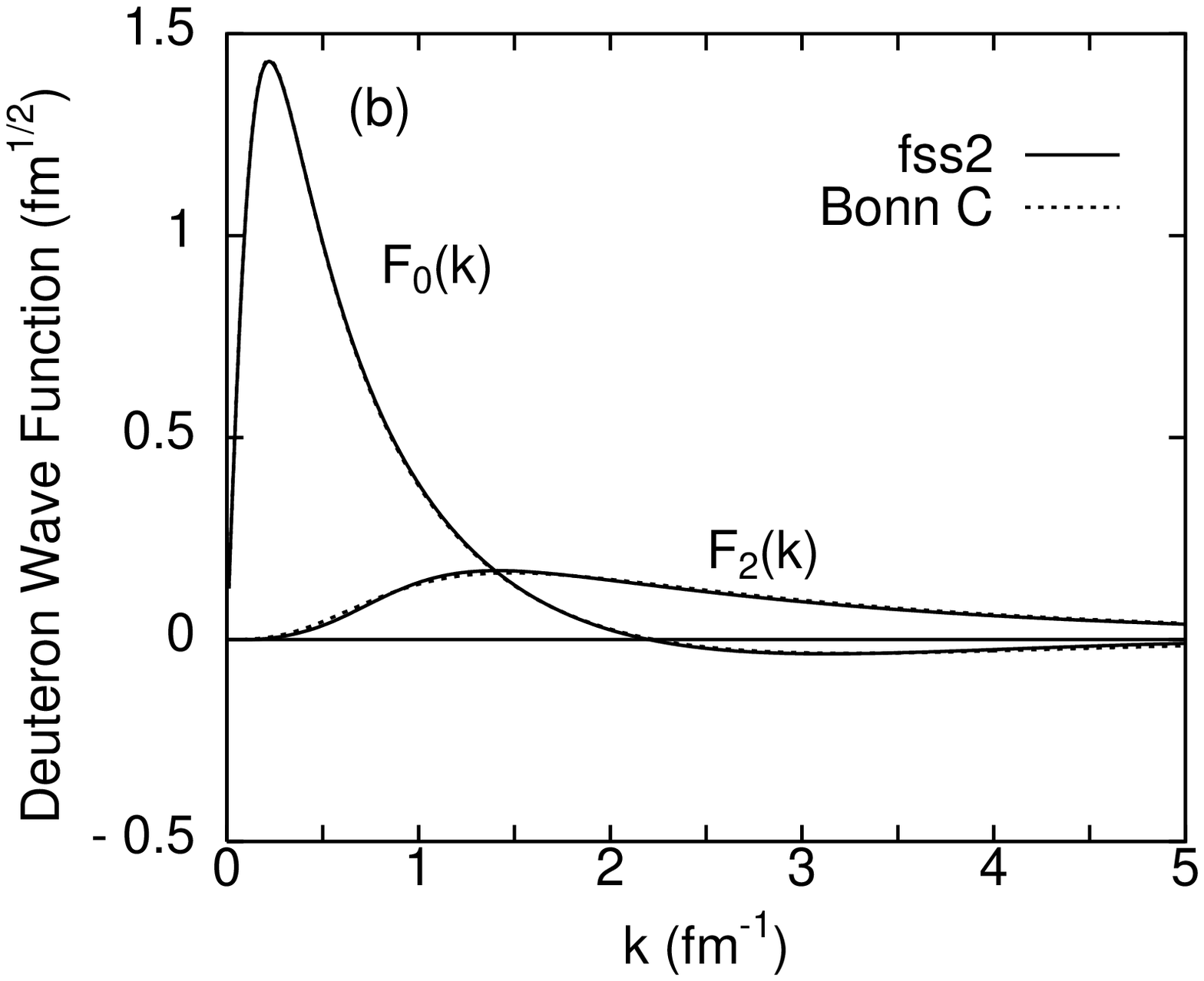}
\end{minipage}
\caption{
The deuteron wave functions predicted by fss2 (solid curves) and
by Bonn model-C \protect\cite{MA89} in the coordinate (a) and
momentum (b) representations.
}
\label{deutf}
\end{center}
\end{figure}

The deuteron properties are calculated
by solving the LS-RGM equation
with respect to the relative wave
functions $f_0(k)$ and $f_2(k)$ in
the momentum representation (see Appendix B of Ref.~\cite{fss2}). 
The properly normalized wave functions
in the Sch{\"o}dinger picture
are not $f_\ell(k)$ but $F_\ell=\sqrt{N}\,f_\ell$,
where $N$ represents
the normalization kernel \cite{FSS}.
The $S$-wave and $D$-wave wave functions in the coordinate
representation, $u(R)$ and $w(R)$, are then
obtained from the inverse Fourier transform of $F_\ell(k)$.
This process is most easily carried out by
expanding $F_\ell(k)$ in a series of Yukawa
functions $\sqrt{2/\pi}k/(k^2+{\gamma_j}^2)$ in the
momentum representation (see Appendix D of Ref.~\cite{MA01}).
We choose $\gamma_j=\gamma+(j-1)\gamma_0$ ($j=1$ - 11) with
$\gamma_0=0.9~\hbox{fm}^{-2}$.
The $\gamma$ is the $S$-matrix pole $q=-i\gamma$,
from which the deuteron energy $\epsilon_d$ is
most accurately calculated by using the relativistic relation
\begin{eqnarray}
M_n+M_p-\epsilon_d=\sqrt{{M_n}^2-\gamma^2}
+\sqrt{{M_p}^2-\gamma^2} \ .
\label{deu1}
\end{eqnarray}
Figure \ref{deutf} shows the deuteron wave function of fss2
in the coordinate and momentum representations,
in comparison with that of the Bonn model-C
potential \cite{MA89} (dotted curves)\footnote{The results
of the Bonn model-C potential in Fig.~\ref{deutf} and
in Table \ref{deutt} are based on the parameterized
deuteron wave functions given in Table C.4
of Ref.~\protect\cite{MA89}.}.
We find that the difference between the two models is very small.
Table \ref{deutt} compares various deuteron properties
calculated in three different schemes: the isospin basis and the
particle basis without or with the Coulomb force.
They are also compared with the empirical values and
the predictions by the Bonn model-C potential.
The final value of the deuteron binding energy
for fss2 is $\epsilon_d=2.2309~\hbox{MeV}$.
If we use the non-relativistic energy
expression,\footnote{In Table \protect\ref{deutt},
the value of $\epsilon_d$ in the isospin basis is calculated
using this non-relativistic formula.}
$\epsilon_d=(\gamma^2/M_N)$ for
$\gamma^2=0.053762~\hbox{fm}^{-2}$ in the full
calculation, we obtain $\epsilon_d=2.2295$ MeV and
the difference is 1.4 keV.
The differences within the deuteron parameters
calculated in the three different schemes
are very small, except for the binding energy $\epsilon_d$.
In particular, the exchange Coulomb kernel
due to the exact antisymmetrization at the quark level
gives an attractive effect to the binding energy,
and increases $\epsilon_d$  by 4.8 keV.
This is even larger than the relativistic
correction included in \eq{deu1}.
The deuteron $D$-state probability is $P_D=5.49~\%$ in fss2,
which is slightly smaller than 5.88 $\%$ in FSS \cite{FSS}.
These values are close to the Bonn model-C
value $P_D=5.60~\%$ \cite{MA89}.
The asymptotic $D/S$ state ratio $\eta$ and
the rms radius are very well reproduced.
On the other hand, the quadrupole moment is
too small by about 5 - 6$\%$.
There are some calculations \cite{AL78,KO83} which claim
that the effect of the meson-exchange currents on the deuteron
quadrupole moment is as large as $\Delta Q_d=0.01~\hbox{fm}^2$.
It is noteworthy that the Bonn model-C almost reproduces
the correct quadrupole moment, in spite of the fact that
the $D$-state probability is very close to ours.
(On the other hand, the quadrupole moment
of CD-Bonn \cite{MA01} is $Q_d=0.270~\hbox{fm}^2$
with a smaller value $P_D=4.85 \%$.)
For the magnetic moment, precise comparison
with the experimental value requires
a careful estimation of various corrections
arising from the meson-exchange currents
and the relativistic effect, etc.

\begin{table}[t]
\caption{Deuteron properties by fss2 in three different
calculational schemes, compared with the predictions of
the Bonn model-C potential \protect\cite{MA89} and the experiment.}
\label{deutt}
%\bigskip
\begin{center}
\renewcommand{\arraystretch}{1.1}
\setlength{\tabcolsep}{2mm}
\begin{tabular}{ccccccc}
%\hline
\hline
& isospin basis & \multicolumn{2}{c}{particle basis} & Bonn C
& Expt. & Ref. \\
\cline{3-4}
& & Coulomb off & Coulomb on & & & \\
\hline
$\epsilon_d$ (MeV) & 2.2250 & 2.2261  & 2.2309 & fitted
& 2.224644 $\pm$ 0.000046 & \protect\cite{DU83} \\
$P_D$ ($\%$)       & 5.490   & 5.490   & 5.494  & 5.60 & & \\
$\eta=A_D/A_S$     & 0.02527 & 0.02527 & 0.02531 & 0.0266 &
0.0256 $\pm$ 0.0004 & \protect\cite{RO90} \\
rms (fm) & 1.9598   & 1.9599 & 1.9582  & 1.968 &
1.9635 $\pm$ 0.0046 & \protect\cite{DU83} \\
$Q_d$ (fm$\hbox{}^2$) & 0.2696 & 0.2696 & 0.2694 & 0.2814 &
0.2860 $\pm$ 0.0015 & \protect\cite{BI79} \\
$\mu_d$ ($\mu_N$) & 0.8485  & 0.8485 & 0.8485 & 0.8479 & 0.85742 & \\
\hline
%\hline
\end{tabular}
\end{center}
\end{table}

\begin{table}[t]
\caption{Effective range parameters of fss2
for the $NN$ interactions.
For the $pp$ and $nn$ systems, the calculation
in the particle basis uses $f_1^S \times 0.9949$,
in order to incorporate the effect of the charge-independence
breaking (CIB). Unit of length is in $\hbox{fm}^{2\ell+1}$ ($a$),
$\hbox{fm}^{-2\ell+1}$ ($r$) and $\hbox{fm}^{-2\ell}$ ($P$)
for the partial wave $\ell$.
The experimental values are taken from \protect\cite{DU83},
\protect\cite{MI90}, \protect\cite{HO98}, \protect\cite{GO99},
\protect\cite{BE88}, and \protect\cite{MA01}.
}
\label{effect}
\begin{center}
\renewcommand{\arraystretch}{1.1}
\setlength{\tabcolsep}{2mm}
\begin{tabular}{cccccc}
%\hline
\hline
 & & isospin basis & \multicolumn{2}{c}{particle basis} & Expt. \\
\cline{4-5}
 & &              & Coulomb off & Coulomb on & \\
\hline
   &  $a$  & & $-17.80$ & \underline{$-7.810$}
& $-7.819 \pm 0.0026$ \\
$pp$ $\hbox{}^1S_0$ & $r$ & & 2.680 & 2.581
& $2.794 \pm 0.0014$ \\
   &  $P$  & & 0.061  & $-0.064$ &  \\
\hline
   &  $a$  & & $-2.876$ & $-3.117$ &
$-4.82 \pm 1.11$, $-2.71 \pm 0.34$ \\
$pp$ $\hbox{}^3P_0$ & $r$ & & 3.834 & 3.735 &
$7.14 \pm 0.93$, $3.8 \pm 1.1$ \\
   &  $P$  & & $-0.011$  & $-0.034$ &  \\
\hline
   &  $a$  & & 1.821 & 1.992 &
$1.78 \pm 0.10$, $1.97 \pm 0.09$ \\
$pp$ $\hbox{}^3P_1$ & $r$ & & $-8.162$ & $-7.5666$ &
$-7.85 \pm 0.52$, $-8.27 \pm 0.37$ \\
   &  $P$  & & 0.001  & 0.001 &  \\
\hline
%\hline
   &  $a$  & & $-18.04$ & $-18.04$ & $-18.05 \pm 0.3$,
$-18.9 \pm 0.4$ \\
$nn$ $\hbox{}^1S_0$ & $r$ & & 2.677 & 2.677 & $2.75 \pm 0.11$ \\
   &  $P$  & & 0.061  & 0.061 &  \\
\hline
   &  $a$  & & $-2.881$ & $-2.881$ &  \\
$nn$ $\hbox{}^3P_0$ & $r$ & & 3.825 & 3.825 &  \\
   &  $P$  & & $-0.011$  & $-0.011$ &  \\
\hline
   &  $a$  & & 1.823 & 1.823 &  \\
$nn$ $\hbox{}^3P_1$ & $r$ & & $-8.154$ & $-8.155$ &  \\
   &  $P$  & & 0.001  & 0.001 &  \\
\hline
%\hline
   &  $a$  & \underline{$-23.76$} & $-27.39$ & $-27.87$
& $-23.748 \pm 0.010$ \\
$np$ $\hbox{}^1S_0$ & $r$ & 2.588 & 2.532 & 2.529 & $2.75 \pm 0.05$ \\
   &  $P$  & 0.059  & 0.052  & 0.052 &  \\
\hline
   &  $a$  & $-2.740$ & $-2.466$ & $-2.466$ &  \\
$np$ $\hbox{}^3P_0$ & $r$ & 3.869 & 3.930 & 3.930 &  \\
   &  $P$  & $-0.013$  & $-0.018$  & $-0.018$ &  \\
\hline
   &  $a$  & 5.399 & 5.400 & 5.395 & $5.424 \pm 0.004$ \\
$np$ $\hbox{}^3S_1$ & $r$ & 1.735  & 1.735 & 1.734
   & $1.759 \pm 0.005$ \\
   &  $P$  & 0.063 & 0.063 & 0.063 &  \\
\hline
   &  $a$  & 2.824 & 2.826 & 2.826 &  \\
$np$ $\hbox{}^1P_1$ & $r$ & $-6.294$ & $-6.300$ & $-6.300$ &  \\
   &  $P$  & $-0.006$  & $-0.006$  & $-0.006$ &  \\
\hline
   &  $a$  & 1.740 & 1.582 & 1.582 &  \\
$np$ $\hbox{}^3P_1$ & $r$ & $-8.198$ & $-8.185$ & $-8.185$ &  \\
   &  $P$  & 0.001 & 0.000  & 0.000 &  \\
\hline
%\hline
\end{tabular}
\end{center}
\end{table}

Table \ref{effect} lists the $S$- and $P$-wave effective range
parameters for the $NN$ system, calculated in the three schemes.
Since the pion-Coulomb correction is not sufficient to explain
the full CIB effect existing in the $np$ and $pp$ $\hbox{}^1S_0$ states,
a simple prescription to multiply
the flavor-singlet $S$-meson coupling
constant $f^S_1$ by a factor 0.9949  is adopted to reduce
too large attraction of the $pp$ central force.
(This prescription is applied only to the calculation
in the particle basis.)
The underlined values of the scattering length $a$ in
Table \ref{effect} indicate
that they are fitted to the experimental values. 
We find that the pion-Coulomb correction
in the $np$ $\hbox{}^1S_0$ state
has a rather large effect on $a$.
The value $a=-23.76~\hbox{fm}$ in the isospin basis changes
to $a=-27.39~\hbox{fm}$ due to the effect of the pion mass
correction and the explicit use of the neutron and proton masses.
It further changes to $a=-27.87~\hbox{fm}$ due to the small effect
of the exchange Coulomb kernel.
These changes should, however, be carefully reexamined by readjusting
the binding energy of the deuteron in Table \ref{deutt}.
We did not carry this out, since the reduction
of $f^{\rm S}_1$ to fit these values to the empirical
value $a=-23.748 \pm 0.010~\hbox{fm}$ does not help
much to reproduce the CIB of the $pp$ channel.
This shortcoming might be related to the insufficient
description of the low-energy $pp$ differential
cross sections around $\theta_{\rm cm}= 90^\circ$,
as seen in  Fig.~\ref{ppdif}.
It was also pointed out by the Nijmegen
group \cite{BE88} that
the Coulomb phase shift should be improved
by the effects of two-photon exchange, vacuum polarization,
and magnetic moment interactions, in order to describe
the $\hbox{}^1S_0$ phase shift precisely at energies
less than 30 MeV. 
Neither of these effects, nor more sophisticated 
charge-symmetry breaking operators due to the interference
of QED and QCD \cite{ST91}, are taken into account
in the present calculation.
The $P$-wave effective range parameters are also
given in Table \ref{effect}, to compare with
a number of empirical predictions.
The parameters of $\hbox{}^3P_2$ state are not given, since
the effective range expansion of this partial wave requires
a correction term related to the accidental $p^5$ low-energy
behavior of OPEP \cite{NA75}.

%\clearpage

\subsection{Hyperon-nucleon interactions}

\subsubsection{The spin-flavor \mib{SU_6} symmetry
in the \mib{B_8 B_8} interactions}

\begin{table}[b]
\caption{
The relationship between the isospin basis
and the flavor-$SU_3$ basis for the $B_8 B_8$ systems. 
The flavor-$SU_3$ symmetry is given by the Elliott
notation $(\lambda \mu)$. The heading $\CP$ denotes
the flavor-exchange symmetry, $S$ the strangeness,
and $I$ the isospin.
}
\label{symmetry}
\vspace{0mm}
\begin{center}
\renewcommand{\arraystretch}{1.1}
\setlength{\tabcolsep}{5mm}
\begin{tabular}{c|c|c|c}
\hline
$S$ & $B_8\,B_8~(I)$ & ${\cal P}=+1$ (symmetric)
    & ${\cal P}=-1$ (antisymmetric) \\
\cline{3-4}
    & & $\hbox{}^1 E$ \quad or \quad $\hbox{}^3 O$
    & $\hbox{}^3 E$ \quad or \quad $\hbox{}^1 O$ \\ 
\hline
$0$ & $NN~(I=0)$ & $-$ & $(03)$  \\
  & $NN~(I=1)$ & $(22)$ & $-$ \\
\hline
  & $\Lambda N$ &  $\frac{1}{\sqrt{10}} [ (11)_s + 3 (22) ]$ &
$\frac{1}{\sqrt{2}} [ -(11)_a + (03) ]$ \\
$-1$ & $\Sigma N~(I=1/2)$ & $\frac{1}{\sqrt{10}}
[ 3 (11)_s - (22) ]$ &
$\frac{1}{\sqrt{2}} [ (11)_a + (03) ]$ \\
  & $\Sigma N~(I=3/2)$ & $(22)$ & $(30)$ \\[1mm]
\hline
  &  $\Lambda\Lambda$ & $\frac{1}{\sqrt{5}}(11)_s
+\frac{9}{2\sqrt{30}}(22)+\frac{1}{2\sqrt{2}}(00)$ & $-$ \\
%\hline
  & $\Xi N~(I=0)$ & $\frac{1}{\sqrt{5}}(11)_s - \sqrt{\frac{3}{10}}(22)
 +\frac{1}{\sqrt{2}}(00)$ & $(11)_a$ \\
  & $\Xi N~(I=1)$ & $\sqrt{\frac{3}{5}}(11)_s+\sqrt{\frac{2}{5}}(22)$
  & $\frac{1}{\sqrt{3}}[-(11)_a+(30)+(03) ]$ \\
%\hline
$-2$ & $\Sigma\Lambda$ & $-\sqrt{\frac{2}{5}}(11)_s
+\sqrt{\frac{3}{5}}(22)$ & $\frac{1}{\sqrt{2}}[(30)-(03)]$ \\
%\hline
  & $\Sigma\Sigma~(I=0)$ & $\sqrt{\frac{3}{5}}(11)_s
-\frac{1}{2\sqrt{10}}(22)-\sqrt{\frac{3}{8}}(00)$ & $-$ \\
  & $\Sigma\Sigma~(I=1)$ & $-$ & $\frac{1}{\sqrt{6}}[2(11)_a+(30)+(03)]$ \\
  & $\Sigma\Sigma~(I=2)$ & $(22)$ & $-$ \\[1mm]
\hline
  & $\Xi \Lambda$ &  $\frac{1}{\sqrt{10}} [ (11)_s + 3 (22) ]$ &
$\frac{1}{\sqrt{2}} [ -(11)_a + (30) ]$ \\
$-3$ & $\Xi \Sigma~(I=1/2)$ & $\frac{1}{\sqrt{10}}
[ 3 (11)_s - (22) ]$ &
$\frac{1}{\sqrt{2}} [ (11)_a + (30) ]$ \\
   & $\Xi \Sigma~(I=3/2)$ & $(22)$ & $(03)$ \\[1mm]
\hline
$-4$ & $\Xi \Xi~(I=0)$ & $-$ & $(30)$  \\
   & $\Xi \Xi~(I=1)$ & $(22)$ & $-$ \\
\hline
\end{tabular}
\end{center}
\end{table}

Our $(3q)$-cluster wave functions in the $(3q)$-$(3q)$ RGM formalism
are the strict $SU_6$ wave functions in the spin-flavor $SU_6$ QM.
This fact and the complete antisymmetrization of quarks
in the RGM framework make it possible to investigate interesting
relationship of the various $B_8 B_8$ interactions
under some plausible assumptions and special roles of pions.
First it should be noted that our phenomenological confinement 
potential does not have any flavor dependence. The contribution of
the $r^2$-type confinement potential to the baryon-baryon
interaction is zero due to the subtraction of the
internal energy contribution.
The FB interaction involves the FSB only through the quark-mass dependence,
but the effect is rather weak. Thus we can assume
that our QM Hamiltonian is approximately $SU_3$ scalar.
On the other hand, the EMEP contribution is also approximately
$SU_3$ scalar, except for the pion contribution.
This is because the $SU_3$ relations of the coupling constants
are automatically incorporated in the QM description. The usage
of the empirical masses for the PS and V mesons breaks
the $SU_3$ symmetry, together with the baryon masses.
The S-meson masses are varied to fit the $NN$ and $YN$ data,
but these are also assumed to be $SU_3$ symmetric
in the first approximation.
After all, our QM Hamiltonian with the EMEP's is approximately
$SU_3$ scalar as a whole, except for the important roles of pions
with the very small mass. If the Hamiltonian is approximately
$SU_3$ scalar, we can expect that the baryon-baryon interactions 
with the same $SU_3$ representation should be very similar,
since the dependence on the internal quantum numbers becomes
negligible. From this discussion, we can correlate different
$B_8 B_8$ interactions according to the $SU_3$ representation
of the $B_8 B_8$ systems.

\begin{figure}[b]
\begin{center}
\begin{minipage}{0.48\textwidth}
\epsfxsize=0.9\textwidth
\epsffile{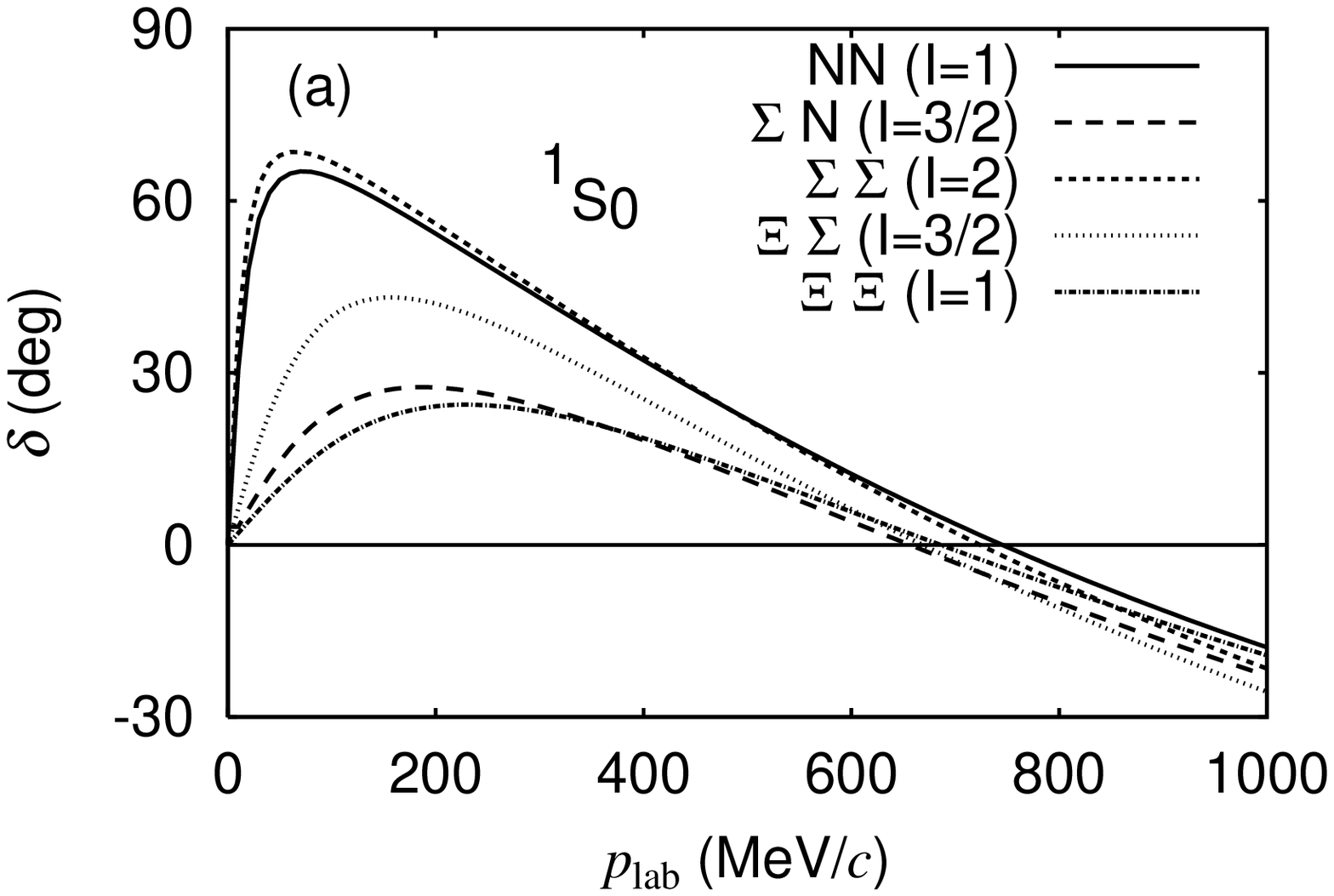}
\end{minipage}~%
%\hfill~%
\begin{minipage}{0.48\textwidth}
\epsfxsize=0.9\textwidth
\epsffile{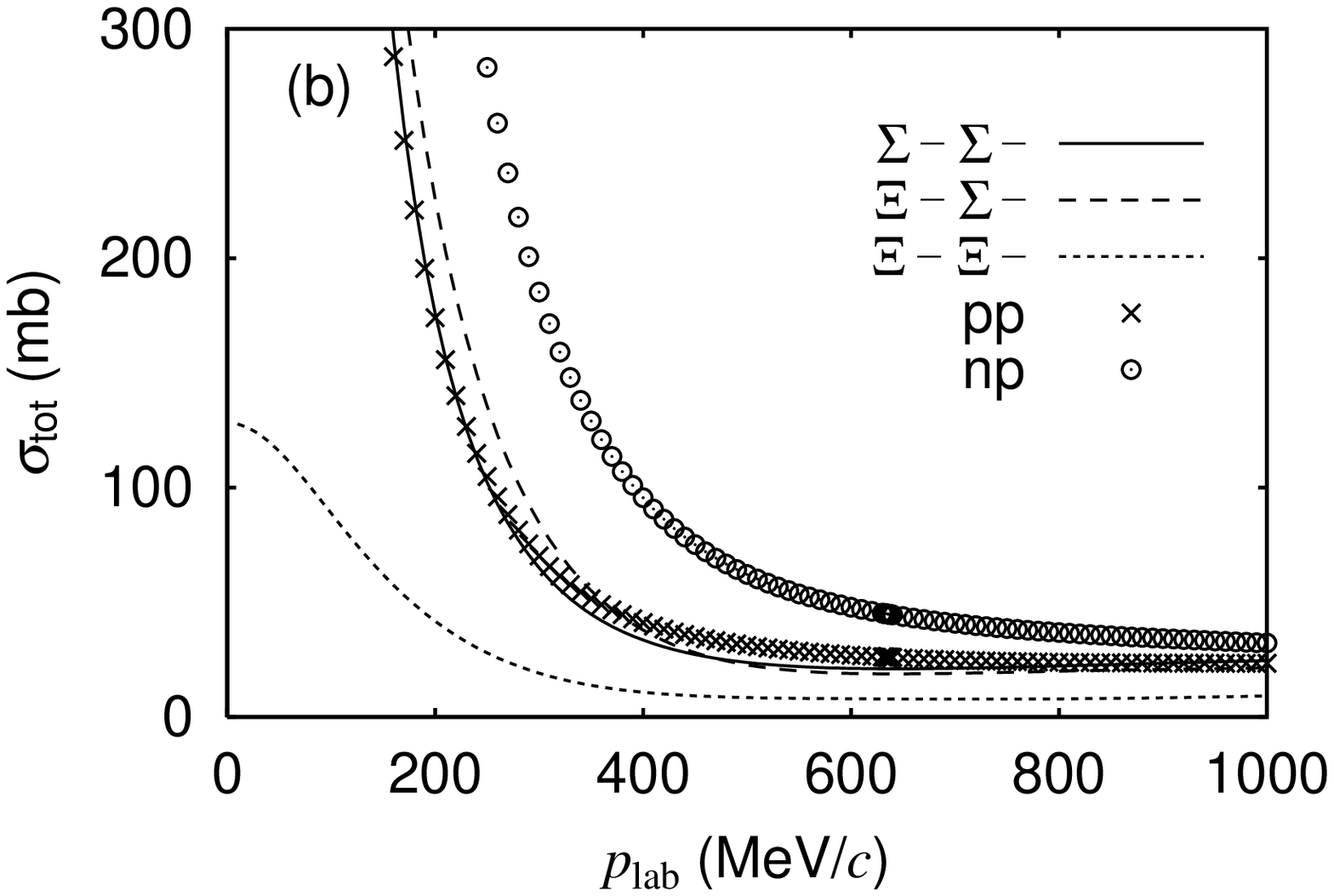}
\end{minipage}
\bigskip
\caption{
(a) $\hbox{}^1S_0$ phase shifts for the $B_8\,B_8$ interactions
with the pure (22) state, predicted by fss2.
(b) Total elastic cross sections for the
pure (22) state ($pp$, $\Sigma^- \Sigma^-$, $\Xi^- \Xi^-$) and
for the (22)+(03) states ($np$, $\Xi^- \Sigma^-$),
predicted by fss2. The Coulomb force is neglected.
}
\label{intro1}
\end{center}
\end{figure}

Table \ref{symmetry} illustrates how the two-baryon systems
in the isospin basis, defined by \eq{form8}, are classified
as a combination of the $SU_3$ bases.
Entirely different structure appears between
the flavor-symmetric and antisymmetric cases.
In the $\hbox{}^1S_0$ state, for example,
we have several systems containing the (22) component dominantly.
The $S$-state $B_8 B_8$ interactions for these systems are
expected to be similar to that for the $NN$ $\hbox{}^1S_0$ state.
The $(11)_s$ component is completely Pauli forbidden
for the most compact $(0s)^6$ configuration,
and is characterized by the strong repulsion
originating from the quark Pauli principle.
The (00) component in the $H$-particle channel
is attractive because of the color-magnetic interaction.
On the other hand, in the $NN$ sector the $\hbox{}^3S_1$ state
with $I=0$ is composed of the pure (03) state,
and the deuteron is bound in this channel
owing to the strong one-pion tensor force.
This $SU_3$ state in the flavor-antisymmetric case
is converted to the (30) state
in the side of more negative strangeness.
Since the (30) state is almost Pauli forbidden
with the eigenvalue $\mu=2/9$ of the normalization kernel,
the interaction is strongly repulsive.
Therefore, the $\Xi \Xi$ interaction is not so attractive
as $NN$, since they are combinations of (22) and (30).
The other $SU_3$ state $(11)_a$ turns out
to have very weak interaction.
After all, the strangeness $S=-2$ sector is most difficult,
since it is a turning point of the strangeness.  
It is also interesting to see that the $\Xi \Sigma$ channel
with $I=3/2$ should be fairly attractive,
since the channel contains the same (22) and (03) $SU_3$ states
as the $NN$ system.

\begin{figure}[t]
\begin{center}
\begin{minipage}{0.48\textwidth}
\epsfxsize=0.9\textwidth
\epsffile{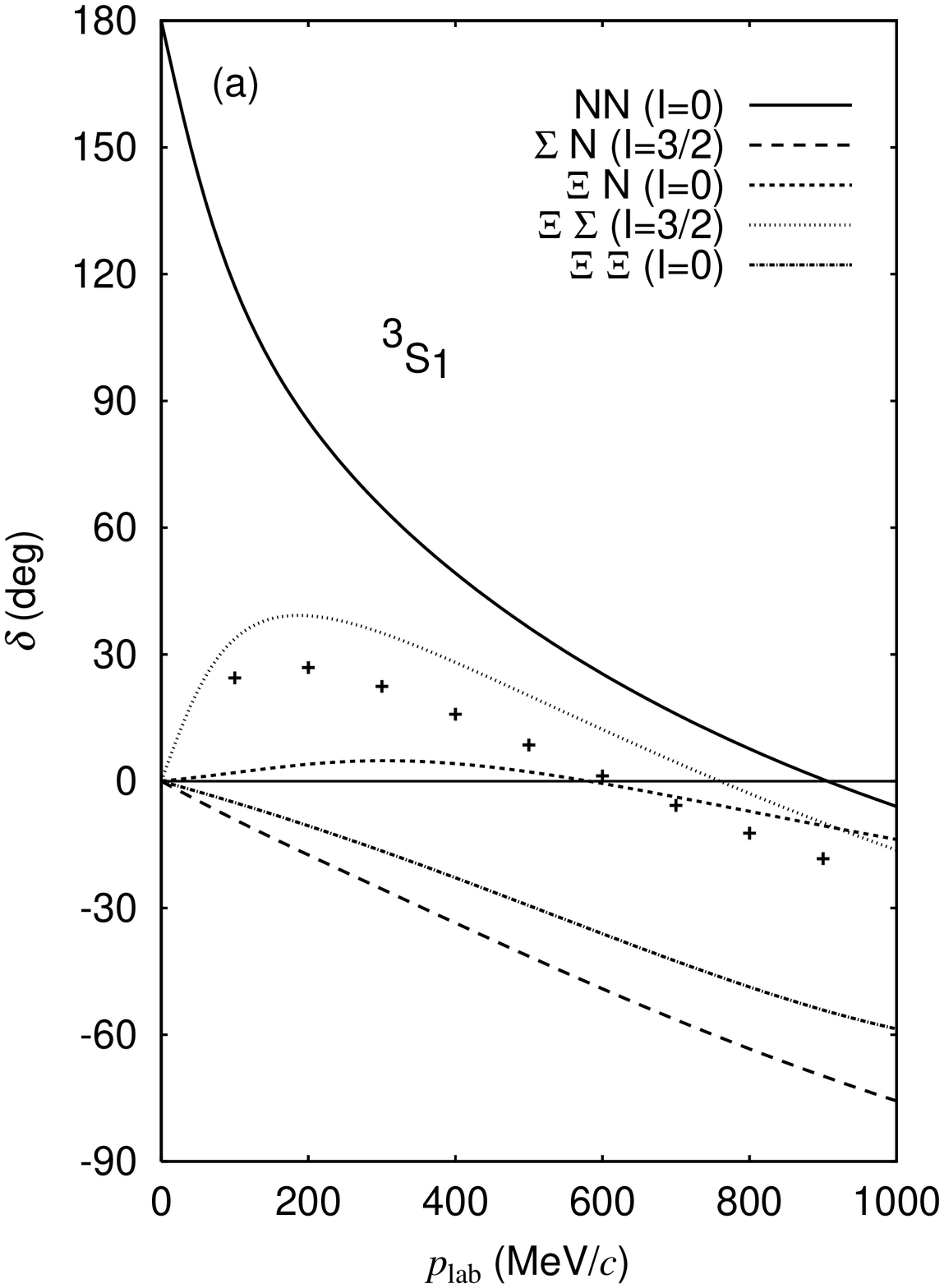}
\end{minipage}~%
%\hfill~%
\begin{minipage}{0.48\textwidth}
\epsfxsize=0.9\textwidth
\epsffile{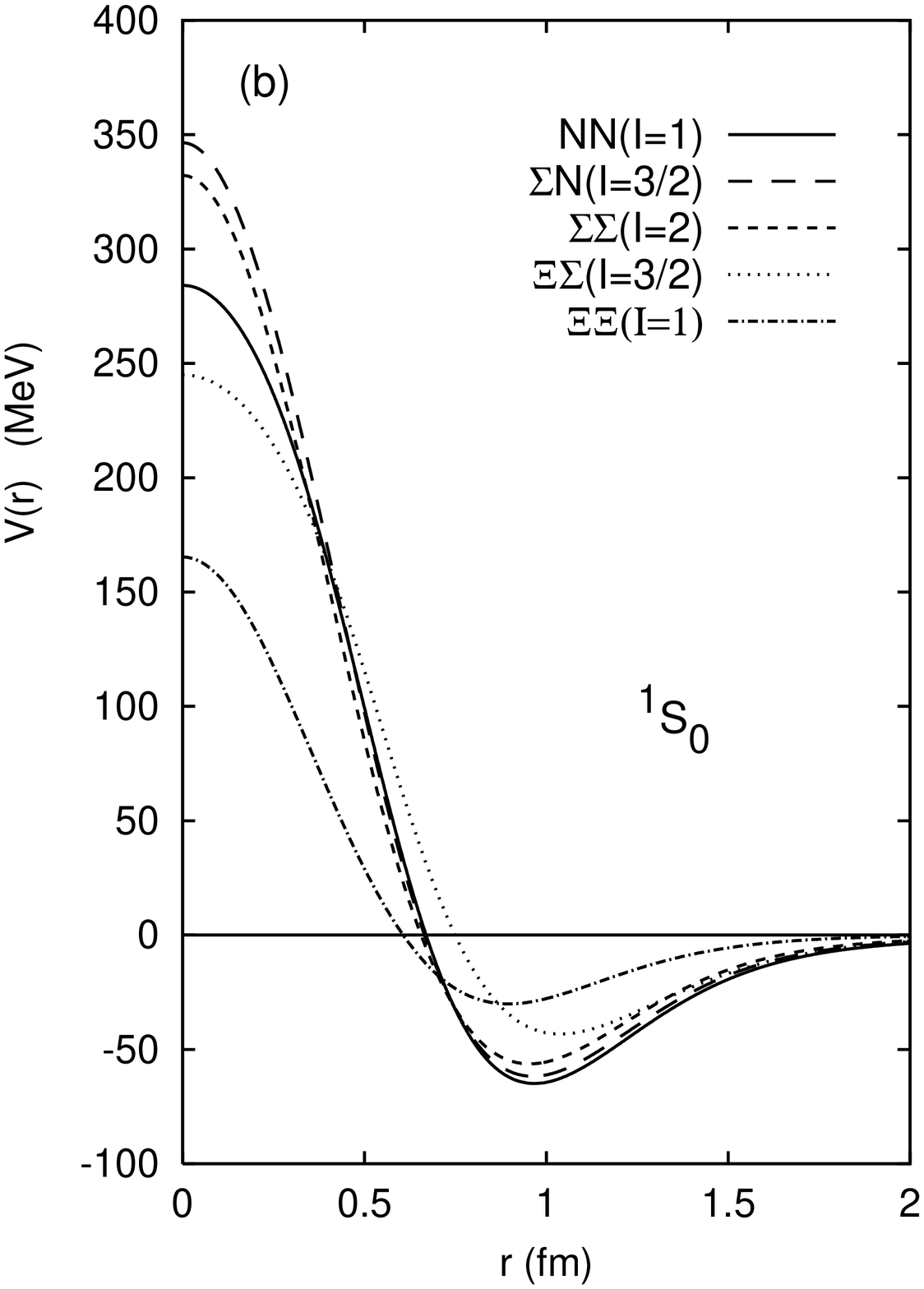}
\end{minipage}
\bigskip
\caption{
(a) $\hbox{}^3S_1$ phase shifts for some $B_8\,B_8$ interactions,
predicted by fss2. The $SU_3$ configurations are (03)
for $NN (I=0)$ and $\Xi \Sigma (I=3/2)$,
$(11)_a$ for $\Xi N (I=0)$, (30) for $\Sigma N (I=3/2)$
and $\Xi \Xi (I=0)$.
The $\hbox{}^3S_1$ $NN (I=0)$ phase shift predicted only
by the central interaction is also shown by crosses.
(b) $^1S_0$ phase-shift equivalent local potentials of $NN (I=1)$, 
$\Sigma N (I=3/2)$ with $\theta_S=33.3295^\circ$ (see
Table \protect\ref{table1}), 
$\Sigma \Sigma (I=2)$, $\Xi \Sigma (I=3/2)$, 
and $\Xi \Xi (I=1)$ systems predicted by fss2 
at $p_{\rm lab}=200~\hbox{MeV}/c$.
}
\label{nn3s}
\end{center}
\end{figure}

Figure \ref{intro1}(a) shows fss2 predictions
for the $\hbox{}^1S_0$ phase shifts
of various $B_8 B_8$ interactions
having the pure (22) configuration.
The corresponding phase-shift equivalent local potentials
at $p_{\rm lab}=200~\hbox{MeV}/c$,
derived by the WKB-RGM technique \cite{SHT83,NA97}, are
depicted in Fig.~\ref{nn3s}(b).
Although the $\Sigma \Sigma$ interaction
with isospin $I=2$ is very similar
to the $NN$ interaction, the other interactions generally
get weaker as more strangeness is involved.
This is a combined effect of both the FSB in the quark sector
and the EMEP contributions.
In particular, the $\Xi \Xi$ interaction shows the lowest
rise of the phase shift, which is less than $30^\circ$.
Accordingly, the $\Xi \Xi$ total cross sections
become much smaller than the other systems,
as seen in Fig.~\ref{intro1}(b).
This figure also shows the total cross sections of
the $np$ and $\Xi^- \Sigma^-$ scatterings, in which 
the flavor-symmetric states have the $SU_3$ (22) representation
and the antisymmetric states (03) representation.
The larger cross sections of the $np$ scattering
are due to the very strong effect of pions.
There is no pion effect in the exchange Feynman diagram
for the $\Xi^- \Sigma^-$ scattering.
More detailed analysis shows that the $NN$ interaction is the
strongest and has the largest cross sections
among any combinations of the $B_8$'s.

This kind of analysis is also applicable to the
flavor-antisymmetric configurations.
Figure \ref{nn3s}(a) shows the phase-shift behavior 
of the $\hbox{}^3S_1$ states for the $NN (I=0)$ system
with (03), $\Xi \Sigma (I=3/2)$ with (03),
$\Xi N (I=0)$ with $(11)_a$, $\Sigma N (I=3/2)$ with (30),
and $\Xi \Xi (I=0)$ with (30). First we find that
the (03) states are attractive similarly to the (22) state.
The existence of the deuteron in the $NN (I=0)$ system is
due to the very strong tensor force of the one-pion exchange.
Next, the $(11)_a$ state represented by the $\Xi N (I=0)$ system
is weakly attractive, and the phase-shift rise is
only $5^\circ$.  Finally, the $(30)$ states represented
by the $\Sigma N (I=3/2)$ and $\Xi \Xi (I=0)$ systems are
both repulsive, but the repulsion of the $\Xi \Xi (I=0)$ state is
slightly weaker than that of the $\Sigma N (I=3/2)$ state
due to the FSB.

It should be kept in mind that the systematics based on the
relationship in Table \ref{symmetry} serves as a guide
and the deviation from the pure $SU_3$ symmetric rules
is very much model dependent. For example, the model FSS
gives $\hbox{}^1S_0$ bound states for the $\Sigma \Sigma (I=2)$,
$\Xi \Sigma (I=3/2)$, and $\Xi \Xi (I=1)$ systems
with the pure (22) representation. This is due to the
imbalanced reduction in the short-range
repulsion of the color-magnetic interaction
and the intermediate-range attraction
of the flavor-singlet scalar-meson exchange EMEP
with increasing strangeness.
Since no such bound states are observed experimentally,
the above discussions based on the model fss2 seem to be more reliable 
and give an important guide line to construct realistic
models for the $B_8 B_8$ interactions.

\subsubsection{\mib{\Sigma^+ p} system}

\begin{figure}[t]
\begin{center}
\begin{minipage}{0.48\textwidth}
\epsfxsize=0.9\textwidth
\epsffile{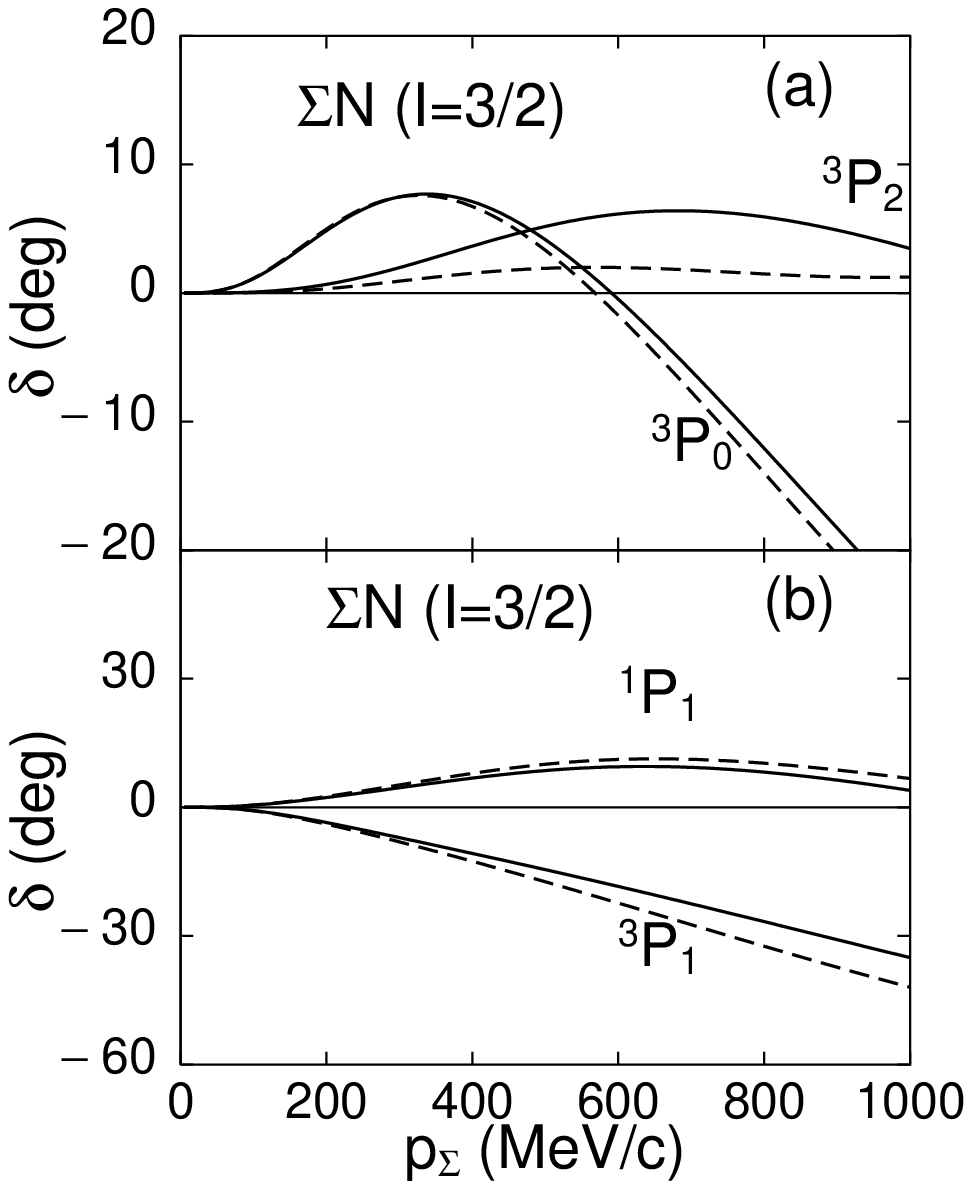}
\vspace{-12mm}
\caption{The $P$-wave phase shifts 
of the $\Sigma N(I=3/2)$ system, predicted by fss2 (solid curves)
and FSS (dashed curves).
}
\label{phsp}
\end{minipage}~%
\qquad
%\hfill
\begin{minipage}{0.48\textwidth}
\epsfxsize=0.9\textwidth
\epsffile{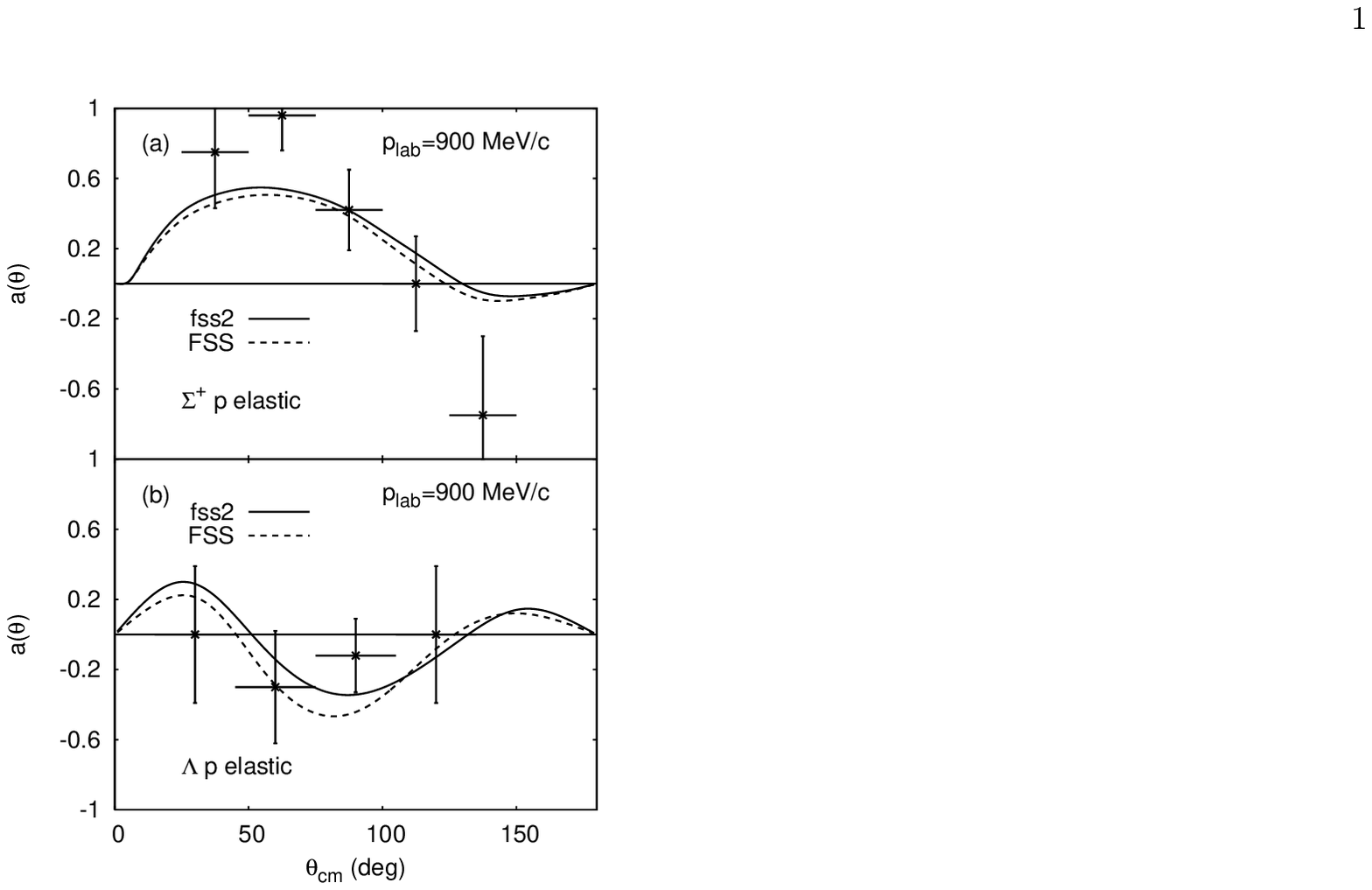}
\caption{
$\Sigma^+p$ (a) and $\Lambda p$ (b) asymmetries
at $p_\Sigma=900~\hbox{MeV}/c$, predicted by fss2 and FSS.
The experimental data are taken from \protect\cite{KA02,KU06}.}
\label{sppol}
\end{minipage}
\end{center}
\end{figure}

\begin{table}[t]
\caption{$\hbox{}^1S_0$ ($a_s$, $r_s$) and $\hbox{}^3S_1$ ($a_t$, $r_t$)
effective range parameters of fss2 for $B_8 B_8$ interactions.
Unit of length is in fm. The experimental values are taken
from \protect\cite{NA73} for $\Sigma^+ p$,
and from \protect\cite{alex68} and \protect\cite{sechi68} for $\Lambda p$.
}
\label{effect2}
\begin{center}
\renewcommand{\arraystretch}{1.1}
\setlength{\tabcolsep}{2mm}
\begin{tabular}{cccccc}
%\hline
\hline
 & & isospin basis & \multicolumn{2}{c}{particle basis} & Expt. \\
\cline{4-5}
 & & & Coulomb off & Coulomb on & \\
\hline
 &  $a_s$  & $-2.51$ & $-2.48$ & $-2.28$ & $-2.42 \pm 0.30$ \\
$\Sigma^+ p$ & $r_s$ & 4.92 & 5.03 & 4.68 & $3.41 \pm 0.30$ \\
 &  $a_t$  & 0.729 & 0.727 & 0.826 & $0.709 \pm 0.001$ \\
 &  $r_t$  & $-1.22$ & $-1.31$ & $-1.52$ & $-0.783 \pm 0.003$ \\
\hline
% &  $a_s$  & & $-2.50$ & $-2.50$ & \\
%$\Sigma^- n$ & $r_s$ & & 5.02 & 5.02 & \\
% &  $a_t$  & & 0.727 & 0.727 & \\
% &  $r_t$  & & $-1.32$ & $-1.32$ & \\
%\hline
%\hline
 &  $a_s$  & $-2.59$ & $-2.59$ & $-2.59$ & $-1.8$, $-2.0$ \\
$\Lambda p$ & $r_s$  &    2.83 &    2.83 & 2.83 & $2.8$, $5.0$ \\
 &  $a_t$  & $-1.60$ & $-1.60$ & $-1.60$ & $-1.6$, $-2.2$ \\
 &  $r_t$  & 3.01    &    3.00 &    3.00 & $3.3$, $3.5$ \\
\hline
% &  $a_s$  & & $-2.56$ & $-2.57$ & \\
%$\Lambda n$ & $r_s$  & &    2.84 & 2.84 & \\
% &  $a_t$  & & $-1.57$ & $-1.57$ & \\
% &  $r_t$  & &    3.04 &    3.04 & \\
%\hline
%\hline
$\Sigma^+ \Sigma^+$ &  $a_s$  & $-85.3$ & $-63.7$ & $-9.72$ & \\
 & $r_s$  & 2.34 & 2.37 & 2.26 & \\
\hline
%\hline
$\Lambda \Lambda$ &  $a_s$  & $-0.821$ & $-0.808$ & $-0.814$ & \\
 &  $r_s$  & 3.78 & 3.83 & 3.80 & \\
\hline
%\hline
 &  $a_s$  & 0.324 & 0.327 & 0.326 & \\
$\Xi^0 p$ & $r_s$  & $-8.93$ & $-9.19$ & $-9.23$ & \\
 &  $a_t$  & $-0.207$ & $-0.203$ & $-0.204$ & \\
 &  $r_t$  &  26.2    &    27.5  &    27.4  & \\
\hline
%\hline
 &  $a_s$  & $-1.08$ & $-1.07$ & $-1.08$ & \\
$\Xi^0 \Lambda$ & $r_s$ & 3.55 & 3.58 & 3.57 & \\
 &  $a_t$  & 0.262 & 0.263 & 0.263 & \\
 &  $r_t$  & 2.15  & 2.14  & 2.14  & \\
\hline
%\hline
 &  $a_s$  & $-4.63$ & $-4.70$ & $-3.46$ & \\
$\Xi^- \Sigma^-$ & $r_s$ & 2.39 & 2.37 & 2.04 & \\
 &  $a_t$  & $-3.48$ & $-3.52$ & $-2.81$ & \\
 &  $r_t$  & 2.52  & 2.52  & 2.24  & \\
\hline
%\hline
 &  $a_s$  & $-1.43$ & $-1.43$ & $-1.43$ & \\
$\Xi^- \Xi^0$ & $r_s$  &  3.20  &  3.17   & 3.17 & \\
 &  $a_t$  & 3.20  &   3.20 &   3.20 & \\
 &  $r_t$  & 0.218 &  0.218 &  0.217 & \\
\hline
$\Xi^- \Xi^-$ &  $a_s$  & & $-1.43$ & $-1.35$ & \\
 & $r_s$      &  & 3.21 & 2.70 & \\
\hline
%\hline
\end{tabular}
\end{center}
\end{table}

Sine the $S$-wave phase shifts of the $\Sigma N(I=3/2)$ system
are already given in Figs.~\ref{intro1}(a) and \ref{nn3s}(a),
we only show the $P$-wave phase shifts in Fig.~\ref{phsp}.
The phase shifts given by FSS are also shown for comparison.
The $\hbox{}^1E$ and $\hbox{}^3O$ states
of the $\Sigma N(I=3/2)$ system belong to 
the (22) representation in the flavor $SU_3$ symmetry,
in exactly the same as the isospin $I=1$ $NN$ system.
The phase-shift behavior of these partial waves
therefore resembles that of the $NN$ system,
as long as the effect of the FSB is not significant.
Figure \ref{intro1}(a) shows that the attraction
in the $\hbox{}^1S_0$ state is much weaker than that
of the $NN$ system, and the phase-shift peak is
about $26^\circ$ around $p_{\Sigma}=200~\hbox{MeV}/c$.
The $^3P_J$ phase shifts show the characteristic
energy dependence observed in the $NN$ phase shifts,
which is caused by different roles of the central,
tensor and $LS$ forces.
The appreciable difference in the two models fss2 and FSS
appears only in the $^3P_2$ state.
It is discussed in Ref.~\cite{FJ96} that the attractive
behavior of the $^3P_2$ phase shift is closely related to
the magnitude of the $\Sigma^+ p$ polarization $P(\theta)$
at the intermediate energies.
The more attractive the phase-shift behavior,
the larger $P(\theta=90^\circ)$.
In Fig.~\ref{sppol}(a), the polarization observable for
the $\Sigma^+ p$ elastic scattering at $p_\Sigma=900~\hbox{MeV}/c$
is shown for the models fss2 and FSS.
The experimental data \cite{KA02} imply
the asymmetry parameter $a^{\rm exp}=0.44 \pm 0.2$ at $p_\Sigma=800
\pm 200~\hbox{MeV}/c$, which is not inconsistent with
our QM predictions provided that the scattering angle
is not specified.  More recent data \cite{KU06}
at $p_\Sigma=700$ - $1,100~\hbox{MeV}/c$ are plotted
in Fig.~\ref{sppol}(a).

On the other hand, the $^3E$ and $^1O$ states
of the $\Sigma N(I=3/2)$ system have the (30) symmetry.
We have no information on the properties of these states
from the $NN$ interaction.
In our QM framework, however,
there is very little ambiguity in the phase
shifts of these states.
Since the $(0s)^6$ configuration in the $^3S_1$ state
is almost Pauli forbidden,
the interaction in the $^3S_1$ state is strongly repulsive.
On the other hand, the $^1P_1$ phase shift
is weakly attractive, which is caused
by the exchange kinetic-energy kernel owing to the Pauli principle.

The effective range parameters of the $YN$ scattering
in the single-channel analysis are given
in Table \ref{effect2} with some empirical values.
For the $\Sigma^+ p$ system, the empirical values given
in Ref.~\cite{NA73} should be compared with the results
in the particle-basis calculation including the
Coulomb force. We find a reasonable agreement
both in the $\hbox{}^1S_0$ and $\hbox{}^3S_1$ states.

\subsubsection{\mib{\Lambda N} system}

\begin{figure}[b]
\begin{center}
\epsfxsize=0.9\textwidth
\epsffile{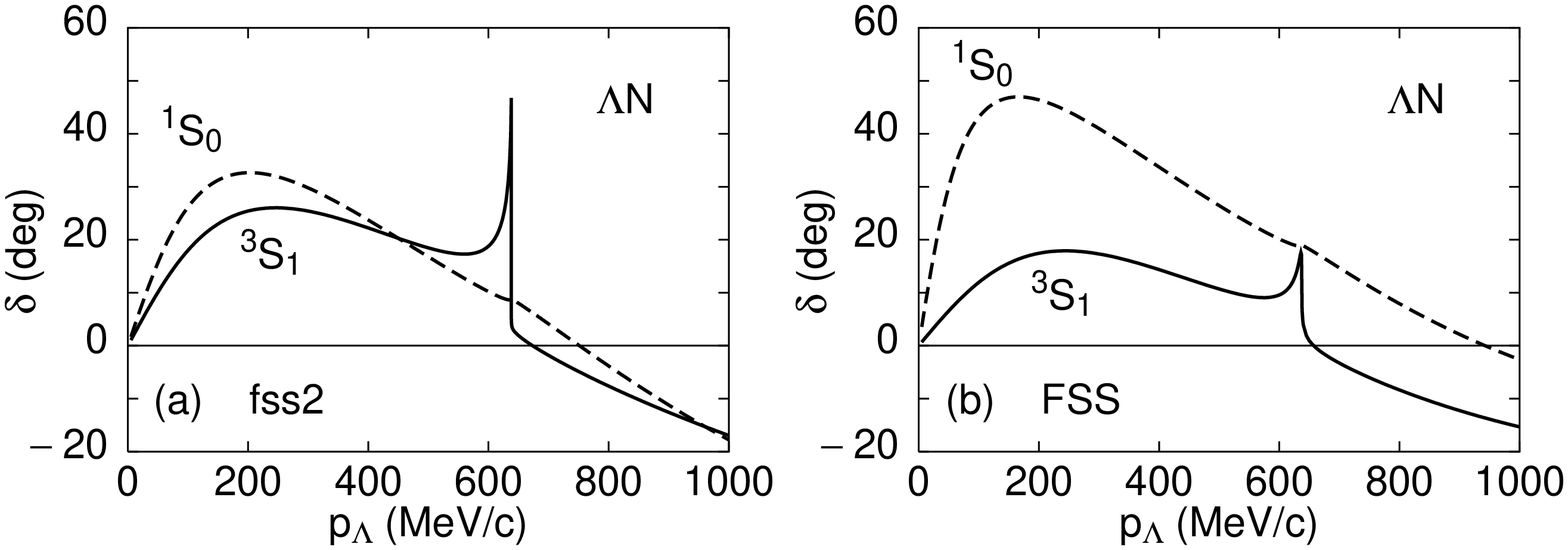}
\bigskip
\caption{The $\Lambda N$ $^3S_1$ and $^1S_0$ phase shifts
calculated by fss2 (a) and by FSS (b).}
\label{phlam1}
\end{center}
\end{figure}

%\clearpage

The total cross section for the $\Lambda p$ elastic scattering
predicted by fss2 in the isospin basis is displayed
in Fig.~\ref{tot}(e) later, together with the previous result
given by FSS. The both models reproduce experimental data
equally well in the low momentum
region $p_{\Lambda} \le 300$ MeV/$c$. 
The total cross section has a cusp structure
at the $\Sigma N$ threshold due to
the strong $\Lambda N$--$\Sigma N (I=1/2)$ $^3S_1$ - $^3D_1$ coupling
caused by the tensor force of the OPEP.
To see the dominant role of the tensor force
of the $\Lambda N$--$\Sigma N$ coupling in more detail,
we show the $^3S_1$ and $^1S_0$ phase shifts
in Fig.~\ref{phlam1} calculated in the isospin basis.
A cusp structure at the $\Sigma N$ threshold is apparent
in the $^3S_1$ channel, while very small in the $^1S_0$ channel.
In both models, the attraction in the $^1S_0$ state is stronger than
that in the $^3S_1$ state.
However, the relative strengths of attraction
in the $^1S_0$ and $^3S_1$ states are different. The model FSS
gives $\delta (\hbox{}^1S_0) - \delta (\hbox{}^3S_1)
\approx 30^\circ$ at $p_{\Sigma} \approx 200$ MeV/$c$,
while fss2 $\delta (\hbox{}^1S_0) - \delta (\hbox{}^3S_1) \approx 10^\circ$.
The difference
of $\delta (\hbox{}^1S_0) - \delta (\hbox{}^3S_1) \lesssim 10^\circ$ is
required from many few-body
calculations \cite{MI99,SH83,YA94,HI00c,NE00}
of the $s$-shell $\Lambda$-hypernuclei.
In Sect. 3.4.2, we will show Faddeev calculations
for the hypertriton \cite{hypt}, using fss2, 
which predict that the desirable difference is only 0 - $2^\circ$.

\begin{figure}[t]
\begin{center}
\epsfxsize=0.9\textwidth
\epsffile{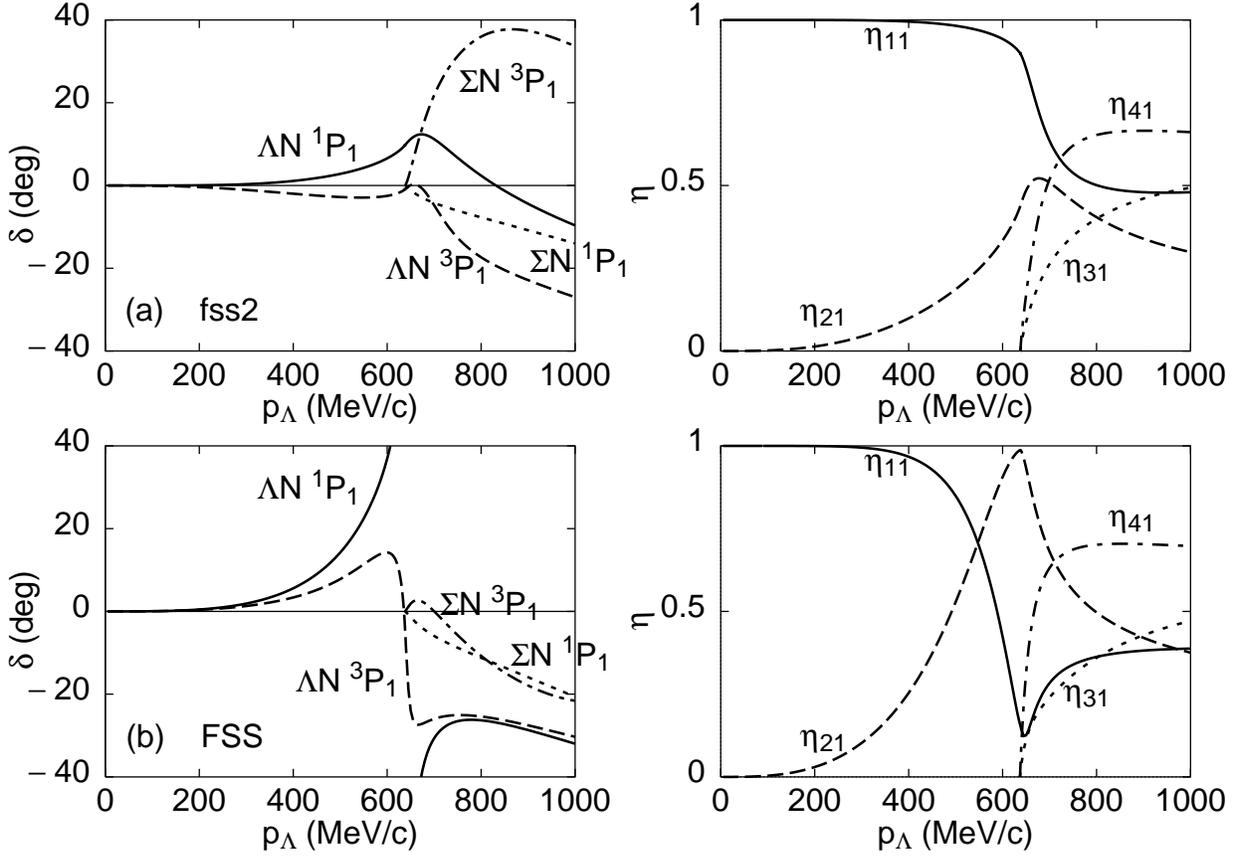}
\bigskip
\caption{The $S$-matrix of the $\Lambda N$--$\Sigma N(I=1/2)$
$\hbox{}^1P_1$ - $\hbox{}^3P_1$ CC system,
predicted by fss2 (upper) and FSS (lower) in the isospin-basis.
The diagonal phase shifts $\delta_{ii}$ (left-hand side),
and the reflection and transmission
coefficients $\eta_{i1}$ (right-hand side) defined
by $S_{ij}=\eta_{ij}e^{2i\delta_{ij}}$ are displayed.
The channels are specified by 1: $\Lambda N$ $\hbox{}^1P_1$,
2: $\Lambda N$ $\hbox{}^3P_1$, 3: $\Sigma N$ $\hbox{}^1P_1$
and 4: $\Sigma N$ $\hbox{}^3P_1$.
}
\label{phlam2}
\end{center}
\end{figure}

As seen in Fig.~\ref{tot}(e) later,
FSS predicts especially large enhancement
of the total cross sections around 
the cusp at the $\Sigma N$ threshold. This is due to a rapid
increase of the $\Lambda N$ $\hbox{}^3P_1$ - $\Lambda N$
$\hbox{}^1P_1$ transition around the threshold.
Figure \ref{phlam2} displays the $S$-matrix
$S_{ij}=\eta_{ij}e^{2i\delta_{ij}}$ 
for the $\Lambda N$--$\Sigma N(I=1/2)$
$^1P_1$ - $^3P_1$ channel coupling for
fss2 (upper) and FSS (lower).
The FSS shows that the transmission
coefficient $\eta_{21}$, which corresponds
to the $\Lambda N$ $\hbox{}^1P_1
\rightarrow \Lambda N$ $\hbox{}^3P_1$ transition,
increases very rapidly around the $\Sigma N$ threshold
as the energy increases. This increase of
the $\eta_{21}$ (and the resultant decrease of the reflection
coefficient $\eta_{11}$) is a common feature of our models.
The strength of the transition, however, has some model dependence. 
The transition is stronger in FSS than in fss2.
The behavior of the diagonal phase shifts is largely
affected by the strength
of the $\hbox{}^1P_1$ - $\hbox{}^3P_1$ channel
coupling (which is directly reflected in $\eta_{21}$).
The model fss2 yields a broad resonance
in the $\Sigma N$ $\hbox{}^3P_1$ channel, while
the FSS with stronger channel
coupling yields no resonance in this channel.
Instead, a step-like resonance appears
in the $\Lambda N$ $\hbox{}^1P_1$ channel.
The location of the resonance is
determined by the strength of the $LS^{(-)}$ force
and the strength of the attractive central force
in the $\Sigma N(I=1/2)$ channel.
In the QM, the central attraction 
by the S mesons is enhanced by the exchange kinetic-energy kernel,
which is attractive in the $\Sigma N$ $\hbox{}^3P_1$ channel
due to the effect of the Pauli principle.
The $LS$ and $LS^{(-)}$ forces also
contribute to increase this attraction.
In FSS, a single channel calculation for
the $\Sigma N (I=1/2)$ system yields a resonance
in the  $\hbox{}^3P_1$ state. When the channel coupling
to the $\Lambda N$ system is introduced, this resonance
moves to the $\Lambda N$ channel due to the
strong $LS^{(-)}$ force generated from the FB interaction. 
On the other hand, the role of the FB $LS^{(-)}$ force becomes
less significant in fss2, since the S-meson exchange
includes only the $LS$ contribution and no $LS^{(-)}$ contribution.
This is the reason the resonance remains in the $\Sigma N$
$\hbox{}^3P_1$ channel in fss2.

In spite of these quantitative differences in the coupling
strength, the essential mechanism
of the $\Lambda N$--$\Sigma N(I=1/2)$
$^1P_1$ - $^3P_1$ channel coupling is the same
for all of our models. It is induced by the strong
FB $LS^{(-)}$ force, which directly connects
the two $SU_3$ configurations
with the $(11)_s$ and $(11)_a$ representations
in the $P$-wave $I=1/2$ channel.
In order to determine the detailed phase-shift behavior
of each channel,
including the position of the $P$-wave resonance,
one has to know the strength of the $LS^{(-)}$ force
and the strength of the attractive central force
in the $\Sigma N(I=1/2)$ channel.
This is possible only by the careful analysis of rich
experimental data for the $\Lambda p$ and $\Sigma^- p$ 
scattering observables in the $\Sigma N$ threshold region.
The asymmetry parameter data for the $\Lambda p$ elastic
scattering at $p_\Lambda=700$ - $1,100~\hbox{MeV}/c$ are plotted
in Fig.~\ref{sppol}(b) \cite{KU06}.

Let us discuss another important feature
of the $\Lambda N$ phase shifts
induced by the $P$-wave coupling due to the $LS^{(-)}$ force.
The $\Lambda N$ $\hbox{}^1P_1$ and $\hbox{}^3P_1$ phase shifts
in Fig.~\ref{phlam2} show weakly attractive behavior
below the $\Sigma N$ threshold.
This is due to the dispersion-like (or step-like
for the $\Lambda N$ $\hbox{}^1P_1$ state in FSS) resonance
behavior with a large width. 
This attraction in the $P$ state can be observed by the forward to
backward ratio ($F/B$) of the $\Lambda p$ differential
cross sections. Both fss2 and FSS
give $F/B>1$ below the $\Sigma N$ threshold \cite{SCAT},
which implies that the $P$-state $\Lambda N$ interaction
is weakly attractive as suggested by Dalitz {\em et al.} \cite{DA72}.
A shell-model analysis of $\hbox{}^{13}_{\Lambda} \hbox{C}$ seems to support
the attractive $P$-wave $\Lambda N$ interaction \cite{MI01}.
In our QM interaction, this attraction originates from
the strong $\Lambda N$--$\Sigma N (I=1/2)$ 
$\hbox{}^1P_1$ - $\hbox{}^3P_1$ coupling
due to the FB $LS^{(-)}$ force \cite{SCAT}.

Some comments are in order with respect to particle-basis
calculations of the $\Lambda p$ and $\Lambda n$ scatterings.
From the energy spectrum
of the $\hbox{}^3_\Lambda\hbox{H}$ - $\hbox{}^3_\Lambda\hbox{He}$ 
isodoublet $\Lambda$-hypernuclei, it is inferred that
the $\Lambda p$ interaction is more attractive than
the $\Lambda n$ interaction. This CSB may have its origin
in the different threshold energies of the $\Sigma N$ particle
channels and the Coulomb attraction
in the $\Sigma^- p$ channel for the $\Lambda n$ system.
The former effect increases the cross sections of
the $\Lambda p$ interaction
and the latter the $\Lambda n$ interaction.
However, the full calculation including the pion-Coulomb
correction using the correct threshold energies yields
very small difference in the $S$-wave phase shifts.
In the energy region up to $p_{\Lambda}=200~\hbox{MeV}/c$,
the $\Lambda p$ phase shift is more attractive
than the $\Lambda n$ phase shift only by less
than $0.2^\circ$ in the $\hbox{}^1S_0$ state, and
by less than $0.4^\circ$ in the $\hbox{}^3S_1$ state.
This can also be seen from the results of 
the $\Lambda p$ and $\Lambda n$ effective range
parameters in the particle basis.
On the other hand, a rather large effect of CSB is
found in the $\Sigma^- p$ channel as discussed in the
next subsection. This is because the CSB effect is enhanced
by the strong $\Lambda N$--$\Sigma N (I=1/2)$ channel-coupling
effect in the $\hbox{}^3S_1$ - $\hbox{}^3D_1$ state.

\subsubsection{\mib{\Sigma^- p} system}

Since the $\Sigma^- p$ system is expressed
as $|\Sigma^- p\rangle=-\sqrt{2/3}|\Sigma N (I=1/2)\rangle
+\sqrt{1/3}|\Sigma N (I=3/2)\rangle$ in the isospin basis,
it is important to know first the phase-shift behavior
of the $\Sigma N (I=1/2)$ system.
The $SU_3$ decomposition of the $\Sigma N (I=1/2)$ state,
that is, $\Sigma N (I=1/2)=\left[ 3(11)_s-(22) \right]/\sqrt{10}$
for $\hbox{}^1E$ and $\hbox{}^3O$
and $\Sigma N (I=1/2)=\left[(11)_a+(03)\right]/\sqrt{2}$
for $\hbox{}^3E$ and $\hbox{}^1O$ (see Table \ref{symmetry}),
is very useful to know the QM prediction for
the phase-shift behavior in the isospin basis.
Since the $(11)_s$ $SU_3$ state
is completely Pauli forbidden,
the $\Sigma N (I=1/2)$ $\hbox{}^1S_0$ phase shift
is strongly repulsive due to the exchange kinetic-energy
kernel \cite{FSS}.
On the other hand, the $\Sigma N (I=1/2)$ $\hbox{}^3S_1$ phase shift
is expected to be attractive similarly
to the $\Lambda N$ $\hbox{}^3S_1$ phase shift, in so far as
the effect of the FSB is not important.
(Note that $\Lambda N=\left[-(11)_a+(03)\right]
/\sqrt{2}$ for $\hbox{}^3E$ and $\hbox{}^1O$ states.)
Unfortunately, the last condition is applicable
only to the central force, since the one-pion tensor force
introduces considerable complexities
in the $\Sigma N (I=1/2)$--$\Lambda N$
$\hbox{}^3S_1$ - $\hbox{}^3D_1$ CC problem.
The strength of the $\Sigma N (I=1/2)$ central attraction,
discussed in the preceding subsection, should therefore
be examined carefully after
this $S$-wave and $D$-wave channel coupling is properly treated. 

\begin{figure}[b]
\begin{center}
\epsfxsize=0.9\textwidth
\epsffile{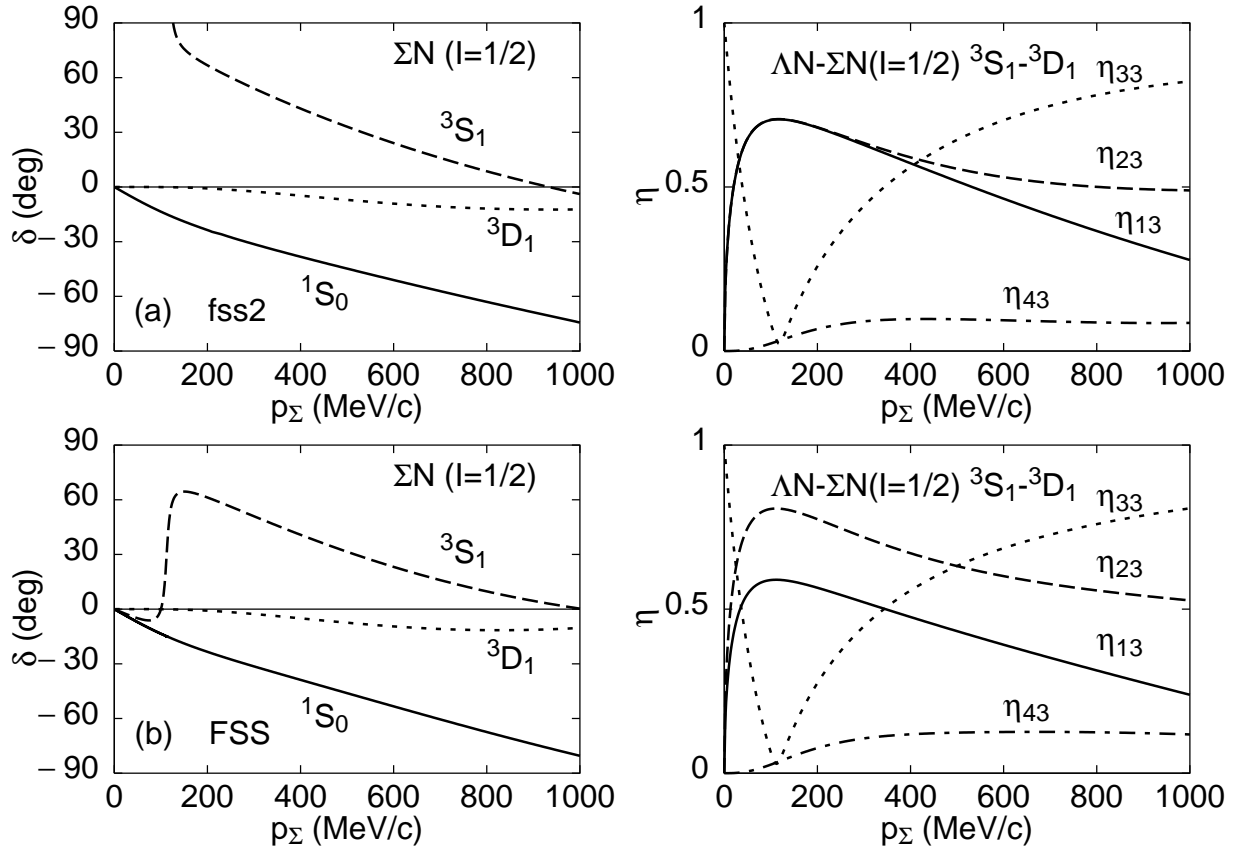}
\bigskip
\caption{The same as Fig.~\protect{\ref{phlam2}} but
for the $^1S_0$ phase shift and for the $S$-matrix
of the $\Lambda N$--$\Sigma N (I=1/2)$
$\hbox{}^3S_1$ - $\hbox{}^3D_1$ CC state.
In the latter system, the channels are specified
by 1: $\Lambda N$ $\hbox{}^3S_1$, 2: $\Lambda N$ $\hbox{}^3D_1$,
3: $\Sigma N$ $\hbox{}^3S_1$ and 4: $\Sigma N$ $\hbox{}^3D_1$.
The upper figures display the results by fss2, while the lower by FSS.
}
\label{phsgn}
\end{center}
\end{figure}

Figure \ref{phsgn} displays the $\hbox{}^1S_0$ phase shift
and the $S$-matrix of the $\Lambda N$--$\Sigma N (I=1/2)$
$\hbox{}^3S_1$ - $\hbox{}^3D_1$ CC state.
The upper figures are the predictions by fss2, while the lower
ones by FSS.
The $\hbox{}^1S_0$ phase shift shows very strong Pauli repulsion,
similarly to the the $\Sigma N (I=3/2)$ $\hbox{}^3S_1$ phase shift
(see Fig.~\ref{nn3s}(a)).
The $\Sigma N (I=1/2)$ $\hbox{}^3S_1$ phase shift
predicted by fss2 starts from $180^\circ$ at $p_{\Sigma}=0$,
decreases moderately down to about $160^\circ$ in
the region $p_{\Sigma}\le 100$ MeV/$c$. 
Then it rapidly decreases to about $80^\circ$ at
around $p_{\Sigma}=120$ MeV/$c$.
Beyond this momentum, a moderate decrease follows again.
The situation in FSS is rather different.
The $\hbox{}^3S_1$ phase shift predicted by FSS starts
from $0^\circ$ at $p_{\Sigma}=0$, and shows a clear resonance behavior
in the region $100 < p_{\Sigma}< 120$ MeV/$c$.
The peak value of the phase shift is about $60^\circ$.
In spite of the very different behavior of the diagonal
phase shifts predicted by fss2 and FSS, the $S$-matrices
are found to be very similar to each other.
This can be seen from the Argand diagram showing the
energy dependence of the $\Sigma N (I=1/2)$ $\hbox{}^3S_1$
$S$-matrix element $S_{33}=\eta_{33}e^{2i\delta_{33}}$.
This implies that the strength of the central attraction
in the $\Sigma N(I=1/2)$ channel is almost the same
as in fss2 and FSS.

The very strong reduction of $\eta_{33}=|S_{33}|$ in
Fig.~\ref{phsgn} is related to the large enhancement
of the transmission coefficients, $\eta_{13}$ and $\eta_{23}$,
from the $\Sigma N (I=1/2)$ $\hbox{}^3S_1$ ($i=3$) to
the $\Lambda N$ $\hbox{}^3S_1$ ($i=1$) and  
$\Lambda N$ $\hbox{}^3D_1$ ($i=2$) channels
around $p_\Sigma= 120~\hbox{MeV}/c$.
These transmission coefficients are directly connected
to the $\Sigma^- p\rightarrow \Lambda n$ reaction cross
sections (which we call $\sigma (\hbox{C})$),
and the driving force for this transition is the one-pion
tensor force.
If we calculate the partial-wave contributions
to this reaction cross section at $p_{\Sigma}=$ 160 MeV/$c$,
the sum of the contributions from the transitions $\hbox{}^3S_1
\rightarrow \hbox{}^3S_1$ and $\hbox{}^3S_1\rightarrow
\hbox{}^3D_1$ amounts to about 120 mb both in fss2 and FSS,
which is much larger than the contribution (3.5 mb)
from the transition $\hbox{}^1S_0 \rightarrow \hbox{}^1S_0$.
This is in accordance with the analysis of the $\Lambda p$ system
in the preceding subsection.
Namely, there is a large cusp structure
in the $\hbox{}^3S_1$ phase shift at the $\Sigma N$ threshold,
while a very small cusp in the $\hbox{}^1S_0$ channel.
There is, however, some quantitative difference
in the detailed feature of this tensor coupling between fss2 and FSS.
In fss2, $\eta_{13}$ and $\eta_{23}$ are almost equal
to each other in the momentum region where the experimental data
exist ($110 \le p_{\Sigma}\le 160$ MeV/$c$),
whereas $\eta_{13}< \eta_{23}$ holds in FSS.
The details of the cross section of the process C indicate that 
$\sigma(\hbox{}^3S_1\rightarrow \hbox{}^3S_1)$ $\approx$
$\sigma(\hbox{}^3S_1\rightarrow \hbox{}^3D_1)$ in fss2,
while $\sigma(\hbox{}^3S_1\rightarrow \hbox{}^3S_1)$
$<$ $\sigma(\hbox{}^3S_1\rightarrow \hbox{}^3D_1)$ in FSS.

It is important to take into account the pion-Coulomb correction
for more detailed evaluation of $\Sigma^- p$ cross sections.
In Ref.~\cite{FU98}, we incorporated the Coulomb force
of the $\Sigma^- p$ channel correctly in the particle basis,
but the threshold energies
of the $\Sigma^0 n$ and $\Lambda n$ channels
were not treated properly. As was discussed in Sec.~2.1,
we can deal with the empirical threshold energies and
the reduced masses in the RGM formalism,
keeping the correct effect of the Pauli principle.
Since the threshold energies and reduced masses are calculated
from the baryon masses, these constitute a part
of the pion-Coulomb correction in the $YN$ interaction.
Although the pion-Coulomb correction may
not be the whole story of the CSB, it is
certainly a first step to improve the accuracy of
the model predictions calculated in the isospin basis.
We can expect that the small difference of the threshold
energies becomes important more and more
for the low-energy $\Sigma^- p$ scattering. In particular,
the charge-exchange total cross section $\Sigma^- p \rightarrow
\Sigma^0 n$ (which we call $\sigma (\hbox{B})$) does not satisfy
the correct $1/v^2$ law in the zero-energy limit,
if the threshold energies of
the $\Sigma^- p$ and $\Sigma^0 n$ channels are assumed to be equal.
We therefore used the prescription \cite{SW62} to multiply
the factor $(k_f/k_i)$, in order
to get $\sigma (\hbox{B})$ from the cross section
calculated by ignoring the difference of the
threshold energies.
Here $k_i$  and $k_f$ are the relative momentum
in the initial and final states, respectively.
We will see below that this prescription is not accurate
and overestimates both $\sigma (\hbox{B})$ and $\sigma (\hbox{C})$.

\subsubsection{The inelastic capture ratio at rest
in the low-energy \mib{\Sigma^- p} scattering}

\begin{table}[t]
\caption{
$\Sp$ zero-energy total cross sections
$k_i \sigma_S$ ($\hbox{mb}~\hbox{fm}^{-1}$)
for each spin-state ($S=0$ and 1) and $k_i \sigma_T$ for reactions
B: $\Sigma^- p \rightarrow \Sigma^0 n$
and C: $\Sigma^- p \rightarrow \Lambda n$.
The divergent factor $C_0$ is taken out in the Coulomb case.
The inelastic capture ratio $r_R$ at rest and $r_F$ in flight
are also given. Two versions of our quark model, FSS and fss2,
are used.}
\label{rest}
\begin{center}
\renewcommand{\arraystretch}{1.1}
\setlength{\tabcolsep}{3mm}
\begin{tabular}{ccccccc}
\hline
& \multicolumn{3}{c}{FSS without Coulomb}
& \multicolumn{3}{c}{fss2 without Coulomb} \\
\cline{2-4}
\cline{5-7}
& $\begin{array}{c} \hbox{B} \\ \end{array}$ &
$\begin{array}{c} \hbox{C} \\ \end{array}$ &
$r_S$, $r_F$ \quad $r_R$ &
$\begin{array}{c} \hbox{B} \\ \end{array}$ &
$\begin{array}{c} \hbox{C} \\ \end{array}$ &
$\begin{array}{c} r_S,~r_F \\ \end{array}$ \quad $r_R$ \\
\hline
$\begin{array}{c}
k_i \sigma_0 \\
k_i \sigma_1 \\
\end{array}$ &
$\begin{array}{c}
56.1  \\ 68.8 \\ \end{array}$ &
$\begin{array}{c}
5.8 \\ 164 \\ \end{array}$ &
$\left. \begin{array}{c}
0.906 \\ 0.296 \\ \end{array}\right\} 0.448$ &
$\begin{array}{c}
61.1 \\ 66.7 \\ \end{array}$ &
$\begin{array}{c}
6.2 \\ 165 \\ \end{array}$ &
$\left. \begin{array}{c}
0.908 \\ 0.289 \\ \end{array} \right\}0.443$ \\
\hline
$k_i \sigma_T$ &
$\begin{array}{c} 65.6 \\ \end{array}$ &
$\begin{array}{c} 124 \\ \end{array}$ &
$\begin{array}{c} 0.346 \\ \end{array}$ \hspace{11mm} &
$\begin{array}{c} 65.3 \\ \end{array}$ &
$\begin{array}{c} 125 \\ \end{array}$ &
$\begin{array}{c} 0.343 \\ \end{array}$ \hspace{11mm} \\
\hline
& \multicolumn{3}{c}{FSS with Coulomb}
& \multicolumn{3}{c}{fss2 with Coulomb} \\
\cline{2-4}
\cline{5-7}
%$(\hbox{mb} \hbox{fm}^{-1})$ &
 &
$\begin{array}{c} \hbox{B} \\ \end{array}$ &
$\begin{array}{c} \hbox{C} \\ \end{array}$ &
$r_S$, $r_F$ \quad $r_R$ &
$\begin{array}{c} \hbox{B} \\ \end{array}$ &
$\begin{array}{c} \hbox{C} \\ \end{array}$ &
$\begin{array}{c} r_S,~r_F \\ \end{array}$ \quad $r_R$ \\
\hline
$\begin{array}{c}
(C_0)^{-1} k_i \sigma_0 \\
(C_0)^{-1} k_i \sigma_1 \\
\end{array}$ &
$\begin{array}{c}
44.4 \\ 39.7 \\ \end{array}$ &
$\begin{array}{c}
4.7 \\ 94.9 \\ \end{array}$ &
$\left. \begin{array}{r}
0.904 \\ 0.295 \\ \end{array} \right\} 0.447$ &
$\begin{array}{c}
48.1 \\ 38.3 \\ \end{array}$ &
$\begin{array}{c}
5.01 \\ 94.6 \\ \end{array}$ &
$\left. \begin{array}{c}
0.906 \\ 0.288 \\ \end{array} \right\} 0.442$ \\
\hline
$(C_0)^{-1} k_i \sigma_T$ &
$\begin{array}{c} 40.9 \\ \end{array}$ &
$\begin{array}{c} 72.3 \\ \end{array}$ &
$\begin{array}{c} 0.361 \\ \end{array}$ \hspace{11mm} &
$\begin{array}{c} 40.7 \\ \end{array}$ &
$\begin{array}{c} 72.2 \\ \end{array}$ &
$\begin{array}{c} 0.361 \\ \end{array}$ \hspace{11mm} \\
\hline
%\hline
\end{tabular}
\end{center}
\end{table}

The largest effect of the pion-Coulomb correction appears
in the calculation of the $\Sigma^- p$ inelastic capture
ratio at rest, $r_R$ \cite{FU98}.
This observable is defined by \cite{SW62}
\begin{eqnarray}
r_R=\frac{1}{4}\,r_{S=0} + \frac{3}{4}\,r_{S=1}
\qquad \hbox{with} \qquad
\left. r_{S=0,1}\equiv \frac{\sigma_{(S=0,1)}({\rm B})}
{\sigma_{(S=0,1)}({\rm B})+\sigma_{(S=0,1)}({\rm C})}
\right|_{p_{\Sigma^-}=0} \ .
\label{sig2}
\end{eqnarray}
This quantity is the ratio of the production rates
of the $\Sigma^0$ and $\Lambda$ particles 
when a $\Sigma^-$ particle is trapped in an atomic orbit
of the hydrogen and interacts with the proton nucleus. 
For the accurate evaluation of $r_R$, we first determine
the effective range parameters of the low-energy $S$-matrix
using the multi-channel effective range theory \cite{FU98}.
The calculations are performed by using the particle basis with
and without the Coulomb force. The scattering-length matrices
are employed in Table \ref{rest} to calculate $r_R$.
Compared to the empirical values $r_R=0.33  \pm 0.05$ \cite{RO58},
$0.474 \pm 0.016$ \cite{HE68} and $0.465 \pm 0.011$ \cite{ST70},
we find that  $r_R=0.442$ predicted by fss2 is
slightly smaller than the recent values
between 0.45 and 0.49.
The contribution from each spin state is also listed in
Table \ref{rest}. We find $r_{S=0}\approx 0.9$, which indicates
that the $\sigma (\hbox{C})$ is very small
in comparison with the $\sigma (\hbox{B})$ in the spin-singlet state.
On the other hand, $r_{S=1} \approx 0.29$ implies that
most of the $\Sigma N$--$\Lambda N$ channel coupling
takes place in the spin-triplet state. 
Here again the one-pion tensor force is very important
in the $\Sigma N (I=1/2)$--$\Lambda N$
$\hbox{}^3S_1$ - $\hbox{}^3D_1$ CC problem.
We also find that the effect of the Coulomb force plays
a minor role for this ratio \cite{FU98}.

\begin{figure}[b]
\begin{center}
\begin{minipage}{0.48\textwidth}
\epsfxsize=0.9\textwidth
\epsffile{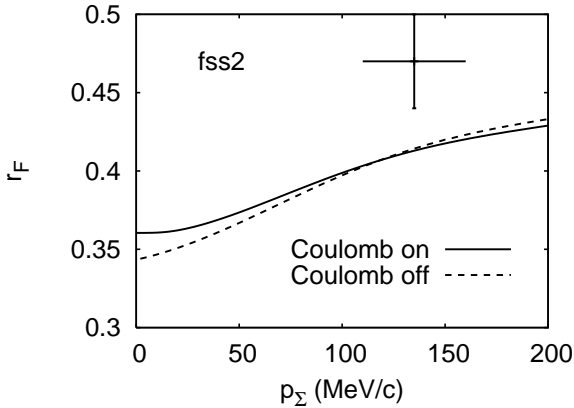}
\end{minipage}
%\bigskip
%\qquad 
\begin{minipage}{0.48\textwidth}
\caption{
$\Sigma^- p$ inelastic capture ratio in flight $r_F$,
predicted by fss2, as a function of the incident
momentum $p_\Sigma$.
Calculation is made in the particle basis
with (solid curve) and without (dashed curve) the
Coulomb force.
}
\label{flight}
\end{minipage}
\end{center}
\end{figure}

The $\Sigma^- p$ inelastic capture ratio in flight $r_F$,
predicted by fss2 in the full calculation,
is illustrated in Fig.~\ref{flight}.
Unlike $r_R$, this quantity is defined directly
using the total cross sections $\sigma_T(\hbox{B})
=(1/4)\sigma_0(\hbox{B})+(3/4)\sigma_1(\hbox{B})$ and 
$\sigma_T(\hbox{C})=(1/4)\sigma_0(\hbox{C})
+(3/4)\sigma_1(\hbox{C})$:
\begin{eqnarray} 
r_F=\frac{\sigma_T(\hbox{B})}{\sigma_T(\hbox{B})
+\sigma_T(\hbox{C})}\ .
\label{sig3}
\end{eqnarray}
This quantity is sensitive to the relative
magnitudes of the $\Sigma^- p \rightarrow \Sigma^0 n$ and
$\Sigma^- p \rightarrow \Lambda n$ total cross sections.
In the momentum region of $p_\Sigma \geq 100~\hbox{MeV}/c$,
the $r_F$ calculated in the particle basis changes little,
irrespective of whether the Coulomb force is included or not.
The empirical value of $r_F$ averaged
in the interval $p_\Sigma=110$ - 160 MeV/$c$
is $r_F=  0.47 \pm 0.03$ \cite{EN66},
as shown in Fig.~\ref{flight} by a cross.
The prediction of fss2, $r_F=0.42$, is too small,
which is the same feature as observed
in FSS ($r_F=0.41$) \cite{FU98}.
The main reason for this disagreement is that
our $\Sigma^- p \rightarrow \Lambda n$ cross sections
are too large.

\subsubsection{\mib{YN} cross sections}

\begin{figure}[t]
\begin{center}
\epsfxsize=0.9\textwidth
\epsffile{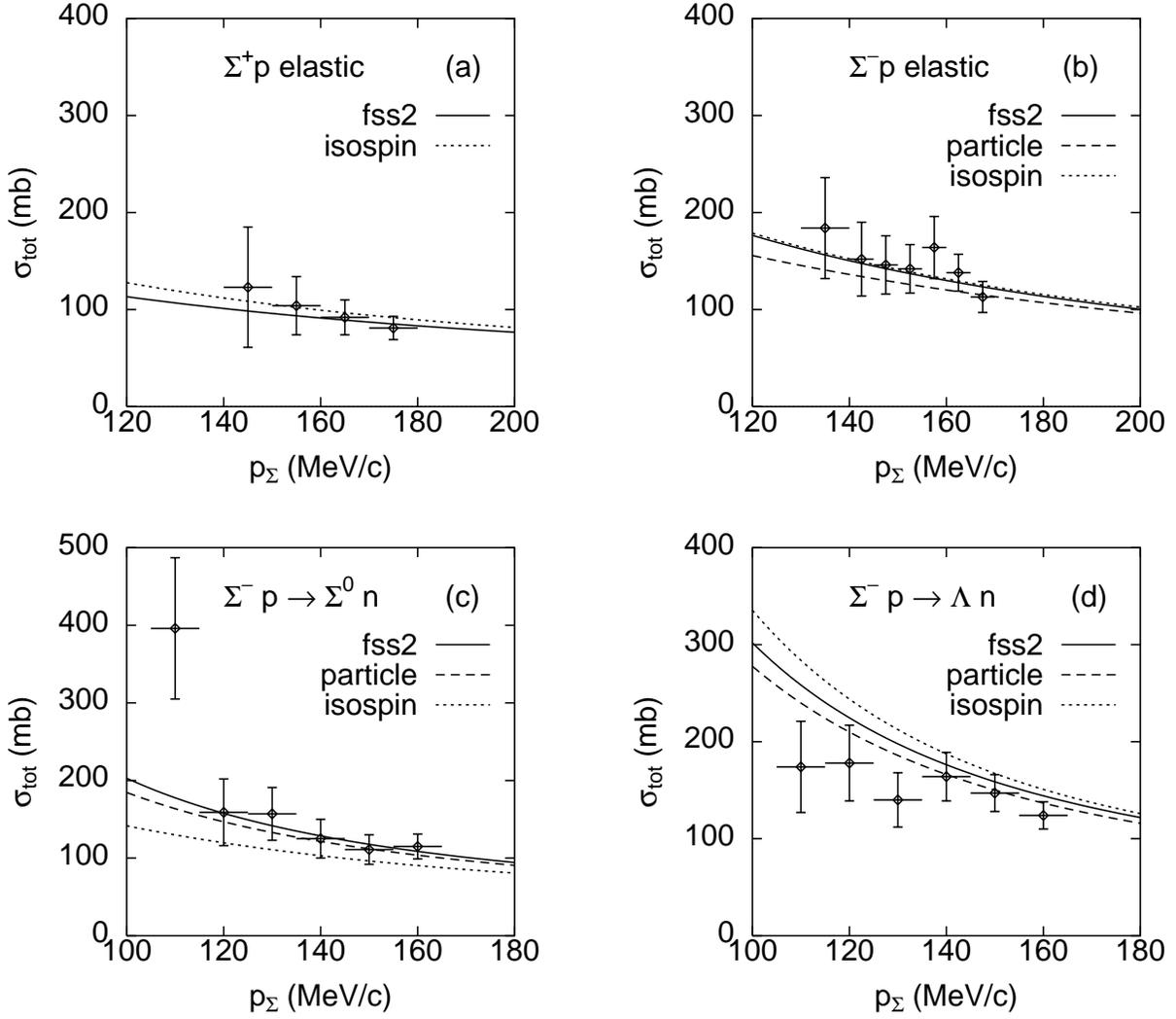}
\caption{
Calculated low-energy $\Sigma^+ p$ and $\Sigma^- p$ scattering
total cross sections by fss2, compared with experimental data:
(a) $\Sigma^+ p$ elastic.
(b) $\Sigma^- p$ elastic. (c) $\Sigma^- p \rightarrow \Sigma^0 n$
charge-exchange. (d) $\Sigma^- p \rightarrow \Lambda n$ reaction
cross sections.
Predictions in the particle basis without the Coulomb
force (dashed curves), and those in the isospin
basis (dotted curves) are also shown.
The experimental data are taken
from \protect\cite{EI71} for (a) and (b),
and from \protect\cite{EN66} for (c) and (d).
}
\label{sgto}
\end{center}
\end{figure}

\begin{figure}[t]
\begin{center}
\epsfxsize=0.9\textwidth
\epsffile{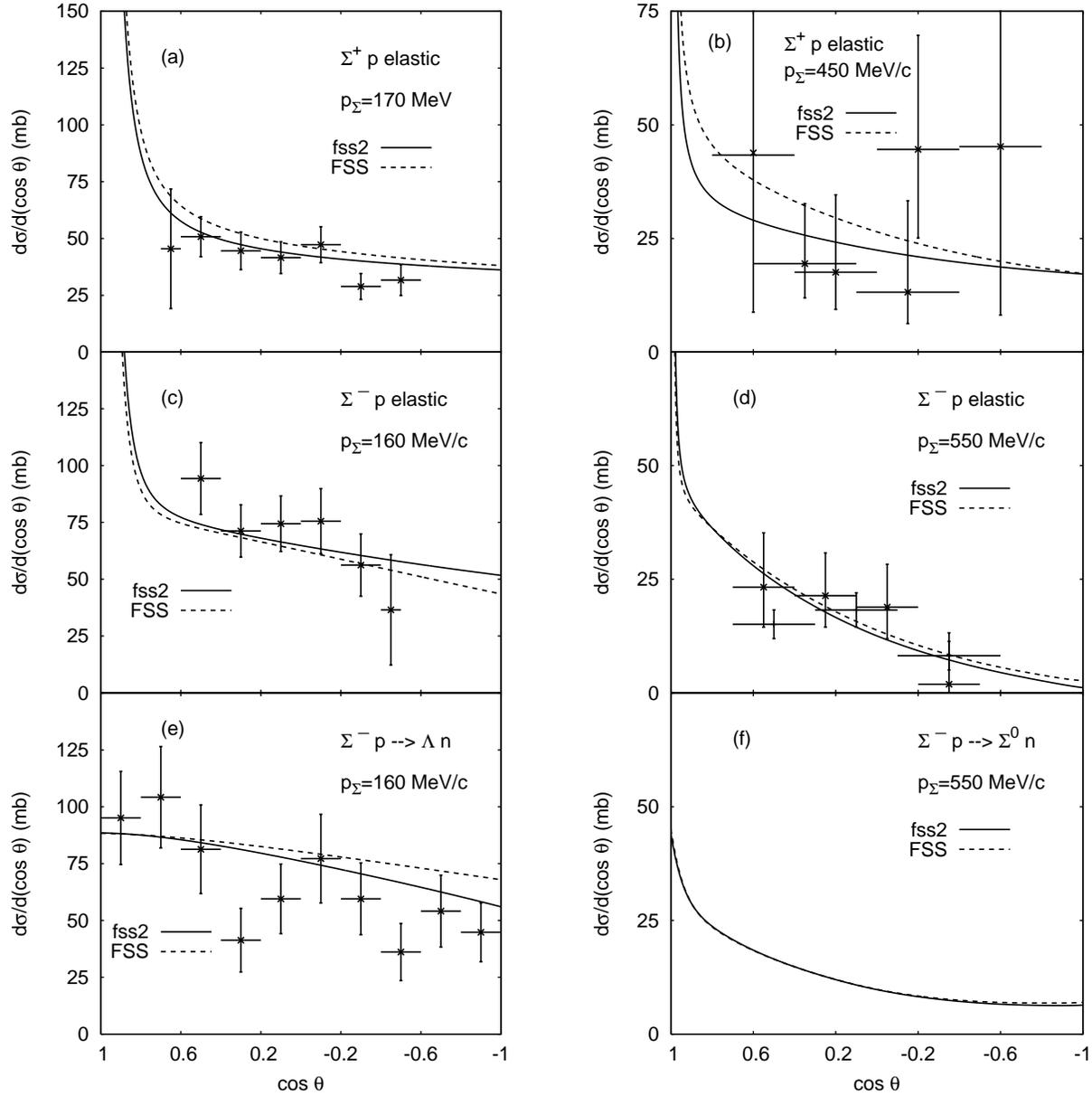}
%\bigskip
\caption{
Calculated $\Sigma^+ p$ and $\Sigma^- p$ differential cross sections
by fss2 (solid curves) and FSS (dashed curves),
compared  with the experimental angular distributions:
(a) $\Sigma^+ p$ elastic scattering at $p_\Sigma=170~\hbox{MeV}/c$.
(b) The same as (a) but at $p_\Sigma=450~\hbox{MeV}/c$.
(c) $\Sigma^- p$ elastic scattering at $p_\Sigma=160~\hbox{MeV}/c$.
(d) The same as (c) but at $p_\Sigma=550~\hbox{MeV}/c$.
(e) $\Sigma^- p$ $\rightarrow$ $\Lambda n$ differential
cross sections at $p_\Sigma=160~\hbox{MeV}/c$.
(f) $\Sigma^- p$ $\rightarrow$ $\Sigma^0 n$ differential
cross sections at $p_\Sigma=550~\hbox{MeV}/c$.
The experimental data are taken
from \protect\cite{EN66} for (a), from \protect\cite{E251,KA05} for (b),
from \protect\cite{EI71} for (c) and (e), and from \protect\cite{E289}
for (d).
}
\label{yndif}
\end{center}
\end{figure}

Figure \ref{sgto} displays the low-energy $\Sigma^- p$ and $\Sigma^+ p$
cross sections predicted by fss2.
The results of three different calculations are shown;
they are in the particle basis including the Coulomb
force (solid curves), in the particle basis
without the Coulomb force (dashed curves),
and in the isospin basis (dotted curves).
The effect of the correct threshold energies and the Coulomb
force is summarized as follows. In the $\Sigma^+ p$ scattering,
the effect of the repulsive Coulomb force reduces the total
cross sections by 11 - 6 mb in the momentum
range $p_\Sigma=140$ - 180 MeV/$c$.
On the other hand, the attractive
Coulomb force in the incident $\Sigma^- p$ channel increases
all the cross sections. An important feature of the present
calculation is the effect of
using the correct $\Sigma^0 n$ threshold energy.
It certainly increases
the $\Sigma^- p \rightarrow \Sigma^0 n$ charge-exchange
cross section, but the prescription to multiply
the factor $(k_f/k_i)$ overestimates this effect.
The real increase is about 1/2 - 2/3 of this estimation.
Furthermore, we find that this change is accompanied with
the fairly large decrease of the $\Sigma^- p$ elastic
and $\Sigma^- p \rightarrow \Lambda n$ reaction cross sections.
Apparently, this effect is due to the conservation of the total
flux. The net effect of the Coulomb and threshold energies
becomes almost zero for the $\Sigma^- p$ elastic scattering.
The charge-exchange reaction cross section largely increases,
and the $\Sigma^- p \rightarrow \Lambda n$ reaction cross
section decreases moderately .
The result of fss2 agrees with the experimental data
reasonably well, although the $\Sigma^- p \rightarrow
\Lambda n$ total reaction cross sections are somewhat too large.

\begin{figure}[t]
\begin{center}
\epsfxsize=0.9\textwidth
\epsffile{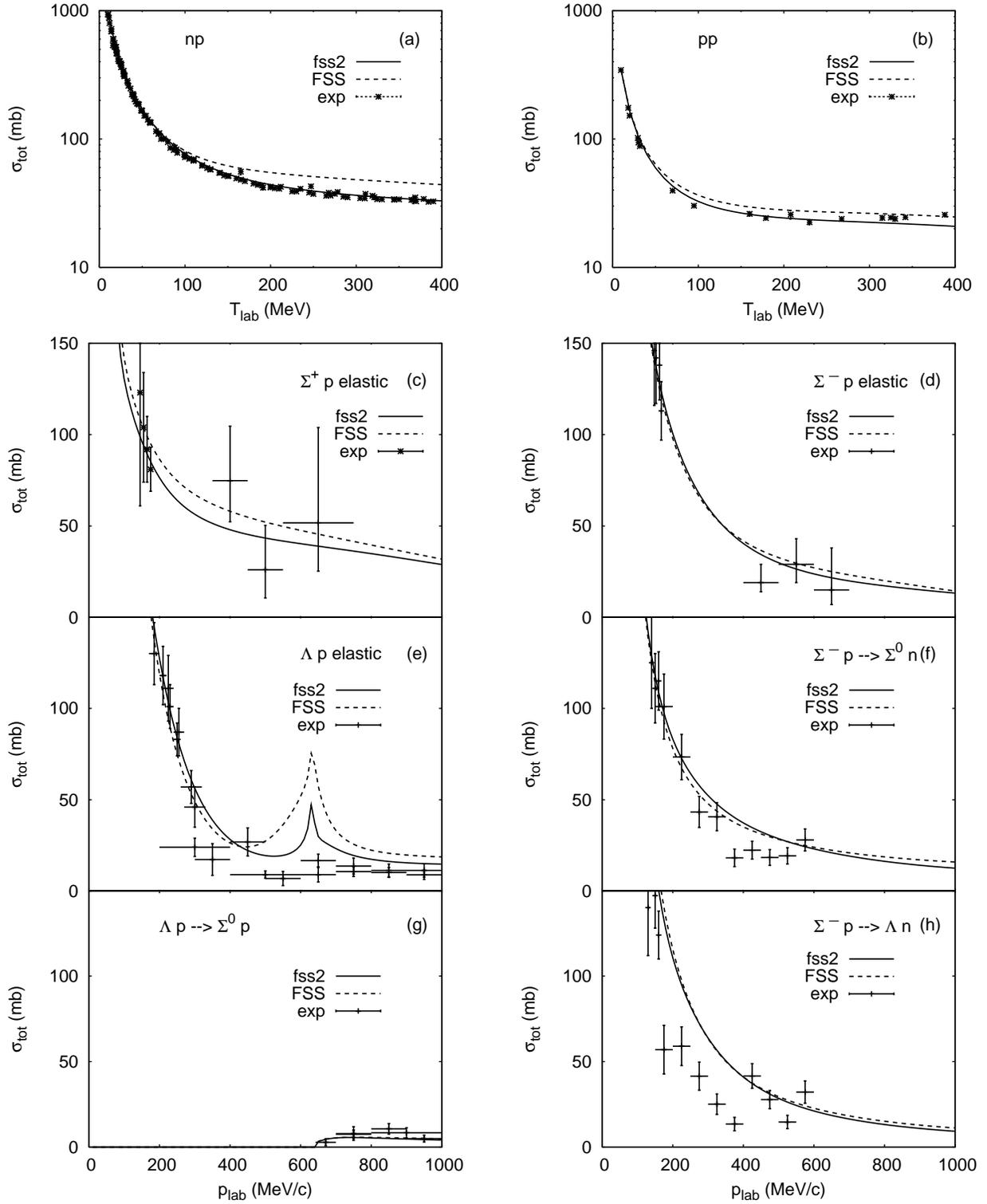}
\caption{
Calculated $NN$ and $YN$ total cross sections
by fss2 (solid curves) and FSS (dashed curves),
compared  with the experimental data.
The calculations are made in the particle basis.
The experimental data are taken
from \protect\cite{SAID} for $NN$,
from \protect\cite{alex68,sechi68,kadyk71} for $\Lambda p$,
from \protect\cite{EN66,E251,KA05} for $\Sigma^+ p$,
and from \protect\cite{ST70,EI71,E289} for $\Sigma^- p$ scattering.
}
\label{tot}
\end{center}
\end{figure}

Figure \ref{yndif} compares the differential cross
sections calculated by fss2 and FSS with experiment. 
For $\Sigma^+ p$ and $\Sigma^- p$ elastic differential
cross sections, the recent data taken
at KEK \cite{E251,KA05,E289} are also compared.
The agreement between theory and experiment is satisfactory.
We display in Fig.~\ref{tot} the total cross sections
of the $NN$ and $YN$ scatterings in a wide energy range.
The solid curves denote the fss2 results,
while the dashed curves the FSS ones.
The ``total'' cross sections in the charged channels,
$pp$, $\Sigma^+ p$, and $\Sigma^- p$, are calculated by
integrating the differential cross sections over the angles
from $\cos \theta_{\rm min}=0.5$ to $\cos \theta_{\rm max}=-0.5$.
In the $NN$ total cross sections, the effect of the inelastic
channel is rather small up to $T_{\rm lab}=400$ MeV.
For the $YN$ total cross sections, the pion-Coulomb correction
is important only for the charged channels and the
low-energy $\Sigma^- p \rightarrow \Sigma^0 n$ reaction
cross sections.

%\clearpage

\subsubsection{\mib{G}-matrix calculations}

\begin{figure}[t]
\begin{center}
\begin{minipage}{0.48\textwidth}
\epsfxsize=0.9\textwidth
\epsffile{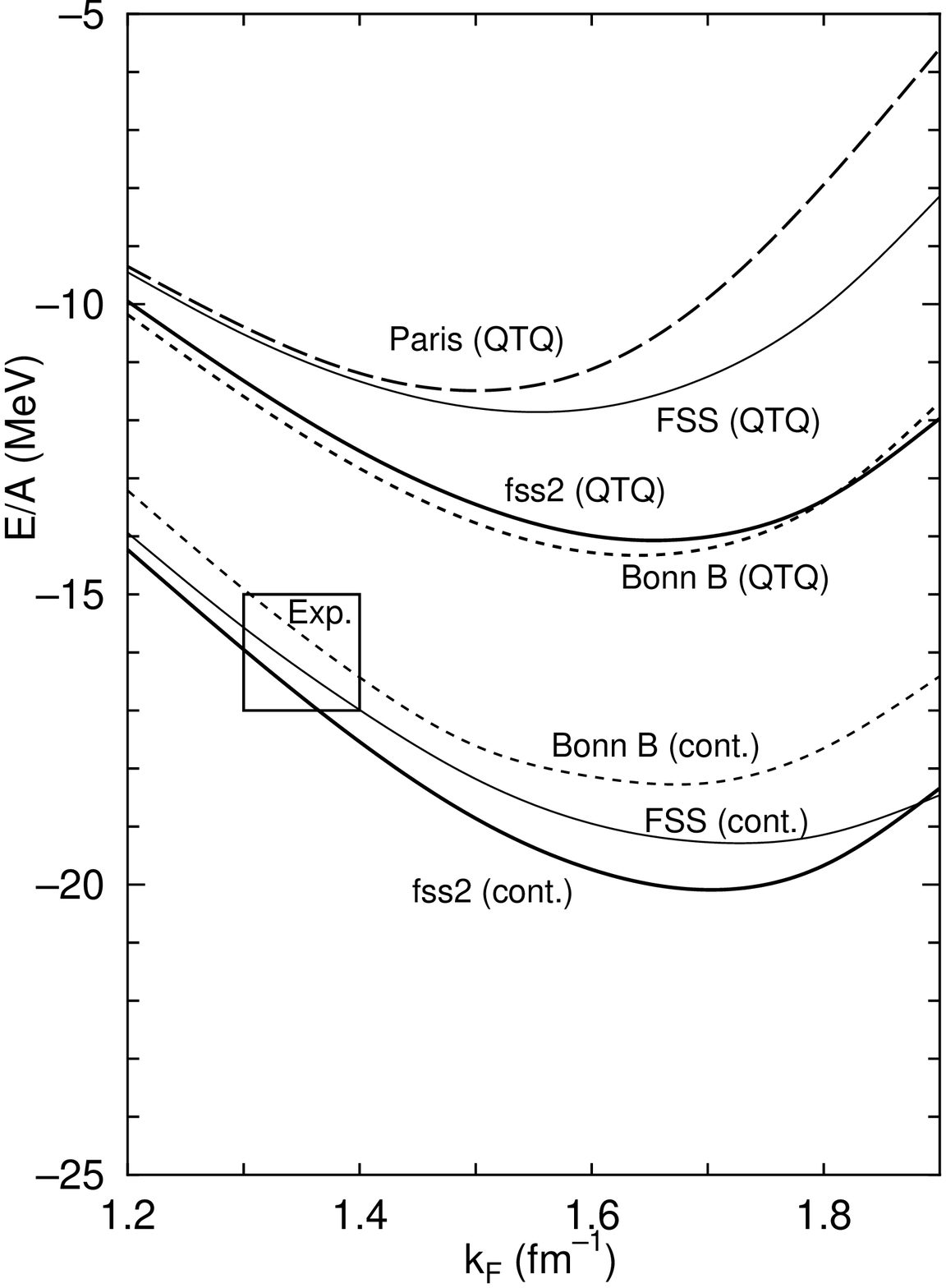}
\end{minipage}
%\qquad
\begin{minipage}{0.48\textwidth}
\caption{
Nuclear matter saturation curves obtained for fss2 and FSS,
together with the results of the Paris
potential \protect\cite{PARI} and
the Bonn model-B (Bonn B) potential \protect\cite{MA89}.
The choice of the intermediate spectra
is specified by ``QTQ'' and ``cont.'' for the $QTQ$ prescription
and the continuous choice, respectively.
The result for the Bonn B potential in the continuous
choice is taken from the non-relativistic calculation
in Ref. \protect\cite{BM90}.
}
\label{matter}
\end{minipage}
\end{center}
\end{figure}

Figure \ref{matter} shows various saturation curves
for symmetric nuclear matter calculated using
the $QTQ$ prescription or the continuous prescription
for intermediate spectra.
The results with the Paris potential \cite{PARI}
and the Bonn B potential \cite{BM90} are also shown for comparison.
Since the short-range part of our QM interaction is
described mainly by the quark-exchange mechanism,
its non-local character is entirely different from
the phenomenological or V-meson exchange pictures in the standard
meson-exchange models.
In spite of this large difference the saturation point
of the QM interaction does not deviate much from the Coester band,
which is similar to other realistic meson-exchange potentials.

\begin{figure}[t]
\begin{center}
\epsfxsize=0.9\textwidth
\epsffile{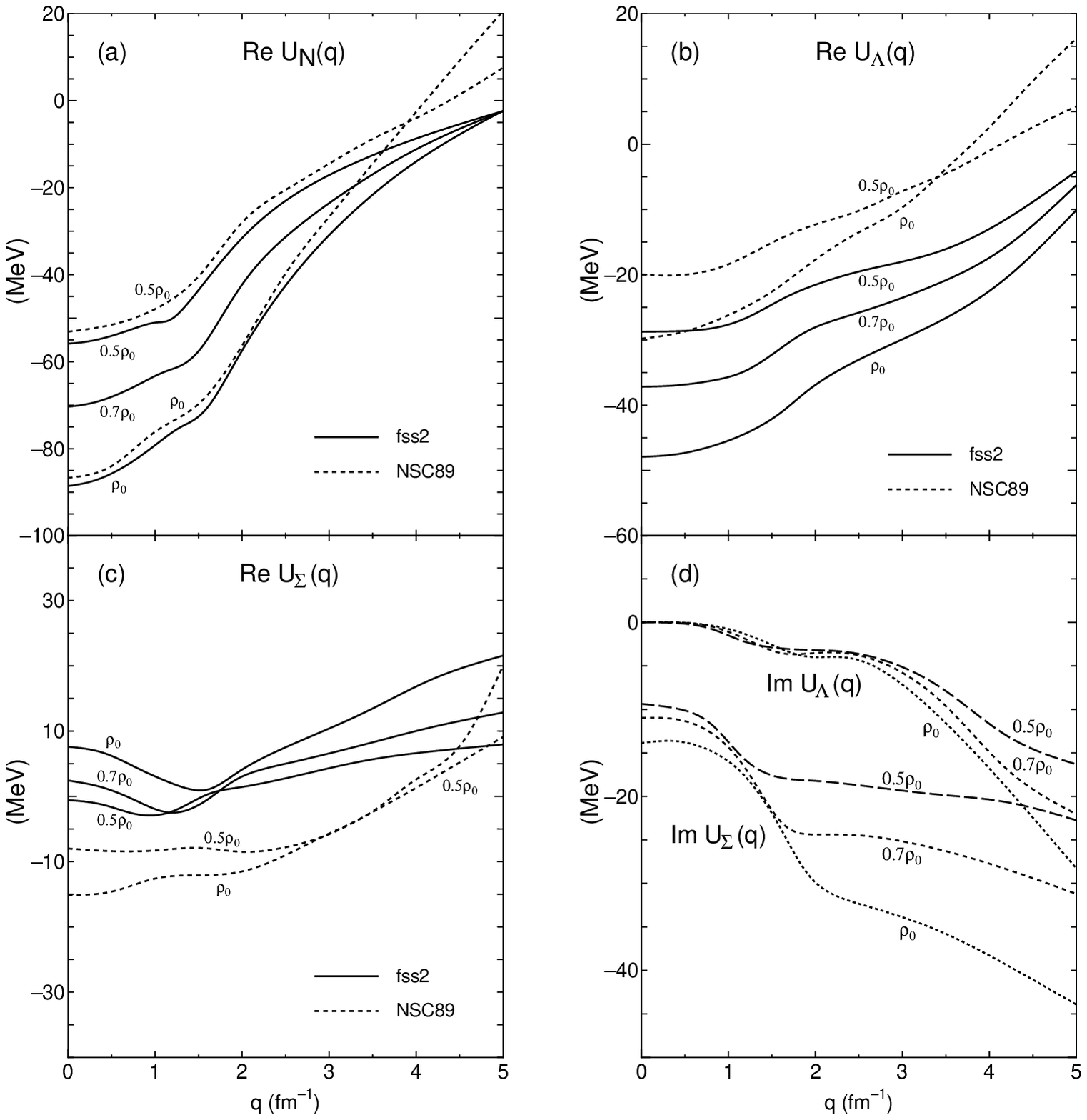}
\caption{
(a) The nucleon s.p.~potential $U_N (q)$ in nuclear
matter in the continuous choice for intermediate spectra.
Predictions by fss2 for three densities $\rho= 0.5\,\rho_0,
~0.7\,\rho_0$ and $\rho_0$ are shown.
Here the normal density $\rho_0$ corresponds
to $k_F =1.35~\hbox{fm}^{-1}$.
The dashed curve is the result achieved by Schulze {\em et al.}
\protect\cite{SCHU} with
the Nijmegen soft-core $NN$ potential NSC89 \protect\cite{NSC89}.
(b) The same as (a) but for the $\Lambda$ s.p.~potential
$U_{\Lambda}(q)$.
(c) The same as (a) but for the $\Sigma$ s.p.~potential
$U_{\Sigma}(q)$.
(d) The same as (b) and (c), but for the imaginary part
of the $\Lambda$ and $\Sigma$ s.p.~potentials $U_B(q)$
predicted by fss2.
}
\label{spnuc}
\end{center}
\end{figure}

The momentum dependence of the nucleon, $\Lambda$,
and $\Sigma$ s.p.~potentials $U_B(q)$ obtained
with the continuous prescription is shown in Fig.~\ref{spnuc} at
three densities $\rho = 0.5\,\rho_0$, $0.7\rho_0$ and $\rho_0$,
where $\rho_0$ = 0.17 fm$^{-3}$ is the normal nucleon density.
(These densities correspond to $k_F =1.07,\;1.2$ and 1.35 fm$^{-1}$,
respectively.)
The results of the Nijmegen soft-core potential
NSC89 \cite{NSC89} calculated by
Schulze {\em et al.} \cite{SCHU} are also shown.
The corresponding figures of the s.p.~potentials calculated by FSS are
given in Figs.\,2 - 5 of Ref.~\cite{GMAT}.
We see that the nucleon s.p.~potential $U_N(q)$ given by fss2 is very
similar to that of FSS except for the higher momentum
region $q \geq 3~\hbox{fm}^{-1}$. As already discussed,
too attractive behavior of FSS in this momentum region
is corrected in fss2, owing to the effect
of the momentum-dependent Bryan-Scott terms
involved in the S-meson and V-meson exchange EMEP's.
The saturation curve in Fig.~\ref{matter} shows that
this improvement of the s.p.~potential in the high-momentum
region is favorable to move the saturation density
to the lower side, as long as the calculation is carried out
with the continuous prescription. 
On the other hand, the saturation curve with the $QTQ$ prescription
alters significantly between FSS and fss2.
The fss2 prediction with the $QTQ$ prescription
is very similar to that of the Bonn model-B potential.
It is interesting to note that the fss2 result is rather close
to Bonn model-C for the deuteron properties (see Table \ref{deutt}),
but to model-B for the nuclear saturation properties.
The model-B has a weaker tensor force than the model-C, which is
a favorable feature for the nuclear saturation properties.

Figures \ref{spnuc}(b) and \ref{spnuc}(c) display the momentum
dependence of the $\Lambda$ and $\Sigma$ s.p.~potentials
in nuclear matter, which are obtained
from the QM $G$-matrices of fss2.
We find that the $U_\Lambda(q)$ predicted by fss2
and FSS (shown in Fig.\,3 of \cite{GMAT}) are again
very similar for $q \leq 2~\hbox{fm}^{-1}$.
The depth of the $U_\Lambda(q)$ at $q=0$ is
48 MeV (fss2) and 46 MeV (FSS) in the continuous prescription.
This value is slightly more attractive
than the value expected from the
experimental data of $\Lambda$-hypernuclei \cite{BMZ}.
On the other hand, the $\Sigma$ s.p.~potential is repulsive in the QM,
reflecting the characteristic repulsion
in the $\hbox{}^3 S_1+\hbox{}^3D_1$ channel
of the isospin $I=3/2$ state (the Pauli repulsion).
The depth of the $U_\Sigma(q)$ at $q=0$ is
8 MeV (fss2) and 20 MeV (FSS).
The repulsive feature of the $\Sigma$ s.p.~potential is
supported by Dabrowski's analysis \cite{DA99} of
the $(K^- ,\pi^{\pm})$ experimental data
at BNL \cite{BNL98}. Recently, inclusive $(\pi^-, K^+)$ spectra
corresponding to the $\Sigma$ formation were measured
at KEK \cite{NU02,SA04} with better accuracy.
These data are satisfactorily reproduced using a repulsive
$\Sigma$--nucleus potential whose strength
is about 30 MeV \cite{KO04}.

\begin{table}[b]
\caption{$\Lambda$ and $\Sigma$ s.p.~potentials
in nuclear matter with $k_F = 1.35~\hbox{fm}^{-1}$,
calculated from fss2 (FSS \protect\cite{GMAT}) $G$-matrices
in the continuous prescription for intermediate spectra.
Predictions of the Nijmegen soft-core potential
NSC89 \protect\cite{NSC89} are also shown
for comparison \protect\cite{SCHU}.}
\label{spcom}
%\vspace{5mm}
\renewcommand{\arraystretch}{1.1}
\setlength{\tabcolsep}{3mm}
\begin{center}
%\begin{tabular}{\textwidth}{@{}c@{\extracolsep{\fill}}rrcrrrr}
%\hline
\begin{tabular}{crrcrrrr}
\hline
 & \multicolumn{2}{c}{$U_\Lambda(0)$ \hspace{1em}[MeV]} &
 & \multicolumn{4}{c}{$U_\Sigma(0)$ \hspace{1em}[MeV]} \\
\cline{2-3} \cline{5-8}
 & fss2 (FSS) & NSC89 & & \multicolumn{2}{c}{fss2 (FSS)}
 & \multicolumn{2}{c}{NSC89} \\ \hline
$I$ & 1/2 & 1/2 &  & 1/2 & 3/2 & 1/2 & 3/2 \\ \hline
$\hbox{}^1S_0$ & $-14.8$ ($-20.1$) & $-15.3$ &  & $6.7$ ($6.1$)
& $-9.2$ ($-8.8$) & $6.7$ & $-12.0$ \\
$\hbox{}^3S_1+\hbox{}^3D_1$ & $-28.4$ ($-21.2$) & $-13.0$ & 
 & $-23.9$ ($-20.2$) & $41.2$ ($48.2$)
 & $-14.9$ & $6.7$\\
$\hbox{}^1P_1+\hbox{}^3P_1$ & $2.1$ ($0.4$) & $3.6$ &
 & $-6.5$ ($-7.0$) & $3.3$ ($4.0$)
 & $-3.5$ & $3.9$ \\
$\hbox{}^3P_0$ & $-0.4$ ($0.5$) & $0.2$ &
  & $2.9$ ($3.0$) & $-2.2$ ($-2.3$) & $2.6$ & $-2.0$ \\
$\hbox{}^3P_2+\hbox{}^3F_2$ & $-5.7$ ($-4.6$) & $-4.0$  &
  & $-1.6$ ($-1.3$) & $-2.5$ ($-1.2$) & $-0.5$ & $-1.9$\\ \hline
 subtotal  &   &  & & $-23.8$ ($-21.0$) & $31.3$ ($40.8$) & $-9.8$ & $-5.5$ \\
total & $-48.2$ ($-46.0$) & $-29.8$ & & \multicolumn{2}{c}{$7.5$ ($19.8$)}
 & \multicolumn{2}{c}{$-15.3$} \\
\hline
\end{tabular}
\end{center}
\end{table}

The partial wave contributions
to $U_{\Lambda}(q=0)$ and $U_{\Sigma}(q=0)$ in symmetric nuclear matter
at $k_F = 1.35~\hbox{fm}^{-1}$ are tabulated
in Table \ref{spcom}, together with the results
of FSS \cite{GMAT} and NSC89 \cite{SCHU}.
For the $\Lambda$ s.p.~potential, the characteristic feature
of fss2 in comparison with FSS appears in the less
attractive $\hbox{}^1S_0$ state
and in the more attractive $\hbox{}^3S_1$ state.
The partial wave contributions of fss2 is very similar
to those of NSC89, except for the $\hbox{}^3S_1+\hbox{}^3D_1$
contribution. The extra attraction of fss2, compared to NSC89,
in the $\Lambda$ s.p.~potential comes mainly
from this channel (15 - 16 MeV).
This is probably because the tensor coupling
is stronger in fss2 than in NSC89.
A minor excess of the attraction comes from
the $\hbox{}^1P_1+\hbox{}^3P_1$ and $\hbox{}^3P_2
+\hbox{}^3F_2$ states (2 - 3 MeV).
In $U_{\Sigma}(q=0)$, the reduction of repulsion
from 20 MeV in FSS to 8 MeV in fss2
is mainly brought about by the 7 MeV reduction
of the $I=3/2$ $\hbox{}^3 S_1+\hbox{}^3D_1$ repulsion
and by the 4 MeV increase
of the $I=1/2$ $\hbox{}^3 S_1+\hbox{}^3D_1$ attraction.
The latter feature is again related to the strong tensor
coupling in fss2. On the other hand, the repulsive contribution
of the $I=3/2$ $\hbox{}^3 S_1+\hbox{}^3D_1$ state in NSC89
is very weak, since this channel has a broad resonance
around $p_\Sigma=500$ - $800~\hbox{MeV}/c$.
(See Fig.\,1 of \cite{FJ96}.)
It is interesting to note that the attractive contributions
to the $\Lambda$ and $\Sigma$ s.p.~potentials
from the $I=1/2$ $\hbox{}^3 S_1+\hbox{}^3D_1$ state
is more than 10 MeV stronger in fss2 than in NSC89.
Although NSC89 is considered to have
a strong $\Lambda N$--$\Sigma N$ coupling,
the $\Lambda N$--$\Sigma N$ coupling of fss2 is even stronger.
%This feature should have some consequence in the energy spectra
%of the $s$-shell $\Lambda$-hypernuclei, if fss2 is used in
%the few-body calculations of these hypernuclei.
The imaginary parts of the $\Lambda$ and $\Sigma$ s.p.~potentials
are shown in Fig.~\ref{spnuc}(d) for fss2.
These results are similar to those of FSS (see Fig.\,4 of ref. \cite{GMAT}).
In particular, $\hbox{Im}~U_\Sigma (q=0)$ at $k_F = 1.35$ fm$^{-1}$ is
$-14.4~\hbox{MeV}$ in fss2 and $-18.4~\hbox{MeV}$ in FSS.
These results are in accord with the calculations
by Schulze {\it et al.} \cite{SCHU} for NSC89.

\begin{table}[t]
\caption{Decomposition of the Scheerbaum factors obtained by fss2
%,$S_{\Lambda}=-11.1~\hbox{MeV}~\hbox{fm}^5$ and $S_{\Sigma}
%= -23.3~\hbox{MeV}~\hbox{fm}^5$, 
at $k_F = 1.35~\hbox{fm}^{-1}$ into various contributions.
The values in parentheses are given by FSS.
The unit is $\hbox{MeV}~\hbox{fm}^5$.}
\label{lscom}
%\bigskip
\begin{center}
\setlength{\tabcolsep}{4mm}
\renewcommand{\arraystretch}{1.1}
%\begin{tabular*}{\textwidth}{@{}c@{\extracolsep{\fill}}cccccc} \hline
\begin{tabular}{@{}c@{\extracolsep{\fill}}cccccc}
\hline
 &  & \multicolumn{2}{c}{$I=1/2$} & &
  \multicolumn{2}{c}{$I=3/2$}\\
 \cline{3-4} \cline{6-7}
 &  & odd & even &
 & odd & even \\ \hline
 & $LS$    & $-19.1~(-17.1)$ & $-0.2~(0.6)$
 & & --- & --- \\
$S_{\Lambda}$ \qquad & $LS^{(-)}$  
 & $7.9~(12.7)$  & $0.3~(0.3)$   & & --- & --- \\
 & total       & \multicolumn{2}{c}{$-11.1~(-3.5)$} &  &  &  \\
\hline
 & $LS$    & $1.3~(2.7)$  & $-0.3~(0.1)$ &
 & $-12.4~(-11.9)$ & $-1.5~(-1.5)$ \\
$S_{\Sigma}$ \qquad & $LS^{(-)}$  & $-9.6~(-10.5)$ & $-0.4~(-0.6)$  &
 & $-0.4~(0.1)$ & $-0.1~(-0.0)$ \\
 & total & \multicolumn{5}{c}{$-23.3~(-21.8)$}  \\
\hline
\end{tabular}
\end{center}
\end{table}

Using the $G$-matrix solution of fss2, we can calculate
the Scheerbaum factor $S_B$ \cite{SC76}, which represents the strength
of the s.p.~spin-orbit potential defined as \cite{SPLS}
\begin{eqnarray}
U_B^{\ell s} (r)= -\frac{\pi}{2}~S_B~\frac{1}{r}
\frac{d\rho (r)}{dr}~\mbox{\boldmath $\ell \cdot \sigma$}\ .
\label{gmat1}
\end{eqnarray}
The value of $S_B$ is momentum-dependent, but here
we only discuss $S_B(q=0)$.
The QM description of the $YN$ interaction
contains the antisymmetric
spin-orbit ($LS^{(-)}$) component, and the large cancellation
between the $LS$ and $LS^{(-)}$ contributions
in the $\Lambda N$ isospin $I=1/2$ channel
leads to a small s.p.~spin-orbit
potential for the $\Lambda$-hypernuclei.
The FSS gives a ratio $S_{\Lambda}/S_N \leq 1/10$ \cite{SPLS}.
The cancellation of these two contributions is less prominent in fss2,
since the S-meson EMEP yields the ordinary $LS$ component
but no $LS^{(-)}$ component (see Sec.~2.2).
Since the total strength of the $LS$ force is fixed in
the $NN$ scattering, the FB contribution
of the $LS$ force is somewhat reduced.
This can easily be seen from the simple formula given in Eq.\,(52)
of \cite{SPLS}, which shows that in the Born approximation
the FB $LS$ contribution to the Scheerbaum factor is
determined only by a single strength
factor $\alpha_S x^3 m_{ud}c^2 b^5$ with $x=(\hbar/m_{ud}cb)$.
The value of this factor is 29.35 $\hbox{MeV}~\hbox{fm}^5$ for fss2,
which is just 3/5 of the FSS value 48.91 $\hbox{MeV}~\hbox{fm}^5$.
The density dependence of $S_B$ is found to be rather weak.
At the normal density $\rho_0$ with $k_F=1.35~\hbox{fm}^{-1}$,
we obtain $S_N=-42.4$, $S_\Lambda=-11.1$,
$S_\Sigma=-23.3$ ($\hbox{MeV}~\hbox{fm}^5$) for fss2, resulting
in $S_{\Lambda}/S_N \approx 0.26$ and $S_{\Sigma}/S_N \approx 0.55$.
The ratios $S_\Lambda/S_N$ and $S_\Sigma/S_N$ become slightly
smaller for lower densities.  
Table \ref{lscom} lists the contributions from the $LS$ and
$LS^{(-)}$ components in the even- and odd-parity states
as well as $I=1/2$ and $I=3/2$ channels for $k_F=1.35~\hbox{fm}^{-1}$.
The parenthesized figures are the predictions by FSS.
A prominent difference between fss2 and FSS
appears only in the $\hbox{}^3O$ contribution of $S_\Lambda$.
Namely, the 5 MeV reduction
of the $\hbox{}^3O$ $LS^{(-)}$ contribution and
the 2 MeV enhancement of the $\hbox{}^3O$ $LS$ contribution,
which explains the change from $S_\Lambda=-3.5$ in FSS
to $S_\Lambda=-11.1$ in fss2.
%More detailed analysis of $S_\Lambda$ for $k_F=1.07~\hbox{fm}^{-1}$ is
%given in Table \ref{tabfad8s}.
In the recent experiment at BNL,
very small spin-orbit splitting is reported
in the energy spectra of $\hbox{}^9_\Lambda \hbox{Be}$
and $\hbox{}^{13}_\Lambda \hbox{C}$ \cite{AK02,TA03}.
A theoretical calculation of these $\Lambda N$ spin-orbit
splittings using OBEP $\Lambda N$ interactions
is carried out in Ref.~\cite{HI00}.
More detailed analysis of the $\alpha \alpha \Lambda$ system
for $\hbox{}^9_\Lambda \hbox{Be}$ is made in Sec. 3.5.1,
in the three-cluster Faddeev formalism
using our QM $\Lambda N$ interaction \cite{2al,2aljj}.

%\clearpage

\subsection{Interactions for other octet-baryons}

\subsubsection{\mib{S=-2} systems}

%\bigskip

{\bf [\mib{\Lambda\Lambda} system]}

\bigskip

Although the direct measurement of $\Lambda \Lambda$ scattering is
not possible, the observation of double-$\Lambda$ hypernuclei gives
an important information on the $\Lambda \Lambda$ interaction.
Several candidates of double-$\Lambda$ hypernuclei are already 
found \cite{TA01,PR66,DA63,AO91}. Among them, the sequential decay
process of $\hbox{}^{\ \,6}_{\Lambda \Lambda}\hbox{He}$ (NAGARA event),
found in the emulsion/scintillating-fiber hybrid experiment at KEK,
has given a definite value for the binding energy
of the two-$\Lambda$ hyperons, $B_{\Lambda \Lambda}$, without ambiguities
arising from the possibilities of excited states.
The $\Lambda \Lambda$ interaction energy extracted from
$\Delta B_{\Lambda \Lambda}(\hbox{}^{\ \,6}_{\Lambda \Lambda}\hbox{He})
=B_{\Lambda \Lambda}(\hbox{}^{\ \,6}_{\Lambda \Lambda}\hbox{He})
-2B_{\Lambda}(\hbox{}^{\,5}_{\Lambda}\hbox{He})$ is
$\Delta B_{\Lambda \Lambda}(\hbox{}^{\ \,6}_{\Lambda \Lambda}\hbox{He})
=1.01 \pm 0.20^{+0.18}_{-0.11}$ MeV
\cite{TA01}. This value is considerably small compared with a similar
measure of the $\Lambda N$ interaction strength,
$\Delta B_{\Lambda N}(\hbox{}^5_\Lambda \hbox{He})
=B_{\Lambda N}(\hbox{}^{\,5}_\Lambda \hbox{He})
-B_{\Lambda}(\hbox{}^{\,4}_{\Lambda}Z)-B_N(\hbox{}^4\hbox{He})
=B_\Lambda(\hbox{}^{\,5}_\Lambda \hbox{He})
-B_\Lambda(\hbox{}^{\,4}_{\Lambda}Z)=1.73 \pm 0.04$ MeV,
where $B_\Lambda(\hbox{}^{\,4}_{\Lambda}Z)$ stands for a proper
spin and charge average over the binding energies of the $0^+$ 
and $1^+$ states in $\hbox{}^{\,4}_\Lambda \hbox{H}$
and $\hbox{}^{\,4}_\Lambda \hbox{He}$ \cite{FI04}.

\begin{figure}[b]
\begin{center}
\begin{minipage}[h]{0.48\textwidth}
\epsfxsize=0.9\textwidth
\epsfbox{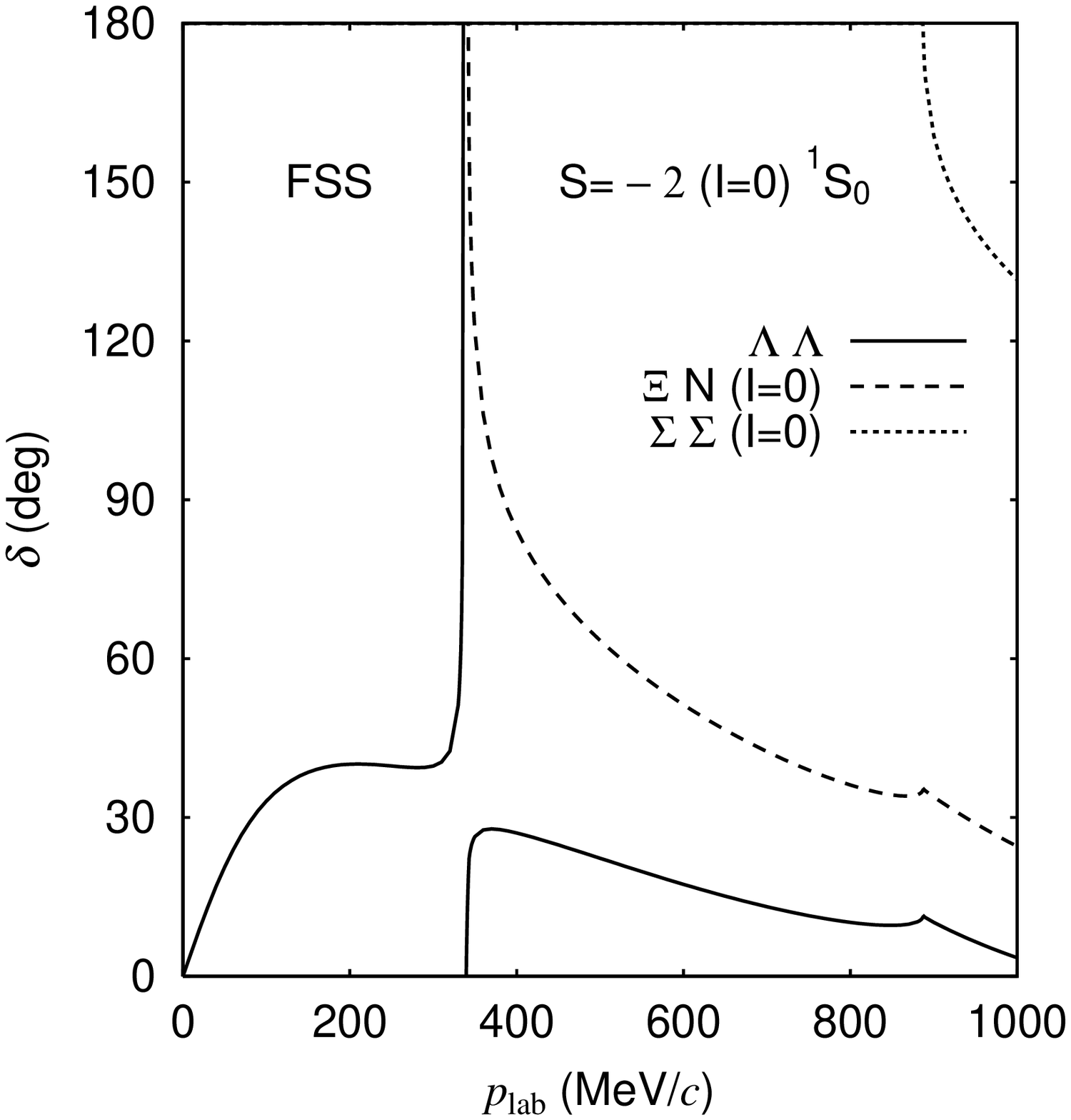}
\caption{$\hbox{}^1S_0 (I=0)$ phase shifts, predicted by FSS,
in the $\Lambda \Lambda$--$\Xi N$--$\Sigma \Sigma$ CC system.}
\label{ll1SFSS}
\end{minipage}
\hfill
\begin{minipage}[h]{0.48\textwidth}
\includegraphics[angle=-90,width=0.9\textwidth]{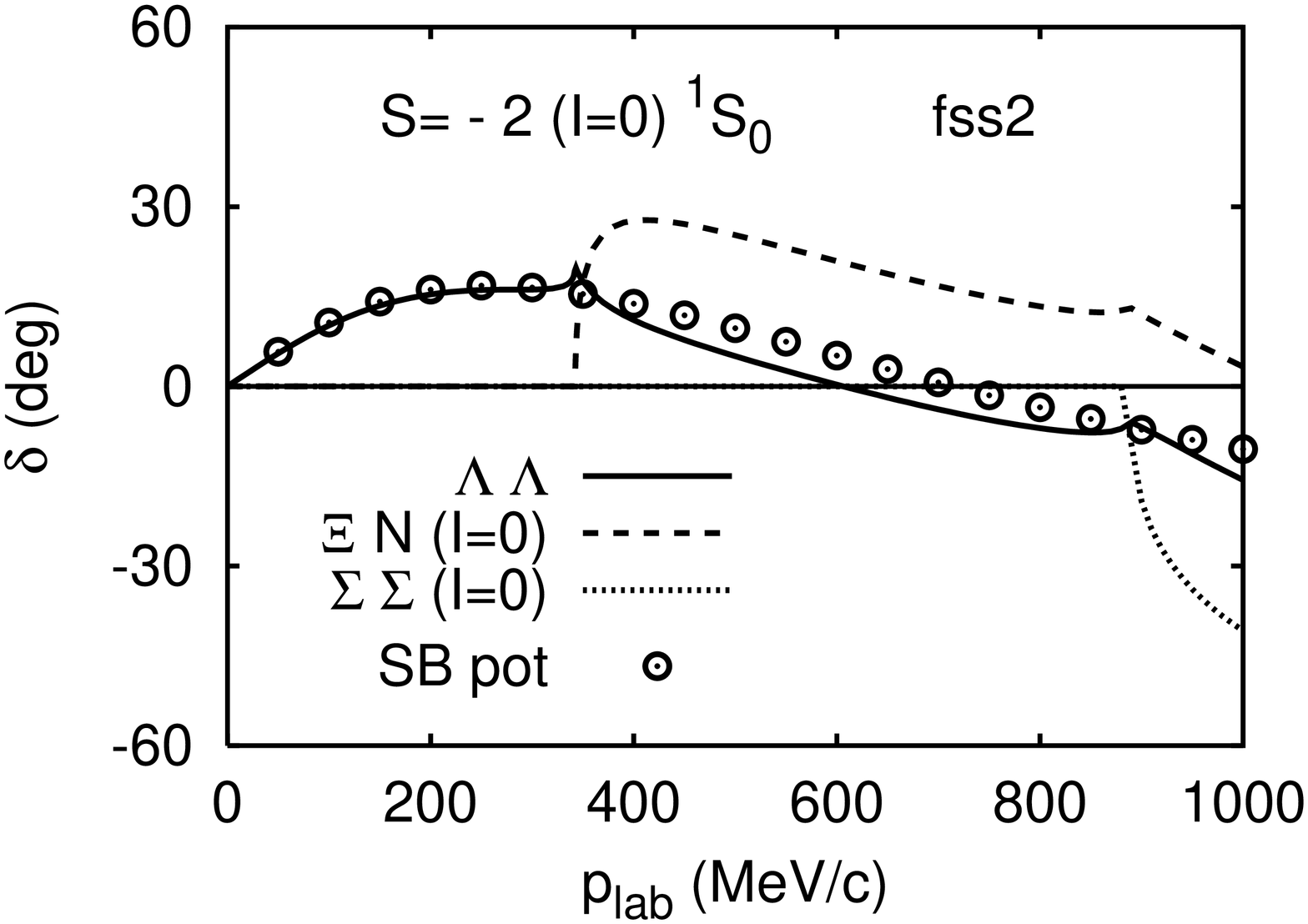}
\caption{
The same as Fig.~\protect\ref{ll1SFSS}, but for fss2.
The single-channel phase shift of the $\Lambda \Lambda$ channel,
predicted by the SB effective potential
in \protect\eq{pot2}, is also shown in circles.
}
\label{ll1Sfss2}
\end{minipage}
\end{center}
\end{figure}

We display in Figs.~\ref{ll1SFSS} and \ref{ll1Sfss2} the $^1S_0$ phase
shifts of the $\Lambda\Lambda$--$\Xi N$--$\Sigma\Sigma$ system 
with isospin $I=0$, calculated by FSS and fss2, respectively.
The $\Lambda\Lambda$--$\Xi N$ CC effect
gives a step-like resonance (FSS) and a cusp (fss2)
at the $\Xi N$ threshold region.
The coupling effect to the $\Sigma\Sigma$ channel also gives cusps 
at the $\Sigma\Sigma$ threshold. 
The $\Lambda\Lambda\,^1S_0$ phase shift rises up to
about $40^\circ$ for FSS and $18^\circ$ for fss2.
The $\Lambda \Lambda$ $T$-matrix of FSS yields
$\Delta B_{\Lambda \Lambda}\approx 3.7$ MeV and that of fss2
$\approx 1.4$ MeV, as will be shown in the Faddeev calculations
of the $\hbox{}_{\Lambda \Lambda}^{\ \,6} \hbox{He}$ system.
(See Table \ref{tabfad9}.)
Another estimate of $\Delta B_{\Lambda \Lambda}
(\hbox{}^{\ \,6}_{\Lambda \Lambda}\hbox{He})$,
using the CC $G$-matrices by fss2, yields 
$\Delta B_{\Lambda\Lambda}=1.12$ - 1.24 MeV,
if the Brueckner rearrangement energy of
the $\alpha$-cluster is properly taken into account \cite{KO03}.
These calculations indicate that fss2 is more appropriate
than FSS, at least for the strength
of the $\Lambda \Lambda$ attraction
in $\hbox{}^{\ \,6}_{\Lambda \Lambda}\hbox{He}$.

\begin{figure}[b]
\begin{center}
%%\vspace{-2cm}
\begin{minipage}[h]{0.48\textwidth}
\vspace{9mm}
\epsfxsize=0.9\textwidth
\epsfbox{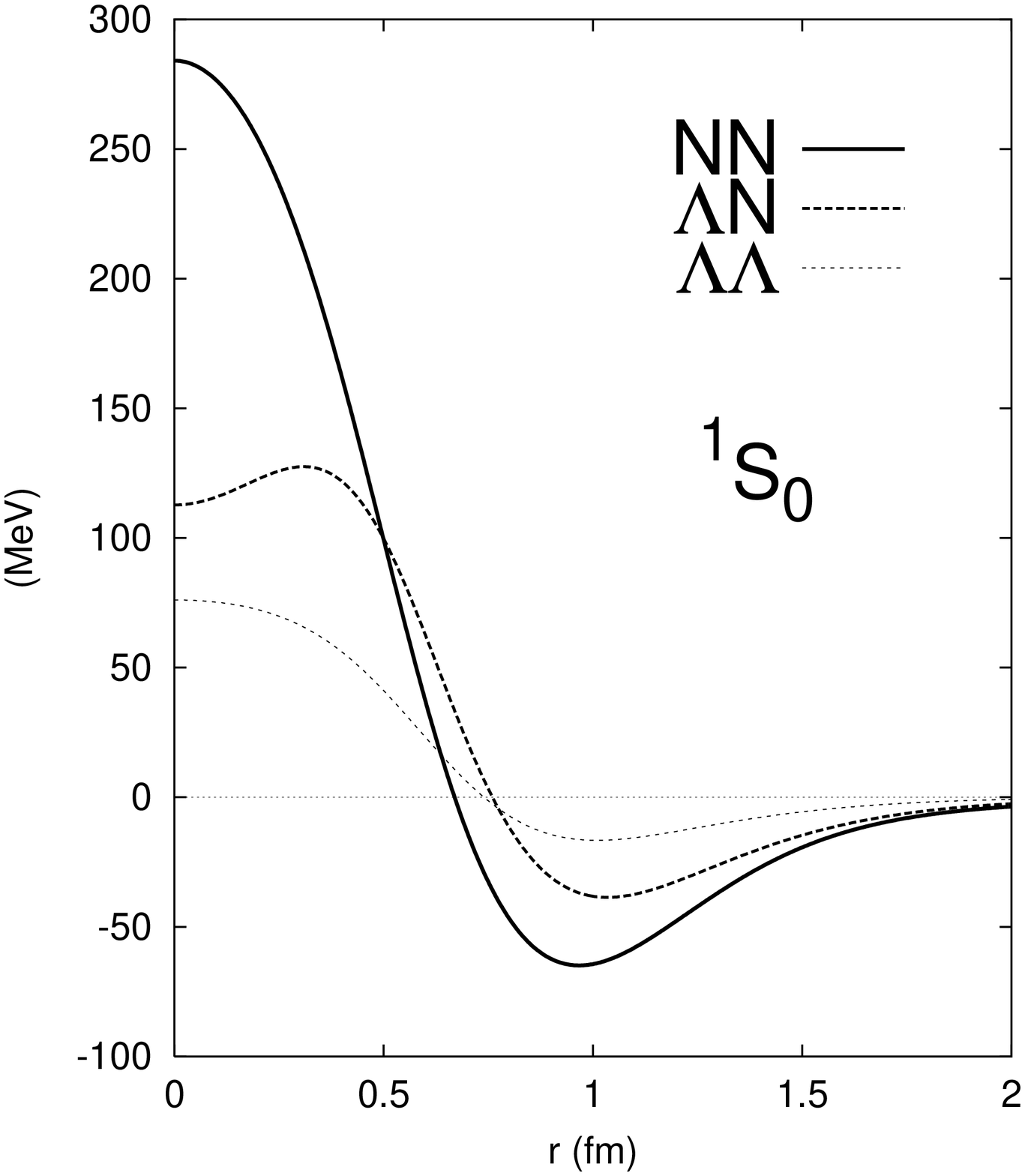}
\caption{The $^1S_0$ phase-shift equivalent
local potentials of single-channel $NN$ (solid curve),
$\Lambda N$ (dashed curve) and $\Lambda \Lambda$ (dotted curve) systems,
predicted by fss2 at $p_{\rm lab}=200~\hbox{MeV}/c$.}
\label{nnlnll-pot}
\end{minipage}
%\qquad
\hfill
\begin{minipage}[h]{0.48\textwidth}
\epsfxsize=0.9\textwidth
\epsfbox{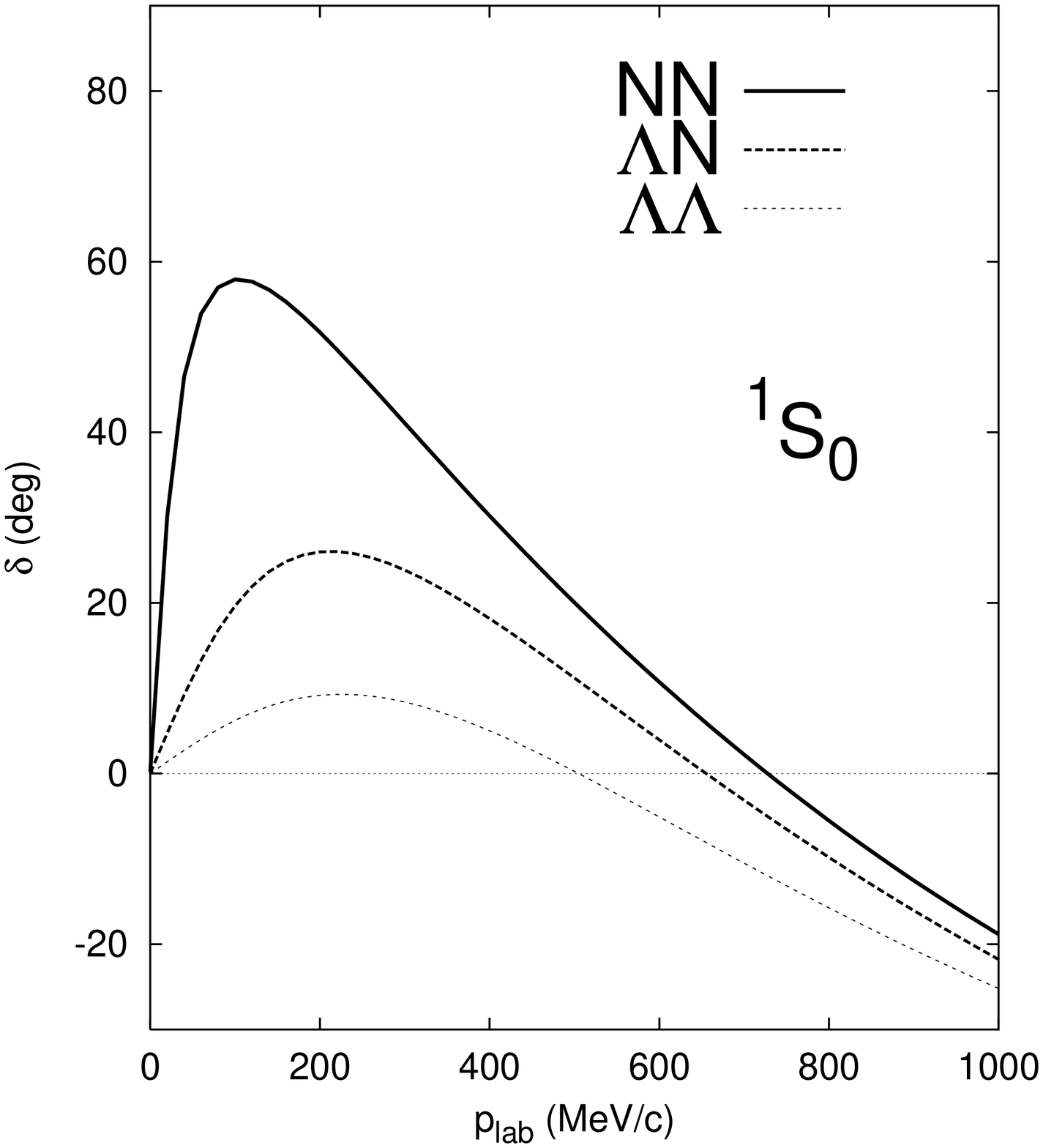}
\caption{$NN$, $\Lambda N$
and $\Lambda\Lambda$ $^1S_0$ phase shifts given 
by the phase-shift equivalent local potentials
in Fig.~\protect\ref{nnlnll-pot} in
the single-channel calculation.}
\label{nnlnll-phs}
\end{minipage}
\end{center}
\end{figure}

It is important to understand the origin of the attraction
of the $\hbox{}^1S_0$ $\Lambda \Lambda$ interaction.
In the single-channel calculation, it is convenient
to display this interaction in terms of the effective local
potential used in Refs.~\cite{SHT83,NA95,NA97},
based on the WKB-RGM approximation.
Figure \ref{nnlnll-pot} shows the phase-shift equivalent
local potentials at $p_{\rm lab}=200\,\hbox{MeV}/c$
for the $\hbox{}^1S_0$ states of $NN$,
$\Lambda N$ and $\Lambda\Lambda$ systems,
obtained by solving the transcendental equation of the 
single-channel RGM kernel of fss2. 
The single-channel $\hbox{}^1S_0$ phase shifts given by these
local potentials are depicted in Fig.~\ref{nnlnll-phs}.
In the low-energy region, our QM gives
the attractive $\Lambda\Lambda$ and $\Lambda N$ interactions
which are similar to the $NN$ one, 
in accordance with the discussion
based on the flavor-$SU_3$ symmetry \cite{NA97}. 
The different strengths of the attraction
in the $NN$, $\Lambda N$ and $\Lambda \Lambda$ systems 
originate from the S-meson nonet exchange, i.e.,
the $SU_3$ relations of the spin-flavor factors. 
On the other hand, the short-range repulsion
arising mainly from the color-magnetic term
of the FB interaction receives a strong effect of
the FSB through the mass ratio
of the $s$- and $ud$-quarks.
Since the color-magnetic term
contains the quark-mass dependence, $1/(m_i m_j)$,
the short-range repulsion is diminished drastically
if the mass ratio $\lambda=(m_s/m_{ud})$ deviates
largely from unity. 
Consequently, the characteristics of the potential
is determined by the subtle balance of the
two reductions; one by the FSB in the short-range repulsion generated
from the color-magnetic term and the other
by the $SU_3$ relations in the medium-range attraction 
generated from the S-meson nonet exchange.
The reduction of the attraction
from the $NN$ to $\Lambda N$ channels suggests 
that the reduction of the medium-range attraction
is larger than that 
of the short-range repulsion in fss2. 
This feature is also reflected when we extend
the present framework from 
the $\Lambda N$ to $\Lambda\Lambda$ systems. 
The relative strength of the attraction for the 
diagonal $\hbox{}^1S_0$ potentials of the $NN$, $\Lambda N$ and 
$\Lambda\Lambda$ systems follows the relationship 
\begin{eqnarray}
|v_{\Lambda\Lambda}|<|v_{\Lambda N}|<|v_{NN}|\ . 
\label{reduction}
\end{eqnarray}
This feature for the two-baryon systems
including the $\Lambda$-particle 
is also seen in NSC97 \cite{SR99}. 

The full $\hbox{}^1S_0$ potentials are obtained by taking into account
the effect of the $\Lambda N$--$\Sigma N$ channel coupling
for the $\Lambda N$ system and the 
$\Lambda\Lambda$--$\Xi N$--$\Sigma\Sigma$ coupling
for the $\Lambda\Lambda$ system. 
First let us discuss the CC effect in the $^1S_0$ state without the EMEP. 
The exchange kinetic-energy, color-Coulombic and 
momentum-dependent Breit retardation terms do not contribute
to the $\Lambda\Lambda$--$\Xi N$ $\hbox{}^1S_0$ transition potential,
because the spin-flavor-color factors of these terms
are exactly zero, as well as that of the
exchange normalization kernel.
Therefore the contribution from the FB interaction
to this coupling potential is solely
from the color-magnetic term. 
This coupling potential gives the sharp cusp structure
for the $\Lambda\Lambda$ phase shift
and the step-like resonance for the $\Xi N$ phase shift,
as seen in Fig.~19 of Ref.~\cite{SH89}.
If the coupling to the $\Sigma \Sigma$ channel
is taken into account, 
the resonance in the $\Xi N$ channel grows up
due to the strong attraction in 
the $\Sigma\Sigma$ channel and moves
to the $\Lambda\Lambda$ channel. 
The actual resonance behavior is, however,
sensitive to the FSB.
If one neglects the FSB and set $\lambda=1$, the CC calculation
of FSS without EMEP yields a bound state of the binding energy 31 MeV,
which corresponds to the $H$-dibaryon predicted
by Ref.~\cite{OSY83}. 
An introduction of the large FSB
reduces the $\Lambda\Lambda$--$\Xi N$ CC effect. 

Next, we introduce EMEP's acting between quarks,
and investigate how they influence the features
of the channel coupling.
The strange-meson exchange contributes
to the $\Lambda\Lambda$--$\Xi N$ diagram
through the direct and exchange kernels in the RGM framework,
whereas the non-strange-meson exchanges
can contribute only through the exchange kernel.
As the result, the dominant contribution from the EMEP's
to the $\Lambda\Lambda$--$\Xi N$ coupling potential
is from the strange-meson exchange. 
This contribution has a sign opposite to that 
of the color-magnetic term \cite{NA97,KEKsympo}
and diminishes the coupling effect by the FB interaction
discussed above.
It turns out that the strange S meson, $\kappa$, gives
the most important contribution
to the $\Lambda\Lambda$--$\Xi N$ $^1S_0$ coupling potential.
The effect of the $\kappa$-meson exchange becomes
more important as the mass becomes smaller. 
It should also be stressed that the spin-flavor factors
of EMEP's are explicitly calculated in our QM baryon-baryon interactions.
The contribution from the exchange RGM kernel
of the EMEP's to the $\Lambda\Lambda$--$\Xi N$ $\hbox{}^1S_0$ coupling
matrix elements is also appreciable, 
as largely as about one third of the corresponding
direct kernel in fss2.
Namely, the following three features of the EMEP reduce the 
$\Lambda\Lambda$--$\Xi N$ CC effect
in the present framework;
1) the introduction of the S-meson nonet
including the $\kappa$ meson, 
2) the comparatively small $\kappa$-meson mass,
and 3) the introduction of the EMEP's not at the baryon level
but at the quark level. 

The above discussion explains
why the $\Lambda\Lambda$--$\Xi N$ CC effect
of fss2 is smaller than 
other several QM's \cite{NA98,OSY83,straub2,KSY90}. 
It is related with the more realistic feature
including the S-meson nonet (with $m_\kappa=936~\hbox{MeV}/c^2$)
and the large FSB (with $\lambda=1.5512$). 
After all, the $\Lambda\Lambda$ interaction of fss2
with the full CC effect satisfies the relationship
of Eq.~(\ref{reduction}). 
This feature of the $\Lambda \Lambda$ interaction
leads to the conclusion that no bound state is possible 
in fss2. (For FSS, the relationship Eq. (\ref{reduction})
is modified to $|v_{\Lambda\Lambda}| \approx
|v_{\Lambda N}| < |v_{NN}|$,
but it is still insufficient to produce a bound state.)
Clearly a realistic QM baryon-baryon interaction
constructed by supplementing EMEP's to the FB interaction
can no longer predict a deeply bound $H$-dibaryon state. 

In Ref.~\cite{FE05}, the $\Lambda \Lambda$ interaction is examined 
in the chiral constituent quark model, which is tightly constrained
by the description of the non-strange sector and the strange meson
and baryon spectra. They found an attractive $\Lambda \Lambda$ interaction
due to the mechanism that the short-range repulsion from the Goldstone-boson
and gluon exchanges are considerably reduced in the $\Lambda \Lambda$
interaction in comparison with the $NN$ interaction. The reduction
of the short-range repulsion produced from the Goldstone-boson exchange
might have some relevance in the systems with more strangeness.

\bigskip

\begin{figure}[b]
\begin{center}
\begin{minipage}{0.48\textwidth}
\epsfxsize=0.9\textwidth
\epsffile{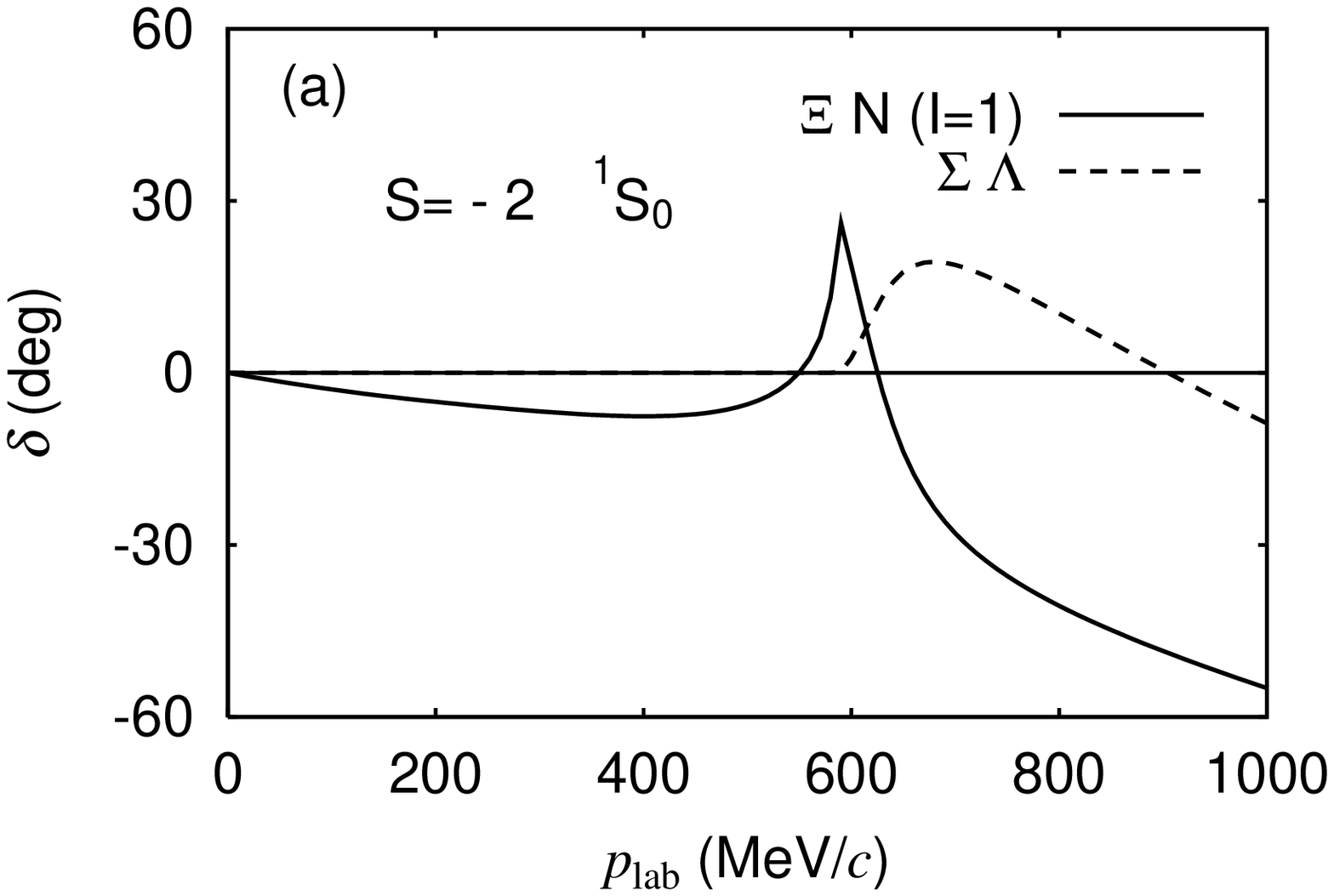}
\end{minipage}~%
%\hfill~%
\begin{minipage}{0.48\textwidth}
\epsfxsize=0.9\textwidth
\epsffile{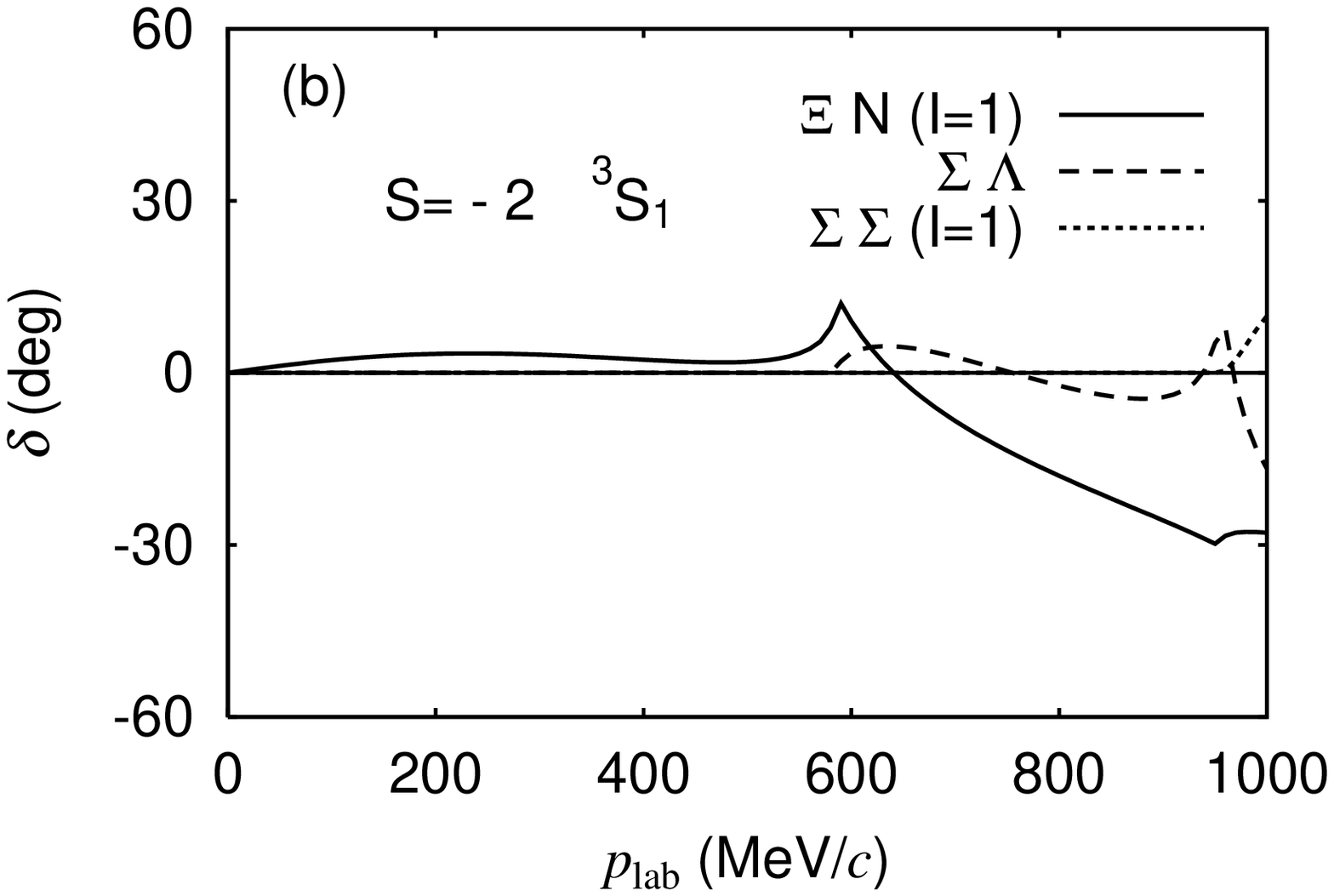}
\end{minipage}
\bigskip
\caption{
(a) $\hbox{}^1S_0$ phase shifts
in the $\Xi N$--$\Sigma \Lambda$ CC system
with $I=1$, predicted by fss2.
(b) $\hbox{}^3S_1$ phase shifts
in the $\Xi N$--$\Sigma \Lambda$--$\Sigma \Sigma$ CC system
with $I=1$, predicted by fss2.
}
\label{fig5}
\end{center}
\end{figure}

\noindent
{\bf [\mib{\Xi N} interaction]}

\bigskip

The $SU_3$ classification in Table \ref{symmetry} is useful
to discuss the characteristics of the $\Xi N$ interaction.
Let us first discuss the phase-shift behavior
of the flavor-symmetric $\hbox{}^1S_0$ state
in the single-channel calculation. 
The color-magnetic term is repulsive
in the (22) state and attractive
in the flavor-singlet (00) state. 
Since the weights of the (22) and (00) components in the 
$\Xi N (I=0)$ $\hbox{}^1S_0$ state are comparable, 
the net contribution of the color-magnetic term is small
in the $I=0$ $\Xi N$ interaction
due to the cancellation between these two components.
The $\Xi N(I=0)$ $\hbox{}^1S_0$ interaction is, therefore,
characterized by the EMEP contribution.
It turns out that the EMEP contribution produces
an attractive $\Xi N(I=0)$ $\hbox{}^1S_0$ interaction.
The situation is different in the $I=1$ state.
In this channel, the leading $SU_3$ component is $(11)_s$,
for which the most compact $(0s)^6$ configuration
gives a Pauli-forbidden state.
This gives a repulsive interaction
in the $\Xi N(I=1)$ $\hbox{}^1S_0$ channel.
As to the flavor-antisymmetric $\hbox{}^3S_1$ - $\hbox{}^3D_1$
state, the $\Xi N (I=0)$ interaction is found to be weakly
attractive, characterizing the property
of the pure $SU_3$ $(11)_a$ state. 
The $\Xi N (I=1)$ interaction is repulsive
because of the almost-forbidden $(0s)^6$ configuration
of the (30) state.
It is concluded that the $\Xi N$ interaction is
attractive in the $I=0$ channel, whereas it is repulsive
in the $I=1$ channel \cite{NA97}. 
The attraction of the $\Xi N (I=0)$
$\hbox{}^1S_0$ state is strongest, and
the repulsion of $\Xi N (I=1)$ $\hbox{}^1S_0$ state
is strongest.

\begin{figure}[b]
\begin{center}
\epsfxsize=\textwidth
\epsffile{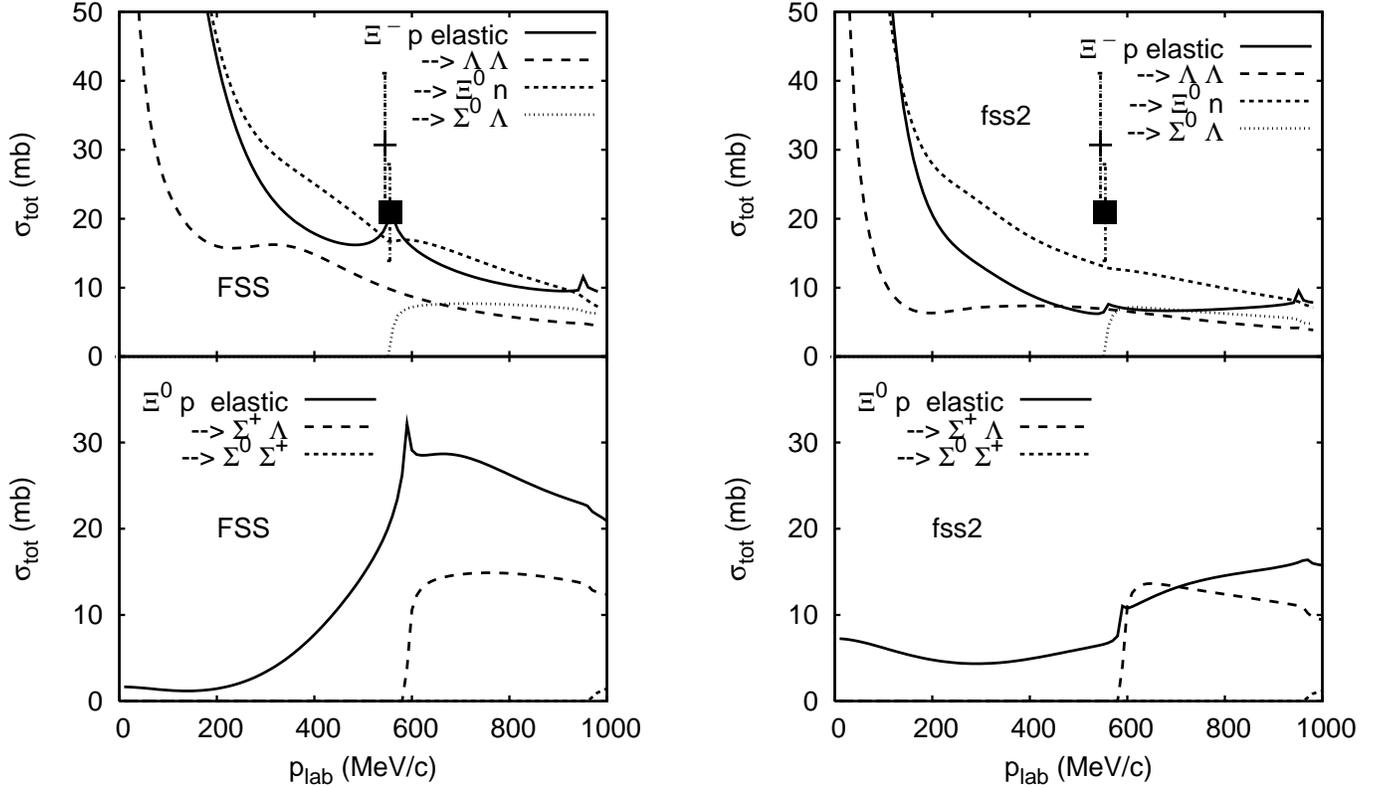}
\caption{$\Xi^- p$ and $\Xi^0 p$ total cross sections
predicted by FSS (left panels) and fss2 (right panels).
The calculation is made in the particle basis, including
the pion-Coulomb corrections.
%For $\Xi^- p$ cross sections both of the
%isospin $I=0$ and 1 channels contribute,
%while for $\Xi^0 p$ (or $\Xi^- n$) only the channel
%with $I=1$ contributes.
}
\label{XNtot}
\end{center}
\end{figure}

The CC effect on the $\Lambda\Lambda$
and $\Sigma\Sigma$ configurations takes palce
in the $I=0$ $\hbox{}^1S_0$ state,
while that on $\Sigma \Lambda$ in the $I=1$ state.
In the former case, the attractive feature
of the $\Xi N$ interaction does not change essentially. 
On the other hand, the effect of the channel coupling
on the $\Sigma \Lambda$ configuration with $I=1$ changes
rather drastically the result of the single-channel
calculation.
Figure \ref{fig5} displays the phase shifts
of the $\hbox{}^1S_0$ and $\hbox{}^3S_1$ states, calculated for
the full $\Xi N$--$\Sigma \Lambda$--$\Sigma \Sigma$ CC system with $I=1$.
In both cases, the phase shifts exhibit a prominent cusp structure
at the $\Sigma\Lambda$ threshold
around $p_{\rm lab}= 600~\hbox{MeV}/c$ and they almost vanish
below the $\Sigma\Lambda$ threshold.
This drastic change is due to the strong CC effect
caused by the large $\Xi N$--$\Sigma\Lambda$ coupling matrix elements.
Although the effect of $\Xi N$--$\Sigma\Lambda$ coupling predicted
by the FB interaction is moderate \cite{OSY87},
the introduction of the EMEP's reinforces it
since the strange-meson exchange can contribute
to the $\Xi N$--$\Sigma\Lambda$ coupling
through both of the direct and exchange
Feynman diagrams \cite{hyp2000}. 

\begin{figure}[b]
\begin{center}
\begin{minipage}{0.48\textwidth}
\epsfxsize=0.9\textwidth
\epsffile{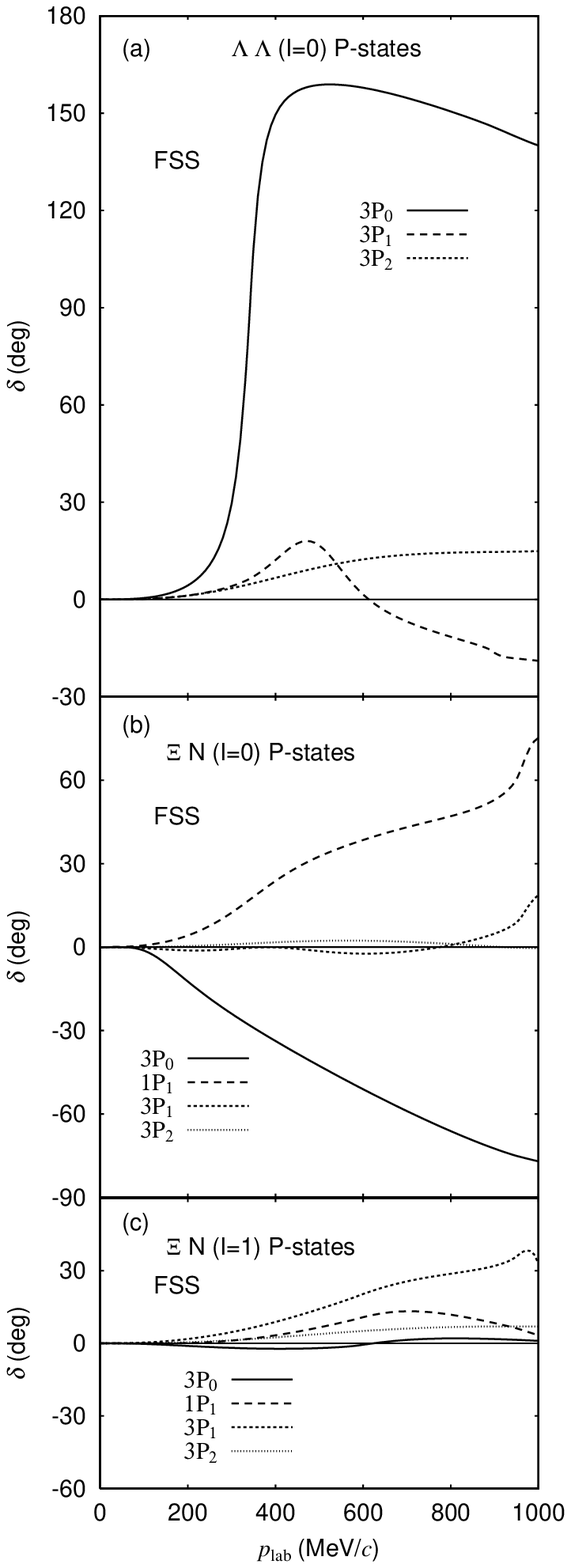}
%\caption{
%The $P$-wave phase shifts
%of the (a) $\Lambda \Lambda (I=0)$,
%(b) $\Xi N (I=0)$ and (c) $\Xi N (I=1)$ scatterings,
%predicted by FSS.}
\end{minipage}~%
%\qquad
\hfill~%
\begin{minipage}{0.48\textwidth}
%\vspace{40mm}
\caption{(left)
The $P$-wave phase shifts
of the (a) $\Lambda \Lambda (I=0)$,
(b) $\Xi N (I=0)$ and (c) $\Xi N (I=1)$ scatterings,
predicted by FSS.}
\label{phxnP_FSS}
\vspace{7mm}
\epsfxsize=0.9\textwidth
\epsffile{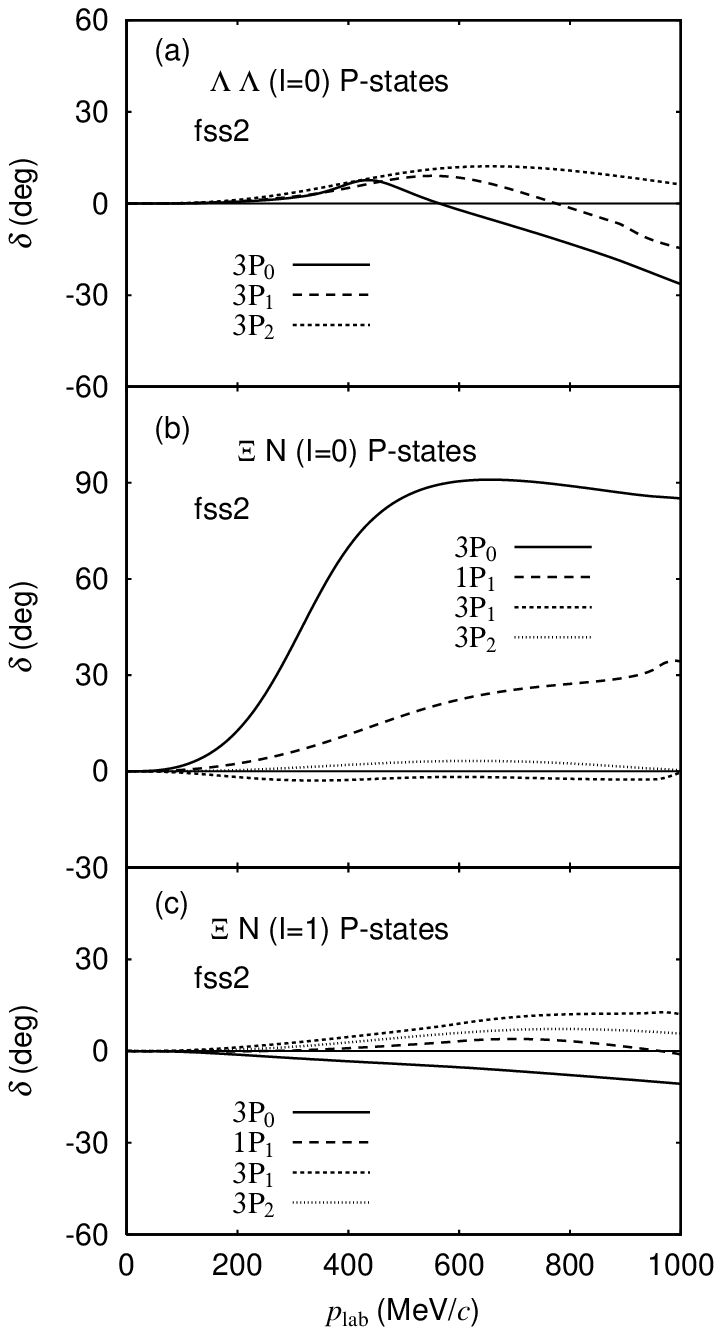}
\caption{(above)
The same as Fig.~\protect\ref{phxnP_FSS} but for fss2.}
\label{phxnP_fss2}
\vspace{7mm}
\end{minipage}
%\bigskip
\end{center}
\end{figure}

Figure \ref{XNtot} shows the $\Xi^- p$ and $\Xi^0 p$ total
cross sections predicted by FSS (left) and fss2 (right).
The calculation is made in the particle basis, including
the pion-Coulomb corrections.
Both of the isospin $I=0$ and 1 channels contribute
to the $\Xi^- p$ cross sections, while only the channel
with $I=1$ contributes to $\Xi^0 p$ (or $\Xi^- n$).
The $\Xi^0 p$ (and $\Xi^- n$) total cross sections
with the pure $I=1$ component are predicted to be very small 
below the $\Sigma \Lambda$ threshold.
This behavior of the $\Xi^- n$ total cross sections
is essentially the same as the Nijmegen result \cite{SR99}.
On the other hand, the $\Xi^- p$ total cross sections
exhibit a typical channel-coupling behavior which is
similar to that of the $\Sigma^- p$ total cross sections.
These features demonstrate that
the $\Sigma \Lambda$ channel coupling
is very important for the correct description
of scattering observables, resulting in the strong
isospin dependence of the $\Xi N$ interaction. 
It is interesting to find that the contribution
of the strange-meson exchange reduces
the $\Lambda \Lambda (I=0)$--$\Xi N (I=0)$ channel coupling,
while it strengthens the $\Xi N (I=1)$--$\Sigma\Lambda (I=1)$ coupling
in the QM including the FB interaction and the EMEP's.
This is quite different from the OBEP models, in which 
both $\Lambda \Lambda$--$\Xi N$ and $\Xi N$--$\Sigma\Lambda$
coupling potentials are generated only from
the strange-meson exchanges. 
Figure \ref{XNtot} also includes
the in-medium experimental $\Xi^- N$ total cross section
around $p_{\rm lab}= 550~\hbox{MeV}/c$ \cite{E906},
where $\sigma_{\Xi^- N}$ (in medium) $=30\pm6.7
\genfrac{}{}{0pt}{1}{+3.7}{-3.6}$\,mb is given.  
Another analysis using the eikonal approximation \cite{YTFM01} 
gives $\sigma=20.9 \pm 4.5 \genfrac{}{}{0pt}{1}{+2.5}{-2.4}~\hbox{mb}$,
as shown by the black square in Fig.~\ref{XNtot}.
The analysis also estimates the cross section ratio 
$\sigma_{\Xi^-p}/\sigma_{\Xi^-n}=1.1
\genfrac{}{}{0pt}{1}{+1.4+0.7}{-0.7-0.4}$\,mb.
A more recent experimental analysis \cite{AH06} for the low-energy
$\Xi^- p$ elastic and $\Xi^- p \rightarrow \Lambda \Lambda$ total
cross sections in the range of 0.2 GeV/$c$ to 0.8 GeV/$c$ shows
that the former is less than 24 mb at 90$\%$ confidence level
and the latter of the order of several mb, respectively.
These results seem to favor the predictions by fss2.
However, we definitely need more experimental data
with high statistics.
A theoretical evaluation of the in-medium cross sections
is also necessary.

\begin{figure}[t]
\begin{center}
\begin{minipage}{0.48\textwidth}
\epsfxsize=0.9\textwidth
\epsffile{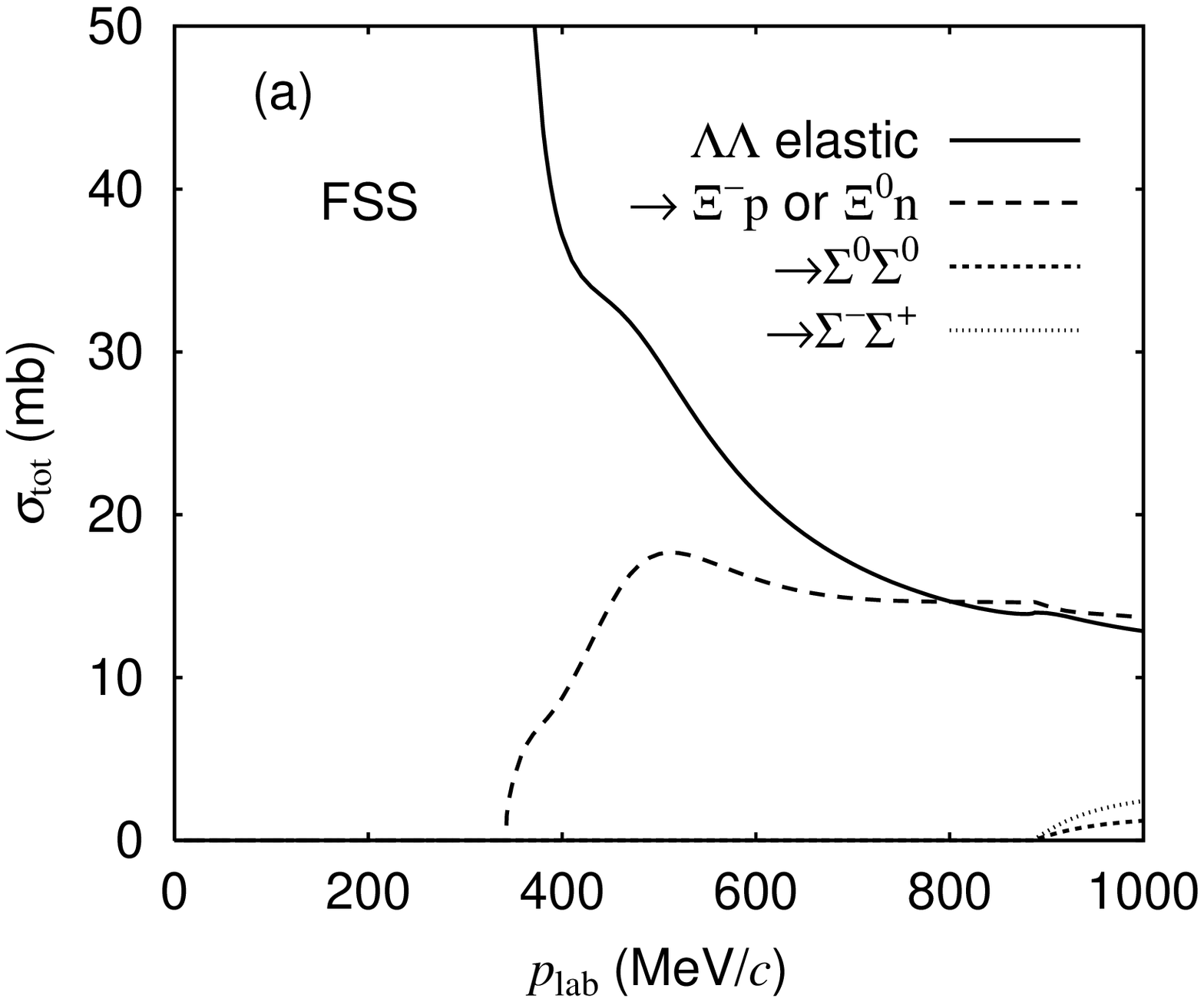}
\end{minipage}~%
\qquad
%\hfill~%
\begin{minipage}{0.48\textwidth}
\epsfxsize=0.9\textwidth
\epsffile{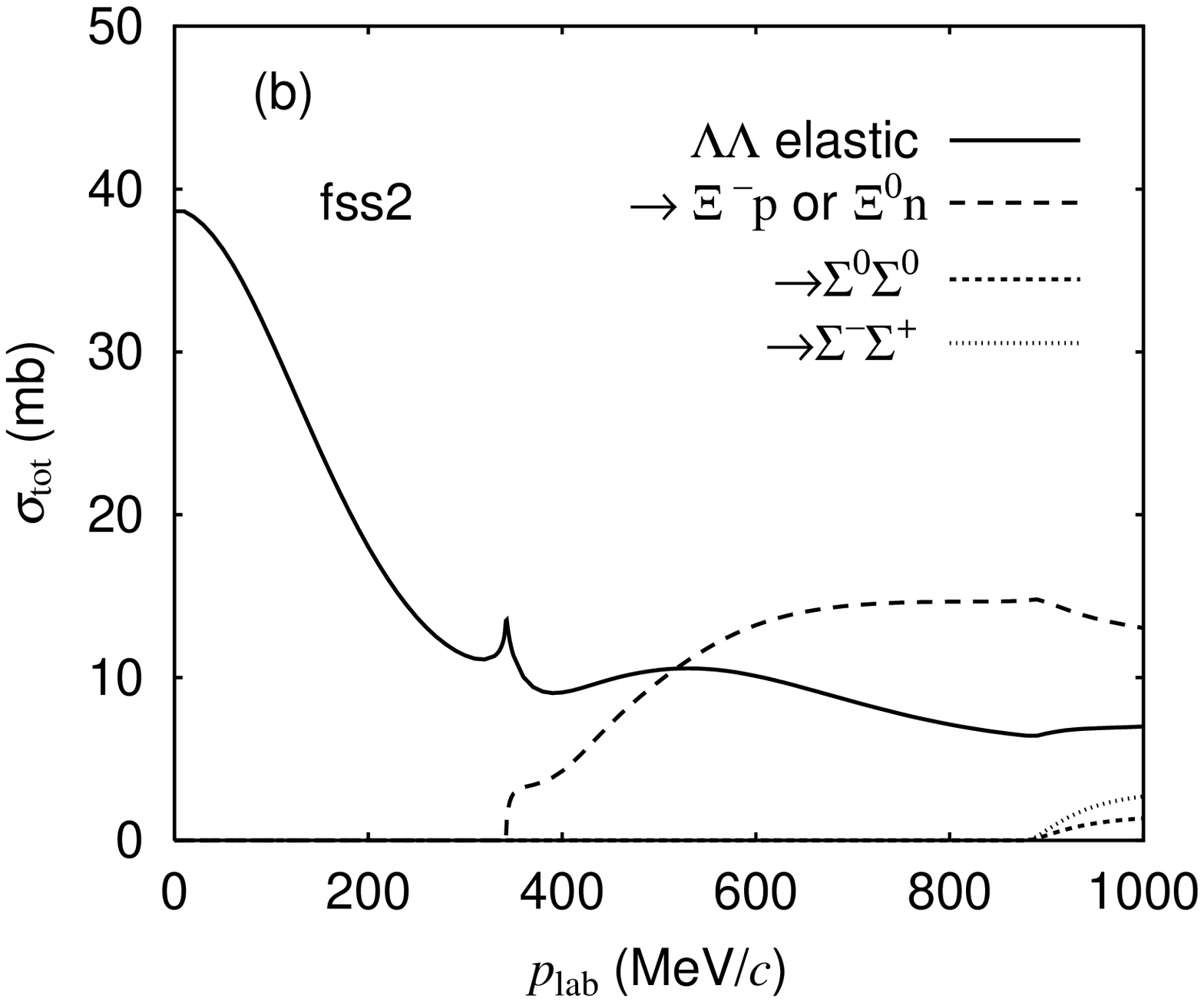}
\end{minipage}
\bigskip
\caption{
Total cross sections for the $\Lambda\Lambda$ scattering,
predicted by FSS (a) and by fss2 (b).
}
\label{LLtot}
\end{center}
\end{figure}

The $P$-wave phase shifts of the $\Lambda \Lambda (I=0)$,
$\Xi N (I=0)$ and $\Xi N (I=1)$ scatterings are
illustrated in Fig.~\ref{phxnP_FSS} for FSS and
in Fig.~\ref{phxnP_fss2} for fss2.
A step-like resonance appears in the $\hbox{}^3P_0$ states
of the $\Lambda \Lambda$ channel (FSS) and the $\Xi N (I=0)$
channel (fss2).
The origin of these resonances is the $\Xi N (I=0)$ $\hbox{}^3P_0$
resonance produced by the non-central forces, i.e.,
the tensor and $LS$ forces. 
This resonance is not clearly seen
in the $\Xi^- p$ total cross sections in Fig.~\ref{XNtot},
due to the small statistical weight of $J=0$.
The shift of the $\Xi N (I=0)$ resonance to
the $\Lambda \Lambda$ channel for FSS resembles the
situation in the $\Sigma N(I=1/2)$ $\hbox{}^3P_1$ resonance,
discussed in Sec.~3.2.3.
If the step-like resonance
in the $\Sigma N(I=1/2)$ $\hbox{}^3P_1$ channel moves 
to the $\Lambda N$ $\hbox{}^1P_1$ channel
by the $\Lambda N$--$\Sigma N$ channel coupling,
the $\Lambda p$ total cross section is enhanced
in the cusp region \cite{RGMFb,FSS}. 
Similarly, the step-like resonance
in the $\Lambda \Lambda$ $\hbox{}^3P_0$ channel
due to the $\Lambda\Lambda$--$\Xi N$ channel coupling
leads to the enhanced $\Lambda\Lambda$ total cross sections
by FSS in the momentum region $p_{\rm lab} \approx 400~\hbox{MeV}/c$.
The total cross section curves for the $\Lambda \Lambda$ scattering
clearly show this situation, as displayed in Fig.~\ref{LLtot}.

\bigskip

\noindent
{\bf [\mib{\Sigma\Sigma} interaction]}

\bigskip

The properties of the single-channel $\Sigma \Sigma$ interactions
are extensively studied in Ref.~\cite{NA97}.
The $\Sigma\Sigma (I=0)$ interaction in the $\hbox{}^1S_0$ state
is strongly attractive
due to the color-magnetic interaction. 
The central force of the $\Sigma \Sigma (I=1)$ interaction 
in the $\hbox{}^3S_1$ state is repulsive due to the Pauli effect. 
Since these $I=0$ and 1 $\Sigma \Sigma$ configurations
couple to other two-baryon continua,
a direct experimental confirmation
of these single-channel properties is not easy.
On the other hand, the $I=2$ $\Sigma\Sigma$ channel does
not couple to any other two-baryon channels. 
Since the $\Sigma\Sigma (I=2)$ $\hbox{}^1S_0$ state consists
of the pure (22) state, the behavior of this interaction
is similar to that of the $NN$ $\hbox{}^1S_0$ state. 
Figure \ref{intro1}(a) shows that this is indeed the case for fss2
and the $\hbox{}^1S_0$ phase shifts
of the $NN$ and $\Sigma \Sigma (I=2)$ channels are very similar.
Consequently, the $\Sigma^- \Sigma^-$ and $pp$ total cross sections
in Fig.~\ref{intro1}(b) are almost equal.

\subsubsection{\mib{S=-3} systems}

\bigskip

\noindent
{\bf [\mib{\Xi\Lambda} interaction]}

\bigskip

\begin{figure}[b]
\begin{center}
\begin{minipage}{0.48\textwidth}
\epsfxsize=0.9\textwidth
\epsffile{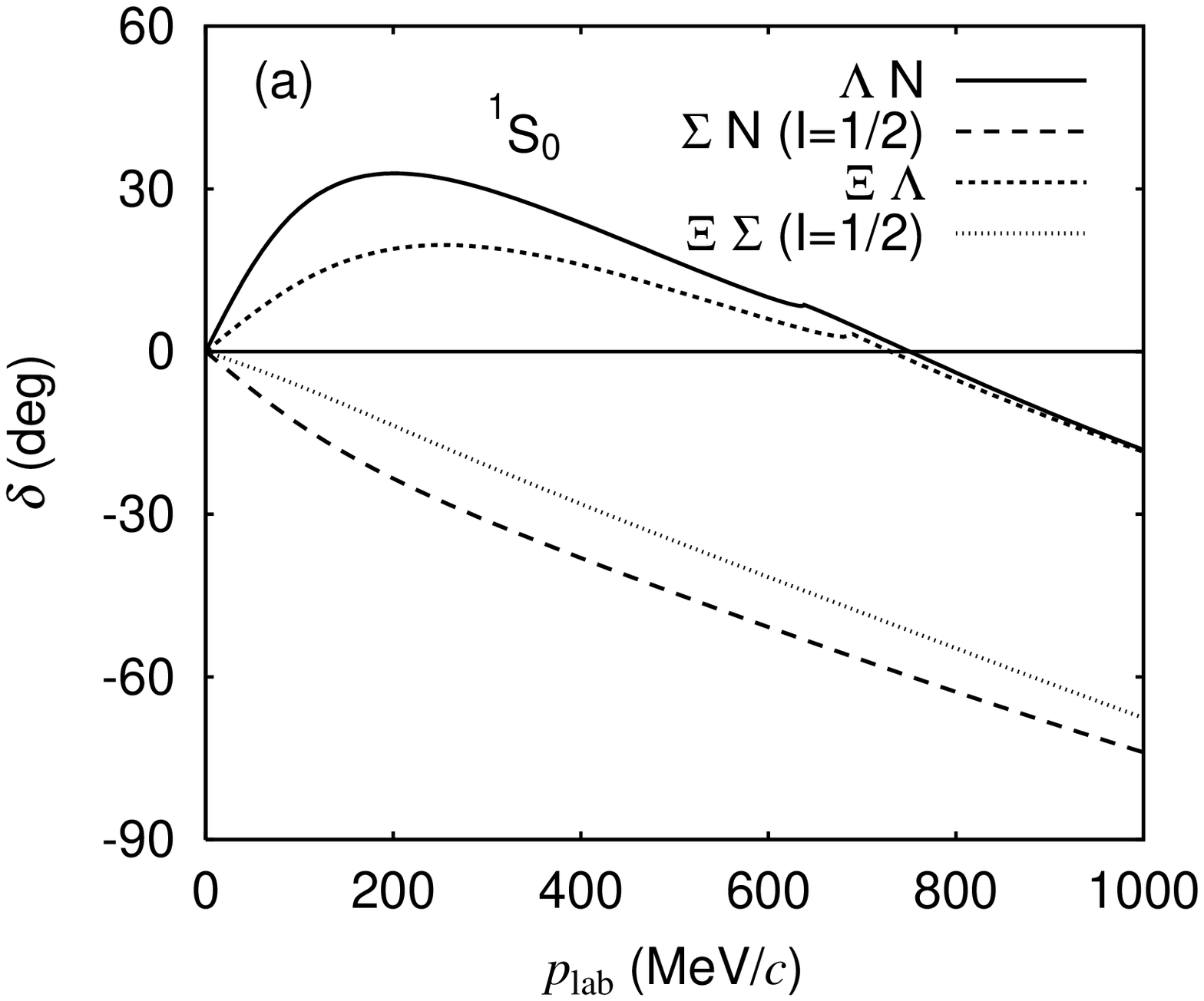}
\end{minipage}~%
\qquad
%\hfill~%
\begin{minipage}{0.48\textwidth}
\epsfxsize=0.9\textwidth
\epsffile{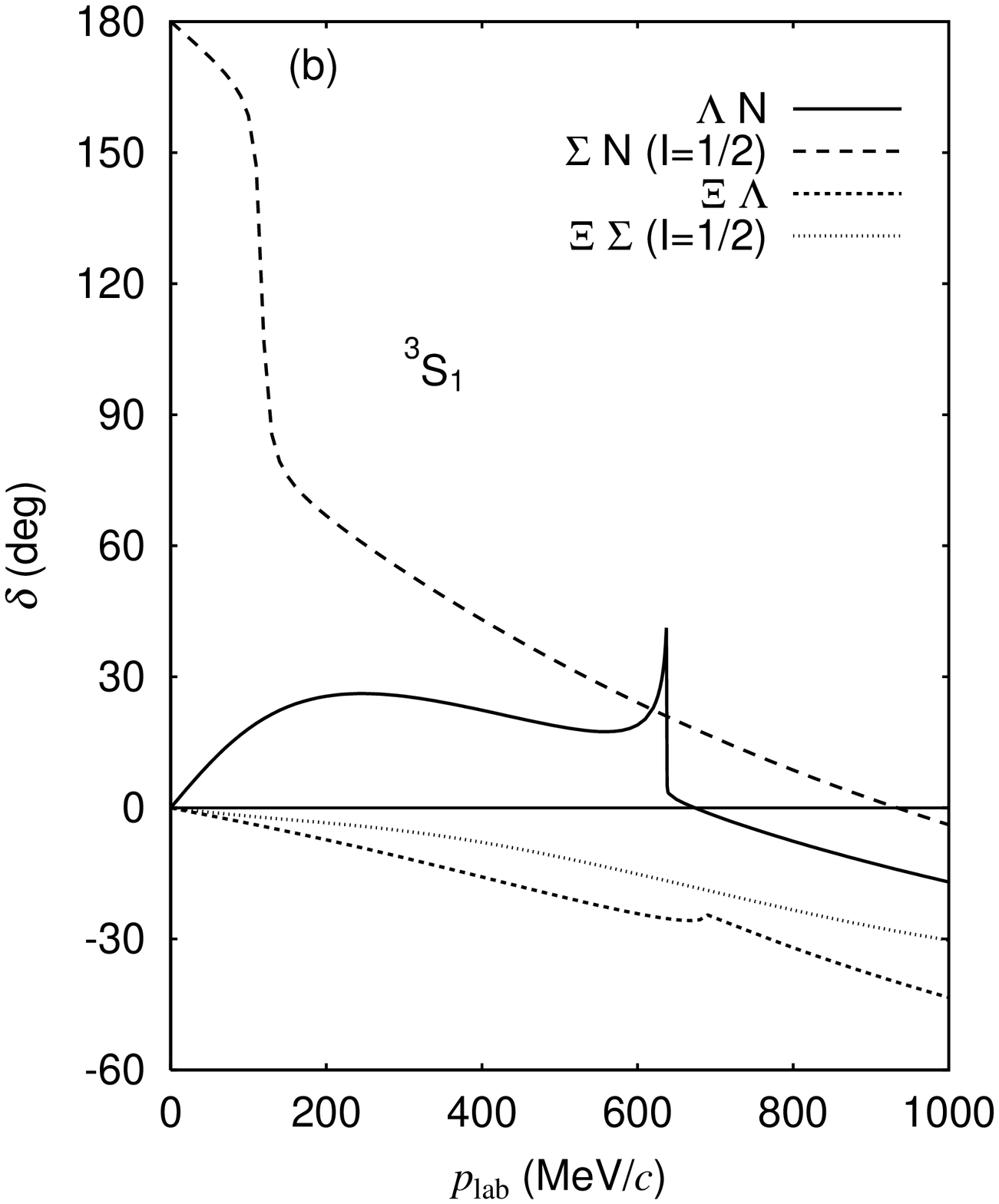}
\end{minipage}
\bigskip
\caption{
(a) $\hbox{}^1S_0$ phase shifts for the $\Lambda N$,
$\Sigma N(I=1/2)$, $\Xi \Lambda$, and $\Xi \Sigma(I=1/2)$ interactions,
predicted by fss2.
(b) The same as (a) but for the $\hbox{}^3S_1$ state.
}
\label{xlxs}
\end{center}
\end{figure}

As discussed in Sec.~3.2.1, the $\Xi \Lambda$--$\Xi \Sigma$
interactions are most easily understood on the analogy of
the $\Lambda N$--$\Sigma N$ interactions. The replacement
of $N$ with $\Xi$ induces the transition
from the (03) to (30) $SU_3$ symmetry. This implies that the attractive
feature of the spin-triplet deuteron channel is no longer
taken over to the $\Xi \Lambda$--$\Xi \Sigma (I=1/2)$ CC system.
Figures \ref{xlxs}(a) and \ref{xlxs}(b) 
illustrate this situation clearly for the $\hbox{}^1S_0$
and $\hbox{}^3S_1$ phase shifts predicted by fss2.
For the flavor-symmetric $\hbox{}^1S_0$ state, the $SU_3$
contents of the $\Xi \Lambda$--$\Xi \Sigma (I=1/2)$ state
are exactly the same as those of the $\Lambda N$--$\Sigma N (I=1/2)$.
Thus the $\Xi \Lambda$ and $\Xi \Sigma (I=1/2)$ phase shifts
are very similar to the $\Lambda N$ and $\Sigma N (I=1/2)$
phase shifts, respectively, with a slight decrease
of the magnitude due to the FSB.
On the other hand, the flavor-antisymmetric $\hbox{}^3S_1$ state
of the $\Xi \Lambda$ and $\Xi \Sigma (I=1/2)$ channels
yields very similar repulsive phase shifts, related to the
strong repulsion of the pure (30) $\Sigma N (I=3/2)$ interaction. 
The fact that the magnitude of the repulsion is almost half
of the latter is consistent with the weakly attractive nature
of the $SU_3$ $(11)_a$ $\Xi N (I=0)$ $\hbox{}^3S_1$ interaction
(see Fig.~\ref{nn3s}(a)).
The $\Xi\Lambda$--$\Xi\Sigma$ channel coupling gives
a very small cusp at the $\Xi \Sigma$ threshold
both in the $\hbox{}^1S_0$ and $\hbox{}^3S_1$ states. 
It should be stressed that, in the $^3S_1$ state,
the $\Xi \Lambda$ interaction does not resemble 
the $\Lambda N$ interaction even qualitatively,
since the $\Lambda N$ $\hbox{}^3S_1$ state consists
of the $(11)_a$ and (03) components, 
while the $\Xi \Lambda$ $\hbox{}^3S_1$ state
consists of $(11)_a$ and (30) components. 

It is reported through Faddeev calculations
of light $\Xi \Lambda$ hypernuclei 
that $\hbox{}_{\Xi \Lambda}^{\ \,6} \hbox{He}$ makes
the onset of nuclear stability for $\Xi$ hyperons \cite{FG02}.
These authors, however, used
an attractive $\Xi \Lambda$ effective force
which simulates the strongly attractive $\Xi \Lambda$ interaction 
predicted by the Nijmegen soft-core NSC97 model \cite{NSC97,SR99}. 
In view of the relationship between the $\Xi \Lambda$
and $\Lambda N$ channels for $\hbox{}^1S_0$
and of the repulsive $\Xi \Lambda$ interaction for $\hbox{}^3S_1$,
it is very unlikely that fss2 predicts
$\hbox{}_{\Xi \Lambda}^{\ \,6} \hbox{He}$ as a 
lightest particle-stable $S=-3$ hypernucleus. 

\bigskip

\noindent
{\bf [\mib{\Xi \Sigma} interaction]}

\bigskip

As discussed above, the $\Xi \Sigma$ interactions with $I=1/2$
are repulsive for both $\hbox{}^1S_0$ and $\hbox{}^3S_1$ states
for the kinematical reason.
The former is due to the $(11)_s$ Pauli-forbidden state
and the latter due to the almost-forbidden (30) component.
One cannot expect the strong $\Xi \Lambda$--$\Xi \Sigma$ coupling
by the one-pion exchange tensor force
unlike the $\Lambda N$--$\Sigma N$ channel coupling,
since the $SU_3$ relation of the meson-baryon coupling
yields $f_{\Xi \Xi \pi}/f_{NN\pi}=2\alpha_m-1=-1/5$ for
the magnetic-type pseudoscalar coupling with $\alpha_m=2/5$
(the $SU_6$ relation).
These features are clearly seen in the phase shifts 
drawn in Figs.~\ref{xlxs}(a) and \ref{xlxs}(b). 

On the other hand, the $\Xi \Sigma$ interactions with $I=3/2$
are attractive for both $\hbox{}^1S_0$ and $\hbox{}^3S_1$ states.
Table \ref{symmetry} shows that the $\hbox{}^1S_0$ state has
the pure (22) symmetry, shared with the $NN$ $\hbox{}^1S_0$ state,
and the $\hbox{}^3S_1$ state has the (03) symmetry
of the $NN$ $\hbox{}^3S_1$ state. Of prime importance are the FSB
and the fact that the $\Xi \Sigma$ system is composed of
two non-identical particles.
For the pion-exchange contributions, the $SU_3$ relations
of the coupling constants are also important.
The $\Xi \Sigma (I=3/2)$ $\hbox{}^1S_0$ phase shift
in Fig.~\ref{intro1}(a) shows that the effect of the
FSB is fairly large for fss2 and
the phase-shift rise is about $40\,\hbox{-}\,50^\circ$. 
On the other hand, the similar magnitude of the attraction
of the $\Xi \Sigma (I=3/2)$ $\hbox{}^3S_1$ phase
shift (Fig.~\ref{nn3s}(a)) is mainly related to the role
of the exchange Feynman diagram, which is entirely 
different from the $NN$ system. 
Unlike the (03) state in $NN$,
the (03) state in $\Xi\Sigma (I=3/2)$ is 
only moderately attractive, since the $\Xi \Sigma$ system
does not allow the strong one-pion exchange
in the exchange Feynman diagram. The reduction
of the attraction in the direct Feynman diagram
is also expected from the $SU_3$ relations
for the $\Xi \Xi \pi$ coupling.
In summary, the $\Xi\Sigma$ interaction predicted
by the QM baryon-baryon interaction has a large 
isospin dependence like the $\Sigma N$ interaction,
namely, the repulsive 
behavior in the $I=1/2$ state and the attractive behavior 
in the $I=3/2$ state. 
In particular, fss2 predicts no bound state
in the $\Xi^-\Sigma^-~^1S_0$ state. 

\subsubsection{\mib{S=-4} systems}

The $SU_3$ classification of the $\Xi \Xi$ system shows that
the $I=0$ system is represented by the pure (30) symmetry,
while the $I=1$ system by the pure (22) symmetry.
We see in Figs.~\ref{nn3s}(a) and \ref{intro1}(a)
that the $\Xi \Xi (I=0)$ interaction is repulsive
like the $\Sigma N (I=3/2)$ one in the $^3S_1$ state 
and the $\Xi \Xi (I=1)$ interaction is attractive
like the $NN (I=1)$ one in the $^1S_0$ state.
In both cases, the effect of the FSB is appreciable.
The prediction by fss2 gives no bound state
in the $\Xi^0 \Xi^0$ $\hbox{}^1S_0$ state.
The $\Xi \Xi$ total cross sections are not so large
as the $NN$ ones, as seen in Fig.~\ref{intro1}(b).

\subsection{Triton and hypertriton Faddeev calculations}

\subsubsection{Three-nucleon bound state}

Since our QM $B_8 B_8$ interaction describes the short-range repulsion
very differently from the meson-exchange potentials, it is interesting
to examine the three-nucleon system predicted by fss2 and FSS.
Here we solve the Faddeev equation for $\hbox{}^3\hbox{H}$ in
the framework discussed in Sec.~2.4 directly
using the QM RGM kernel in the isospin basis \cite{triton}.
For the $NN$ system, we have no Pauli-forbidden state.
However, the energy dependence of the two-cluster
RGM kernel should be treated properly
through the self-consistency condition (\ref{three1-2}).
Our Faddeev calculation is the full 50-channel calculation,
including all the angular momentum up to $J=6$ for
the $NN$ interaction, and the triton energy
converges almost completely as seen
in Table \ref{tabfad3} \cite{PANIC02}.
The fss2 prediction, $-8.52$ MeV, seems to be overbound
compared to the experimental
value $E^{\rm exp}(\hbox{}^3\hbox{H})=-8.482$ MeV.
In fact, this is not the case,
since the present calculation is carried out
using the $np$ interaction.
The charge dependence of the $NN$ interaction indicates 
that the $\hbox{}^1S_0$ interaction of the $nn$ system
is less attractive than that of the $np$ system.
The effect of the charge dependence is estimated
to be $-0.19$ MeV for the triton binding
energy \cite{MA89,Bra88a}.
If we take this into account,
our result is still 0.15 MeV less bound.

\begin{table}[b]
\caption{
The three-nucleon bound state properties predicted by the Faddeev
calculation with fss2 and FSS. The $np$ interaction is used
in the isospin basis. 
The heading ``No. of channels'' implies the number of two-nucleon
channels included; $n_{\rm max}$ is the dimension
of the diagonalization for the Faddeev equation;
$\varepsilon_{NN}$ the $NN$ expectation value
determined self-consistently;
$E(\hbox{}^3\hbox{H})$ the ground-state energy; and 
$\protect\sqrt{\langle r^2\rangle_{\hbox{}^3{\rm H}}}$ 
($\protect\sqrt{\langle r^2\rangle_{\hbox{}^3{\rm He}}}$)
the charge rms radius for $\hbox{}^3{\rm H}$ ($\hbox{}^3{\rm He}$),
including the proton and neutron size corrections
through \protect\eq{rms1}. The Coulomb force and
the relativistic corrections are neglected.}
\renewcommand{\arraystretch}{1.1}
\setlength{\tabcolsep}{4mm}
\begin{center}
\begin{tabular}{ccccccc}
\hline
model & No. of & $n_{\rm max}$ & $\varepsilon_{NN}$
 & $E(\hbox{}^3\hbox{H})$
 & $\protect\sqrt{\langle r^2\rangle_{\hbox{}^3{\rm H}}}$
 & $\protect\sqrt{\langle r^2\rangle_{\hbox{}^3{\rm He}}}$ \\
 & channels & & (MeV) & (MeV)  & (fm) & (fm) \\
\hline
     &  2 ($S$)       &   2,100 & 2.361 & $-7.807$ & 1.80 & 1.96 \\
     &  5 ($SD$)      &   5,250 & 4.341 & $-8.189$ & 1.75 & 1.92 \\
fss2 & 10 ($J \le 1$) &  10,500 & 4.249 & $-8.017$ & 1.76 & 1.94 \\
     & 18 ($J \le 2$) &  18,900 & 4.460 & $-8.439$ & 1.72 & 1.90 \\
     & 34 ($J \le 4$) &  35,700 & 4.488 & $-8.514$ & 1.72 & 1.90 \\
     & 50 ($J \le 6$) & 112,500 & 4.492 & $-8.519$ & 1.72 & 1.90 \\
\hline
     &  2 ($S$)       &   2,100 & 2.038 & $-7.675$ & 1.83 & 1.99 \\
     &  5 ($SD$)      &   5,250 & 3.999 & $-8.034$ & 1.78 & 1.95 \\
FSS  & 10 ($J \le 1$) &  10,500 & 3.934 & $-7.909$ & 1.78 & 1.97 \\
     & 18 ($J \le 2$) &  18,900 & 4.160 & $-8.342$ & 1.74 & 1.93 \\
     & 34 ($J \le 4$) &  35,700 & 4.175 & $-8.390$ & 1.74 & 1.92 \\
     & 50 ($J \le 6$) & 112,500 & 4.177 & $-8.394$ & 1.74 & 1.92 \\
\hline
\end{tabular}
\end{center}
\label{tabfad3}
\end{table}

For a realistic calculation of the $\hbox{}^3\hbox{H}$ binding
energy, it is essential to use such an $NN$ interaction that
reproduces both the correct $D$-state probability $P_D$ of
the deuteron and the effective range parameters
of the $\hbox{}^1S_0$ scattering \cite{Bra88b}.
Since all the realistic $NN$ interactions reproduce the $NN$ phase
shifts more or less correctly, the strength of the central
attraction is counterbalanced with that of the tensor force.
Namely, if the interaction has a weaker tensor force, then
it should have a stronger central attraction.
Generally speaking, the effect of the tensor force is
reduced in the nuclear many-body systems compared to
the bare two-nucleon collision.
Namely, the $NN$ interaction with a weaker tensor force
is favorable in order to obtain sufficient binding energies
of the nuclear many-body systems.
The weak tensor force, however, causes various problems
such as a too small deuteron quadrupole moment $Q_d$ and
the disagreement of the mixing parameter $\varepsilon_1$ of
the $\hbox{}^3S_1+\hbox{}^3D_1$ coupling.

\begin{figure}[t]
\begin{center}
\begin{minipage}{0.48\textwidth}
\epsfxsize=0.9\textwidth
\epsffile{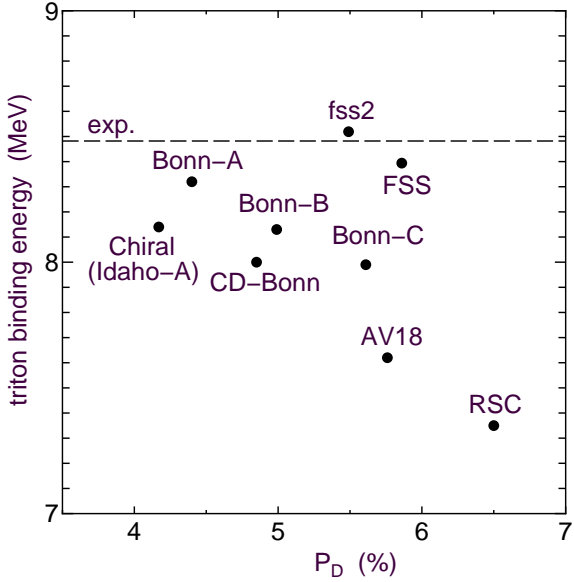}
\end{minipage}~%
\qquad
%\hfill~%
\begin{minipage}{0.48\textwidth}
\caption{Calculated $\hbox{}^3\hbox{H}$ binding
energies $B_t$ as a function
of the deuteron $D$-state probability $P_D$.
All the calculations are made in the isospin basis, using the 
$np$ interaction, except for the CD-Bonn potential.
The calculated values are taken
from \protect\cite{MA89} (RSC, Bonn-A, B, C),
from \protect\cite{NO00} (AV18 \protect\cite{WI95},
CD-Bonn \protect\cite{MA01}), and
from \protect\cite{EM02} (Chiral).
The experimental value, $B_t=8.482$ MeV, is shown by the dashed
line.}
\label{tritbe}
\end{minipage}
\end{center}
\end{figure}

The correlation between the triton binding
energy, $B_t=-E(\hbox{}^3\hbox{H})$, and
the $D$-state probability of the deuteron is plotted
in Fig.~\ref{tritbe} for fss2 and FSS as well as many other
Faddeev calculations using modern realistic meson-exchange potentials.
In the traditional meson-exchange potentials,
the calculated points are located on a straight line,
which is similar to the Coester line for the saturation
point of symmetric nuclear matter.
For example, the Reid soft-core potential
(RSC) \cite{Rei68} gives $P_D=6.5~\%$ and
predicts too small $\hbox{}^3\hbox{H}$ binding energy,
$B_t=7.35~\hbox{MeV}$.
A series of the Bonn potentials reproduce
the $NN$ phase shifts very accurately, but they tend to have
a rather weak tensor force \cite{MA89}.
The model C has the strongest tensor force, $P_D=5.61~\%$,
yielding $B_t=7.99~\hbox{MeV}$.
The value $P_D$ becomes smaller for models B and A,
and the value of $B_t$ becomes larger accordingly.
The following results are given in Ref.~\cite{MA89}:
model-B ($P_D=5.0~\%,~B_t=8.13~\hbox{MeV}$),
model-A ($P_D=4.4~\%,~B_t=8.32~\hbox{MeV}$).
These results are all obtained in the 34 channel calculations
(including the $NN$ total angular momentum $J\leq 4$), 
and by using the $np$ interaction.
In fact, the effects of the charge dependence and
the charge asymmetry are important for the detailed discussion.
The most recent Faddeev calculation employing the CD-Bonn
potential \cite{MA01} incorporates these effects,
and predicts $B_t=8.014~\hbox{MeV}$ \cite{NO97}
for $P_D=4.85~\%$.
The present status of the $\hbox{}^3\hbox{H}$ binding energy
calculation is summarized as follows: more than 0.5 MeV is missing
for any realistic two-nucleon interactions \cite{NO00}.

On the other hand, our result of the $P_D$ value
is 5.49 $\%$ for fss2 and 5.86 $\%$ for FSS.
The deuteron quadrupole moment $Q_d$ is $0.270~\hbox{fm}^2$ for
fss2 and $0.283~\hbox{fm}^2$ for FSS.
Our results of the effective range parameters
listed in Tables \ref{effect} are not as perfect as
those of the meson-exchange potentials,
but the deuteron binding energy and the scattering length $a_s$ for
the $\hbox{}^1S_0$ state are fitted in determining
our QM parameters in the isospin basis.
Under these circumstances, the results of fss2 and FSS in
Fig.~\ref{tritbe} seem to indicate that
they form another $B_t$ - $P_D$ line
deviating from the traditional meson-exchange potentials.
After the correction by the charge dependence
of the $NN$ interaction, our $\hbox{}^3\hbox{H}$ binding energies
are smaller than the experimental value by merely 0.15 MeV for fss
and 0.28 MeV for FSS.
If we attribute these differences to the effect
of the three-nucleon force, it is by far smaller than the
generally accepted values, 0.5 - 1 MeV \cite{NO00},
obtained with the meson-theoretical $NN$ interactions.
A remarkable point of our results is that
we can reproduce enough binding energy of the triton
without reducing the deuteron $D$-state probability.

Table \ref{tabfad3} lists the charge rms radii
of $\hbox{}^3{\rm H}$ and $\hbox{}^3{\rm He}$,
predicted by fss2 and FSS. The finite size corrections
of the nucleons are made through
\begin{eqnarray}
& & \langle r^2\rangle_{\hbox{}^3{\rm H}}
=\left[R_C(\hbox{}^3{\rm H})\right]^2
+(0.8502)^2-2\times (0.3563)^2\ ,\nonumber \\
& & \langle r^2 \rangle_{\hbox{}^3{\rm He}}
=\left[R_C(\hbox{}^3{\rm He})\right]^2
+(0.8502)^2-\frac{1}{2}\times (0.3563)^2\ ,
\label{rms1}
\end{eqnarray}
where ${R_C}^2$ stands for the squared charge radius
for the point nucleons which is obtained from the Faddeev calculations.
Since our three-nucleon bound state wave functions
are given in the momentum representation,
we first calculate the charge form factors $F_C(Q^2)$,
according to the formulation given in Ref.~\cite{GL82b}.
The values ${R_C}^2$ are then extracted from
the power series expansion of $F_C(Q^2)$ with respect to $Q^2$.
%We have employed 20 points, $Q=0.05 \times n~\hbox{fm}^{-1}$
%with $n=1~\hbox{-}~20$, for the extrapolation to $Q=0$.
In the present calculation, the Coulomb force and the relativistic
correction terms \cite{KI88} of the charge current operator
are entirely neglected.
The experimental values are difficult to
determine, as discussed in Ref.~\cite{KI88}. Here we compare
our results with two empirical values
\begin{eqnarray}
\sqrt{\langle r^2\rangle_{\hbox{}^3{\rm H}}}
=\left\{ \begin{array}{c}
1.70\pm 0.05~\hbox{fm} \quad \protect\cite{CO65} \\
1.81\pm 0.05~\hbox{fm} \quad \protect\cite{MA86} \\
\end{array}\right.\ \ ,\qquad
\sqrt{\langle r^2\rangle_{\hbox{}^3{\rm He}}}
=\left\{ \begin{array}{c}
1.87\pm 0.05~\hbox{fm} \quad \protect\cite{CO65} \\
1.93\pm 0.03~\hbox{fm} \quad \protect\cite{MA86} \\
\end{array}\right.\ \ .
\label{rms2}
\end{eqnarray}
The agreement with the experiment is satisfactory
both for fss2 and FSS.  

\begin{table}[b]
\caption{
Decomposition of the total triton energy $E$ into
the kinetic-energy and potential-energy contributions:
$E=\langle H_0 \rangle+\langle V \rangle$.
The unit is in MeV. In the present framework,
this is given by the expectation
value $\varepsilon_{NN}$ of the two-cluster Hamiltonian
with respect to the Faddeev solution, which is determined
self-consistently. The results of CD-Bonn \protect\cite{MA01}
and AV18 \protect\cite{WI95} are
taken from Ref. \protect\cite{NO00}.
}
\renewcommand{\arraystretch}{1.1}
\setlength{\tabcolsep}{9mm}
\begin{center}
\begin{tabular}{ccccc}
\hline
model & $\varepsilon_{NN}$ & $E$ & $\langle H_0\rangle$
      & $\langle V \rangle$ \\
\hline
fss2     &  4.492 & $-8.519$ & 43.99 & $-52.51$ \\
FSS      &  4.177 & $-8.394$ & 41.85 & $-50.25$ \\
CD-Bonn  &  3.566 & $-8.012$ & 37.42 & $-45.43$ \\
AV18     &  5.247 & $-7.623$ & 46.73 & $-54.35$ \\
\hline
\end{tabular}
\label{tabfad4}
\end{center}
\end{table}

For the systems composed of three identical particles,
the self-consistent energy of the two-cluster RGM kernel,
$\varepsilon_{NN}$, has a clear physical meaning
related to the decomposition of the total
triton energy $E$ into the kinetic-energy and
potential-energy contributions.
A simple manipulation of the Faddeev equations and
the self-consistency condition yields
\begin{eqnarray}
\langle H_0 \rangle=2 (3 \varepsilon_\alpha-E)\ \ ,\qquad
\langle V \rangle=3 (E-2 \varepsilon_\alpha)\ ,
\label{rms3}
\end{eqnarray}
where the expectation value is calculated
with respect to the three-nucleon total wave function.
From the first equation, the $\varepsilon_{NN}$
is obtained from the kinetic-energy contribution
through $\varepsilon_{NN}=E/3+\langle H_0 \rangle/6$.
Table \ref{tabfad4} lists this decomposition,
together with the results
of CD-Bonn \cite{MA01} and AV18 \cite{WI95} potentials \cite{NO00}.
We see that our QM results of $\varepsilon_{NN}$ by
FSS and fss2 are just in between these potentials,
which have very different strengths of the tensor force.
The energy-dependence of the $NN$ RGM kernel with
the explicit structure $+\varepsilon K$ has some effects
favorable to give a larger binding energy of the triton,
since the value of the exchange
normalization kernel $K$ is $-1/9$ for the most
compact $(0s)$ wave function between the two nucleon-clusters.
The RGM kernel becomes more attractive for a larger value
of $\varepsilon_{NN}$. Table \ref{tabfad4} indicates
a larger $\langle V \rangle$ for a larger $\varepsilon_{NN}$,
whereas the CD-Bonn and AV18 potentials do not have
such an energy dependence.

The Faddeev calculations for $\hbox{}^3\hbox{H}$ using
the QM $NN$ potentials have also been carried out
by Takeuchi, Cheon and Redish \cite{TA92}
and by the Salamanca-J{\" u}lich group \cite{SA01}.
In the former calculation, the model QCM-A gives the $NN$ phase
shifts with almost the same accuracy as our model FSS,
and predicts $P_D=5.58~\%$ and $B_t=8.01~\hbox{-}~ 8.02~\hbox{MeV}$ in
the 5-channel calculation.
These are very similar to our FSS results. 
On the other hand, the Salamanca-J{\" u}lich group
predicts $B_t=7.72~\hbox{MeV}$, in spite of the small $D$-state
probability $P_D=4.85~\%$. It is not clear to us how they
treated the energy dependence of the RGM kernel
when the separable expansion is introduced for solving the Faddeev
equations. We think that the fit of the $NN$ phase shifts
for higher partial waves, especially for the $P$ waves,
has to be improved in order to extend their calculation
to more than 5 channels. 

\subsubsection{The hypertriton}

\begin{table}[b]
\caption{Results of the hypertriton Faddeev
calculations by fss2 and FSS.
The heading $E$ is the hypertriton
energy measured from the $\Lambda NN$ threshold, $B_\Lambda$
the $\Lambda$ separation energy, $\varepsilon_{NN}$
($\varepsilon_{\Lambda N}$) the $NN$ ($\Lambda N$) expectation
value determined self-consistently,
and $P_\Sigma$ the $\Sigma NN$ probability in percent.
The calculated deuteron binding energy
is $\varepsilon_d=2.2247$ MeV for fss2 and 2.2561 MeV for FSS
($\varepsilon^{\rm exp}_d=2.224644 \pm 0.000046$ MeV \cite{DU83}).
The norm of admixed redundant components is less than $10^{-9}$.}
\renewcommand{\arraystretch}{1.1}
\setlength{\tabcolsep}{5mm}
\begin{center}
\begin{tabular}{@{}ccccccc}
\hline
model & No. of & $E$ & $B_\Lambda$
& $\varepsilon_{NN}$ & $\varepsilon_{\Lambda N}$
& $P_\Sigma$ \\
      & channels & (MeV) & (keV) & (MeV) & (MeV) & $(\%)$ \\
\hline
 &   6 ($S$)  & $-2.362$ & 137 & $-1.815$ & 5.548 & 0.450 \\
 &  15 ($SD$) & $-2.423$ & 198 & $-1.762$ & 5.729 & 0.652 \\
fss2 &  30 ($J \le 1$) & $-2.403$ & 178 & $-1.786$ & 5.664 & 0.615 \\
 & 54 ($J \le 2$) & $-2.498$ & 273 & $-1.673$ & 5.974 & 0.777 \\
%&  78 ($J \le 3$) & $-2.510$ & 285 & $-1.660$ & 6.014 & 0.800 \\
 & 102 ($J \le 4$) & $-2.513$ & 288 & $-1.658$ & 6.022 & 0.804 \\
%& 126 ($J \le 5$) & $-2.514$ & 289 & $-1.657$ & 6.024 & 0.805 \\
 & 150 ($J \le 6$) & $-2.514$ & 289 & $-1.657$ & 6.024 & 0.805 \\
\hline
 &   6 ($S$)  & $-2.910$ & 653 & $-1.309$ & 3.984 & 1.022 \\
 &  15 ($SD$) & $-2.967$ & 710 & $-1.433$ & 6.171 & 1.200 \\
FSS &  30 ($J \le 1$) & $-2.947$ & 691 & $-1.427$ & 6.143 & 1.191 \\
 & 54 ($J \le 2$) & $-3.121$ & 865 & $-1.323$ & 6.467 & 1.348 \\
%&  78 ($J \le 3$) & $-3.128$ & 872 & $-1.320$ & 6.480 & 1.357 \\
 & 102 ($J \le 4$) & $-3.134$ & 877 & $-1.317$ & 6.488 & 1.360 \\
%& 126 ($J \le 5$) & $-3.134$ & 878 & $-1.317$ & 6.488 & 1.361 \\
 & 150 ($J \le 6$) & $-3.134$ & 878 & $-1.317$ & 6.488 & 1.361 \\
\hline
\end{tabular}
\end{center}
\label{tabfad5}
\end{table}

Next, we apply our QM $NN$ and $YN$ interactions
to the hypertriton ($\hbox{}^3_\Lambda \hbox{H}$) which has
the small separation energy of the $\Lambda$-particle,
${B_\Lambda}^{\rm exp}=130 \pm 50~\hbox{keV}$ \cite{hypt}.
Since the $\Lambda$-particle is far apart from
the two-nucleon subsystem, the on-shell properties
of the $\Lambda N$ and $\Sigma N$ interactions are expected to be
well reflected in this system. 
In particular, this system is very useful to learn the relative
strength of $\hbox{}^1S_0$ and $\hbox{}^3S_1$ attractions
of the $\Lambda N$ interaction, since the $\hbox{}^1S_0$ component
plays a more important role than $\hbox{}^3S_1$ in this system
and the available low-energy $\Lambda p$ total cross section
data cannot discriminate many possible combinations
of the $\hbox{}^1S_0$ and $\hbox{}^3S_1$ interactions.
In fact, Refs.~\cite{MI95,MI00,NO02} showed that
most meson-theoretical interactions fail to bind the hypertriton
except for the Nijmegen soft-core potentials NSC89 \cite{NSC89},
NSC97f and NSC97e \cite{NSC97}. It is also pointed out 
in Refs.~\cite{MI95} and \cite{NO02} that a small admixture
of the $\Sigma NN$ components less than 1\% is
very important for this binding.
We therefore carry out the $\Lambda NN$--$\Sigma NN$ CC Faddeev
calculation, by properly taking into account the existence
of the $SU_3$ Pauli-forbidden state $(11)_s$ at the quark level.
The necessary formulation is given in Sec.~2.4.

\begin{table}[t]
\caption{Correlation between $\hbox{}^1S_0$ and $\hbox{}^3S_1$
effective range parameters of various $\Lambda N$ interactions
and the $\Lambda$ separation energy $B_\Lambda$ of the
hypertriton. The $B_\Lambda$ value for NSC89 is taken
from Ref.~\protect\cite{NO02}. The experimental $B_\Lambda$ value
is $130 \pm 50~\hbox{keV}$ \protect\cite{BMZ}.
}
\renewcommand{\arraystretch}{1.1}
\setlength{\tabcolsep}{5mm}
\begin{center}
\begin{tabular}{@{}cccccc}
\hline
model & $a_s$ (fm) & $r_s$ (fm) & $a_t$ (fm) & $r_t$ (fm)
& $B_\Lambda$ (keV) \\
\hline
FSS \protect\cite{FSS} & $-5.41$ & 2.26 & $-1.02$ & 4.20
& 878 \\
fss2 \protect\cite{fss2,B8B8} & $-2.59$ & 2.83 & $-1.60$ & 3.01
& 289 \\
NSC89 \protect\cite{NSC89} & $-2.59$ & 2.90 & $-1.38$ & 3.17
& 143 \\
\hline
\end{tabular}
\label{tabfad6}
\end{center}
\end{table}

Table \ref{tabfad5} shows the results of the Faddeev calculations
using fss2 and FSS.
In the 15-channel calculation including the $S$ and $D$ waves
of the $NN$ and $YN$ interactions,
we have already obtained $B_\Lambda=-\varepsilon_d
-E(\hbox{}^3_\Lambda \hbox{H}) \approx 200$ keV for fss2.
The convergence in the partial-wave expansion is very rapid,
and the total angular-momentum $J \leq 4$ of the baryon pairs
is enough to obtain 1 keV accuracy.
With 150-channel $\Lambda NN$ and $\Sigma NN$ configurations
we obtain $B_\Lambda=289$ keV with
the $\Sigma NN$ component $P_\Sigma=0.80\,\%$ for the
fss2 prediction,
and $B_\Lambda=878$ keV with $P_\Sigma=1.36\,\%$ for FSS.

Table \ref{tabfad6} shows the correlation
between the $\Lambda$ separation
energy $B_\Lambda$ and the $\hbox{}^1S_0$ and $\hbox{}^3S_1$
$\Lambda N$ effective range parameters of FSS, fss2 and NSC89.
Although all of these $\Lambda N$ interactions reproduce
within the experimental error bars
the low-energy $\Lambda N$ total cross section data
below about $p_\Lambda= 300~\hbox{MeV}/c$,
our QM interactions seem to be slightly more
attractive than the NSC89 potential \cite{NSC89}.
The model FSS gives a large overbinding
since the $\hbox{}^1S_0$ $\Lambda N$ interaction is strongly
attractive. The phase-shift difference
of the $\hbox{}^1S_0$ and $\hbox{}^3S_1$ states
at about $p_\Lambda= 200~\hbox{MeV}/c$ is $\Delta \delta
=\delta(\hbox{}^1S_0)-\delta(\hbox{}^3S_1) \approx 30^\circ$ for FSS,
while $\Delta \delta \approx 10^\circ$ for fss2. 
Since the fss2 result is still slightly overbound,
this difference should be made smaller in order to reproduce
the experimental value.
From the two results given by fss2 and FSS, we extrapolate
the desired difference to be $0^\circ$ - $2^\circ$ in our QM.
The same conclusion is also obtained in Ref.~\cite{NE02}, using
simulated interactions of the Nijmegen models.

\begin{table}[b]
\caption{Decomposition of the $NN$ and $YN$ expectation
values ($\varepsilon_{NN}$ and $\varepsilon_{YN}$), the
deuteron energy ($-\varepsilon_d$) and the total three-body
energy $E$ to the kinetic-energy and potential-energy
contributions. The unit is in MeV. 
The results for NSC89 are taken from Ref.~\protect\cite{MI95}.}
\renewcommand{\arraystretch}{1.1}
\setlength{\tabcolsep}{10mm}
\begin{center}
\begin{tabular}{@{}cc@{}cc}
\hline
model & $h_{NN}+v_{NN}=\varepsilon_{NN}$ &
& $h_d+v_d=-\varepsilon_d$ (deuteron) \\
\hline
FSS   & $19.986-21.303=-1.317$ & & $16.982-19.238=-2.256$ \\
fss2  & $19.371-21.029=-1.657$ & & $17.495-19.720=-2.225$ \\
NSC89 & $20.48 -22.25 =-1.77$  & & $19.304-21.528=-2.224$ \\
\hline
model & $h_{YN}+v_{YN}=\varepsilon_{YN}$ &
& $\langle H \rangle+\langle V \rangle=E$ \\
\hline
FSS   & $10.036-4.602=5.435$ & & $27.372-30.506=-3.134$ \\
fss2  &  $8.066-2.667=5.400$ & & $23.849-26.362=-2.514$ \\
NSC89 &  $7.44 -3.54 =3.90$  & & $23.45 -25.79 =-2.34$  \\
\hline
\end{tabular}
\end{center}
\label{tabfad7}
\end{table}

In order to make sure that this extrapolation gives
a good estimation,
we modify the $\kappa$-meson mass of the model fss2
from the original value,
$m_\kappa=936$ MeV (see Table \ref{table1}), to 1,000 MeV,
and repeat the whole calculation.
This modification makes the $\hbox{}^1S_0$ $\Lambda N$ interaction
less attractive and the $\hbox{}^3S_1$ more attractive.
We obtain $B_\Lambda=145$ keV with $P_\Sigma=0.53\,\%$,
which is very close to the NSC89 prediction 
$B_\Lambda=143$ keV with $P_\Sigma=0.5\,\%$ \cite{MI95,NO02}.
The effective range parameters of this modified fss2 interaction
are $a_s=-2.15$ fm, $r_s=3.05$ fm,
and $a_t=-1.80$ fm, $r_t=2.87$ fm. The phase-shift difference is
only $1.3^\circ$ and the increase of the total cross section
of the $\Lambda N$ scattering is at most 10 mb at
$p_{\rm \Lambda}=100~\hbox{MeV}/c$ (from 286 mb to 296 mb),
which is still within the experimental error bars.
It should be kept in mind that the effective range parameters
or the $S$-wave phase-shift values determined in this way is
very much model dependent, since the $B_\Lambda$ value is not
determined by these quantities alone. It depends on how higher
partial waves contribute and also on the details
of the $\Lambda N$--$\Sigma N$ coupling of a particular model.
The reliability of the extrapolation shown here is based
on the fact that the models fss2 and FSS have a common framework
for the quark sector and EMEP's. 

Table \ref{tabfad5} also shows that the expectation value
of the $NN$ Hamiltonian, $\varepsilon_{NN}$, determined
self-consistently is close
to the deuteron energy $-\varepsilon_d$, especially in fss2.
This feature is even more conspicuous if these energies are decomposed
to the kinetic-energy and potential-energy contributions.
Table \ref{tabfad7} shows this decomposition with respect
to fss2, FSS and NSC89. (For this comparison,
we use the definition of the kinetic-energy part of the deuteron,
given by $h_d=\langle \chi_d|h_{NN}|\chi_d \rangle/\langle \chi_d|
\chi_d \rangle$, where $\chi_d$ is the RGM relative wave function
between the neutron and the proton.)
In fss2, the kinetic energy of the $NN$ subsystem
is 1.88 MeV larger than that of the deuteron,
which suggests that the $NN$ subsystem
shrinks by the effect of the outer $\Lambda$-particle, 
in comparison with the deuteron in the free space.
In NSC89, this difference is smaller, i.e., 1.18 MeV.
These results are consistent with the fact
that the hypertriton in NSC89 is more loosely
bound ($B_\Lambda=143$ keV) \cite{NO02} than in fss2 (289 keV),
and the $\Lambda$-particle is very far apart from the $NN$ subsystem.
Table \ref{tabfad7} lists the kinetic-energy and
potential-energy decompositions
for the $\Lambda N$--$\Sigma N$ averaged $YN$ expectation
value $\varepsilon_{YN}$ and the total energy $E$.
The kinetic energies of $\varepsilon_{YN}$ are much smaller
than those of $\varepsilon_{NN}$, which indicates that the
relative wave function between the hyperon and the nucleon is
widely spread in the configuration space.
The comparison of the total-energy decomposition shows that
the wave functions of fss2 and NSC89 may be very similar.
A clear difference between them, however, appears
in the roles of the higher partial waves. The energy increase
due to the partial waves higher than the $S$ and $D$ waves
is 91 keV in fss2 and 168 keV in FSS, respectively.
On the other hand, the results
in Refs.~\cite{MI95} and \cite{NO02} indicate that
this is only 20 - 30 keV in the case of NSC89.
This difference originates from both
of the $NN$ and $YN$ interactions. Since the characteristics 
of the meson-theoretical $YN$ interactions in higher partial
waves are {\em a priori} unknown, more detailed analysis of the
fss2 results sheds light on the adequacy
of the QM baryon-baryon interactions.

\subsection{Application to \mib{\hbox{}_\Lambda^9 \hbox{Be}} and
\mib{\hbox{}_{\Lambda \Lambda}^{\ \,6} \hbox{He}} systems}

\subsubsection{The \mib{\alpha \alpha \Lambda} system
for \mib{\hbox{}^9_{\Lambda}\hbox{Be}}}

\begin{figure}[t]
\begin{center}
\begin{minipage}[ht]{0.48\textwidth}
\includegraphics[angle=-90,width=0.9\textwidth]{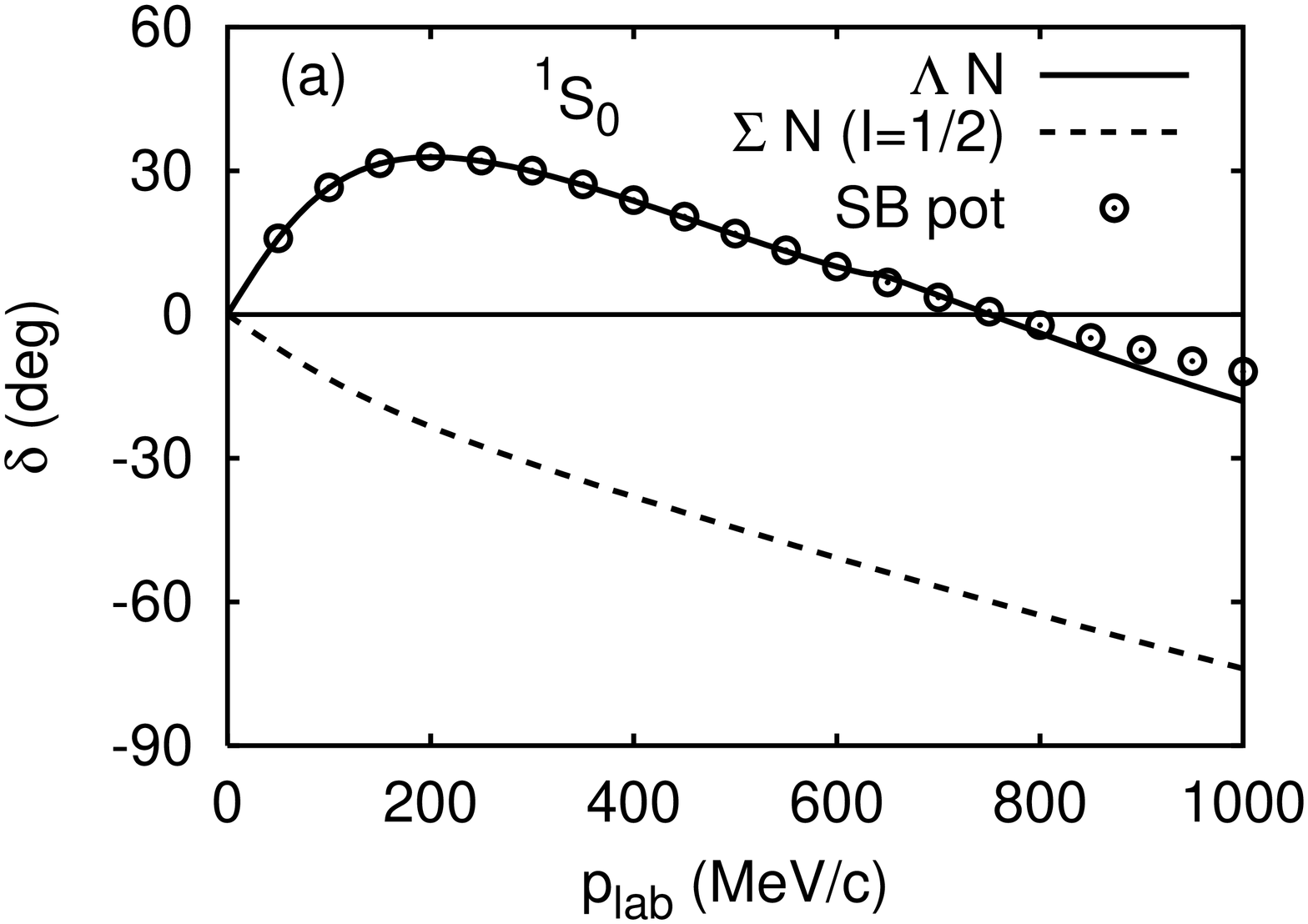}
\end{minipage}
%
%\qquad
\hfill
\begin{minipage}[ht]{0.48\textwidth}
\includegraphics[angle=-90,width=0.9\textwidth]{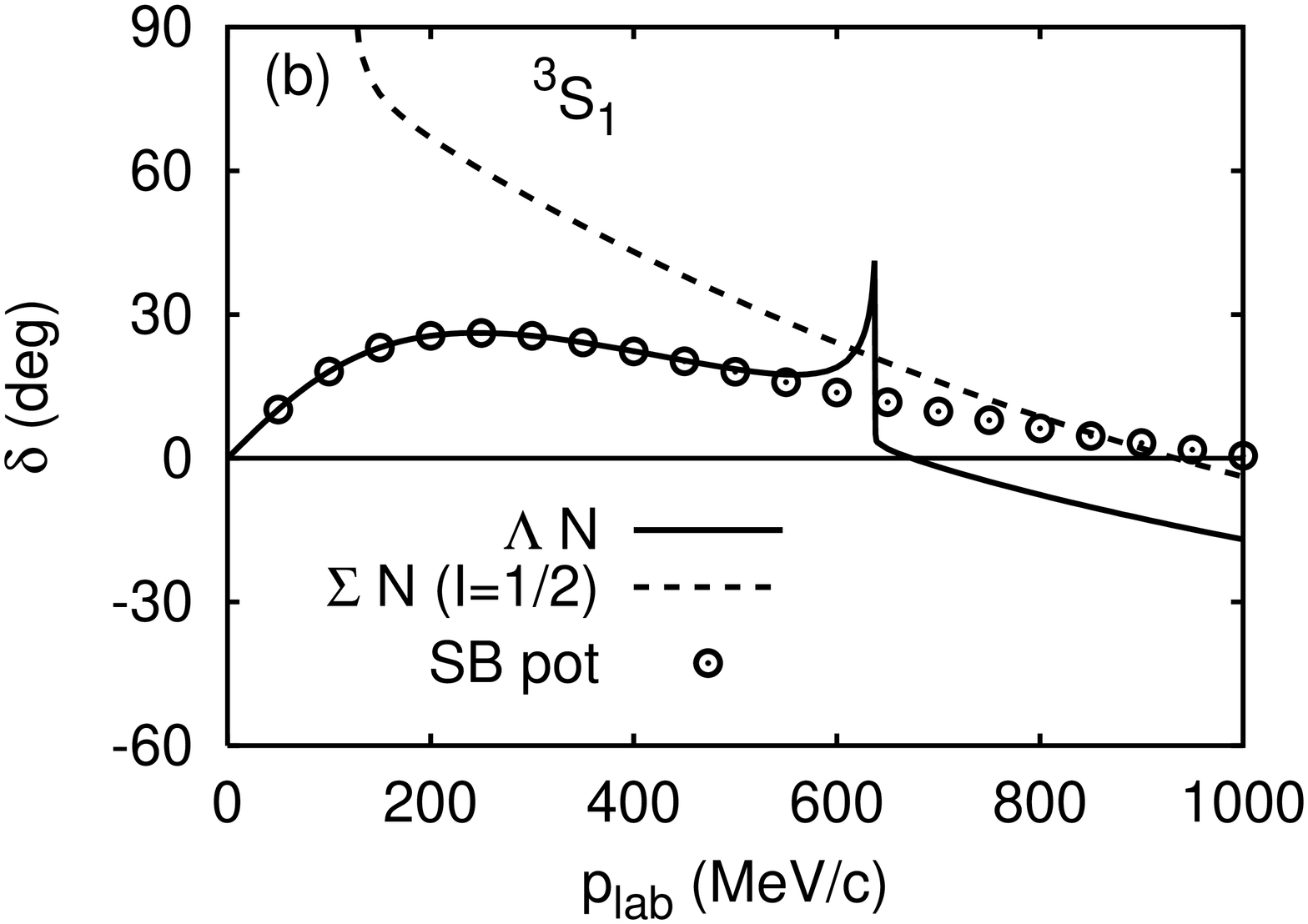}
\end{minipage}
\caption{$\Lambda N$--$\Sigma N$ $\hbox{}^1S_0$ (a)
and $\hbox{}^3S_1$ (b) phase shifts
for the isospin $I=1/2$ channel,
calculated with fss2 (solid and dashed curves)
and the single-channel SB potentials ($f=1$)
in \protect\eq{la5} (circles).}
\label{fig12}
\end{center}
\end{figure}

As a typical example of three-cluster systems composed of
two identical clusters, we apply the present formalism
to the $\alpha \alpha \Lambda$ Faddeev calculation
for $\hbox{}^9_{\Lambda}\hbox{Be}$, using the $\alpha \alpha$
RGM kernel and the $\Lambda \alpha$ folding potential generated 
from a simple $\Lambda N$ effective interaction \cite{2al,2aljj}.
For the $\alpha \alpha$ RGM kernel,
we use the three-range Minnesota
force \cite{TH77} with the Majorana exchange mixture $u=0.94687$,
and the h.o. size parameter $\nu=0.257~\hbox{fm}^{-2}$.
The effective $\Lambda N$ central interaction,
denoted SB in Table \ref{tabfad8}, is a Minnesota-type potential
\begin{eqnarray}
v_{\Lambda N}& = & \left[\,v(\hbox{}^1E) \frac{1-P_\sigma}{2}
+v(\hbox{}^3E) \frac{1+P_\sigma}{2}\,\right]
\left[\,\frac{u}{2}+\frac{2-u}{2}P_r\,\right]\ ,
\label{la4}
\end{eqnarray}
where $v(\hbox{}^1E)$ and $v(\hbox{}^3E)$ are simple two-range
Gaussian potentials generated from the $\hbox{}^1S_0$ and
$\hbox{}^3S_1$ phase shifts predicted
by the QM $\Lambda N$ interaction fss2.
The inversion method based on supersymmetric quantum
mechanics is used to derive phase-shift equivalent
local potentials \cite{SB97}. These potentials are
then fitted by two-range Gaussian functions:
\begin{eqnarray}
v(\hbox{}^1S_0) & = & -128.0~\exp(-0.8908~r^2)
+1015~\exp(-5.383~r^2)\qquad (\hbox{MeV})\ ,\nonumber \\
v(\hbox{}^3S_1) & = & -56.31\,f\,\exp(-0.7517~r^2)
+1072~\exp(-13.74~r^2)\qquad (\hbox{MeV})\ ,
\label{la5}
\end{eqnarray}
where $f$ is an adjustable parameter and $r$ is the relative distance
between $\Lambda$ and $N$ measured in fm.
Figure \ref{fig12} shows that these potentials fit
the $\hbox{}^1S_0$ and $\hbox{}^3S_1$ $\Lambda N$ phase shifts
obtained by the full $\Lambda N$--$\Sigma N$ CCRGM calculations of fss2.
In the $\hbox{}^3S_1$ state, the phase shifts are fitted only
in the low-momentum region with $p_{\rm lab} < 600~\hbox{MeV}/c$,
since the cusp structure is never reproduced
in the single-channel calculation.
Since any central and single-channel effective $\Lambda N$ force 
leads to the well-known overbinding
problem of $\hbox{}^5_{\Lambda}\hbox{He}$ by about 2 MeV (in the
present case, it is 1.63 MeV) \cite{DA72}, the attractive part
of the $\hbox{}^3S_1$ $\Lambda N$ potential is modified to reproduce
the correct binding energy,
$E^{\rm exp}(\hbox{}^5_\Lambda \hbox{He})
=-3.12\pm 0.02$ MeV with $f=0.8923$.
This overbinding problem is mainly attributed
to the Brueckner rearrangement effect of
the $\alpha$-cluster, originating from the starting energy
dependence of the bare two-nucleon interaction
due to the addition of an extra $\Lambda$-particle \cite{BA80}. 
The odd-state $\Lambda N$ interaction is
introduced in \eq{la4} with the Majorana exchange
parameter $u \neq 1$.
All partial waves up to $\lambda_{\rm Max}={\ell_1}_{\rm Max}
=6$ for the $\alpha \alpha$ and $\Lambda \alpha$ pairs are included.
The direct and exchange Coulomb kernel between
the two $\alpha$-clusters is introduced at the nucleon level
with the cut-off radius, $R_C=10~\hbox{fm}$.
In the first calculation using only the central force,
the SB potential with the pure Serber character ($u=1$
in \eq{la4}) can reproduce the energies of the ground
and excited states within 100 - 200 keV accuracy \cite{2al}.

As already discussed in Sec.~3.2.7, one of the interesting problems
in $\hbox{}^9_{\Lambda}\hbox{Be}$ is
the very small spin-orbit ($\ell s$) splitting,
$\Delta E_{\ell s}=E_{\rm x}(3/2^+)-E_{\rm x}(5/2^+)$,
of the $5/2^+$ and $3/2^+$ excited states.
The recent Hyperball $\gamma$-ray spectroscopy experiment
predicts $\Delta E^{\rm exp}_{\ell s}=43 \pm 5$ keV \cite{AK02,TA03}.
In the non-relativistic models of the $YN$ interaction,
this phenomenon is usually described by the strong cancellation
of the ordinary $LS$ component and the antisymmetric $LS$
component ($LS^{(-)}$ force), the latter of which is
a characteristic feature of baryon-baryon interactions
between non-identical baryons. For example,
our FSS interaction yields
such a strong $LS^{(-)}$ component \cite{NA93} that is about one half
of the ordinary $LS$ component with the opposite sign.
We performed the $G$-matrix calculation in symmetric nuclear
matter, using this QM interaction \cite{GMAT},
and calculated the Scheerbaum factor $S_B$ \cite{SPLS}. 
The ratio of $S_B$ to the nucleon
strength $S_N \approx -40~\hbox{MeV} \hbox{fm}^5$ is
$S_\Lambda/S_N \approx 1/5$ and $S_\Sigma/S_N \approx 1/2$ in the Born
approximation, while it changes to $S_\Lambda/S_N \approx 1/12$ in
the $G$-matrix calculation of the model FSS at
the normal density $\rho_0=0.17~\hbox{fm}^{-3}$.
The significant reduction of $S_\Lambda$ in the latter
is due to the enhancement of the $LS^{(-)}$ component
in the diagonal $\Lambda N$ channel,
owing to the $P$-wave $\Lambda N$--$\Sigma N$ coupling.

\begin{table}[t]
\caption{The ground-state energy $E_{\rm gr}(1/2^+)$,
the $5/2^+$ and $3/2^+$ excitation energies, $E_{\rm x}(5/2^+)$
and $E_{\rm x}(3/2^+)$, and the spin-orbit
splitting, $\Delta E_{\ell s}=E_{\rm x}(3/2^+)-E_{\rm x}(5/2^+)$,
calculated by solving the Faddeev equations
for the $\alpha \alpha \Lambda$ system
in the $jj$ coupling scheme.
The exchange mixture parameter of the SB $\Lambda N$ force
is assumed to be $u=0.82$. The $\Lambda \alpha$ spin-orbit force
is generated from the Born kernel of the FSS and
fss2 $\Lambda N$ $LS$ interactions.
For the fss2 $LS$ interaction, the $LS$ component
from EMEP's is also included.
The $\Lambda N$ central potentials $v^C_{\Lambda N}$ used are 
Eqs.~(\ref{la4}) and (\ref{la5}) (SB),
the simulated Nijmegen models (NS, ND, NF)
and J{\"u}lich model-A and B potentials (JA, JB) \cite{HI97}.
}
\label{tabfad8}
\begin{center}
\renewcommand{\arraystretch}{1.1}
\setlength{\tabcolsep}{6mm}
\begin{tabular}{cccccc}
\hline
$v^{LS}_{\Lambda N}$ & $v^C_{\Lambda N}$
& $E_{\rm gr}(1/2^+)$
& $E_{\rm x}(5/2^+)$ & $E_{\rm x}(3/2^+)$
& $\Delta E_{\ell s}$ \\
& & (MeV) & (MeV) & (MeV) & (keV) \\
\hline
     & SB & $-6.623$ & 2.854 & 2.991 & 137 \\ 
     & NS & $-6.744$ & 2.857 & 2.997 & 139 \\
FSS  & ND & $-7.485$ & 2.872 & 3.024 & 152 \\
     & NF & $-6.908$ & 2.877 & 3.002 & 125 \\
     & JA & $-6.678$ & 2.866 & 2.991 & 124 \\
     & JB & $-6.476$ & 2.858 & 2.980 & 122 \\
\hline
     & SB & $-6.623$ & 2.828 & 3.026 & 198 \\ 
     & NS & $-6.745$ & 2.831 & 3.033 & 202 \\
fss2 & ND & $-7.487$ & 2.844 & 3.064 & 220 \\
     & NF & $-6.908$ & 2.853 & 3.035 & 182 \\
     & JA & $-6.679$ & 2.843 & 3.024 & 181 \\
     & JB & $-6.477$ & 2.834 & 3.012 & 178 \\
\hline
\multicolumn{2}{c}{Exp't~\protect\cite{AK02,TA03}}
 & $-6.62(4)$ & 3.024(3) & 3.067(3)
 & $43(5)$ \\
\hline
\end{tabular}
\end{center}
\end{table}

In the $\alpha \alpha \Lambda$ three-cluster model,
the origin of the $\ell s$ splitting    
of the $5/2^+$ and $3/2^+$ excited states is
only from the $\Lambda \alpha$ $\ell s$ interaction.
We therefore carry out the Faddeev calculation
in the $jj$-coupling scheme, by using the central plus
spin-orbit $\Lambda \alpha$ interactions \cite{2aljj}.
The $\Lambda \alpha$ spin-orbit interaction is generated
from the Born kernel of the $\Lambda N$ $LS$ QM interaction
\eq{form32} by the $\alpha$-cluster folding.
The three possible $LS$ types, $LS$, $LS^{(-)}$,
and $LS^{(-)}\sigma$, in \eq{form33} are all included.
The derivation of the $\Lambda \alpha$ Born kernel
from the $\Lambda N$ Born kernel can be carried out
analytically, yielding the final form
\begin{eqnarray}
& & V^{LS}_{\Lambda \alpha}(\bq_f, \bq_i)
=\sum_\CT \left[\,X^d_\CT~V^{LS\,d}_\CT (\bq_f, \bq_i)
+X^e_\CT~V^{LS\,e}_\CT (\bq_f, \bq_i)\,\right]
~i \bn \cdot \bS_\Lambda\ .
\label{ls2}
\end{eqnarray}
Here $X^{d, e}_\CT$ and $V^{LS\,d, e}_\CT (\bq_f, \bq_i)$
are the spin-flavor factors and the spatial integrals,
respectively, for the QM interaction
types, $\CT=D_-$, $D_+$ and $S$ ($S^\prime$),
in the $\Lambda N$ interaction.
Since the QM $\Lambda N$ interaction contains
the strangeness exchange term, both the direct ($d$)
and exchange ($e$) Feynman diagrams between $\Lambda$ and
the nucleon in the $\alpha$-cluster contribute,
and they are extremely non-local. The $LS$ components from EMEP's
are also included for the model fss2.

Table \ref{tabfad8} shows the results of
Faddeev calculations in the $jj$-coupling scheme,
predicted by our QM $\Lambda N$ $LS$ interaction plus the SB and
other various $\Lambda N$ central potentials \cite{HI97}.
The ground-state energy does not change much
from the $LS$-coupling calculation, indicating the
dominant $s$-wave coupling of the $\Lambda$-hyperon.
We can reproduce the ground-state energy using
the SB central force with the Majorana exchange
mixture $u=0.82$. The final values for the $\ell s$ splitting
of the $5/2^+$ - $3/2^+$ excited states
are $\Delta E_{\ell s}=137$ keV (FSS) and 198 keV (fss2).
Compared to the experimental value of $43 \pm 5$ keV,
these predictions are three to five times too large.
For the $G$-matrix simulated NSC97f $LS$ potential
in Ref.~\cite{HI00}, we obtain 209 keV for the same SB force.
The deviation from 0.16 MeV in Ref.~\cite{HI00} is due to
the model dependence
of the $\alpha \alpha$ and $\Lambda \alpha$ interactions.
Our FSS prediction for $\Delta E_{\ell s}$ is
less than 2/3 of the NSC97f prediction,
while fss2 gives almost the same result as NSC97f. 
If we switch off the EMEP contribution
in the fss2 calculation,
we find $\Delta E_{\ell s}=86$ keV.
This results from the dominant $LS$ component (and the very
small $LS^{(-)}$ component) generated from the EMEP of fss2.  

\begin{table}[b]
\caption{The Scheerbaum factor $S_\Lambda$ in
symmetric nuclear matter with $k_F=1.07~\hbox{fm}^{-1}$,
predicted by $G$-matrix calculations of FSS and fss2
in the continuous prescription for intermediate spectra.
Decompositions into various contributions are shown,
together with the cases
when the $\Lambda N$--$\Sigma N$ coupling by
the $LS^{(-)}$ and $LS^{(-)}\sigma$ forces is
switched off (coupling off).
The unit is in $\hbox{MeV}~\hbox{fm}^5$.}
\label{tabfad8s}
\begin{center}
\renewcommand{\arraystretch}{1.1}
\setlength{\tabcolsep}{6mm}
\begin{tabular}{ccrrrr}
\hline
model & & \multicolumn{2}{c}{full}
 & \multicolumn{2}{c}{coupling off} \\
\hline
      & & odd & even & odd & even \\
\hline
     & $LS$       & $-17.36$ & 0.38 & $-19.70$ & 0.30 \\
FSS  & $LS^{(-)}$ &  14.83   & 0.22 &    8.37  & 0.26 \\
     & total & \multicolumn{2}{c}{$-1.93$}
     & \multicolumn{2}{c}{$-10.77$} \\
\hline
     & $LS$       & $-19.97$ & $-0.14$ & $-21.04$ & $-0.20$ \\
fss2 & $LS^{(-)}$ &    8.64  &    0.21 &     6.12 &    0.23 \\
     & total & \multicolumn{2}{c}{$-11.26$}
     & \multicolumn{2}{c}{$-14.89$} \\
\hline
\end{tabular}
\end{center}
\end{table}

We see that the direct use of the
QM Born kernel for the $\Lambda N$ $LS$ component is
not good enough to reproduce the small experimental
value $43 \pm 5$ keV for the $\ell s$ splitting.
In order to clarify the origin of the problem,
we have carried out an analysis
of the $\Lambda \alpha$ s.p. spin-orbit potential,
using the Scheerbaum factor, $S_\Lambda$ \cite{2aljj}.
Table \ref{tabfad8s} lists $S_\Lambda$ in
symmetric nuclear matter, obtained by the $G$-matrix
calculations in the continuous prescription.
The Fermi momentum, $k_F=1.07~\hbox{fm}^{-1}$,
corresponding to the half of the normal
density $\rho_0=0.17~\hbox{fm}^{-3}$, is assumed.
The table also shows the decompositions into various contributions
and the results obtained by turning off
the $\Lambda N$--$\Sigma N$ coupling through
the $LS^{(-)}$ and $LS^{(-)}\sigma$ forces.
For FSS, we find a large reduction of $S_\Lambda$ from
the Born value $-7.8~\hbox{MeV}~\hbox{fm}^5$ \cite{SPLS},
especially when the $P$-wave $\Lambda N$--$\Sigma N$ coupling
is properly taken into account.
When all the $\Lambda N$--$\Sigma N$ couplings
including those by the pion tensor force are turned off,
the $LS^{(-)}$ contribution is just a half
of the $LS$ contribution with the opposite
sign (in the dominant odd partial wave),
which is the same result as in the Born approximation.
The $\hbox{}^3P_2+\hbox{}^3F_2$ $\Lambda N$--$\Sigma N$ coupling
slightly enhances the attractive $LS$ contribution,
while the $\hbox{}^1P_1+\hbox{}^3F_1$
$\Lambda N$--$\Sigma N$ coupling largely enhances
the repulsive $LS^{(-)}$ contribution.
If we use this reduction of the $S_\Lambda$ factor
from $-7.8~\hbox{MeV}~\hbox{fm}^5$ to $-1.9~\hbox{MeV}~\hbox{fm}^5$ in
the realistic $G$-matrix calculation,
the $\Delta E_{\ell s}$ value is reduced from $-137$ MeV to
an almost correct value $-33$ keV.
However, such a reduction of the Scheerbaum factor due to
the $\Lambda N$--$\Sigma N$ coupling is
supposed to be hindered in the $\Lambda \alpha$ system
in the lowest-order approximation from the isospin consideration.
On the other hand, the situation of fss2 in
Table \ref{tabfad8s} is rather different,
although the cancellation mechanism
between the $LS$ and $LS^{(-)}$ components and the reduction
effect of $S_\Lambda$ factor in the full calculation are
equally observed.
When all the $\Lambda N$--$\Sigma N$ couplings are neglected,
the ratio of the $LS^{(-)}$ and $LS$ contributions
in the quark sector is still one half.
Since the EMEP contribution is mainly for the $LS$ type,
it amounts to about $-6~\hbox{MeV}~\hbox{fm}^5$,
which is very large and remains with the same magnitude
even after the $P$-wave $\Lambda N$--$\Sigma N$ coupling
is included. Furthermore, the increase of the $LS^{(-)}$ component
is rather moderate in comparison with the FSS case.
This is because the model fss2 contains an appreciable EMEP
contribution (about $40\%$) which has very
little $LS^{(-)}$ contribution.
As a result, the total $S_\Lambda$ value
in fss2 $G$-matrix calculation is 3 - 6 times
larger than the FSS value, depending on the Fermi
momentum $k_F=1.35$ - $1.07~\hbox{fm}^{-1}$.
Such an appreciable EMEP contribution
to the $LS$ component of the $YN$ interaction is
not favorable to reproduce the negligibly
small $\ell s$ splitting of $\hbox{}^9_\Lambda \hbox{Be}$.
In conclusion, the failure of the present Faddeev calculation
is probably because the $P$-wave $\Lambda N$--$\Sigma N$ coupling
is not properly taken into account.
The QM baryon-baryon interaction with a large spin-orbit
contribution from the S-meson exchange potentials is not appropriate
to reproduce the very small $\ell s$ splitting observed
in $\hbox{}^9_\Lambda \hbox{Be}$.
A way of incorporating the $P$-wave $\Lambda N$--$\Sigma N$ coupling
in the cluster model calculations is a future problem. 

\subsubsection{The \mib{\Lambda \Lambda \alpha} system 
for \mib{\hbox{}^{\ \,6}_{\Lambda \Lambda} \hbox{He}}}

\begin{table}[b]
\caption{Comparison of $\Delta B_{\Lambda \Lambda}$ values
in MeV, predicted by various $\Lambda \Lambda$ interactions
and $\Lambda N$ potentials.
The $\Lambda \Lambda$ potential $v_{\Lambda \Lambda}$(Hiyama) is
the three-range Gaussian potential
used in Ref.~\protect\cite{HI97},
and $v_{\Lambda \Lambda}$(SB) the two-range Gaussian potential
given in \protect\eq{pot2}.
FSS and fss2 use the $\Lambda \Lambda$ RGM
$T$-matrix ($\widetilde{T}$-matrix)
with $\varepsilon_{\Lambda \Lambda}$, which is 
the $\Lambda \Lambda$ expectation value determined self-consistently.
The $\Lambda N$ central potentials $v^C_{\Lambda N}$ used are 
Eqs.~(\ref{la4}) and (\ref{la5}) (SB),
the simulated Nijmegen models (NS, ND, NF)
and J{\"u}lich model-A and B potentials (JA, JB) \cite{HI97}.
The experimental value is $\Delta B^{\rm exp}_{\Lambda \Lambda}
=1.01 \pm 0.20$ MeV \protect\cite{TA01}.
}
\renewcommand{\arraystretch}{1.1}
\setlength{\tabcolsep}{4mm}
\begin{center}
\begin{tabular}{ccccccc}
\hline
$v_{\Lambda \Lambda}$ & Hiyama
& \multicolumn{2}{c}{FSS} & \multicolumn{2}{c}{fss2}
& SB \\
\hline
$v^C_{\Lambda N}$ & $\Delta B_{\Lambda \Lambda}$
 & $\Delta B_{\Lambda \Lambda}$ & $\varepsilon_{\Lambda \Lambda}$
 & $\Delta B_{\Lambda \Lambda}$ & $\varepsilon_{\Lambda \Lambda}$
 & $\Delta B_{\Lambda \Lambda}$ \\
\hline
SB & 3.618 &  3.657 & 5.124 & 1.413 & 5.938 & 1.910 \\
NS & 3.548 &  3.630 & 5.151 & 1.366 & 5.947 & 1.914 \\
ND & 3.181 &  3.237 & 4.479 & 1.288 & 5.229 & 1.645 \\
NF & 3.208 &  3.305 & 4.622 & 1.271 & 5.407 & 1.713 \\
JA & 3.370 &  3.473 & 4.901 & 1.307 & 5.702 & 1.824 \\
JB & 3.486 &  3.599 & 5.141 & 1.327 & 5.952 & 1.911 \\
\hline
\end{tabular}
\label{tabfad9}
\end{center}
\end{table}

As another application of the $\Lambda \alpha$ $T$-matrix in
the $\alpha \alpha \Lambda$ Faddeev calculation, we discuss
the ground-state energy 
of $\hbox{}^{\ \,6}_{\Lambda \Lambda} \hbox{He}$ \cite{HE6LL}.
The full CC $T$-matrices of fss2 and FSS
with strangeness $S=-2$ and isospin $I=0$ are employed for
the $\Lambda \Lambda$ RGM $T$-matrix ($\widetilde{T}$-matrix). 
Table \ref{tabfad9} shows the two-$\Lambda$ separation
energy $\Delta B_{\Lambda \Lambda}$ (defined
by $\Delta B_{\Lambda \Lambda}
=B_{\Lambda \Lambda}(\hbox{}^{\ \,6}_{\Lambda \Lambda}\hbox{He})
-2B_{\Lambda}(\hbox{}^{\,5}_{\Lambda}\hbox{He})$),
predicted by various combinations
of the $\Lambda N$ and $\Lambda \Lambda$ interactions.
The results of a simple three-range Gaussian potential,
$v_{\Lambda \Lambda}$(Hiyama), used in Ref.~\cite{HI97} are also shown.
This $\Lambda \Lambda$ potential and
our Faddeev calculation using the FSS $\widetilde{T}$-matrix yield
very similar results with large $\Delta B_{\Lambda \Lambda}$ values
of about 3.6 MeV, in accordance with the fact that
the $\Lambda \Lambda$ phases shifts predicted by these interactions
increase up to about $40^\circ$ (see Fig.~\ref{ll1SFSS}).
The improved QM fss2
yields $\Delta B_{\Lambda \Lambda}=1.41$ MeV.
(For the $\Lambda \Lambda$ single-channel $T$-matrix,
this value is reduced to 1.14 MeV.)
In Table \ref{tabfad9}, results are also shown
for $v_{\Lambda \Lambda}$(SB) which is generated
from the fss2 $\hbox{}^1S_0$ $\Lambda \Lambda$ phase
shift (see Fig.~\ref{ll1Sfss2}) in the full-channel calculation,
using the supersymmetric inversion method \cite{SB97}.
This single-channel effective potential is given by
\begin{eqnarray}
v_{\Lambda \Lambda}({\rm SB})=-103.9~\exp(-1.176~r^2)
+658.2~\exp(-5.936~r^2)\qquad (\hbox{MeV})\ .
\label{pot2}
\end{eqnarray}
We think that the 0.5 MeV difference between our fss2 result
and the $v_{\Lambda \Lambda}$(SB) result is probably because
we neglected the CC effects of the $\Lambda \Lambda \alpha$ channel
to $\Xi N \alpha$ and $\Sigma \Sigma \alpha$ channels.
We should also keep in mind that in all of these
three-cluster calculations
the Brueckner rearrangement effect of the $\alpha$-cluster
is very important, producing
about $-1$ MeV (repulsive) contribution \cite{KO03}.
It is also reported in Ref.~\cite{SU99} that
the quark Pauli effect between the $\alpha$-cluster
and the $\Lambda$-hyperon yields a non-negligible
repulsive contribution of 0.1 - 0.2 MeV for
the two-$\Lambda$ separation energy
of $\hbox{}^{\ \,6}_{\Lambda \Lambda} \hbox{He}$,
even when a rather compact $(3q)$ size of $b=0.6$ fm is assumed
as in our QM interactions.
Taking all of these effects into consideration, we can conclude
that the present results by fss2 are in good agreement
with the recent experimental value,
$\Delta B^{\rm exp}_{\Lambda \Lambda}
=1.01 \pm 0.20$ MeV, deduced from the NAGARA event \cite{TA01}.

\section{Summary}

Since the advent of the Yukawa's meson theory, an enormous amount of efforts
have been devoted to understand the fundamental
nucleon-nucleon ($NN$) interaction and related hadronic interactions.
The present view of these interactions is a non-perturbative
realization of inter-cluster interactions, governed by the
fundamental theory of the strong interaction,
quantum chromodynamics (QCD), in which the gluons are 
the field quanta exchanged between quarks.
On this basis, the meson ``theory'' can be understood as 
an effective description of the quark-gluon dynamics
in the low-energy regime. The short-range part of
the $NN$ and hyperon-nucleon ($YN$) interactions are
still veiled with unsolved mechanism of quark
confinement and multi-gluon effects.

In this review article, we applied a constituent
quark model to the study of the baryon-baryon interactions,
in which some of the essential features of QCD are explicitly
taken into account in the non-relativistic framework.
For example, the color degree of freedom of quarks is explicitly
included, and the full antisymmetrization of quarks
is carried out in the resonating-group method (RGM) formalism.
The gluon exchange effect is represented in the form
of the quark-quark interaction,
for which a color analogue of the Fermi-Breit (FB)
interaction is used with an adjustable parameter of the
quark-gluon coupling constant $\alpha_S$.
The confinement potential is a phenomenological $r^2$-type potential,
which has no contributions to the baryon-baryon interactions
in the present framework.
Since the meson-exchange effects are the non-perturbative
aspect of QCD, these are described by the effective-meson
exchange potentials (EMEP's) acting between quarks.

Our purpose is to construct
a realistic model of the $NN$ and $YN$ interactions,
which describes not only the baryon-baryon scattering
quantitatively in the wide energy region, but also reproduces
rich phenomena observed in few-baryon systems
and various types of finite nuclei and nuclear matter. 
The present framework incorporates both the quark and mesonic
degrees of freedom into the model explicitly.
This framework is very versatile,
since it is based on the natural
picture that the quarks and gluons are the most economical  
ingredients in the short-range region,
while the meson-exchange processes dominate
in the medium- and long-range regions of the interaction.
The short-range repulsion is described by the color-magnetic
term of the FB interaction and the Pauli effect at the quark level. 
In this article, we have focused on our
recent versions called FSS \cite{PRL,FSS,SCAT}
and fss2 \cite{fss2,B8B8}.
In FSS only the scalar-meson and pseudoscalar-meson
exchanges are introduced as the EMEP's, while in fss2
the vector-meson exchanges are introduced as well.
All possible standard terms used in the non-relativistic
one-boson exchange potentials (OBEP) can in principle be included
in the final version. As a first step to study the charge
dependence of the baryon-baryon interactions, the pion-Coulomb
correction is also taken into account in the particle basis.
After these improvements mainly related to the EMEP's,
the most recent model fss2 has achieved an accurate
description of the $NN$ and $YN$ interactions.

An advantage of using the quark model for the study
of the baryon-baryon interactions is that the flavor $SU_3$ symmetry
(and also the spin-flavor $SU_6$ symmetry) is automatically
incorporated in the framework, so that all the interactions between
the octet-baryons ($B_8$) are treated in a unified manner
in close connection with the well-known $NN$ interaction.
We extended the $(3q)$-$(3q)$ RGM study of the
the $NN$ and $YN$ interactions 
to the strangeness $S=-2,~-3$ and $-4$ sectors
with no {\em ad hoc} parameters, and clarified the characteristic
features of the $B_8 B_8$ interactions \cite{B8B8}.
The results appear to be reasonable,
if we consider i) the spin-flavor $SU_6$ symmetry of $B_8$,
ii) the weak pion effect in the strangeness sector,
and iii) the effect of the flavor-symmetry breaking (FSB) of
the underlying quark Hamiltonian.
In particular, the existence of the Pauli-forbidden state
in the $SU_3$ $(11)_s$ channel (for the flavor-symmetric
configuration) and the almost Pauli-forbidden state
in the $(30)$ channel (for the flavor-antisymmetric configuration)
is of prime importance, in order to understand the
origin of the Pauli repulsion and the relative strength
among various baryon-baryon interactions.
This is the most prominent feature
of the spin-flavor $SU_6$ symmetry. The second feature,
the weak pion effect in the strangeness sector,
is a direct consequence of the $SU_3$ relationship
in the model framework. The last item is important
even in the kinetic-energy term, since the heavier baryon masses
reduce the kinetic energies appreciably. In fact, the light pion mass
itself is a reflection of the chiral symmetry, which gives
the most important source of the FSB.

These $B_8 B_8$ interactions are now used for the detailed study
of the few-body systems, as well as medium-weighted hypernuclei
and baryonic matter through the $G$-matrix calculations.
In particular, the $G$-matrix calculations are very useful
to clarify the characteristics of the model interactions.
Both fss2 and FSS are found to produce the nucleon
single-particle (s.p.) potentials and
nuclear-matter saturation curves, which are similar to those
obtained with the standard meson-exchange potentials.
For example, fss2 gives results very similar to 
the Bonn model-B potential.
It is interesting to note that the deuteron properties
calculated with fss2 are close to those of model-C, which has
a larger $D$-state probability of the deuteron than model-B.
Since fss2 reproduces the $NN$ phase shifts at non-relativistic
energies quite well, the difference of the off-shell effect
between our quark model and the other OBEP models does not
seem to appear prominently, as far as the nuclear saturation
properties are concerned.
Some interesting features of our quark model appear in the predictions
for hyperon properties in nuclear medium.
The $\Lambda$ s.p.~potential has a depth
of 48 MeV (fss2) and 46 MeV (FSS) if the continuous
prescription is used for intermediate energy spectra.
This value is slightly more attractive
than the value expected from the
experimental data of $\Lambda$-hypernuclei \cite{BMZ}.
However, the recent analysis of the $\Lambda$ s.p.~potentials
in finite nuclei, using the local-density
and Thomas-Fermi approximations, seems to suggest that
this is an appropriate strength of attraction.
% \cite{Ko05}.
The $\Sigma$ s.p.~potential is repulsive, with the strength
of about 8 MeV (fss2) and 20 MeV (FSS).
The origin of this repulsion is due to the strong Pauli effect
of the (30) component in the $\Sigma N (I=3/2)$ $\hbox{}^3S_1$ state.
This result is consistent with the recent analysis \cite{DA99,KO04}
of the $(K^{-}, \pi^{\pm})$ and $(\pi^-, K^+)$ experimental
data \cite{BNL98,NU02,SA04}.
Future experiments are expected to settle the problem
of the $\Sigma$ s.p.~potential and the isospin dependence
of the $\Sigma N$ interaction.
One of the characteristic features of fss2 is the $LS$ force
generated from the scalar-meson EMEP. If this contribution is large,
the cancellation of the $LS$ and $LS^{(-)}$ components from the
FB interaction becomes less prominent.
The model fss2 predicts the ratio of the Scheerbaum factors
$S_\Lambda/S_N \approx 1/4$, which is larger
than the FSS value $1/12$.
Though these ratios are for symmetric nuclear matter with the normal
density $\rho_0=0.17~\hbox{fm}^{-3}$,
the density dependence of $S_\Lambda/S_N$ is
actually weak in fss2.

We also discussed some applications
of the $NN$, $YN$ and $YY$ interactions
to the Faddeev calculations of the three-nucleon bound state,
the $\Lambda NN$--$\Sigma NN$ system for the hypertriton,
the $\alpha \alpha \Lambda$ system
for $\hbox{}^9_{\Lambda}\hbox{Be}$,
and the $\Lambda \Lambda \alpha$ system
for $\hbox{}^{\ \,6}_{\Lambda \Lambda} \hbox{He}$.
For these applications, we developed a new three-cluster
Faddeev formalism which uses two-cluster RGM kernels
for the interacting pairs. The model fss2 gives the triton
binding energy close enough to compare with the experiment.
The charge root-mean-square radii
of $\hbox{}^3\hbox{H}$ and $\hbox{}^3\hbox{He}$ are
also correctly reproduced.
The application to the hypertriton calculation shows that
fss2 gives a result similar to the Nijmegen
soft-core model NSC89, except for the appreciable
contributions of higher partial waves.
We find that the $\hbox{}^1S_0$ state
of the $\Lambda N$ interaction should be only slightly more
attractive than the $\hbox{}^3S_1$ state, in order to
reproduce the empirical hypertriton binding energy.
A desirable difference of the $\hbox{}^1S_0$ and $\hbox{}^3S_1$
phase shifts around $p_{\Lambda}= 200~\hbox{MeV}/c$
is only $0^\circ$ - $2^\circ$, although fss2 is still too attractive,
giving about $10^\circ$.
In the application to the $\alpha \alpha \Lambda$ system,
the $\alpha \alpha$ RGM kernel and some appropriate $\Lambda N$ forces
generated from the low-energy phase-shift behavior of fss2 reproduce
the ground-state and excitation energies
of $\hbox{}^9_{\Lambda}\hbox{Be}$ within 100 - 200 keV accuracy.
The very small spin-orbit splitting
of the excited $5/2^+$ and $3/2^+$ states
of $\hbox{}^9_{\Lambda}\hbox{Be}$ is studied in the
Faddeev calculation in the $jj$-coupling scheme,
where $\Lambda \alpha$ spin-orbit potential is generated
from the $\Lambda N$ Born kernel of the FSS and fss2 RGM kernels.
It was argued that the $P$-wave $\Lambda N$--$\Sigma N$ coupling
by the $LS^{(-)}$ and $LS^{(-)}\sigma$ forces is very
important to achieve the extremely small spin-orbit splitting
as an almost complete cancellation between
the $LS$ and $LS^{(-)}$ contributions.
The EMEP $LS$ force generated from the scalar-meson exchange
in fss2 reduces this cancellation.
As deduced from the NAGARA event \cite{TA01}
for $\hbox{}^{\ \,6}_{\Lambda \Lambda} \hbox{He}$,
the weak $\Lambda \Lambda$ force
is reasonably well reproduced by the Faddeev calculation
of the $\Lambda \Lambda \alpha$ system,
using the present $\Lambda \alpha$ $T$-matrix
and the full CC $\Lambda \Lambda$--$\Xi N$--$\Sigma \Sigma$
$\widetilde{T}$-matrix of fss2.

The interactions of the octet-baryons, derived from the quark model 
fss2 in the RGM formalism, enable us to reproduce 
a large number of $NN$ and $YN$ scattering observables. 
To further test these interactions, it is vital to apply them to 
the systems of more than two particles.  
The three-cluster formalism presented here will serve as a basic tool 
for this purpose and open a way to solve the few-body systems
which interact via the quark-model baryon-baryon interactions,
while keeping the essential features of the RGM kernels,
i.e., the non-locality, the energy dependence
and a possible existence of the pairwise Pauli-forbidden state.
Since this formalism makes it possible
to relate the underlying $NN$ and $YN$ interactions directly
to the structure of the few-baryon systems, we expect to learn the 
interplay of the ingredients of the quark model, the baryon-baryon 
interactions, and the structure and binding mechanism of the nuclear 
systems. 

\bigskip

%\begin{center}
{\Large\bf Acknowledgments}
%\end{center}

\bigskip

This research is supported by Grants-in-Aid for Scientific
Research from the Japanese Ministry of Education, Science,
Sports and Culture (Nos. 04640296, 07640397, 08239203, 09225201,
12640265), Grants-in-Aid for Scientific
Research (C) (Nos. 15540270, 15540284, 15540292 and 18540261) and
for Young Scientists (B) (No.~15740161) from
the Japan Society for the Promotion of Science (JSPS),
and a Grant for Promotion of Niigata University Research
Project (2005 - 2007).
Some of the results presented in this article are
based on the collaborations
with T. Fujita (Japan Meteorological Agency),
M. Kohno (Kyushu Dental College),
and K. Miyagawa (Okayama Science University).

\bigskip

%\begin{center}
{\Large\bf Appendix}
%\end{center}

%\bigskip

\begin{table}[htb]
\caption{
List of acronyms and abbreviations used in the text
}
\label{acronyms}
%\bigskip
\begin{center}
\renewcommand{\arraystretch}{1.2}
\setlength{\tabcolsep}{2mm}
\begin{tabular}{rlrl}
\hline
Term & Meaning & Term & Meaning \\
\hline
$B_8$ & octet-baryons &   QM    & quark model \\
$NN$  & nucleon-nucleon & FB    & Fermi-Breit \\
$YN$  & hyperon-nucleon & FSB   & flavor-symmetry breaking \\
$YY$  & hyperon-hyperon & CIB   & charge-independence breaking \\
$B_8 B_8$ & baryon-octet baryon-octet &
CSB   & charge-symmetry breaking \\
RGM   & resonating-group method  &
LS-RGM & Lippmann-Schwinger RGM \\
GCM   & generator-coordinate method & CC    & coupled-channels \\
OCM   & orthogonality condition model &
$\widetilde{T}$-matrix & RGM $T$-matrix in \protect\eq{fad9} \\
OPEP  & one-pion exchange potential &  $LS$  & spin-orbit \\
OBEP  & one-boson exchange potential & $LS^{(-)}$ & antisymmetric $LS$ \\
EMEP  & effective-meson exchange potential &
$LS^{(-)}\sigma$  & antisymmetric $LS$ with $P_\sigma$
in \protect\eq{form33} \\
s.p. & single-particle &
$QLS$ & quadratic $LS$ in \protect\eq{form34} \\
h.o. & harmonic-oscillator & 
$\ell s$ & one-body $LS$ \\
\hline
%\hline
\end{tabular}
\end{center}
\end{table}

\addtocounter{table}{-1}

\begin{table}[htb]
\caption{-continued
}
\label{acronyms-2}
%\bigskip
\begin{center}
\renewcommand{\arraystretch}{1.2}
\setlength{\tabcolsep}{3mm}
\begin{tabular}{rll}
\hline
Term & Meaning & Reference \\
\hline
RGM-F & quark model including effective-meson exchange  & \\
      & potentials at the baryon level
      & \protect\cite{NA95,RGMFa,RGMFb}  \\
FSS   & quark model including scalar- and pseudoscalar- & \\
      & meson interactions
      & \protect\cite{PRL,FSS,SCAT} \\
RGM-H & an alternative version of FSS & \protect\cite{FSS,SCAT} \\
fss2  & quark-model including scalar-, pseudoscalar- and & \\
      & vector-meson interactions
      & \protect\cite{fss2,B8B8} \\
QMPACK & quark-model package homepage & \\
       & http://qmpack.homelinux.com/$\sim$qmpack/index.php & \\
SAID   & Scattering Analysis Interactive Dial-up & \\
SP99   & phase shift analysis in the SAID program
       & \protect\cite{SAID} \\
PWA93  & phase shift analysis by the Nijmegen group
       & \protect\cite{ST94} \\
NSC**  & Nijmegen soft-core potential
       & \protect\cite{NSC89,NSC97,SR99} \\
ESC**  & Nijmegen extended soft-core potential
       & \protect\cite{ESC04a,ESC04b} \\
CD-Bonn & charge-dependent Bonn & \protect\cite{MA01} \\
RSC    & Reid soft-core potential & \protect\cite{Rei68} \\
AV18   & Argonne potential with 18 operators including
CIB and CSB & \protect\cite{WI95} \\
SB     & Sparenberg and Baye potential
in Eqs.~(\ref{la5}),~(\ref{pot2}) & \protect\cite{2al,HE6LL} \\
\hline
%\hline
\end{tabular}
\end{center}
\end{table}

\vspace{20mm}

%%%%%%%%%%%%%%%%%%%%%%%%% References


\begin{thebibliography}{99}
\itemsep -2pt
\newcommand{\etal}{{\em et al.}}
% 1
\bibitem{VA05} A. Valcarce, H. Garcilazo,
F. Fern{\'a}ndez and P. Gonz{\'a}lez,
\Journal{\em Rep. Prog. Phys.} {68}{965}{2005}
% 2
\bibitem{LI77} D. A. Liberman,
\Journal{\PRD} {16}{1542}{1977}
% 3
\bibitem{RU75} A. De R{\'u}jula, H. Georgi and S.L. Glashow,
\Journal{\PRD} {12}{147}{1975}
% 4
\bibitem{OK80} M. Oka and K. Yazaki,
\Journal{\PL} {90B}{41}{1980};
\Journal{\PRO} {66}{556, 572}{1981}
% 5
\bibitem{HA80} M. Harvey,
\Journal{\NPA} {352}{301, 326}{1981}
% 6
\bibitem{FA82} Amand Faessler, F. Fernandez, G. L{\"u}beck
and K. Shimizu,
\Journal{\PLB} {112}{201}{1982}
% 7
\bibitem{SU84} Y. Suzuki,
\Journal{\NPA} {430}{539}{1984}
% 8
\bibitem{WS84} See, for example, M. Oka and K. Yazaki,
in {\em Quarks and Nuclei}, ed. W. Weise
(World Scientific, Singapore, 1984), p.489
% 9
\bibitem{WO86} C.W. Wong,
\Journal{\PREP} {136}{1}{1986}
% 10
\bibitem{OSY87}
M. Oka, K. Shimizu and K. Yazaki,
\Journal{\NPA} {464}{700}{1987}
% 11
\bibitem{SH89} K. Shimizu,
\Journal{\em Rep. Prog. Phys.} {52}{1}{1989}
% 12
\bibitem{OK83} M. Oka and K. Yazaki,
\Journal{\NPA} {402}{477}{1983}
% 13
\bibitem{YA86} Y. Yamauchi and M. Wakamatsu,
\Journal{\NPA} {457}{621}{1986}
% 14
\bibitem{TA89} S. Takeuchi, K. Shimizu and K. Yazaki,
\Journal{\NPA} {504}{777}{1989}
% 15
\bibitem{SH84} K. Shimizu,
\Journal{\PL} {148B}{418}{1984}
% 16
\bibitem{ST88} U. Straub, Zhang Zong-Ye, K. Br{\" a}uer,
Amand Faessler, S.B. Khadkikar and G. L{\"u}beck,
\Journal{\NPA} {483}{686}{1988};
\Journal{} {508}{385c}{1990}
% 17
\bibitem{BR90} K. Br{\" a}uer, Amand Faessler,
F. Fern{\'a}ndez and K. Shimizu,
\Journal{\NPA} {507}{599}{1990}
% 18
\bibitem{FE93} F. Fern{\'a}ndez, A. Valcarce, U. Straub
and A. Faessler,
\Journal{\em J. Phys. G} {19}{2013}{1993}
% 19
\bibitem{VA94} A. Valcarce, A. Buchmann, F. Fern{\'a}ndez
and Amand Faessler,
\Journal{\PRC} {50}{2246}{1994};
\Journal{} {51}{1480}{1995}
% 20
\bibitem{ZH94} Zong-ye Zhang, Amand Faessler, U. Straub
and L. Ya. Glozman,
\Journal{\NPA} {578}{573}{1994}
% 21
\bibitem{QI95} L. J. Qi, J. H. Zhang, P. N. Shen, Z. Y. Zhang
and Y. W. Yu,
\Journal{\NPA} {585}{693}{1995}
% 22
\bibitem{YU95} Y. W. Yu, Z. Y. Zhang, P. N. Shen and L. R. Dai,
\Journal{\PRC} {52}{3393}{1995}
% 23
\bibitem{SH99} P. N. Shen, Z. Y. Zhang, Y. W. Yu, X. Q. Yuan
and S. Yang,
\Journal{\JPG} {25}{1807}{1999}
% 24
\bibitem{HA05} J. Haidenbauer and Ulf-G. Mei{\ss}ner,
\Journal{\PRC} {72}{044005}{2005}
% 25
\bibitem{ESC04a} Th. A. Rijken,
\Journal{\PRC}{73}{044007}{2006}
% 26
\bibitem{ESC04b}
Th. A. Rijken and Y. Yamamoto,
\Journal{\PRC} {73}{044008}{2006}
% 27
\bibitem{FU93} Y. Fujiwara,
\Journal{\PRO} {90}{105}{1993}
% 28
\bibitem{SHT84} Y. Suzuki and K. T. Hecht,
\Journal{\NPA} {420}{525}{1984}; \Journal{} {446}{749 (E)}{1985}
% 29
\bibitem{NA93} C. Nakamoto, Y. Suzuki and Y. Fujiwara,
\Journal{\PLB} {318}{587}{1993}
% 30
\bibitem{FU85} Y. Fujiwara and K. T. Hecht,
\Journal{\NPA}{444}{541}{1985};
\Journal{} {451}{625}{1986};
\Journal{} {456}{669}{1986}
% 31
\bibitem{FU87} Y. Fujiwara and K. T. Hecht,
\Journal{\NPA} {462}{621}{1987}
% 32
\bibitem{FA81} A. Faessler, G. L{\"u}beck and K. Shimizu,
\Journal{\PRD} {26}{3280}{1981}
% 33
\bibitem{OK85} M. Oka,
\Journal{\PRD} {31}{2274}{1985}
% 34
\bibitem{BU89} J. Burger, R. M{\"u}ller, K. Tregel and H. M. Hofmann,
\Journal{\NPA} {493}{427}{1989}
% 35
\bibitem{NA95} C. Nakamoto, Y. Suzuki and Y. Fujiwara,
\Journal{\PRO} {94}{65}{1995}
% 36
\bibitem{RGMFa} Y. Fujiwara, C. Nakamoto and Y. Suzuki,
\Journal{\PRO} {94}{215}{1995}
% 37
\bibitem{RGMFb} Y. Fujiwara, C. Nakamoto and Y. Suzuki,
\Journal{\PRO} {94}{353}{1995}
% 38
\bibitem{NA79} M. M. Nagels, T. A. Rijken and J. J. de Swart,
\Journal{\PRD} {20}{1633}{1979}
% 39
\bibitem{PRL} Y. Fujiwara, C. Nakamoto and Y. Suzuki,
\Journal{\PRL} {76} {2242}{1996}
% 40
\bibitem{FSS} Y. Fujiwara, C. Nakamoto, and Y. Suzuki,
\Journal{\PRC} {54}{2180}{1996}
% 41
\bibitem{SCAT} T. Fujita, Y. Fujiwara, C. Nakamoto and Y. Suzuki,
\Journal{\PRO} {100}{931}{1998}
% 42
\bibitem{fss2} Y. Fujiwara, T. Fujita, M. Kohno, C. Nakamoto
and Y. Suzuki,
\Journal{\PRC} {65}{014002}{2002}
% 43
\bibitem{B8B8} Y. Fujiwara, M. Kohno, C. Nakamoto and Y. Suzuki, 
\Journal{\PRC} {64}{054001}{2001}
% 44
\bibitem{LSRGM} Y. Fujiwara, M. Kohno, C. Nakamoto and Y. Suzuki,
\Journal{\PRO} {103}{755}{2000}
% 45
\bibitem{GMAT} M. Kohno, Y. Fujiwara, T. Fujita,
C. Nakamoto and Y. Suzuki,
\Journal{\NPA} {674}{229}{2000}
% 46
\bibitem{SPLS} Y. Fujiwara, M. Kohno, T. Fujita,
C. Nakamoto and Y. Suzuki,
\Journal{\NPA} {674}{493}{2000}
% 47
\bibitem{MA01} R. Machleidt,
\Journal{\PRC} {63}{024001}{2001}
% 48
\bibitem{NA73}
M. M. Nagels, T. A. Rijken and J. J. de Swart,
\Journal{\ANNP} {79}{338}{1973}
% 49
\bibitem{SP87} Y. Fujiwara,
\Journal{\em Prog. Theor. Phys. Suppl.} {No. 91}{160}{1987}
% 50
\bibitem{KA77} M. Kamimura,
\Journal{\em Prog. Theor. Phys. Suppl.} {No. 62}{236}{1977}
% 51
\bibitem{KU93} K. Kume and S. Yamaguchi,
\Journal{\PRC} {48}{2097}{1993}
% 52
\bibitem{SAL00} D. R. Entem, F. Fern{\'a}ndez and A. Valcarce,
\Journal{\PRC} {62}{034002}{2000}
% 53
\bibitem{NO65} H. P. Noyes,
\Journal{\PRL} {15}{538}{1965}
% 54
\bibitem{KO65} K. L. Kowalski,
\Journal{\PRL} {15}{798, 908 (E)}{1965}
% 55
\bibitem{GRGM} Y. Fujiwara, M. Kohno, C. Nakamoto and Y. Suzuki,
\Journal{\PRO} {104}{1025}{2000}
% 56
\bibitem{KI94} Y. Fujiwara and Y. C. Tang,
\Journal{\em Memoirs of the Faculty of Science, Kyoto University,
Series A of Physics, Astrophysics, Geophysics and Chemistry}
{Vol. XXXIX, No. 1, Article 5}{91} {1994}
% 57
\bibitem{SU83} Y. Suzuki,
\Journal{\NPA} {405}{40}{1983}
% 58
\bibitem{FU92}
Y. Fujiwara,
\Journal{\PRO} {88}{933}{1992}
% 59
\bibitem{FU97}
Y. Fujiwara, C. Nakamoto, Y. Suzuki and Zhang Zong-ye,
\Journal{\PRO} {97}{587}{1997}:
The $S$-type $LS^{(-)}\sigma$ factor in Eq.~(C.4) of this paper should
be read as 
$X^{LS^{(-)\sigma}}_S=-\frac{1}{108}\left(1-\frac{1}{\lambda}\right)
\left[\frac{5}{3}\left(1+\frac{1}{\lambda}\right)+\left(1-\frac{1}{\lambda}
\right)P_F\right]$ for $B_3 B_1=\Sigma \Lambda$.
Accordingly, the second expression in Eq.\,(5.17) is changed to
$\left(X^{LS^{(-)\sigma}}_{S^\prime}\right)_{\Lambda \Sigma}
=\frac{1}{108}\left(1-\frac{1}{\lambda}\right)
\left[\frac{5}{3}\left(1+\frac{1}{\lambda}\right)-\left(1-\frac{1}{\lambda}
\right)P_F\right]$.
% 60
\bibitem{YA90}
K. Yazaki,
\Journal{\em Prog. Part. Nucl. Phys.} {24}{353}{1990}
% 61
\bibitem{SAID} Scattering Analysis Interactive Dial-up (SAID),
Virginia Polytechnic Institute, Blacksburg, Virginia, R. A. Arndt:
Private Communication
% 62
\bibitem{ST94}
V. G. J. Stoks, R. A. M. Klomp, C. P. F. Terheggen
and J. J. de Swart,
\Journal{\PRC} {49}{2950}{1994}
% 63
\bibitem{SC76} R. R. Scheerbaum,
\Journal{\NPA} {257}{77}{1976}
% 64
\bibitem{VP74} C. M. Vincent and S. C. Phatak,
\Journal{\PRC} {10}{391}{1974}
% 65
\bibitem{apfb05} Y. Fujiwara, in {\em Proceedings
of the Third Asia-Pacific Conference on Few-Body
Problems in Physics} (APFB05), Nakhon Ratchasima, Thailand,
July 26 - 30, 2005.
% 66
\bibitem{HO74} H. Horiuchi,
\Journal{\PRO} {51}{1266}{1974}; \Journal{} {53}{447}{1975}
% 67
\bibitem{TRGM} Y. Fujiwara, H. Nemura, Y. Suzuki, K. Miyagawa
and M. Kohno, \Journal{\PRO} {107}{745}{2002}
% 68
\bibitem{RED} Y. Fujiwara, Y. Suzuki, K. Miyagawa, M. Kohno
and H. Nemura, \Journal{\PRO} {107}{993}{2002}
% 69
\bibitem{TH77} D. R. Thompson, M. LeMere and Y. C. Tang,
\Journal{\NPA} {286}{53}{1977}
% 70
\bibitem{BE9L} Y. Fujiwara, K. Miyagawa, M. Kohno, Y. Suzuki,
D. Baye and J.-M. Sparenberg,
\Journal{\PRC} {70}{024002}{2004}
% 71
\bibitem{post} Y. Fujiwara, K. Miyagawa, M. Kohno,
Y. Suzuki, D. Baye and J.-M. Sparenberg,
\Journal{\NPA} {738}{495}{2004}
% 72
\bibitem{OR94} S. Oryu, K. Samata, T. Suzuki,
S. Nakamura and H. Kamada,
\Journal{{\em Few-Body Systems}} {17}{185}{1994}
% 73
\bibitem{OCMFAD} Y. Fujiwara, M. Kohno and Y. Suzuki,
\Journal{{\em Few-Body Systems}} {34}{237}{2004}
% 74
\bibitem{KU78} V. I. Kukulin and V. N. Pomerantsev,
\Journal{\ANNP} {111}{330}{1978}
% 75
\bibitem{BF77} B. Buck, H. Friedrich and C. Wheatley,
\Journal{\NPA} {275}{246}{1977}
% 76
\bibitem{AMRED} Y. Fujiwara, Y. Suzuki and M. Kohno,
\Journal{\PRC} {69}{037002}{2004}
% 77
\bibitem{tur01} E. M. Tursunov,
\Journal{\JPG} {27}{1381}{2001}
% 78
\bibitem{TB03} E. M. Tursunov, D. Baye and P. Descouvemont,
\Journal{\NPA} {723}{365}{2003}
% 79
\bibitem{DE03} P. Descouvemont, C. Daniel and D. Baye,
\Journal{\PRC} {67}{044309}{2003}
% 80
\bibitem{BR67} R. A. Bryan and B. L. Scott,
\Journal{\PREV} {164}{1215}{1967}
% 81
\bibitem{QMPACK} http://qmpack.homelinux.com/$\sim$qmpack/index.php
% 82
\bibitem{ST93}
V. G. J. Stoks, R. A. M. Klomp, M. C. M. Rentmeester
and J. J. de Swart,
\Journal{\PRC} {48}{792}{1993}
% 83
\bibitem{VA95}
A. Valcarce, Amand Faessler and F. Fern{\'a}ndez,
\Journal{\PLB} {345}{367}{1995}
% 84
\bibitem{PARI} M. Lacombe, B. Loiseau, J. M. Richard, R. Vinh Mau,
J. C\^{o}t\'{e}, P. Pir\`{e}s and R. de Tourreil,
\Journal{\PRC} {21}{861}{1980}
% 85
\bibitem{MA89} R. Machleidt,
\Journal{\em Adv. Nucl. Phys.} {19}{189}{1989}
% 86
\bibitem{DU83}
O. Dumbrajs, R. Koch, H. Pilkuhn, G. C. Oades, H. Behrens,
J. J. de Swart and P. Kroll,
\Journal{\NPB} {216}{277}{1983}
% 87
\bibitem{AL78}
L. J. Allen, H. Fiedeldey and N. J. McGurk,
\Journal{\em J. Phys. G} {4}{353}{1978}
% 88
\bibitem{KO83}
M. Kohno,
\Journal{\em J. Phys. G} {9}{L85}{1983}
% 89
\bibitem{RO90}
N. L. Rodning and L. D. Knutson,
\Journal{\PRC} {41}{898}{1990}
% 90
\bibitem{BI79}
David M. Bishop and Lap M. Cheung,
\Journal{\PRA} {20}{381}{1979}
% 91
\bibitem{MI90}
G. A. Miller, B. M. K. Nefkens and I. {\v S}laus,
\Journal{\PREP} {194}{1}{1990}
% 92
\bibitem{HO98}
C. R. Howell \etal,
\Journal{\PLB} {444}{252}{1998}
% 93
\bibitem{GO99}
D. E. Gonz{\'a}lez Trotter \etal,
\Journal{\PRL} {83}{3788}{1999}
% 94
\bibitem{BE88}
J. R. Bergervoet, P. C. van Campen, W. A. van der Sanden
and J. J. de Swart,
\Journal{\PRC} {38}{15}{1988}
% 95
\bibitem{ST91}
G. T. Stephenson, Jr., Kim Maltman and T. Goldman,
\Journal{\PRD} {43}{860}{1991}
% 96
\bibitem{NA75}
M. M. Nagels, T. A. Rijken and J. J. de Swart,
\Journal{\PRD} {12}{744}{1975}
% 97
\bibitem{SHT83} Y. Suzuki and K. T. Hecht,
\Journal{\PRC} {27}{299}{1983} 
% 98
\bibitem{NA97} C. Nakamoto, Y. Suzuki and Y. Fujiwara,
\Journal{\PRO} {97}{761}{1997}
% 99
\bibitem{FJ96} T. Fujita, Y. Fujiwara, C. Nakamoto and Y. Suzuki,
\Journal{\PRO} {96}{653}{1996}
% 100
\bibitem{KA02}
T. Kadowaki \etal,
% \etal~(KEK-PS E452 collaboration),
\Journal{\em Eur. Phys. J.} {A15}{295}{2002}
% 101
\bibitem{KU06}
M. Kurosawa, KEK-Report 2005-104
% 102
\bibitem{MI99} K. Miyagawa and H. Yamamura,
\Journal{\PRC} {60}{024003}{1999}
% 103
\bibitem{SH83} S. Shinmura, Y. Akaishi and H. Tanaka,
\Journal{\PRO} {71}{546}{1984}
% 104
\bibitem{YA94} Y. Yamamoto, T. Motoba, H. Himeno, K. Ikeda
and S. Nagata,
\Journal{\em Prog. Theor. Phys. Suppl.} {No. 117}{361}{1994} 
% 105
\bibitem{HI00c} E. Hiyama,
\Journal{\NPA} {670}{273c}{2000}
% 106
\bibitem{NE00} H. Nemura, Y. Suzuki, Y. Fujiwara
and C. Nakamoto,
\Journal{\PRO} {103}{929}{2000}
% 107
\bibitem{hypt} Y. Fujiwara, K. Miyagawa, M. Kohno and Y. Suzuki,
\Journal{\PRC} {70}{024001}{2004}
% 108
\bibitem{alex68} G. Alexander, U. Karshon, A. Shapira, G. Yekutieli,
R. Engelmann, H. Filthuth and W. Lughofer,
\Journal{\PREV} {173}{1452}{1968}
% 109
\bibitem{sechi68} B. Sechi-Zorn, B. Kehoe, J. Twitty and R. A. Burnstein,
\Journal{\PREV} {175}{1735}{1968}
% 110
\bibitem{kadyk71} J. A. Kadyk, G. Alexander, J. H. Chan, P. Gaposchkin 
and G. H. Trilling,
\Journal{\NPB} {27}{13}{1971}
% 111
\bibitem{DA72} R. H. Dalitz, R. C. Herndon and Y. C. Tang,
\Journal{\NPB} {47}{109}{1972}
% 112
\bibitem{MI01} D. J. Millener, 
\Journal{\NPA} {691}{93c}{2001}
% 113
\bibitem{FU98} Y. Fujiwara, T. Fujita, C. Nakamoto and Y. Suzuki,
\Journal{\PRO} {100}{957}{1998}
% 114
\bibitem{SW62} J. J. de Swart and C. Dullemond,
\Journal{\ANNP} {19}{458}{1962}
% 115
\bibitem{RO58} R. R. Ross,
\Journal{\em Bull. Ann. Phys. Sic.} {3}{335}{1958}
% 116
\bibitem{HE68} V. Hepp and H. Schleich,
\Journal{\em Z. Phys.} {214}{71}{1968}
% 117
\bibitem{ST70} D. Stephen, Ph. D. thesis, Univ. of Massachusetts, 1970
(unpublished)
% 118
\bibitem{EN66}
R. Engelmann, H. Filthuth, V. Hepp and E. Kluge,
\Journal{\PL} {21}{587}{1966}
% 119
\bibitem{EI71}
F. Eisele, H. Filthuth, W. F{\"o}hlisch, V. Hepp and G. Zech,
\Journal{\PL} {37B}{204}{1971}
% 120
\bibitem{E251}
J. K. Ahn \etal,
\Journal{\NPA} {648}{263}{1999}
% 121
\bibitem{KA05}
J. K. Ahn \etal,
\Journal{\NPA} {761}{41}{2005}
% 122
\bibitem{E289}
Y. Kondo \etal,
\Journal{\NPA} {676}{371}{2000}
% 123
\bibitem{BM90} R. Brockmann and R. Machleidt,
\Journal{\PRC} {42}{1965}{1990}
% 124
\bibitem{NSC89} P. M. M. Maessen, Th. A. Rijken and J. J. de Swart,
\Journal{\PRC} {40}{2226}{1989}
% 125
\bibitem{SCHU} H.-J. Schulze, M. Baldo, U. Lombardo, J. Cugnon and
A. Lejeune,
\Journal{\PRC} {57}{704}{1998}
% 126
\bibitem{BMZ} H. Bando, T. Motoba and J. \u{Z}ofka,
\Journal{\em Int. J. of Mod. Phys. A} {5}{4021}{1990}
% 127
\bibitem{DA99} J. D\c{a}browski,
\Journal{\PRC} {60}{025205}{1999}
% 128
\bibitem{BNL98} R. Sawafta,
\Journal{\NPA} {585}{103c}{1995};
\Journal{} {639}{103c}{1998}
% 129
\bibitem{NU02} H. Noumi \etal,
\Journal{\PRL} {89} {072301}{2002}; \Journal{} {90}{049902 (E)}{2003}
% 130
\bibitem{SA04} P. K. Saha \etal,
\Journal{\PRC} {70} {044613}{2004}
% 131
\bibitem{KO04}
M. Kohno, Y. Fujiwara, Y. Watanabe, K. Ogata and M. Kawai,
\Journal{\PRO} {112}{895}{2004};
to appear in {\em Phys. Rev.} C (2006)
% 132
\bibitem{AK02} H. Akikawa \etal,
\Journal{\PRL} {88}{082501}{2002}
% 133
\bibitem{TA03} H. Tamura \etal,
\Journal{\NPA} {754}{58c}{2005}
% 134
\bibitem{HI00} E. Hiyama, M. Kamimura, T. Motoba, T. Yamada
and Y. Yamamoto,
\Journal{\PRL} {85}{270}{2000}
% 135
\bibitem{2al} Y. Fujiwara, K. Miyagawa, M. Kohno, Y. Suzuki,
D. Baye and J.-M. Sparenberg,
\Journal{\PRC} {70}{024002}{2004}
% 136
\bibitem{2aljj} Y. Fujiwara, M. Kohno, K. Miyagawa
and Y. Suzuki,
\Journal{\PRC} {70}{047002}{2004}
% 137
\bibitem{TA01}
H.~Takahashi \etal,
%(KEK-PS E373 collaboration), 
\Journal{\PRL} {87}{212502}{2001}
% 138
\bibitem{PR66} D. J. Prowse,
\Journal{\PRL} {17}{782}{1966}
% 139
\bibitem{DA63} M. Danysz \etal,
\Journal{\PRL} {11}{29}{1963}
% 140
\bibitem{AO91} S. Aoki \etal,
\Journal{\PRO} {85}{1287}{1991}
% 141
\bibitem{FI04} I. N. Filikhin, A. Gal and V. M. Suslov,
\Journal{\NPA} {743}{194}{2004}
% 142
\bibitem{KO03}
M. Kohno, Y. Fujiwara and Y. Akaishi, 
\Journal{\PRC} {68}{034302}{2003}
% 143
\bibitem{SR99}
V. G. J. Stoks and Th. A. Rijken,
\Journal{\PRC}{59}{3009}{1999}
% 144
\bibitem{OSY83}
M. Oka, K. Shimizu and K. Yazaki,
\Journal{\PL} {130B}{365}{1983}
% 145
\bibitem{KEKsympo}
C. Nakamoto, Y. Fujiwara and Y. Suzuki,
\Journal{\NPA} {670}{315c}{2000} 
% 146
\bibitem{NA98} C. Nakamoto, Y. Fujiwara and Y. Suzuki,
\Journal{\NPA} {639}{51c}{1998}
% 147
\bibitem{straub2}
U. Straub, Zong-Ye Zhang, K. Br{\"a}uer, Amand Faessler
and S. B. Khadkikar,
\Journal{\PLB} {200}{241}{1988} 
% 148
\bibitem{KSY90}
Y. Koike, K. Shimizu and K. Yazaki,
\Journal{\NPA} {513}{653}{1990}
% 149
\bibitem{FE05} T. Fern{\'a}ndez-Caram{\'e}s, A. Valcarce
and P. Gonz{\'a}lez,
\Journal{\PRD} {72}{054008}{2005}
% 150
\bibitem{hyp2000}
C. Nakamoto, Y. Fujiwara and Y. Suzuki,
\Journal{\NPA} {691}{238c}{2001}
% 151
\bibitem{E906}
T. Tamagawa \etal,
%$et\,al.$ (BNL-E906 collaboration),
\Journal{\NPA} {691}{234c}{2001}
% 152
\bibitem{YTFM01}
Y. Yamamoto, T. Tamagawa, T. Fukuda and T. Motoba,
\Journal{\PRO} {106}{363}{2001}
% 153
\bibitem{AH06} J. K. Ahn \etal, 
\Journal{\PLB} {633}{214}{2006}
% 154
\bibitem{FG02}
I. N. Filikhin and A. Gal,
\Journal{\NPA} {707}{491}{2002};
\Journal{\PRC} {65}{041001}{2002}
% 155
\bibitem{NSC97}
Th. A. Rijken, V. G. J. Stoks and Y. Yamamoto, 
\Journal{\PRC}{59}{21}{1999}
% 156
\bibitem{triton} Y. Fujiwara, K. Miyagawa, M. Kohno, Y. Suzuki
and H. Nemura, 
\Journal{\PRC} {66}{021001}{2002}
% 157
\bibitem{PANIC02} Y. Fujiwara, K. Miyagawa, Y. Suzuki,
M. Kohno and H. Nemura, \Journal{\NPA} {721}{983c}{2003}
% 158
\bibitem{Bra88a} R. A. Brandenburg, G. S. Chulick,
Y. E. Kim, D. J. Klepacki, R. Machleidt, A. Picklesimer
and R. M. Thaler,
\Journal{\PRC} {37}{781}{1988}
% 159
\bibitem{Bra88b} R. A. Brandenburg, G. S. Chulick, R. Machleidt,
A. Picklesimer and R. M. Thaler,
\Journal{\PRC} {37}{1245}{1988}
% 160
\bibitem{Rei68} 
R. V. Reid, \Journal{\ANNP} {50}{411}{1968}
% 161
\bibitem{WI95} R. B. Wiringa, V. G. J. Stoks and R. Schiavilla,
\Journal{\PRC} {51}{38}{1995} 
% 162
\bibitem{NO97} A. Nogga, D. H{\" u}ber, H. Kamada
and W. Gl{\" o}ckle,
\Journal{\PLB} {409}{19}{1997}
% 163
\bibitem{NO00} A. Nogga, H. Kamada and W. Gl{\" o}ckle,
\Journal{\PRL} {85}{944}{2000}
% 164
\bibitem{EM02} D. R. Entem and R. Machleit,
\Journal{\PLB} {524}{93}{2002}
% 165
\bibitem{GL82b}
W. Gl{\" o}ckle,
\Journal{\NPA} {381}{343}{1982}
% 166
\bibitem{KI88}
Kr. T. Kim, Y. E. Kim, D. J. Klepacki, Richard A. Brandenburg,
E. P. Harper and R. Machleidt,
\Journal{\PRC} {38}{2366}{1988}
% 167
\bibitem{CO65}
H. Collard, R. Hofstadter, E. B. Hughes, A. Johansson,
M. R. Yearian, R. B. Day and R. T. Wagner,
\Journal{\PREV} {138}{B57}{1965}
% 168
\bibitem{MA86}
J. Martino, in {\em The Three-Body Force
in the Three-Nucleon System},
Vol. 260 of {\em Lecture Note in Physics}, edited by
B. L. Berman and B. F. Gibson (Springer-Verlag, Berlin, 1986),
p. 129
% 169
\bibitem{TA92}
S. Takeuchi, T. Cheon and E. F. Redish,
\Journal{\PLB} {280}{175}{1992};
\Journal{\NPA} {508}{247c}{1990}
% 170
\bibitem{SA01}
B. Juli{\' a}-D\'{\i}az, J. Haidenbauer, A. Valcarce
and F. Fern{\' a}ndez,
\Journal{\PRC} {65}{034001}{2002}
% 171
\bibitem{MI95} K. Miyagawa, H. Kamada, W. Gl{\"o}ckle and V. Stoks,
\Journal{\PRC} {51}{2905}{1995}
% 172
\bibitem{MI00} K. Miyagawa, H. Kamada, W. Gl{\"o}ckle, H. Yamamura,
T. Mart and C. Bennhold,
\Journal{{\em Few-Body Systems Suppl.}} {12}{324}{2000}
% 173
\bibitem{NO02} A. Nogga, H. Kamada and W. Gl{\"o}ckle,
\Journal{\PRL} {88}{172501}{2002}
% 174
\bibitem{NE02} H. Nemura, Y. Akaishi and Y. Suzuki,
\Journal{\PRL} {89}{142504}{2002}
% 175
\bibitem{SB97} J.-M. Sparenberg and D. Baye,
\Journal{\PRC} {55}{2175}{1997}
% 176
\bibitem{BA80} H. Bando and I. Shimodaya,
\Journal{\PRO} {63}{1812}{1980}
% 177
\bibitem{HI97} E. Hiyama, M. Kamimura, T. Motoba, T. Yamada
and Y. Yamamoto, \Journal{\PRO} {97}{881}{1997}
% 178
\bibitem{HE6LL} Y. Fujiwara, M. Kohno, K. Miyagawa, Y. Suzuki
and J.-M. Sparenberg,
\Journal{\PRC} {70}{037001}{2004}
% 179
\bibitem{SU99}
Y. Suzuki and H. Nemura, \Journal{\PRO} {102}{203}{1999}
%
\end{thebibliography}
\end{document}